\documentclass[numberedappendix]{emulateapj}
\submitted{Accepted by \apj}

\def\arcdeg{\hbox{$^\circ$}}

\def\as{\hbox{$^{\prime\prime}$}}
\def\deg{\hbox{$^\circ$}}
\def\mbh{M$_{BH}$}
\def\Msun{M$_{\odot}$}
\def\mic{$\mu$m~}
\def\micns{$\mu$m}
\def\lam{$\lambda$}
\def\sig{$\sigma$~}
\def\signs{$\sigma$}
\def\it{\textit{}}
\def\htwo{H$_{2}$~}
\def\htwons{H$_{2}$}
\def\bg{Br$\gamma$~}
\def\bgns{Br$\gamma$}
\def\kms{km s$^{-1}$~}
\def\kmsns{km s$^{-1}$}
\def \oiii{[O III]}
\def \sm{$\sim$}
\def \kb{{\em K}-band}
\def \ch{$\chi^{2}$}
\def \rch{$\chi^{2}$/dof}
\def \n{{\em n}}
\def \ml{$\Upsilon$}
\def \mlu{M$_{\odot}$/L$_{\odot,H}$}
\def \inc{{\em i}}
\def \t{$\theta$}
\def \rfit{$r_{fit}$}
\def \s{~}

\begin{document}

\title{Circumnuclear Gas in Seyfert 1 Galaxies: Morphology, Kinematics, and Direct Measurement of Black Hole Masses}

\author{Erin K. S. Hicks\altaffilmark{1,2} and Matthew A. Malkan\altaffilmark{1}}

\altaffiltext{1}{Department of Physics and Astronomy, University of California,   Los Angeles, CA, 90095-1562}
\altaffiltext{2}{Current address: Max-Planck-Institut f\"{u}r extraterrestrische Physik, Postfach 1312, 85741 Garching, Germany}

\begin{abstract}
The two-dimensional distribution and kinematics of the molecular, ionized, and highly ionized gas in the nuclear regions of Seyfert 1 galaxies have been measured using high spatial resolution ($\sim$0\as.09) near-infrared spectroscopy from NIRSPEC with adaptive optics on the Keck telescope.  Molecular hydrogen, \htwons, is detected in all nine Seyfert 1 galaxies and, in the majority of galaxies, has a spatially resolved flux distribution.  In contrast, the narrow component of the \bg emission has a distribution consistent with that of the \kb\s continuum.  In general, the kinematics of the molecular hydrogen are consistent with thin disk rotation, with a velocity gradient of over 100 \kms measured across the central 0\as.5 in three galaxies, and a similar gradient across the central 1\as.5 in an additional two galaxies.  The kinematics of \bg are in agreement with the \htwo rotation, except that in all four cases the central 0\as.5 is either blue- or redshifted by more than 75 \kmsns.  The highly ionized gas, measured with the [Ca VIII] and [Si VII] coronal lines, is spatially and kinematically consistent with \bg in the central 0\as.5.  In addition, the velocity dispersion of both the coronal and \bg emission is greater than that of \htwo (by 1.3-2.0 times), suggesting that both originate from gas that is located closer to the nucleus than the \htwo line emitting gas.  Dynamical models have been fitted to the two-dimensional \htwo kinematics, taking into account the stellar mass distribution, the emission line flux distribution, and the point spread function.  For NGC 3227 the modeling indicates a black hole mass of \mbh=2.0$^{+1.0}_{-0.4}\times10^{7}$ \Msun, and for NGC 4151 \mbh=3.0$^{+0.75}_{-2.2}\times10^{7}$ \Msun.  In NGC 7469 the best fit model gives \mbh$<5.0\times10^{7}$ \Msun.  In all three galaxies, modeling suggests a near face-on disk inclination angle, which is consistent with the unification theory of active galaxies.  The direct black hole mass estimates verify that masses determined from the technique of reverberation mapping are accurate to within a factor of three with no additional systematic errors.
\end{abstract}


\section{Introduction}

Central dark masses, which are assumed to be black holes (BHs), have now been measured in several quiescent galaxies using either stars or gas to trace the BH gravitational potential.  These studies have revealed correlations between black hole mass (\mbh) and properties of the host galaxy, such as bulge luminosity and \sig (the stellar velocity dispersion; \citealt{ferrarese00}; \citealt{gebhardt00}).  These relationships imply that even though a BH's gravitational influence is limited to within only tens of parsecs, it is coupled with global properties of the host galaxy at scales of kiloparsecs; the BH and host galaxy therefore coevolve. 

The presence of BHs in quiescent galaxies implies that these galaxies experienced an active phase, during which the mass of the BH was accumulated.  An assessment of BH demographics in current active galactic nuclei (AGN), both locally and through cosmic time, will therefore provide constraints for galaxy formation and evolution models.  In addition, measurement of \mbh\s in AGN is essential to understanding the AGN phenomenon itself, since \mbh\s is the fundamental parameter in AGN theory and models.  

Indirect methods exist for measuring \mbh\s in AGN, but their validity has not yet been demonstrated, nor have their systematic uncertainties been assessed  (\citealt{sun89}).  Reverberation mapping (\citealt{peterson04}) can measure \mbh, both locally and at higher redshifts, but the unknown broad line region geometry and kinematics limits its precision.  Secondary methods, such as the use of rest-frame ultraviolet luminosity and emission line width, which have been shown to correlate with reverberation mapped \mbh\s (e.g. \citealt{wandel99}, \citealt{vest02}), are powerful in that they can measure \mbh\s with relative ease in large samples of AGN out to large redshifts.  However, they are calibrated against the reverberation mapping results, and are thus even more uncertain.  Independent validation and calibration of the reverberation mapping method, even if for only a few galaxies, would provide a foundation from which a `\mbh\s scale' can be built and extended to higher redshift and larger samples.

It is especially challenging to measure \mbh\s directly in active galaxies because of the bright non-stellar emission from the nucleus, which dilutes the gas and stellar spectral signatures.  Until recently, the only directly measured \mbh\s in AGN were for galaxies with weak activity, such as NGC 4258 (\citealt{pastorini07}), or for Seyfert 2 galaxies in which the nucleus is obscured (e.g. \citealt{greenhill96}, \citealt{schinnerer00}, \citealt{silge05}, \citealt{capetti05}).  In the case of NGC 4258, water masers can be used as a tracer of the central potential, which gives an exceptionally accurate \mbh\s estimate (\citealt{miyoshi95}).  Unfortunately, this method is not applicable to other AGNs because the required favorable inclination and `well behaved' maser disks are extremely rare.  The glare of the unobscured nuclear emission in Seyfert 1 galaxies has restricted most dynamical studies of these galaxies to only placing upper limits on \mbh\s (e.g. \citealt{winge99}). 

With the advent of adaptive optics, which compensates for the distortions of the Earth's atmosphere, it has become possible to achieve the spatial resolution necessary to overcome some of the challenges presented by the non-stellar emission from bright AGN.  At infrared wavelengths, the non-stellar emission is lessened in comparison to the stellar and gas spectral features, and with a high spatial resolution (i.e. $<$0\as.1) these features can now be used to trace the gravitational potential down to small enough radii to measure the influence of a central BH.  This technique has been successfully demonstrated with a few individual galaxies.  For example, molecular hydrogen emission has been used to study the gas dynamics and to place constraints on \mbh\s in two galaxies, NGC 7469 and Mrk 231 (\citealt{davies04} and \citealt{davies04b}, respectively).  In addition, using stellar kinematics, \citet{davies06} have been able to estimate \mbh\s in the galaxy NGC 3227.  Increasing the number of AGN \mbh\s estimates to be statistically significant is critical to the calibration of indirect methods, such as the reverberation mapping technique, and to our understanding of the role BHs play in galaxy formation and evolution.

In this paper we present the first high spatial resolution (tens of parsecs) 2-D study of the nuclear regions in a sample of active galaxies with a direct view of the central engine, in this case Seyfert 1s (galaxies with active nuclei of M$_{B} >$ -22 and permitted lines with a broad component of 10$^{3}$-10$^{4}$ \kmsns).  The velocity, velocity dispersion, and flux distribution of molecular hydrogen, ionized hydrogen, and coronal gas is measured in a 2-D field using single slit observations. The 2-D kinematics of the nuclear molecular hydrogen is modeled and from this \mbh\s estimates are obtained.  

Selection of the sample of Seyfert 1 galaxies and the observations are described in $\S$ 2.  The data reduction and methods for basic analysis, as well as the construction of the 2-D maps, are discussed in $\S$ 3, and $\S$ 4 presents the 2-D kinematic and flux distribution maps for each Seyfert 1 galaxy.  The general technique used to model the nuclear gas kinematics is discussed in $\S$ 5 and the results of the kinematic modeling are presented in $\S$ 6.  The last section, $\S$7, gives our general conclusions.  The electronic edition of the Journal contains color versions of the figures throughout this paper, as well as all of the 2-D maps of the flux distributions and kinematics for each of the Seyfert 1 galaxies.

\section{Near-Infrared Spectra of Seyfert 1 Galaxies}

\subsection{Seyfert 1 Sample Selection}

Targets were selected from all of Seyfert 1 galaxies with reverberation mapped \mbh\s estimates using three criteria.  The first is simply the visibility of the galaxies from the Keck II telescope\footnote[1]{The W.M. keck Observatory is operated as a scientific partnership among the California Institute of Technology, the University of California, and the National Aeronautics and Space Administration.  The Observatory was made possible by the generous financial support of the W. M. Keck Foundation.} such that they could be observed with an air mass above 2.0.  The second criterion is the ability to measure the velocity field at a spatial resolution sufficient to detect the gravitational influence of the BH.  To achieve the best possible spatial resolution, natural guide star (NGS) adaptive optics (AO) is used to compensate for the distortions from the Earth's atmosphere, providing near diffraction limited observations.  Thus, the third criterion for the sample is that an adequate guide source for NGS AO system on Keck II is available, which, for virtually all Seyfert 1 galaxies, is limited to the AGN nucleus itself.

The spatial resolution necessary to detect the gravitational influence of the BH on the measured velocity field depends on two things: the mass of the BH and the distance of the galaxy.  The mass of the BH must be high enough to alter the steepness of the central velocity gradient such that it is distinguishable from the gradient expected when no BH is present.  The sphere of influence, the range within which the BH dominates the gravitational potential over that of the stars, is defined as
\begin{equation}
R_{h} \, = \, \frac{G M_{BH}}{\sigma^{2}}
\end{equation}
\noindent where $\sigma$ is the velocity dispersion of the host galaxy bulge.  To resolve the diameter of this sphere at a given wavelength it is required that
\begin{equation}
2R_{h} \, > \, \frac{d \, 1.22 \, \lambda}{D}
\end{equation}
\noindent where D is the telescope aperture, d is the distance of the galaxy, and \lam\s is the wavelength of the observation.  However, the above relationship assumes the observations are diffraction limited, which is often not the case.  In practice the spatial resolution with the Keck II NGS AO system in the {\em K}-band, during good conditions, is 0.06\as, giving the following criteria
\begin{equation}
2R_{h} \, > \, R_{res} = \, \frac{0.06\as \, d(pc)}{206265}
\end{equation}

\begin{figure}[!t] 
\epsscale{1.0}
\plotone{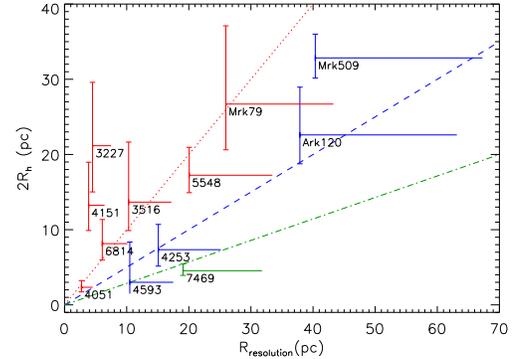}
\caption[R$_{resolved}$ versus R$_{h}$]{ R$_{resolved}$ versus R$_{h}$ for the selected sample of Seyfert 1 galaxies. The dotted, dashed, and dash-dotted lines correspond to resolving 1, 2, and 3.5 times the sphere of influence, respectively. Labels are the NGC number or the full galaxy name. \label{RresRh}}
\end{figure}

\begin{deluxetable*}{lcrrccccccr}
\tabletypesize{\scriptsize}
\tablecaption{Selected Sample of Seyfert 1 Galaxies \label{t_sample}} 
\tablewidth{0pt}
\tablehead{
\colhead{Galaxy} & 
\colhead{Nucleus\tablenotemark{a}} &
\colhead{$\alpha$(2000.0)\tablenotemark{b}} & 
\colhead{$\delta$(2000.0)\tablenotemark{b}} & 
\colhead{z} & 
\colhead{\mbh (10$^{6}$ \Msun)\tablenotemark{c}} & 
\colhead{$\sigma$ (km s$^{-1}$)} & 
\colhead{Reference\tablenotemark{d}} & 
\colhead{$R_{h}$ (pc)} &
\colhead{$R_{res}$ (pc)\tablenotemark{e}} \\ 
}
\startdata

Ark 120	& Sy1		&	05 16 11.4	&	- 00 08 59.4	&	0.0323	&	150 $\pm$ 19	&	239 $\pm$ 36	&	1	&	11.3 $^{+ 6.3 }_{- 3.8 }$	&	37.9	\\
Mrk 79	& Sy1.5	&	07 42 32.8	&	+ 49 48 34.8	&	0.0222	&	52.4 $\pm$ 14.4	&	130 $\pm$ 20	&	1	&	13.3 $^{+ 10.4 }_{- 6.1 }$	&	26.0	\\
NGC 3227	& Sy1.5 	&	10 23 30.6	&	+ 19 51 54.0	&	0.0039	&	42.2 $\pm$ 21.4	&	131 $\pm$ 11	&	1	&	10.6 $^{+ 8.4 }_{- 6.1 }$	&	4.5	\\
NGC 3516	& Sy1.2	&	11 06 47.5	&	+ 72 34 06.9	&	0.0088	&	42.7 $\pm$ 14.6	&	164 $\pm$ 35	&	1	&	6.8 $^{+ 8.0 }_{- 3.8 }$	&	10.3	\\
NGC 4051	& Sy1.2	&	12 03 09.6	&	+ 44 31 52.8	&	0.0023	&	1.91 $\pm$ 0.78	&	84 $\pm$ 9		&	1	&	1.2 $^{+ 0.9 }_{- 0.6 }$	&	2.7	\\
NGC 4151	& Sy1.5	&	12 10 32.6	&	+ 39 24 20.6	&	0.0033	&	45.7$^{+5.7}_{-4.7}$	&	93 $\pm$ 14	&	1	&	6.6 $^{+ 5.7 }_{- 3.3 }$	&	3.9	\\
NGC 4253	& Sy1.2	&	12 18 26.5	&	+ 29 48 46.0	&	0.0129	&	10 $\pm$ 5		&	108.3 $\pm$ 12.4	&	2	&	3.7 $^{+ 3.4 }_{- 2.2 }$	&	15.1	\\
NGC 4593	& Sy1		&	12 39 39.4	&	- 05 20 39.3	&	0.0090	&	9.8 $\pm$ 2.1	&	124 $\pm$ 29	&	1	&	1.5 $^{+ 5.3 }_{- 2.2 }$	&	10.5	\\
NGC 5548	& Sy1.5	&	14 17 59.5	&	+ 25 08 12.4	&	0.0172	&	65.4$^{+2.6}_{-2.5}$	&	183 $\pm$ 27	&	1	&	8.6 $^{+ 3.7 }_{- 2.3 }$	&	20.1	\\
NGC 6814	& Sy1		&	19 42 40.6	&	+ 10 19 25.0	&	0.0052	&	12 $\pm$ 5		&	112.7 $\pm$ 12.4	&	2	&	4.1 $^{+ 3.2 }_{- 2.1 }$	&	6.1	\\
Mrk 509	& Sy1		&	20 44 09.7	&	- 10 43 24.5	&	0.0344	&	143 $\pm$ 12	&	193.6 $\pm$ 9.0	&	2	&	16.4 $^{+ 3.2 }_{- 2.7 }$	&	40.4	\\
NGC 7469	& Sy1		&	23 03 15.6	&	+ 08 52 26.4	&	0.0163	&	12.2 $\pm$ 1.4	&	152 $\pm$ 16	&	1	&	2.3 $^{+ 0.9 }_{- 0.6 }$	&	19.1	\\

\enddata
\tablenotetext{a}{Classification of the AGN nucleus is from \citet{ho97} and \citet{whittle92}.}
\tablenotetext{b}{Units of right ascension are hours, minutes, and seconds, and units of declination are degrees, arcminutes, and arcseconds.}
\tablenotetext{c}{\mbh\s estimates are from \citet{peterson04} for all, except NGC 6814 and NGC 4253 are from \citet{wandel02}, NGC 4151 from \citet{bentz06}, NGC 4593 from \citet{denney06}, and NGC 5548 from \citet{bentz07}.}
\tablenotetext{d}{The \sig measurements are from (1) Onken et al. 2004 or (2) determined using the \mbh-\sig relationship as given by Ferrarese 2002.}
\tablenotetext{e}{Value of $R_{res}$ assumes a spatial resolution of 0\as.06.}

\end{deluxetable*}

The sphere of influence for each reverberation mapped Seyfert 1 is estimated using the reverberation mapped \mbh\s results (\citealt{peterson04}) and \sig measurements based on Ca II triple stellar absorption (\citealt{onken04}; see Table \ref{t_sample}).  It is essential to remember that systematic error in the reverberation mapping method has yet to be assessed, and \mbh\s estimates could be in error by as much as several times.  In addition, it is not necessary to resolve the sphere of influence to be able to estimate \mbh\s based on nuclear gas kinematics.  If the velocity field measurements are accurate enough (error less than $\sim$ 20 km s$^{-1}$), then it is possible to distinguish the rotation expected without a BH from the rotation expected with a BH of a mass less than that needed to produce a resolvable sphere of influence.  The combination of the uncertainty of the reverberation mapped \mbh\s estimates and the ability to measure \mbh\s even when the sphere of influence is not quite resolved, allows for the selection of all galaxies for which at least 3.5 times the estimated sphere of influence is resolved (see Fig. \ref{RresRh}).  Table \ref{t_sample} lists all twelve Seyfert 1 galaxies selected for the sample based on the feasibility of detecting their BH.  The Hubble constant is assumed to be H$_{0}$ = 75 \kms Mpc$^{-1}$.

The ability of the AO system to use the Seyfert nucleus as the natural guide source is not as easy to asses due to the varying brightness of the AGN on timescales of days to months (\citealt{rosenblatt92}, \citealt{webb00}).  The AO system requires a point source of {\em V}-band magnitude brighter than 13-14 magnitudes.  The AGN must therefore be at least this bright, as well as bright enough in comparison to the stellar light of its host galaxy to be seen as a point source.  If the AGN is in a bright flux state then it will often meet this criterion, but if seeing conditions are poor and/or the AGN is in a low flux state, then no guide source will be available.  The final selection of galaxies observed was therefore determined at the telescope by choosing those that provided the best AO correction, and thus the highest achievable spatial resolutions, during each observing run.  The final selection includes nine Seyfert 1 galaxies, as indicated in Table \ref{t_obs}.

\begin{deluxetable*}{lcccccccl}
\tablecaption{NIRSPEC/AO Observations \label{t_obs}} 
\tablewidth{0pt}
\tablehead{
\colhead{} & 
\colhead{} & 
\colhead{No. of} &
\colhead{Exp.} &
\colhead{PA} & 
\colhead{Slit} & 
\colhead{PSF} & 
\colhead{} & 
\colhead{} \\ 
\colhead{Seyfert 1} & 
\colhead{Date} & 
\colhead{Spec.} &
\colhead{(s)} &
\colhead{($^o$)} & 
\colhead{Width (\as)} & 
\colhead{FWHM (\as)} & 
\colhead{WFSC\tablenotemark{a}} & 
\colhead{Notes} \\ 
}
\startdata


Ark 120	&	12 Dec 2003	&	5	&	600	&	92	&	0.037	&	0.10	&	60	&		\\
		&	13 Dec 2003	&	5	&	600	&	0	&	0.037	&	0.12	&	55	&		\\
Mrk 79	&	13 Dec 2003	&	\nodata	&	\nodata	&	\nodata	&	\nodata	&	\nodata	&	15-20	&	WFSC too low	\\
NGC 3227	&	20 Feb 2001	&	12	&	400	&	163	&	0.054	&	0.12	&	25-40	&		\\
		&	19 Feb 2003	&	10	&	600	&	43	&	0.037	&	0.11	&	60-75	&		\\
		&	21 Apr 2003	&	9	&	600	&	103	&	0.037	&	0.07	&	65	&		\\
		&	12 Dec 2003	&	4	&	600	&	55	&	0.037	&	0.11	&	70	&		\\
NGC 3516	&	20 Feb 2001	&	10	&	400	&	28	&	0.054	&	0.12	&	50	&		\\
		&	20 Feb 2003	&	10	&	600	&	48	&	0.037	&	0.17	&	60-65	&		\\
NGC 4051	&	20 Feb 2001	&	8	&	400	&	-47	&	0.054	&	0.12	&	20-35	&		\\
		&	20 Feb 2003	&	9	&	600	&	-47	&	0.037	&	0.12	&	55-60	&		\\
		&	22 Apr 2003	&	3	&	600	&	43	&	0.037	&	0.06	&	20-30	&		\\
		&	12 Dec 2003	&	6	&	600	&	43	&	0.037	&	0.12	&	60	&		\\
NGC 4151	&	19 Feb 2001	&	34	&	400	&	40	&	0.054	&	0.12	&	20-40	&	Variable Seeing \& Clouds	\\
		&	19 Feb 2003	&	8	&	600	&	141	&	0.037	&	0.12	&	85-100	&	AO at 146 Hz, gains 0.4, 0.35	\\
		&	22 Apr 2003	&	8	&	600	&	90	&	0.037	&	0.08	&	75-107	&	AO at 58 Hz, gains 0.35, 0.45	\\
		&	13 Dec 2003	&	10	&	600	&	40	&	0.037	&	0.16	&	120	&	AO at 76 Hz, gains 0.4,0.6	\\
NGC 4593	&	22 Apr 2003	&	2	&	600	&	90	&	0.037	&	0.08	&	70-85	&		\\
NGC 4253	&	21 Apr 2003	&	\nodata	&	\nodata	&	\nodata	&	\nodata	&	\nodata	&	10-15	&	WFSC too low	\\
		&	13 Dec 2003	&	\nodata	&	\nodata	&	\nodata	&	\nodata	&	\nodata	&	15-20	&	WFSC too low	\\
NGC 5548	&	24 Aug 2001	&	6	&	400	&	90	&	0.054	&	0.08	&	20	&		\\
		&	25 Aug 2001	&	8	&	400	&	0	&	0.054	&	0.07	&	20	&		\\
NGC 6814	&	25 Aug 2001	&	8	&	400	&	35	&	0.054	&	0.07	&	20	&		\\
Mrk 509	&	\nodata	&	\nodata	&	\nodata	&	\nodata	&	\nodata	&	\nodata	&	\nodata	&	Never available	\\
NGC 7469	&	24 Aug 2001	&	6	&	400	&	90	&	0.054	&	0.08	&	40	&		\\
		&	25 Aug 2001	&	7	&	400	&	150	&	0.054	&	0.07	&	35-40	&		\\
		&	12 Dec 2003	&	6	&	600	&	47	&	0.037	&	0.17	&	55	&		\\
		&	13 Dec 2003	&	4	&	600	&	-47	&	0.037	&	0.12	&	50	&		\\

\enddata
\tablenotetext{a}{WFSC with AO correction at 55 Hz and gains of 0.25 and 0.35 for the tip-tilt and deformable mirrors, respectively, except where noted in the table.}

\end{deluxetable*}

\subsection{NIRSPEC/AO Keck Observations}

Data for this project were obtained during ten half nights with NIRSPEC/AO (\citealt{mclean98}) on Keck II, which provides long-slit spectra with a pixel scale of 0$''$.0185 with an Aladdin InSb $1024 \times 1024$ chip.  For the first four half nights a slit width of 0$''$.056 was used, while the remaining nights used a 0$''$.037 wide slit, and both had a slit length of 3$''$.93.  The spectral resolution with the 0$''$.056 and 0$''$.037 slits is $R \simeq 2000$ and 3300, respectively.  The NIRSPEC-7 filter is used, providing coverage from 1.9 to 2.5 \mic (exact wavelength coverage varied for each run depending on the setup).  Exposure times were typically 400 or 600 seconds, giving an average total exposure time per galaxy of 2.5 hours, ranging from 20 minutes to 5.1 hours; see Table \ref{t_obs} for details. 

Indispensable to the project is the slit viewing camera, SCAM.  This camera has a 0$''$.017 pixel scale, and with the 256 $\times$ 256 PICNIC HgCdTe chip, a field of view (FOV) of $4''.4 \times 4''.4$.  Images are taken approximately every minute throughout spectroscopic exposures, resulting in a accurate determination of the slit position with respect to the galaxy nucleus and any drift of the slit throughout the spectroscopic exposure (see $\S$ 3.2).  In addition, the SCAM images were used to monitor and estimate the PSF (see $\S$ 5.1.1).  Another near-infrared spectrograph with high throughput (NIRC2) was available at the Keck telescope during the time of this observing campaign, but the ability to simultaneously monitor the slit position and PSF during the spectroscopic observations made NIRSPEC with SCAM more favorable for this project. 

Over the course of the ten half nights, the AO system was able to lock onto the nuclei of nine of the Seyfert galaxies in the sample.  Typically the AO system was corrected at a rate of 55 Hz.  The wave front sensor counts (WFSC) obtained during each run, for each galaxy, are listed in Table \ref{t_obs}.  Of the remaining galaxies, two were attempted, but were found to have less than 20 WFSC, which is not sufficient to provide a significant AO correction, and one galaxy was never available from the Keck telescope during the nights scheduled.

With AO correction the average full-width-at-half-maximum (FWHM) resolution was 0$''$.09, with a range from 0$''.$06 to 0$''$.17 (see Table \ref{t_obs} for the conditions during each run).  Slit widths and positions were chosen so the inner arcsecond of the galaxy was, at a minimum, Nyquist sampled, and the target was dithered along the slit by at least 100 pixels for sky subtraction.  Immediately after obtaining the set of spectra for each galaxy, neon and argon arc lamp spectra were taken for wavelength calibration.

For removal of atmospheric absorption features spectra of A-type stars were obtained during each run with the same instrumental set up and observing procedures as described above.  A neutral density filter is used to reduce the flux going to the AO camera for stars with a {\em V}-band magnitude brighter than the galaxies.  This enables the use of a lower AO correction, one that is more similar to that achieved with the Seyfert 1s, so atmospheric templates are more closely matched to the galaxy spectra.  In addition, M-type stellar spectra were obtained in the same fashion as templates for measuring the stellar velocity dispersion (see $\S$ 4.3.1).

\section{2-D Maps: Data Reduction and Analysis}
Until recently, single slit observations were the only way to obtain high spatial resolution near-infrared spectra of the nuclear regions of AGN.  Although it is time intensive to obtain enough single slit observations to measure a significant fraction of a 2-D field, it is necessary to be able to constrain parameters such as position angle of the major axis of rotation and disk inclination, and for measuring the flux distribution of the line emitting gas, all of which can significantly alter the observed rotation field.  Even more importantly, it is essential to measure the 2-D field to determine the likelihood that the gas motion is dominated by non-gravitational forces, which is not always evident in the velocities measured in a single slit position, or even in a few slits.  In addition, the position of the slit must be known to high precision for accurate modeling of the steep nuclear velocity gradient (which can change by as much as 100 km s$^{-1}$ in just 0.\as1), to which \mbh\s estimates are highly sensitive.  The high resolution available with NIRSPEC/AO, combined with the slit viewing camera SCAM, make it possible to measure the nuclear regions of many Seyfert 1 galaxies at the level of accuracy needed to estimate \mbh\s directly from gas kinematics for the first time.

\subsection{Reduction and Analysis of NIRSPEC/AO Spectra}

\subsubsection{Reduction of Near-Infrared Spectra}

\begin{deluxetable}{lcccc}
\tabletypesize{\scriptsize}
\tablecaption{Summary of Extraction Apertures for Spectra \label{t_aps}} 
\tablewidth{0pt}
\tablehead{
\colhead{} & 
\colhead{} &
\colhead{Rest \lam} & 
\colhead{} &
\colhead{E.W.} \\
\colhead{Seyfert 1} & 
\colhead{Transition} &
\colhead{(\micns)} & 
\colhead{Measured (\as)} &
\colhead{Width (\as)\tablenotemark{a}} \\
}
\startdata


NGC 3227	&	\htwo 1-0 S(1)  &	 2.1218	&	1.4	&	0.037	\\
		&	\bg  		&	 2.1661	&	1.4	&	0.056	\\
                &       [Ca VIII]       &        2.3213 &       1.4     &       0.074   \\
		&	\htwo 1-0 Q(1)  &        2.4066	&	1.4	&	0.056	\\
		&	\htwo 1-0 Q(3)  &	 2.4237	&	1.4	&	0.056	\\
                &       [Si VII]        &        2.4833 &       1.4     &       0.074   \\
NGC 3516	&	\htwo	1-0 S(1)&	 2.1218	&	1.3	&	0.148	\\
                &       [Ca VIII]       &        2.3213 &       1.3     &       0.148   \\
NGC 4051	&	\htwo 1-0 S(1)  &	 2.1218	&	1.0	&	0.056	\\
		&	\bg  		&	 2.1661	&	1.0	&	0.111	\\
                &       [Ca VIII]       &        2.3213 &       1.0     &       0.056   \\
                &       [Si VII]        &        2.4833 &       1.0     &       0.056   \\
NGC 4151	&	\htwo 1-0 S(1)  &        2.1218	&	1.3	&	0.056	\\
		&	\bg  		&	 2.1661	&	1.3	&	0.056	\\
                &       [Ca VIII]       &        2.3213 &       1.3     &       0.037   \\
                &       [Si VII]        &        2.4833 &       1.3     &       0.037   \\
NGC 4593	&	\htwo 1-0 S(1)  &	 2.1218	&	1.0	&	0.185	\\
		&	\htwo 1-0 Q(1)  &	 2.4066	&	1.0	&	0.111	\\
NGC 5548	&	\htwo 1-0 S(1)  &	 2.1218 &	1.4	&	0.148	\\
                &       [Ca VIII]       &        2.3213 &       1.4     &       0.148   \\
NGC 6814	&	\htwo 1-0 S(1)  &	 2.1218	&	1.0	&	0.148	\\
		&	\htwo 3-2 S(3)  &	 2.2014	&	1.0	&	0.185	\\
		&	\htwo 1-0 S(0)  &	 2.2235	&	1.0	&	0.185	\\
		&	\htwo 2-1 S(1)  &	 2.2477	&	1.0	&	0.185	\\
NGC 7469	&	\htwo 1-0 S(3)  &	 1.9576	&	1.2	&	0.074	\\
		&	\htwo 1-0 S(1)  &	 2.1218	&	1.4	&	0.037	\\
		&	\bg  		&	 2.1661	&	1.4	&	0.056	\\
                &       [Ca VIII]       &        2.3213 &       1.4     &       0.037   \\
Ark 120	        &	\htwo 1-0 S(1)  &	 2.1218	&	1.3	&	0.370	\\

\enddata
\tablenotetext{a}{The sizes and locations of the extraction windows (E.W.) varied for each spectroscopic exposure depending on the feature strength at the particular aperture location.  Listed here is the typical size of the extraction window.  In regions of low emission line flux the windows sizes are as much as three times larger.  The aperture on the sky of a particular spectrum is determined by the extraction window size as well as the slit width, which is either 0\as.037 or 0\as.056 (see Table. \ref{t_obs}).} 

\end{deluxetable}

Reduction of the NIRSPEC/AO data is done using IRAF software\footnote[1]{IRAF is distributed by NOAO, which is operated by AURA Inc., under contract to the NSF}.  First cosmic rays are removed using the COSMICRAYS package.  Then the spectral images are rectified so wavelength and spatial dispersion run parallel to the rows and columns, respectively, using the WMKONSPEC reduction package.  Next the pairs of spectra that were dithered along the slit are subtracted, removing the sky emission to first order as well as subtracting the bias.  The spectra are then extracted using the IRAF task APALL.  Three important steps optimized the quality of the extracted spectra: 1) choice of placement and sizes of the extraction windows, 2) locating the center of the object light profile, and 3) placement of background subtraction windows.  The best light profile center and background windows are chosen for each spectroscopic exposure individually, while the extraction windows used are optimized for each emission line of interest in each spectrum.

2-D coverage of each galaxy is achieved by obtaining multiple single slit observations across the FOV of the galaxy from which multiple spectra are extracted along the spatial dimension (the length of the slit).  To achieve as high a spatial resolution as possible, while maintaining at least a 3$\sigma$ detection of each emission line, the location and size of the extraction windows are selected independently for each emission line at each slit position (Table \ref{t_aps}).  Aperture location is established by the slit position and the location of the extraction window along the slit length.  Aperture size is determined by the width of the slit (0\as.037 or 0\as.056; Table \ref{t_obs}) and the size of the extraction window along the slit length, which in cases of high emission line flux is as small as 0\as.037 (two pixels).  Placement of the extraction windows is with respect to the continuum peak and is held constant with wavelength. 

In extracting the spectra, APALL places extraction windows with respect the center of the galaxy, which is determined by the continuum peak.  To ensure an accurate determination of the extraction window locations the small distortion (less than one pixel across the 1024 pixels from one end of the chip to the other) of the spatial and wavelength axes that remains after the data reduction must be accounted for during the spectral extraction.  This is achieved in APALL by doing a trace of the galaxy light profile to determine the position of the nucleus for each wavelength interval (each pixel).  A smooth function, either a linear spline or cubic spline, both of order one, is then fit to the position of the galaxy light profile center as a function of wavelength, and all aperture placement is based on this center.  In less than 20\% of the extracted spectra the continuum peak is poorly measured (often due to poor seeing or AO correction) and in these cases a trace from another exposure is used.  

The last parameter that requires careful consideration is the location of the two windows used to determine the background that is subtracted from the spectra during extraction.  Although this is a second order background subtraction, with the first order subtraction accomplished by differencing the dithered pairs, the additional step is necessary to account for differences in the background between the subtracted exposures.  After extraction, the spectra are wavelength calibrated using spectra from neon and argon arc lamps, obtained immediately following the completion of the set of spectra obtained for each target, and the calibration is confirmed by checking the wavelengths of strong sky emission lines.  The spectra have 4.33 $\times 10^{-4}$\mic per pixel, and spectral resolutions with the 0\as.056 and 0\as.037 slits of $R \simeq 2000$ (with FWHM $\simeq 1.1 \times10^{-3}$\mic or 150 km s$^{-1}$) and $R \simeq 3300$ (with FWHM $\simeq 0.7 \times 10^{-3}$\mic or 90 km s$^{-1}$), respectively.  The M- and A-type stellar spectra are reduced using the same procedure described above, except only one spectrum is extracted for each single slit observation. 

\begin{figure}[!t]
\epsscale{1}
\plotone{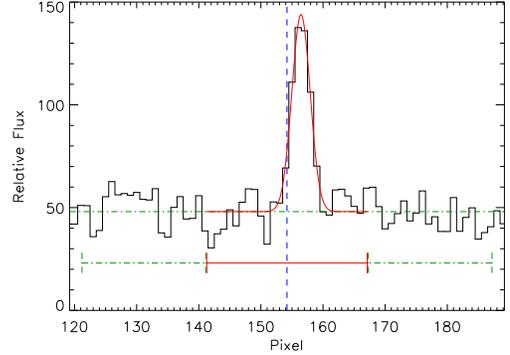}
\caption[Example Gaussian Fit to an Emission Line]{ Example Gaussian fit to an emission line in NGC 3227.  The solid horizontal bar indicates the region of the spectrum (black histogram) that is fit by the Gaussian (solid curve).  The windows used to determine the continuum (dashed line) are indicated by the dashed horizontal bars.  The expected wavelength of the \htwo emission feature is shown by the vertical dashed line.  \label{gauss}}
\end{figure}

\subsubsection{Measurement of Velocity, Velocity Dispersion, and Flux Distribution}

Once the spectra have been extracted and calibrated, the emission lines are measured to obtain the gas velocity, velocity dispersion, and emission line flux using code written in IDL.  A single Gaussian line profile is fit to the emission features and two key parameters are varied to optimize the fit to the data: the number of pixels included in the fit and the region(s) of the spectra used to determine the continuum level.  Typically the Gaussian is fit to the 26 pixels (113 \AA) centered on the expected wavelength of the feature, and the continuum level is determined from 10 pixel (43 \AA) windows on either side of the fitting window (Fig. \ref{gauss}).  Velocity measurements are typically accurate to 20 \kmsns, while measurement of the velocity dispersion is accurate to 30 \kmsns.  The accuracy of these measurements for the emission lines considered will be discussed further in $\S$ 4. 

It is not known a priori that a single Gaussian is a reasonable representation of the emission line profiles.  However, non-Gaussian profiles are more common in highly ionized gas such as \oiii\s (e.g. \citealt{winge99}) rather than in molecular and ionized gas.  In the sample of Seyfert 1 galaxies selected for this study, it is found that, at least at our spectral resolution, the line profiles of the molecular hydrogen, and even of ionized hydrogen (\bgns), are well represented by a single Gaussian.

\subsection{Reduction and Analysis of SCAM Direct Images}

\subsubsection{Reduction of SCAM Images}
After removal of cosmic rays (preformed with the IRAF package COSMICRAYS), the SCAM images were flat fielded. Since neither dome nor sky flats are available for many of the observing runs, flat fields were created from the galaxy images themselves.  For each set of observations the galaxy was placed off slit and several images taken with the object dithered to the four quadrants of the chip.  At least four images are selected, each with the target in a different location on the chip by more than 50 pixels compared to the others.  These images were then averaged using a rejection of the two highest valued pixels, eliminating the galaxy light from the average.  The resulting averaged sky frame is then normalized and used for flat fielding.  The location of the slit in the flat field is set to one to prevent its altering the slit in the flat fielded images.  This method of creating a flat field is adequate for the purpose of accurately determining the slit position.  

\subsubsection{Determining Slit Positions from SCAM Images}

\begin{figure}[!t]
\epsscale{1}
\plotone{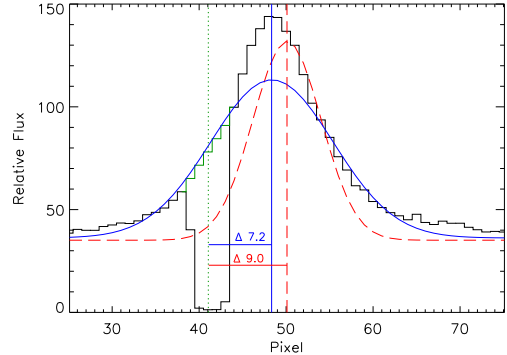}
\caption[Example Slit Position Fit]{An example of the slit determined from a fit to a corrected galaxy light profile (histogram).  The correction to the light profile is shown by the histogram across the slit location (dotted vertical line).  Gaussian fits to the corrected (solid Gaussian with a center indicated by the solid vertical line) and uncorrected (dashed Gaussian and vertical line) light profile demonstrate the correction is critical to accurately determine the slit position.  This is a cut through the 2-D fit. \label{spfit}}
\end{figure}

The SCAM images are critical in that they provided the means to accurately determine the position of the spectroscopic slit, which has been a challenge in previous studies, resulting in significant uncertainty in \mbh\s measurements.  Typically 8-15 images are taken simultaneously with each spectroscopic exposure and the slit position, with respect to the galaxy center, is determined using code written in IDL.  A 2-D Gaussian is fit to the galaxy {\em K}-band continuum light profile to find its center.  Before a Gaussian is fit, the slit (which is seen in the SCAM images as a clearly defined strip of no flux) must be removed from the galaxy light profile to prevent the fit from being biased due to the artificial drop in the galaxy light.  The missing galaxy light is replaced by a straight line fit to the flux on either side of the slit.  Fig. \ref{spfit} shows a cut through a typical SCAM image illustrating the method used to determine the slit position.  As indicated in the figure, a slit position of nine pixels would be estimated if no correction were made to the galaxy light profile, whereas the actual slit position is 7.2 pixels

\begin{figure}[!htt]
\epsscale{1}
\plotone{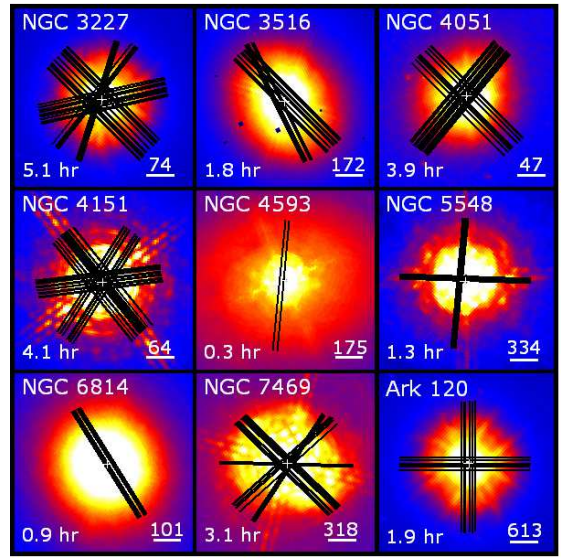}
\caption[Slits Overlaid on HST Images]{NIRSPEC/AO slit positions superimposed on HST images (F160W images for all except Ark 120 and NGC 4051, which are WFPC2 images) for each Seyfert 1 galaxy in the sample.  Each thin black line is a single slit position; thicker lines are overlapping slit positions.  Also shown are the total exposure times of the data obtained for each galaxy in hours, a bar representing 1\as, and the corresponding distance in parsecs assuming H$_{0}$ = 75 km s$^{-1}$ Mpc$^{-1}$. \label{overlay}}
\end{figure}

The slit position as a function of time is determined for each spectroscopic exposure by fitting a straight line.  The slit is found to drift no more than a pixel during the 400-600 second exposures.  A summary of the slit positions measured for each Seyfert 1 galaxy is given in Table \ref{t_sp} and the exact slit positions can be seen in Fig. \ref{overlay} overlaid on Hubble Space Telescope (HST) images. 

\subsection{Construction of 2D Maps}
2-D maps were made for each significant emission line present in the spectra of each Seyfert 1 galaxy (see Table \ref{t_aps} for a list of the lines measured).  The accuracy of placing each measurement on the 2-D maps depends on the precision of determining the galaxy center for both the slit position and the extraction window location, and is typically better than 0.2 pixels, or 0\as.004.  The single Gaussian fit to the emission line in each spectrum is used to construct 2-D maps of the gas velocity field, velocity dispersion, and relative flux distribution using the centroid, the width, and the integrated flux, respectively.  In addition, the maps presented have been smoothed by a 0\as.07 Gaussian, which is similar to the size of the PSF under the best seeing conditions.  Maps of the Pearson correlation coefficient (PCC) of the Gaussian fit to the emission line spectrum were also constructed.  This correlation coefficient measures the quality of the Gaussian fit to the emission line profile, with a value of one indicating a perfect fit and low values (0.2-0.4) indicating a poor fit (see e.g. \citealt{babu96}).  Since the emission line profiles in this sample are well represented by a single Gaussian, a low PCC served as an indicator of a weak signal-to-noise measurement of the emission line, rather than a non-Gaussian profile, and is found to correlate well with the relative flux.  The PCC values are therefore used in further analysis when only the highest quality data are required.

\begin{deluxetable}{lccccc}
\tablecaption{NIRSPEC/AO Slit Positions \label{t_sp}} 
\tablewidth{0pt}
\tablehead{
\colhead{} & 
\colhead{} & 
\colhead{No. of} &
\colhead{PA} & 
\colhead{Slit Pos.} & 
\colhead{\% 2-D} \\ 
\colhead{Seyfert 1} & 
\colhead{Date} & 
\colhead{Spec.} &
\colhead{($^o$)} & 
\colhead{(pix)\tablenotemark{a}} & 
\colhead{(0\as.5,1\as)\tablenotemark{b}} \\ 
}
\startdata


NGC 3227	&	20 Feb 2001	&	12	&	163	&	-8 to 0	&	87, 60	\\
		&	19 Feb 2003	&	10	&	43	&	-19 to 24	&		\\
		&	21 Apr 2003	&	9	&	103	&	-14 to 10	&		\\
		&	12 Dec 2003	&	4	&	55	&	9 to 21	&		\\
NGC 3516	&	20 Feb 2001	&	10	&	28	&	-8 to 0	&	65, 39	\\
		&	20 Feb 2003	&	10	&	48	&	-6 to 27	&		\\
NGC 4051	&	20 Feb 2001	&	8	&	-47	&	-4 to -8	&	84, 52	\\
		&	20 Feb 2003	&	9	&	-47	&	-8 to 20	&		\\
		&	22 Apr 2003	&	3	&	43	&	-8 to -3	&		\\
		&	12 Dec 2003	&	6	&	43	&	-19 to 14	&		\\
NGC 4151	&	19 Feb 2001	&	34	&	40	&	-10 to 5	&	90, 66	\\
		&	19 Feb 2003	&	8	&	141	&	-5 to 16	&		\\
		&	22 Apr 2003	&	8	&	90	&	-8 to 8	&		\\
		&	13 Dec 2003	&	10	&	40	&	-22 to 20	&		\\
NGC 4593	&	22 Apr 2003	&	2	&	90	&	-9,-6		&	14, 7	\\
NGC 5548	&	24 Aug 2001	&	6	&	90	&	-1 to 3	&	42, 22	\\
		&	25 Aug 2001	&	8	&	0	&	-10 to -2	&		\\
NGC 6814	&	25 Aug 2001	&	8	&	35	&	0 to 7	&	26, 13	\\
NGC 7469	&	24 Aug 2001	&	6	&	90	&	0 to 3	&	71, 42	\\
		&	25 Aug 2001	&	7	&	150	&	0 to 6	&		\\
		&	12 Dec 2003	&	6	&	47	&	-13 to 12	&		\\
		&	13 Dec 2003	&	4	&	-47	&	-13 to -4	&		\\
Ark 120	&	12 Dec 2006	&	5	&	92	&	-11 to 12	&	62, 34	\\
		&	13 Dec 2006	&	5	&	0	&	-11 to 11	&		\\

\enddata
\tablenotetext{a}{Slit position is defined as positive to the right and negative to the left with PA in degrees east of north.} 
\tablenotetext{b}{Percentage of pixels measured within a 0\as.5 and 1\as FOV, respectively.} 

\end{deluxetable}

For each galaxy, 2-D maps of the mean continuum flux are also constructed for three wavelength windows free of line emission: the blue end of the spectrum (2.14-2.15\micns), the middle (2.23-2.24\micns), and the red end (2.448-2.453\micns).  To measure the continuum flux at high spatial resolution, spectra extraction with apertures determined for the strongest emission line, typically \htwo \lam 2.1218\micns, were used (see Table \ref{t_aps}).

\section{Circumnuclear Gas in Seyfert 1 Galaxies: Hydrogen and Coronal Gas}

\subsection{Distribution and Kinematics of Molecular and Ionized Hydrogen}

Based on the nine galaxies in this sample, molecular hydrogen is spatially resolved in the nuclear regions of Seyfert 1 galaxies on scales of tens of parsecs.  Molecular hydrogen is detected at 1-0 S(1) \lam2.1218 in all galaxies, and in four of the galaxies more than one transition is seen (Table \ref{t_nsigma}).  In the majority of galaxies (six of nine) the distribution of the \htwo emission is spatially resolved and the position angle (PA) of the major axis is consistent with that of the larger scale optical emission.  In the remaining galaxies the distribution is consistent with that of the \kb\s continuum.  Furthermore, in two of the galaxies, several knots of \htwo emission are detected throughout the 2\as\s FOV.  Broad \bg is seen in the spectra of each Seyfert 1 galaxy and four of the galaxies have a distinguishable narrow \bg component.  In contrast to \htwons, the flux distribution of narrow \bg (including the major axis PA) is found to be consistent with that of the \kb\s continuum, which is assumed to have a peak flux at the location of the AGN.

The kinematics of both \htwo and narrow \bg are consistent with thin disk rotation in five of the galaxies, with the exception of \bg in the central 0\as.5, which is found to be either blue- or redshifted with respect to the \htwo kinematics.  Organized rotation is seen in three galaxies with a velocity gradient of over 100 \kms measured across the central 0\as.5, and in two additional galaxies a similar velocity gradient is seen across the central 1\as.5.  In general the PA of the kinematic major axis agrees with that of the major axis of both the \htwo flux distribution and large-scale optical emission.  Although no rotation is detected in the velocity field of the remaining four galaxies, in two of these galaxies data along only one PA was obtained, and it is possible that the selected PA is the minor axis (which would not show rotation).  Furthermore, it is possible that a rotating gas disk is still present, but is viewed face-on, and thus no rotation along the line of sight could be detected.

For the sample of galaxies, the velocity dispersion (\signs, where FWHM$\approx$2.35\signs) of the narrow \bg emission is found to be, on average, 2.0$\pm$0.7 times greater than that of \htwo (Table \ref{t_nsigma_coronal}; measured from 1\as.4 aperture spectra, see $\S$ 4.3).  The velocity dispersion reported for \oiii\s (\citealt{peterson04}) is consistent with that measured for \bgns, with an average dispersion 2.0$\pm$0.9 times that of \htwons.  If the velocity dispersions measured are assumed to be a result of rotation, then they suggest that narrow \bg and \oiii\s originate from gas at similar distances from the nucleus, and that this distance is less than that of the \htwo line emitting gas.  Contamination of the narrow \bg velocity dispersion measurements from broad line \bgns, which is only an issue close to the nucleus, is estimated to be within the measurement error (typically 30 \kmsns) in most cases, and in the worst cases to be less than 20\%.  The impact of the broad \bg emission, which typically has a velocity dispersion of thousands of \kmsns, is mitigated by restricting the number of pixels used in the single Gaussian fit to 15-20 pixels (65-87 \AA).  The continuum is determined on either side of this narrow spectral range, which effectively removes the broad line component.  This method was tested by comparing the velocity dispersions measured from a careful fit of both the broad and narrow components in three galaxies, and the single Guassian fits are found to overestimate the velocity dispersion by no more than 20\%.  Assuming a generous overestimate of 10-20\% for each galaxy (depending on how reliably the two components can be separated) gives an average narrow \bg velocity dispersion that is 1.7$\pm$0.6 time wider than \htwons. 

In summary, although \htwo and \bg share similar kinematics outside of a radius of 0\as.5, they are spatially and kinematically distinct.  Evidence supporting this includes (1) that \htwo is spatially resolved while narrow \bg is consistent with the \kb\s continuum, (2) the difference in major axis position angle of the flux distributions, (3) the blue- or redshifted central 0\as.5 of \bg compared to the ordered rotation of \htwo down to the smallest scales, and (4) the greater velocity dispersion measured for the narrow \bgns.

\subsection{Size and Morphology of the Coronal Line Region}

A common feature in the Seyfert 1 spectra is coronal emission, detected in [Ca VIII] \lam2.3213 and [Si VI] \lam2.4833.  [Ca VIII] is detected in all eight of the galaxies for which wavelength coverage permits its detection, and it is sufficiently strong in six of them for 2-D maps to be constructed of the flux distribution and kinematics.  [Si VII] is detected in each of the three galaxies for which it is available in the spectra, and 2-D maps of [Si VII] are also constructed.  

Measurement of both coronal emission lines mentioned is challenging because of neighboring spectral features.  For [Ca VIII] the measurement is difficult because of a stellar absorption feature (the $^{12}$CO(3,1) \lam 2.3226 bandhead), which is only 12\AA\s to the red of the expected wavelength.  However, in Seyfert 1 galaxies this feature is often weak.  For [Si VII] the measurement is problematic because of the strong atmospheric absorption at the longer wavelengths of the \kb\s window.  Of course the degree to which measurement of the [Si VII] coronal line is impacted by the telluric features depends on the redshift of the galaxy.  However, for each of the three galaxies for which [Si VII] is available in the spectra the Gaussian fit to the line profile is affected by a telluric feature and any result based on the [Si VII] emission is thus uncertain.

In galaxies for which both the [Ca VIII] and [Si VII] coronal lines are measured, the two are found to have similar flux distributions.  The distribution of the coronal emission varies from galaxy to galaxy, with extended emission out to a couple tens of parsecs up to 130 parsecs which is consistent with the range of distributions found for other active galaxies (e.g. \citealt{prieto06}).  In all but one galaxy, the coronal line flux distribution is consistent with the distributions of both \bg and the \kb\s continuum.  The exception is NGC 4151, for which the coronal line distribution is found to be significantly greater.  

Contrary to previous studies, a comparison of the coronal emission with the radio and \oiii\s emission in the Seyfert galaxies finds no general correlation in their spatial distributions (e.g. \citealt{reunanen03}, \citealt{prieto06}).  Out of the six galaxies for which the 2-D coronal emission was measured, NGC 4151 was the only galaxy to have coronal emission spatially coincident with \oiii\s and radio emission.  This apparent lack of correlation between the coronal emission and the radio emission could indicate that, in general, the excitation of the coronal gas is not a result of shocks from a nuclear jet.

The average velocity of the coronal gas, measured in both [Ca VIII] and [Si VII], agrees with the blue- or redshift seen in the central region for \bg in each galaxy for which the narrow component of \bg is measured.  Furthermore, the velocity dispersion of the more reliably measured [Ca VIII] coronal line is found to be, on average, 1.3$\pm$0.5 times greater than the \htwo dispersion, and 0.8$\pm$0.4 times the velocity dispersions reported for \oiii\s (\citealt{peterson04}; see Table \ref{t_nsigma_coronal}).  If this broadening of the lines is assumed to be rotational, then the velocity dispersion of the coronal gas indicates that this gas is located at a greater distance from the AGN than both \oiii\s and \bg, and closer to the nucleus than \htwons.  The coronal gas is kinematically and spatially more consistent with the narrow \bg emission than it is with \htwo based on its velocity shift, velocity dispersion, and flux distribution.

\subsection{Circumnuclear Gas in Individual Galaxies}

\begin{figure*}[!htt]
\epsscale{0.5}
\plotone{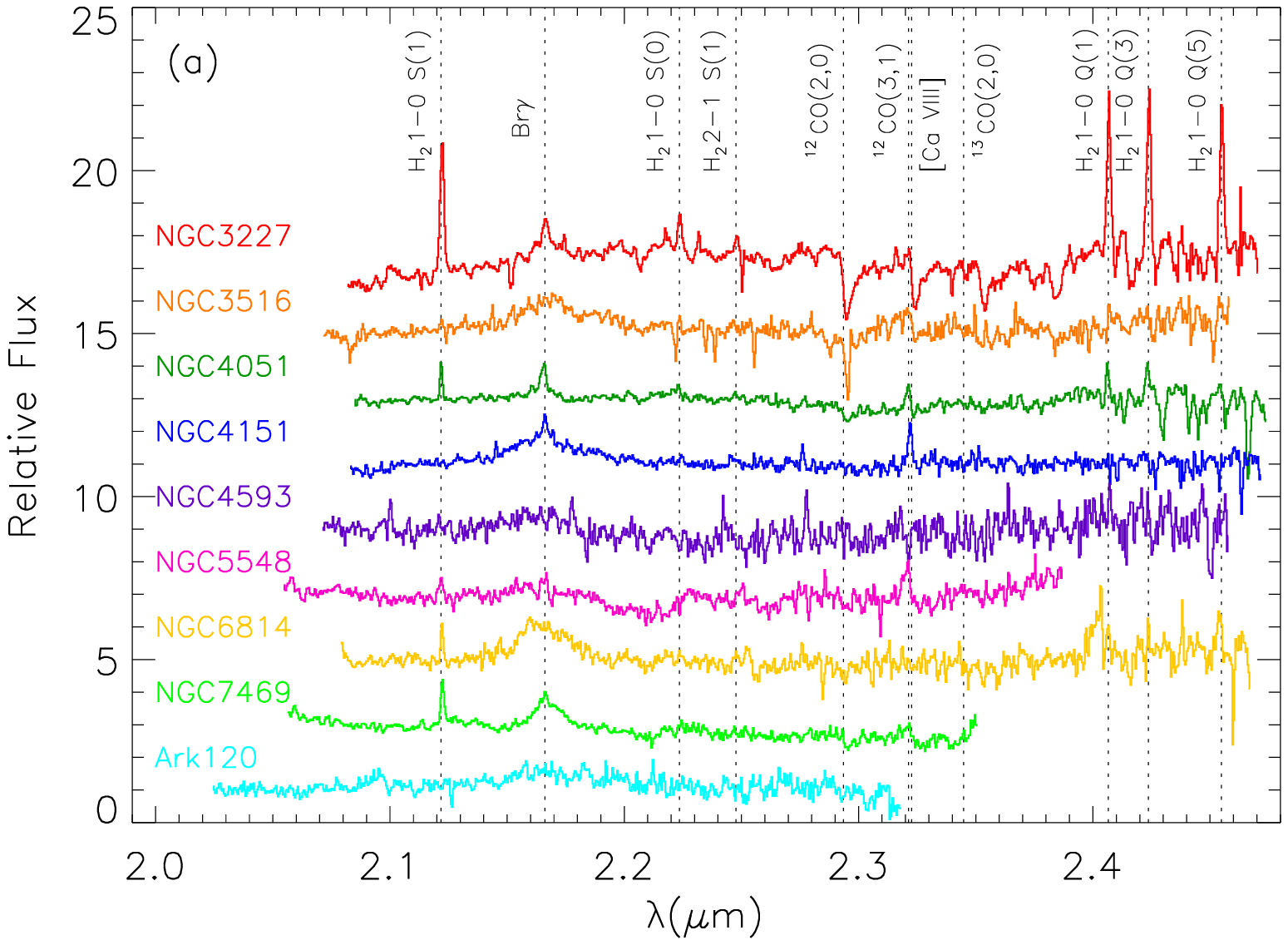}
\plotone{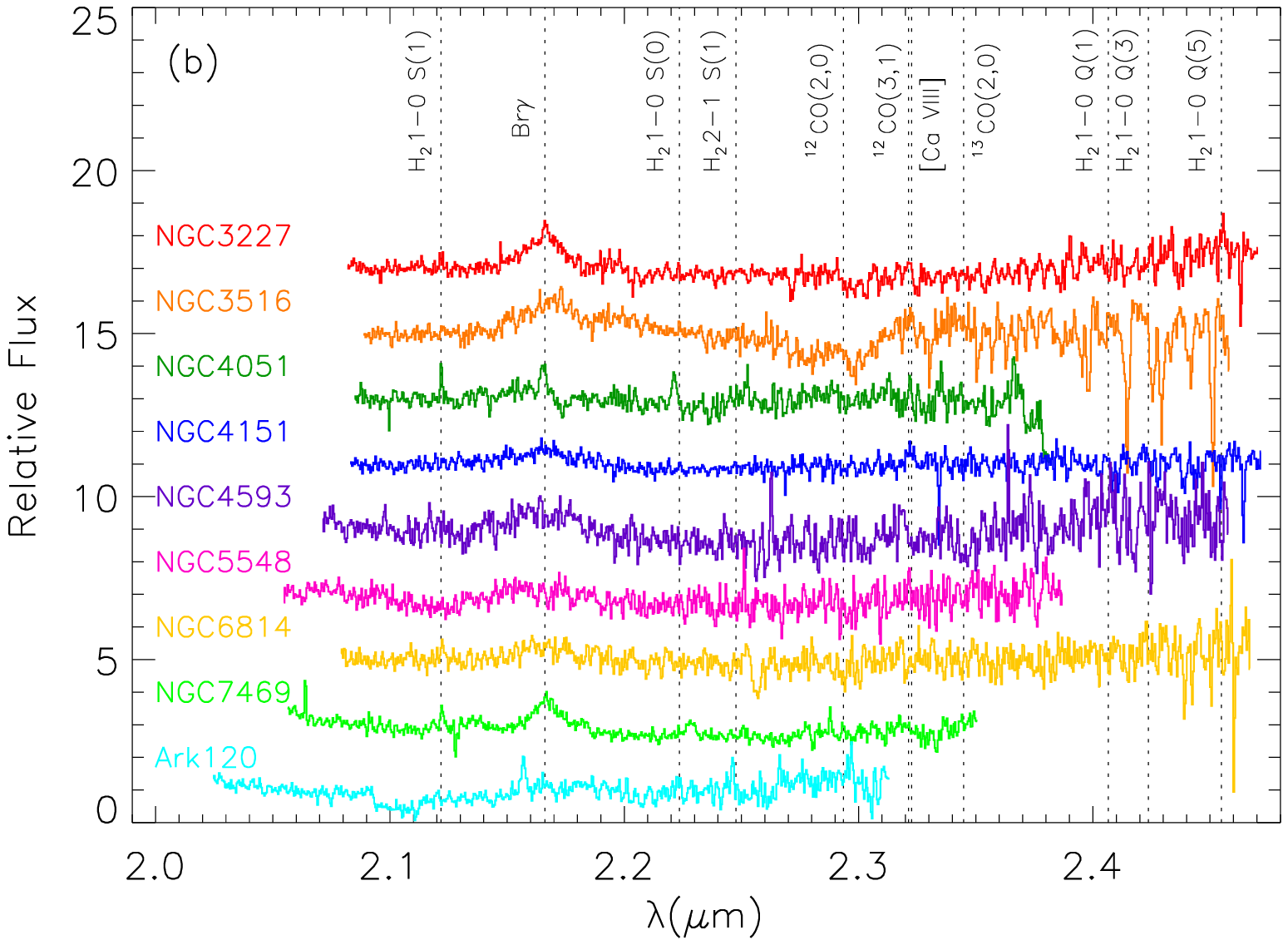}
\plotone{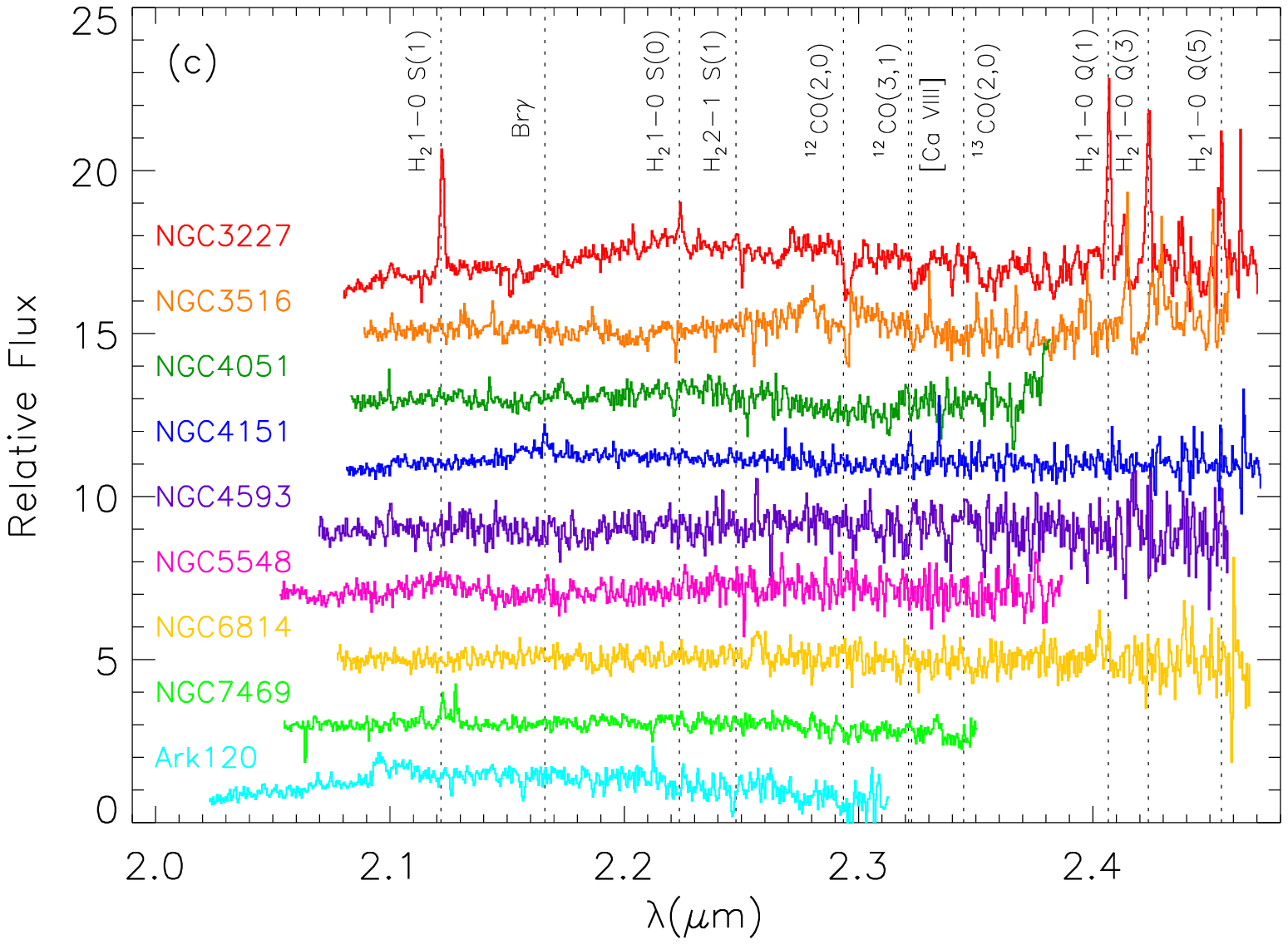}
\caption[Composite Spectra]{ Seyfert 1 galaxy composite spectra from (a) a 1\as.4 diameter aperture, (b) a 0\as.2 diameter aperture, and (c) an annulus with radius 0\as.1-0\as.7.  Significant spectral features are indicated by the dotted lines and the galaxy names are indicated on the left.  \label{spectra}}
\end{figure*}

Nuclear and off-nuclear spectra of all nine Seyfert 1 galaxies are shown in Figures \ref{spectra}a-\ref{spectra}c.  Fig. \ref{spectra}a contains composite spectra of all slit positions for each galaxy extracted with a 1\as.4 diameter aperture centered on the \kb\s continuum peak, which is assumed to be the AGN location.  Since the number of slit positions for each galaxy varied, the combined exposure time of each composite spectrum is different, which can be seen in the signal-to-noise ratio (SNR) of the spectra.  Fig. \ref{spectra}b shows nuclear spectra extracted with a 0\as.2 diameter aperture from a single slit position selected to be located as near to the nucleus as possible.  The spectra from an annulus of 0\as.1-0\as.7 (the difference of the other spectra) are shown in Fig. \ref{spectra}c.  These spectra, in one plot for each galaxy, are also available in Appendix A (Fig. \ref{indspec}) for the purpose of comparing the spectral features to atmospheric features.  As mentioned, the relative significance of the detected emission in these spectra, as well as the continuum SNR, are listed in Tables \ref{t_nsigma} and \ref{t_nsigma_coronal} for the hydrogen and coronal lines, respectively.  In addition, Table \ref{t_nsigma_coronal} gives the velocity dispersion, or line width, of the coronal emission, narrow \bgns, and \htwo 2.1218, as well as of \oiii\s taken from the literature.

\begin{figure*}[!ht]
\epsscale{0.4}
\plotone{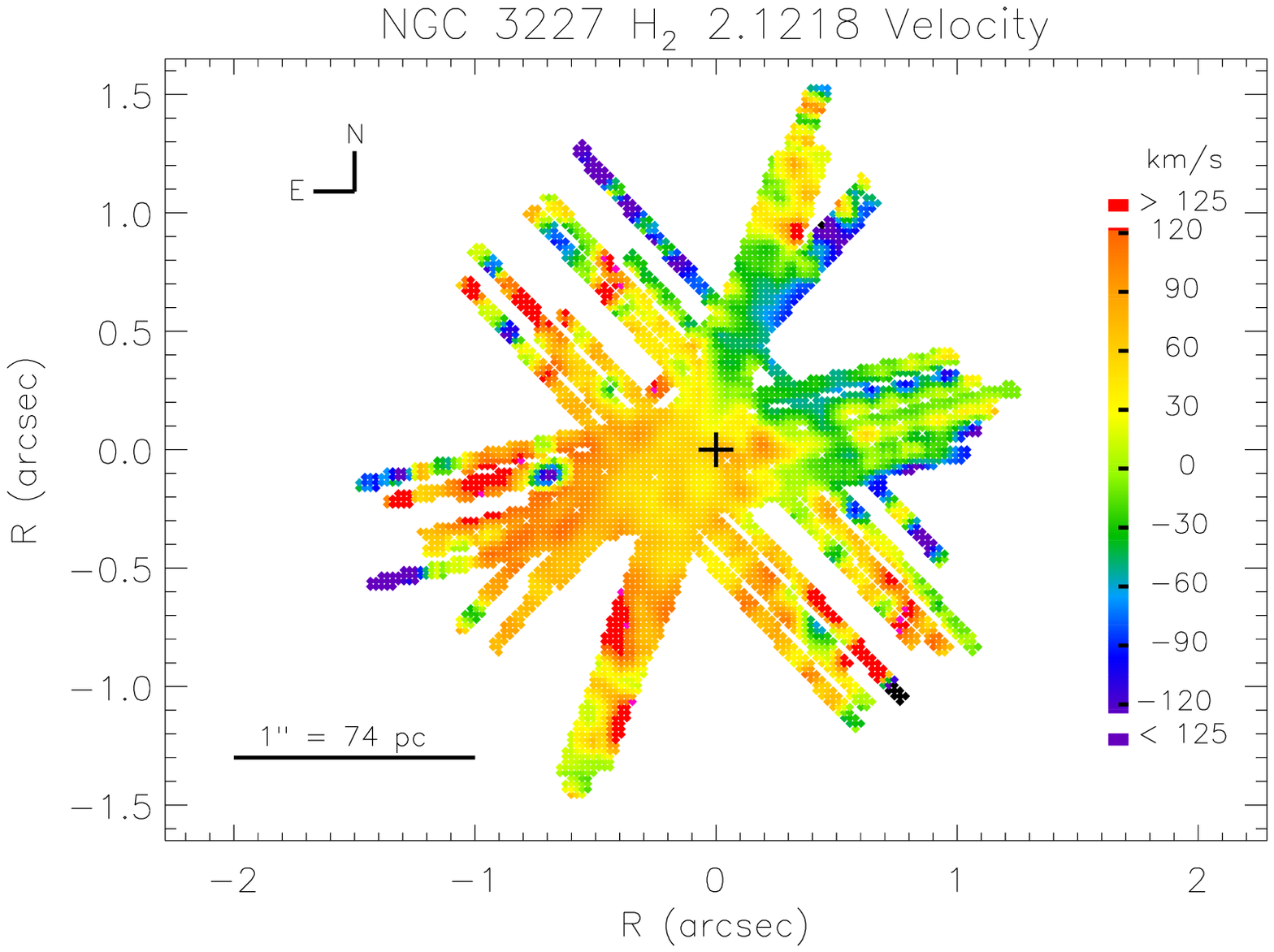}
\plotone{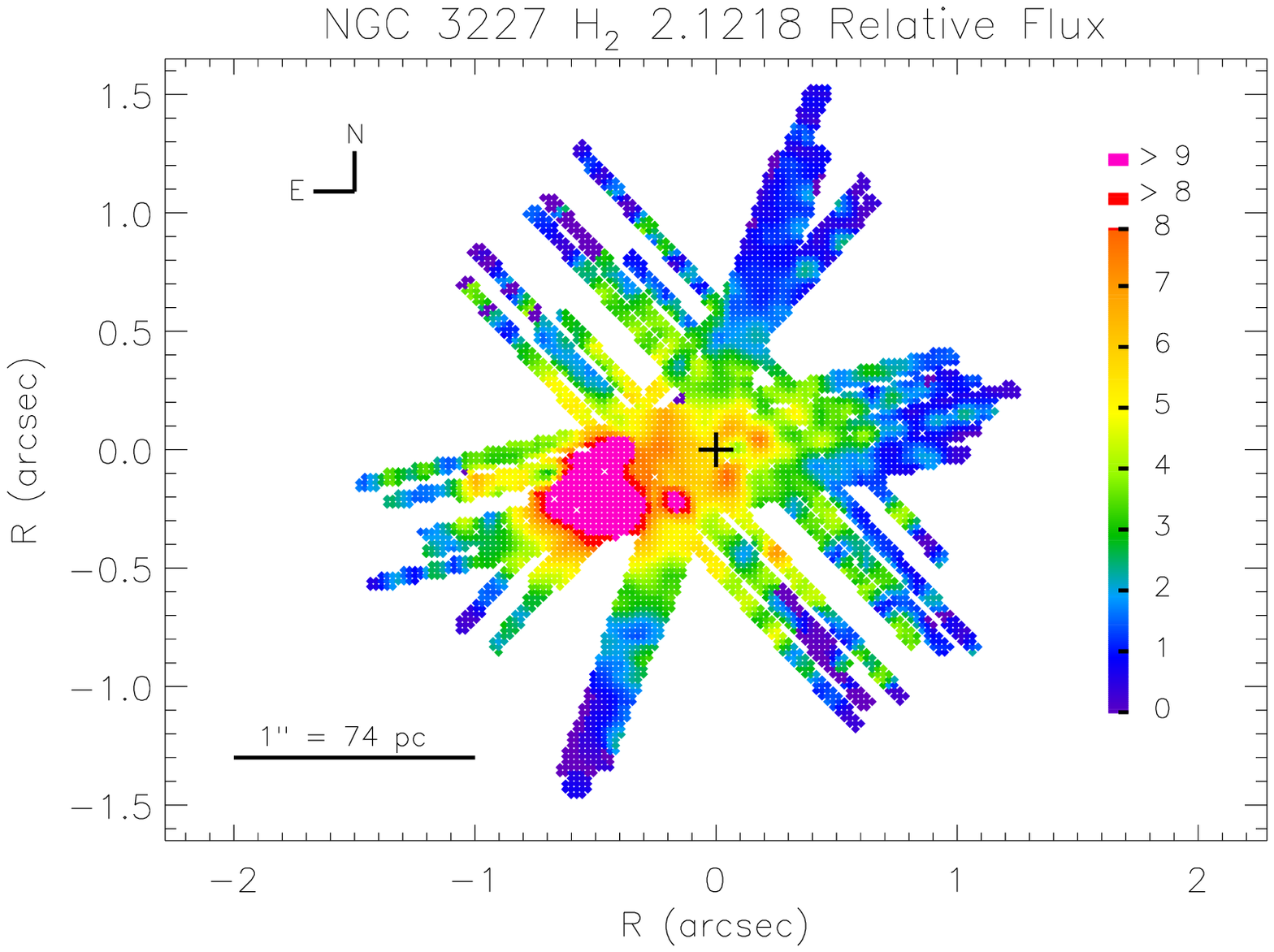}
\plotone{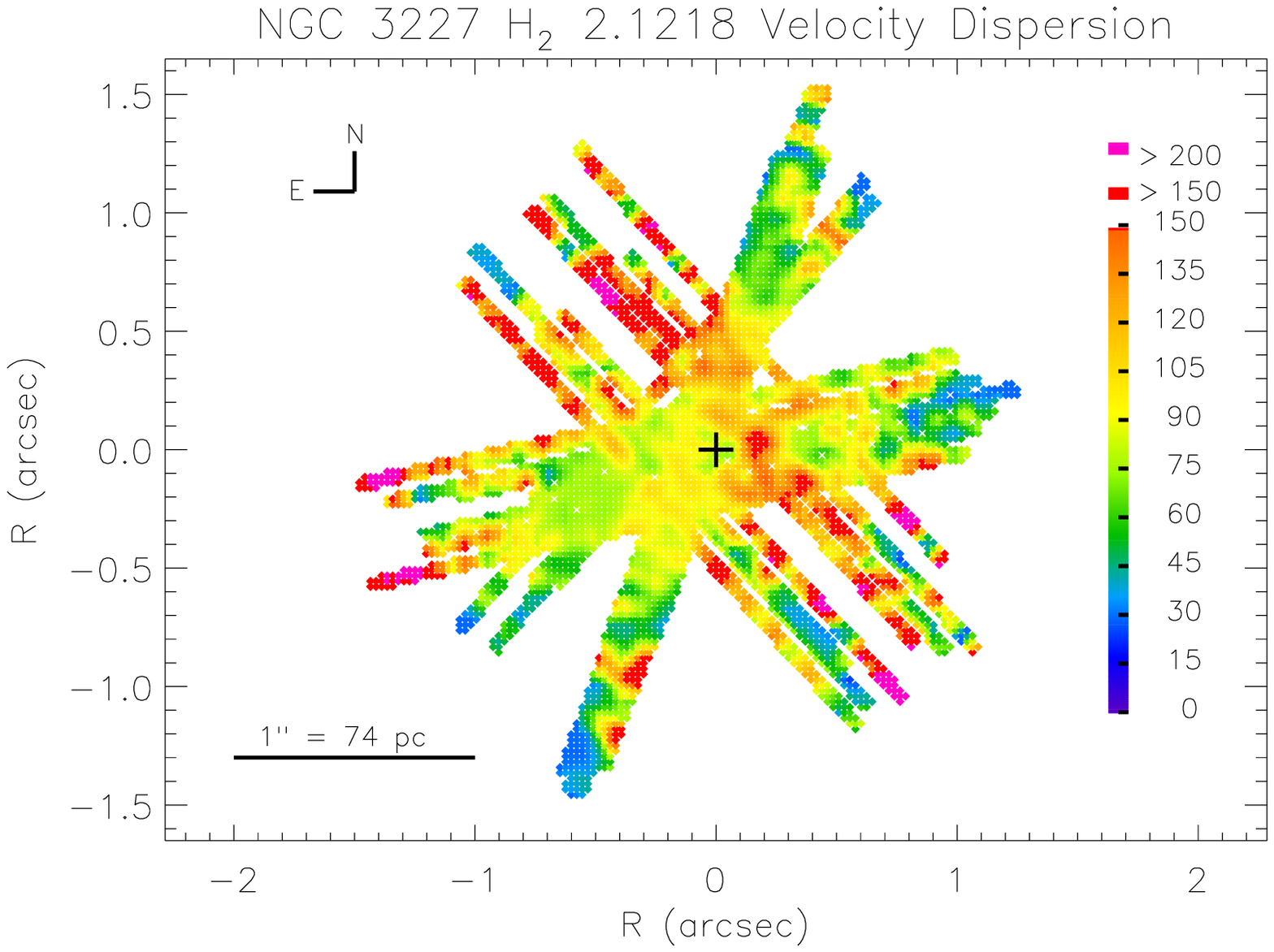}
\plotone{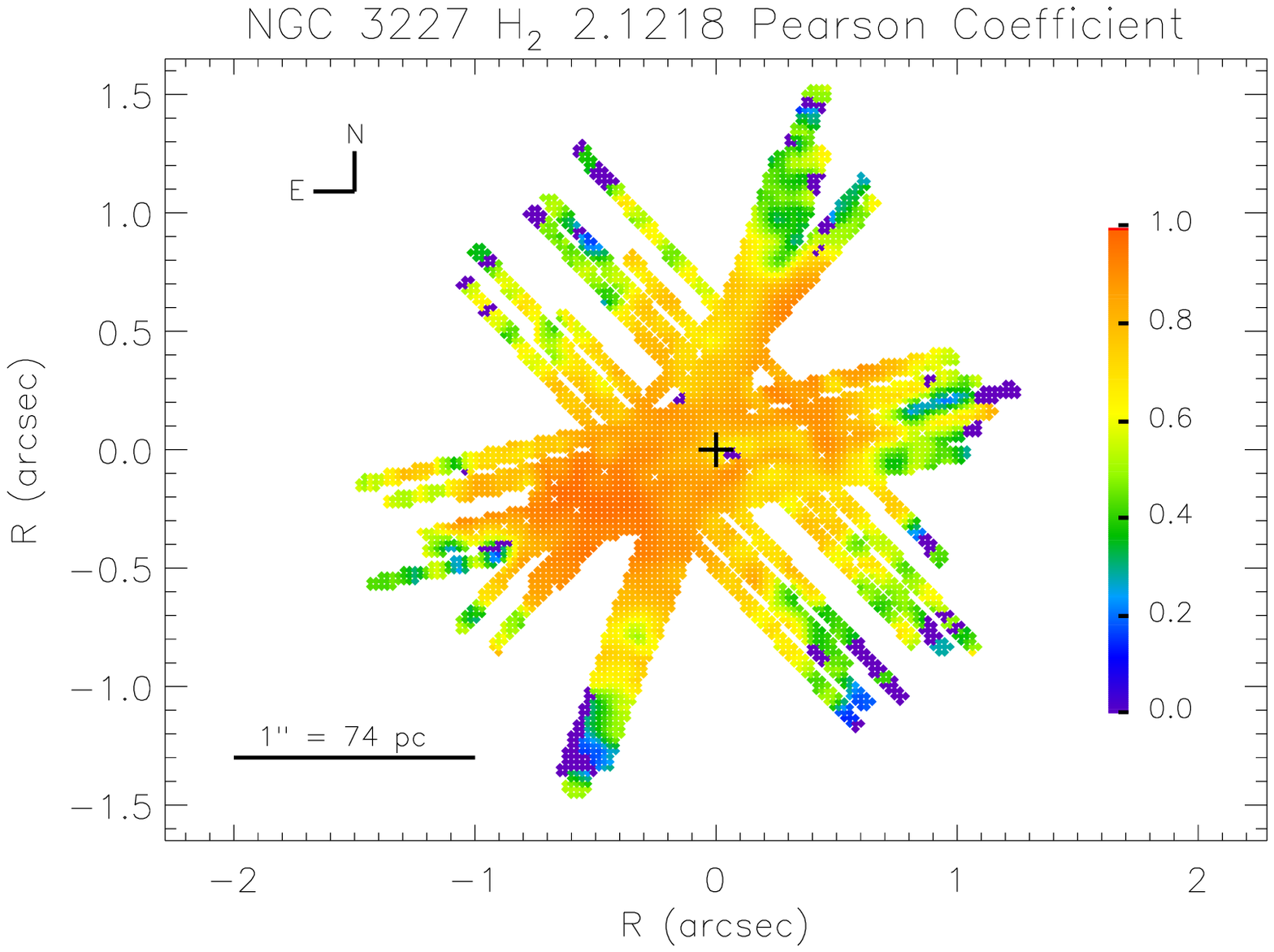}
\caption[]{2-D maps of the flux distribution, velocity, velocity dispersion, and Pearson correlation coefficient for all significant emission lines (see Table \ref{t_aps}) and the continuum flux distribution in three wavelength bands (see $\S$ 3.3) for each Seyfert 1 galaxy.  Each map has a title that includes the galaxy name, the emission line or continuum wavelength band, and the quantity measured, and the maps are presented in the same order as listed in Table \ref{t_aps}.  The bar in the left corner of the maps indicates 1\as\s and the corresponding distance in parsecs is given.   All maps have been smoothed by a Gaussian with a FWHM of 0\as.07, which is representative of the PSF under the best seeing conditions.  Two sets of maps are shown for NGC 4151.  The first set contains all data except those from 19 Feb 2001 for which clouds and a marginal AO correction resulted in poor data, and the second set contains data from only 22 Apr 2003, which has the highest spatial resolution.  The printed edition of the Journal contains only a sample of the 2-D maps.  All of the maps for each Seyfert 1 galaxy are available in the electronic edition of the Journal.  All maps are also available at www.astro.ucla.edu/$\sim$ehicks. \label{allmaps}}
\end{figure*}

For each galaxy, 2-D maps of the flux distribution, velocity, and velocity dispersion of \htwo 1-0 S(1) \lam2.1218, as well as other significant emission such as other \htwo transitions, the narrow component of \bgns, and coronal emission, are presented in Fig. \ref{allmaps}.  In addition, 2-D maps of the \kb\s continuum are presented.  Since the maps are constructed from observations on several different nights (sometimes separated by many months) the flux distribution of unresolved, or marginally resolved, emission varied due to changes in the seeing conditions and the quality of the AO correction. This should be kept in mind when examining the 2-D flux distribution maps, as some asymmetric emission might be explained by a difference in the quality of the data at different PAs.  The flux maps are scaled as a percentage of the peak flux.

\begin{deluxetable*}{lrrrrrrrrrrr}
\tablecaption{\htwo and \bg N$_{sigma}$ Values
 \label{t_nsigma}} 
\tablewidth{0pt}
\tablehead{
\colhead{} & 
\colhead{\htwo} &
\colhead{\htwo} &
\colhead{\htwo} &
\colhead{\bg} &
\colhead{\htwo} &
\colhead{\htwo} &
\colhead{\htwo} &
\colhead{\htwo} &
\colhead{\htwo} &
\colhead{\htwo} &
\colhead{Cont.} \\
\colhead{} & 
\colhead{1-0 S(3)} &
\colhead{1-0 S(2)} &
\colhead{1-0 S(1)} &
\colhead{} &
\colhead{3-2 S(3)} &
\colhead{1-0 S(0)} &
\colhead{2-1 S(1)} &
\colhead{1-0 Q(1)} &
\colhead{1-0 Q(3)} &
\colhead{1-0 Q(5)} &
\colhead{SNR} \\  
\colhead{Galaxy} & 
\colhead{1.9576} &
\colhead{2.0338} &
\colhead{2.1218} &
\colhead{2.1661} &
\colhead{2.2014} &
\colhead{2.2235} &
\colhead{2.2477} &
\colhead{2.4066} &
\colhead{2.4237} &
\colhead{2.4548} &
\colhead{2.23-2.24} \\ 
}
\startdata

NGC3227  &  \nodata  &  41.6  &  140.  &  32.3  &  14.4  &  13.0  &  35.2  &  
92.4 &  98.4  &  89.9   & 196	\\
         &  \nodata  &  \nodata  &  7.4  &  27.4  &   0.0   &  2.5  &  7.5
  &   0.0   &  3.8  &  6.9   & 126 \\
         &  \nodata  &  \nodata  &  77.6  &  7.5  &  13.5  &  16.1  &  12.8
  &  72.0  &  70.6  &  55.9  & 113 \\
NGC3516  &  \nodata  &  \nodata  &  8.3  &  5.3  &   0.0   &  6.0  &  6.9
  &  6.8  &  0.0  &  0.0  & 167	\\
         &  \nodata  &  \nodata  &  2.0  &  4.2  &   0.0   &  3.8  &  4.8
  &  4.6  &  0.0  &  0.0  &  125 \\
         &  \nodata  &  \nodata  &  1.8  &  2.9  &   0.0   &   0.0   &  1.1
  &   0.0   &  0.0  &   0.0  & 96  \\
NGC4051  &  \nodata  &  12.2  &  57.0  &  48.9  &  22.4  &  25.1  &  15.1  &  
17.3  &  34.0  &  7.9   & 192	\\
         &   \nodata   &  34.4  &  20.6  &  36.5  &  5.2  &  26.1  &  2.0  &  
8.2  &  7.3  &  0.0  & 108 \\
         &  \nodata  &   0.0   &  2.4  &   0.0   &  3.9  &  7.7  &  1.5  &  
0.0  &  0.0  &  0.0  & 89 \\
NGC4151  &   \nodata   &  0.0  &   4.4   &  66.5  &  9.7  &  10.6  &  16.2  &  
0.0  &  0.0  &  0.0   & 300	\\
         &  \nodata  &  0.0 &  7.5  &  13.1  &  4.3  &  2.7  &  2.8
  &  0.0  &  0.0  &   0.0  & 143  \\
         &  \nodata  &  0.0  &   2.8   &  24.7  &  5.0  &  9.4  &  8.8
  &   0.0   &  0.0   &  0.0 & 105  \\
NGC4593  &  \nodata  &  \nodata  &  6.1  &   0.0   &  2.2  &   0.0   &  
 0.0   &  7.0  &  \nodata  &  2.5   & 100 \\
         &  \nodata  &  \nodata  &  4.0  &   0.0   &  1.4  &   0.0   &  0.0
  &  4.7  &  \nodata  &   0.0  & 43  \\
         &  \nodata  &  \nodata  &  2.1  &  0.0  &  1.7  &  0.0  &   0.0 
  &  0.0  &   \nodata   &  0.0  & 40 \\
NGC5548  &  \nodata  &  \nodata  &  19.7  &  22.5  &  6.5  &   0.0   &  12.4
  &  6.5  &  14.1  &  0.0   & 125 \\
         &  \nodata  &  \nodata  &  1.2  &   0.0   &  3.8  &   0.0   &  6.9
  &  0.0  &  2.8  &  0.0  & 27 \\
         &  \nodata  &  \nodata  &  5.5  &  9.2  &   0.0   &  0.0  &  0.0
  &  8.5  &  1.4  &  \nodata  & 23 \\
NGC6814  &  \nodata  &  \nodata  &  37.3  &  4.1  &  13.7  &  4.9  &  4.7
  &  3.9  &  \nodata  &  \nodata  & 111  \\
         &  \nodata  &  \nodata  &  8.8  &  2.3  &  5.8  &   0.0   &  3.1
  &   0.0   &  \nodata  &  \nodata  & 33 \\
         &  \nodata  &  \nodata  &  0.0  &  2.5  &  1.5  &  2.5  &  3.5
  &  3.0  &  \nodata  &   \nodata  & 27  \\
NGC7469  &  14.6  &  \nodata  &  46.1  &  36.6  &  10.7  &  5.4  &  3.8  &  
\nodata  &  \nodata  &  \nodata  &  173 \\
         &  9.0  &  \nodata  &  11.1  &  23.4  &  1.9  &  0.0  &   0.0 
  &  \nodata  &  \nodata  &  \nodata  & 154  \\
         &  3.3  &  \nodata  &  39.5  &  1.7  &  8.6  &  4.0  &  2.1
  &  \nodata  &  \nodata  &  \nodata  & 132 \\
Ark120  &  0.0  &  0.0  &  4.7   &  4.8  &  3.5  &  2.7  &   \nodata   &  
\nodata  &  \nodata  &  \nodata   & 48	\\
         &   0.0   &  0.0  &  6.6  &  5.5  &  8.5  &  7.6  &  \nodata  &  
\nodata  &  \nodata  &  \nodata  & 23 \\
         &  0.0  &  0.0  &  2.7  &  0.0  &   0.0   &   0.0   &   \nodata   &  
\nodata  &  \nodata  &  \nodata & 25  \\

\enddata
\tablecomments{The lines of N$_{sigma}$ values, top, middle, and bottom, for each Seyfert 1 galaxy are for the spectra from a large 1\as.4 aperture, a nuclear 0\as.2 aperture, and an annulus aperture (r=0\as.1-0\as.7; the difference of the 0\as.2 and 1\as.4 aperture spectra), respectively.}

\end{deluxetable*}

\begin{deluxetable*}{lrrrrrrr}
\tablecaption{Velocity Dispersion and Coronal Line N$_{sigma}$ 
\label{t_nsigma_coronal}} 
\tablewidth{0pt}
\tablehead{
\colhead{} & 
\multicolumn{2}{c}{[Ca VIII] 2.3213} &
\multicolumn{2}{c}{[Si VII] 2.4833} &
\colhead{\bg} &
\colhead{\htwo 2.1218} &
\colhead{\oiii \tablenotemark{a}} \\ 
\cline{2-5} \\
\colhead{Galaxy} & 
\colhead{N$_{sigma}$} &
\colhead{FWHM} &
\colhead{N$_{sigma}$} &
\colhead{FWHM} &
\colhead{FWHM} &
\colhead{FWHM} &
\colhead{FWHM} \\
\colhead{ } & 
\colhead{ } &
\colhead{(\kmsns)} &
\colhead{} &
\colhead{(\kmsns)} &
\colhead{(\kmsns)} &
\colhead{(\kmsns)} &
\colhead{(\kmsns)} \\
}
\startdata

NGC3227  &  11.8  	&  126	&  5.1      &   31	& 146     & 122    &  206  \\ 
         &  8.1  	&  124	&  5.3      & 44	& 135     & 89     &  \\  
         &  5.9  	&  142	&  3.1      &  21       & 120	  & 120    &  \\    
NGC3516  &  16.8  	&  90	&  \nodata  &  \nodata  & 97	  & 72     &  163  \\ 
         &  9.3  	&  60	&  \nodata  & \nodata   & 177	  & 65:    &  \\     
         &  2.3  	&  68: &  \nodata  & \nodata    & 58:     & 51:    &  \\     
NGC4051  &  58.0  	&  120	&  8.4      &   66	& 177	  & 69     &  81  \\ 
         &  4.9  	&  75	&  2.7      &   57:     & 178	  & 57     &  \\     
         &  9.2  	&  119	&  4.3      &   41	& \nodata & 45:    &  \\     
NGC4151  &  29.9  	&  106	&  12.3     &  94	& 160	  & 83     &  181  \\ 
         &  13.2  	&  166	&  6.6      &  67	& 157	  & 60     &  \\     
         &  13.0  	&  81	&  3.8	    &  65	& 106	  & 43:    &  \\    
NGC4593  &  0.0   &  \nodata	&  \nodata  &  \nodata  & \nodata & 63     &  109  \\
         &  0.0   &  \nodata	&  \nodata  &  \nodata  & \nodata & 60     &  \\	
         &  0.0   &  \nodata	&  \nodata  &  \nodata  & \nodata & 43:    &  \\	
NGC5548  &  35.4  	&  163	&  \nodata  &  \nodata  & 141     & 119    &  174  \\
         &  6.4  	&  60	&  \nodata  & \nodata   & \nodata & 43:    &  \\
     	 &  0.0   &  \nodata	&  \nodata  & \nodata   &   83	  & 45     &  \\
NGC6814  &  0.0   &  \nodata	&  \nodata  &  \nodata  &  177    & 71     &  \nodata \\
    	 &  0.0   &  \nodata	&  \nodata  &  \nodata  &  96:	  & 64     &  \\
         &  0.0   &  \nodata	&  \nodata  & \nodata   &  72:	  & \nodata &  \\
NGC7469  &  29.0  	&  71	&  \nodata  &  \nodata  & 176     & 102    &  153  \\
         &  9.7  	&  65	&  \nodata  & \nodata   & 175  	  & 80     &  \\
         &  7.6  	&  52	&  \nodata  & \nodata   & 57: 	  & 110    &  \\
Ark120   &  \nodata  &  \nodata	&  \nodata  &  \nodata  & 174     & 68     &  209  \\
         &  \nodata  &  \nodata	&  \nodata  &  \nodata  & 174	  & 92     &  \\
         &  \nodata  &  \nodata	&  \nodata  &  \nodata  & \nodata & 90:    &  \\

\enddata
\tablecomments{The three lines of N$_{sigma}$ and FWHM values given in the top, middle, and bottom positions for each galaxy are the measurements of spectra from a large 1\as.4 aperture, a nuclear 0\as.2 aperture, and an annulus aperture (r=0\as.1-0\as.7; the difference of the 0\as.2 and 1\as.4 aperture spectra), respectively.  Measurements of lines with less than a 2.0$\sigma$ detection are indicated by a colon.}
\tablenotetext{a}{The FWHM values for \oiii \lam5077 \AA\s are from \citet{peterson04}.}

\end{deluxetable*}

\begin{deluxetable*}{lccccccc}
\tablecaption{Gaussian Fits to Flux Distribution \label{t_fluxgfits}} 
\tablewidth{0pt}
\tablehead{
\colhead{Seyfert 1} & 
\colhead{Trasition} & 
\colhead{Rest \lam(\mic)} &
\colhead{FWHM (\as)} &
\colhead{Offset $\Delta\alpha$(\as), $\Delta\delta$(\as)} & 
\colhead{Axis Ratio} & 
\colhead{PA (\deg)}  &
\colhead{Method \tablenotemark{a}}  \\ 
}
\startdata

NGC 3227	&	\htwo 1-0 S(1)  &	 2.1218	&	0.538$\pm$0.034	&		-0.359,-0.155	&	\nodata	&	\nodata	&	2-D, C\tablenotemark{b}	\\
		&	\bg  		    &	 2.1661	&	0.055$\pm$0.006	& 0.043,-0.032	&	0.67$\pm$0.22	&	90$\pm$20	&	2-D	\\
		&	\htwo 1-0 Q(1)  &  2.4066	&	0.266$\pm$0.034	&		-0.542,-0.218	&	\nodata	&	\nodata	&	2-D, C\tablenotemark{b} \\
		&	\htwo 1-0 Q(3)  &	 2.4237	&	0.460$\pm$0.066	&		-0.487,-0.198	&	\nodata	&	\nodata	&	2-D, C\tablenotemark{b} \\
NGC 3516	&	\htwo 1-0 S(1)  &	 2.1218	&	0.210$\pm$0.062	&	-0.120,0.031	&	\nodata	&	\nodata	&	S	\\
NGC 4051	&	\htwo 1-0 S(1)  &	 2.1218	&	1.097$\pm$2.133	&		-0.530,0.005	&	0.57$\pm$0.22	&	90$\pm$30	&	2-D	\\
		&	\bg  		    &	 2.1661	&	0.172$\pm$0.029	&	0.0,0.0	&	0.75$\pm$0.18	&	119$\pm$22	&	2-D	\\
NGC 4151	&	\htwo 1-0 S(1)  &	 2.1218	&	0.13$\pm$0.20	&	-0.871,-0.160	&	0.67$\pm$0.21	&	145$\pm$38	&	2-D\tablenotemark{c}	\\
		&	  &	 &	0.11$\pm$0.25	&	-0.000,-0.080	&	1.00$\pm$0.15	&	160$\pm$25 	&	2-D\tablenotemark{c}	\\
		&	  &	 &	0.23$\pm$0.25	&	0.501,0.000	&	0.20$\pm$0.16	&	123$\pm$31 	&	2-D\tablenotemark{c}	\\
		&	  &	 &	0.25$\pm$0.21	&	1.021,0.000	&	0.26$\pm$0.25	&	95$\pm$17 	&	2-D\tablenotemark{c}	\\
		&	\bg  		    &	 2.1661	&	0.048$\pm$0.001	&		0.0,0.0	&	\nodata	&	\nodata	&	2-D, C	\\
NGC 4593	&	\htwo 1-0 S(1)  &	 2.1218	&	0.132$\pm$0.056	&	0.0,0.0	&	\nodata	&	\nodata	&	S	\\
		&	\htwo 1-0 Q(1)  &	 2.4066	&	0.179$\pm$0.072	&	0.0,0.0	&	\nodata	&	\nodata	&	S	\\
NGC 5548	&	\htwo 1-0 S(1)  &	 2.1218	&	0.131$\pm$0.083	&	0.0,0.0	&	\nodata	&	\nodata	&	S	\\
NGC 6814	&	\htwo 1-0 S(1)  &	 2.1218	&	0.389$\pm$0.067	&	0.0,0.0	&	\nodata	&	\nodata	&	S	\\
		&	\htwo 3-2 S(3)  &	 2.2014	&	0.984$\pm$0.243	&	0.0,0.0	&	\nodata	&	\nodata	&	S	\\
		&	\htwo 1-0 S(0)  &	 2.2235	&	1.137$\pm$0.395	&	0.0,0.0	&	\nodata	&	\nodata	&	S	\\
		&	\htwo 2-1 S(1)  &	 2.2477	&	0.102$\pm$0.341	&	0.0,0.0	&	\nodata	&	\nodata	&	S	\\
NGC 7469	&	\htwo 1-0 S(1)  &	 1.9576	&	0.327$\pm$0.016	&	0.0,0.0	&	\nodata	&	\nodata	&	2-D, C	\\
		&	\htwo 1-0 S(1)  &	 2.1218	&	0.448$\pm$0.055	&	0.0,0.0	&	\nodata	&	\nodata	&	2-D, C	\\
		&	\bg  		    &	 2.1661	&	0.085$\pm$0.010	&	0.0,0.0	&	0.91$\pm$0.16	&	90$\pm$20	&	2-D	\\
Ark 120	&	\htwo 1-0 S(1)  &	 2.1218	&	0.134$\pm$0.017	&	0.0,0.0	&	\nodata	&	\nodata	&	2-D, C	\\

\enddata
\tablenotetext{a}{Fits were done with a single 2-D Gaussian (2-D) with no constraints, a single 2-D Gaussian assuming circular symmetry (2-D,C), or a fit to the flux distribution along a single slit position (S; done in cases of little 2-D coverage).}
\tablenotetext{b}{The \htwo flux distribution in NGC 3227 is very patchy and elongated along a PA of about 135\deg.  The fits reported here are circularly symmetric fits to the peak of the flux distribution.} 
\tablenotetext{c}{For NGC 4151 the distribution is very patchy and a single 2-D Gaussian is not a good representation of the flux distribution.  Instead four 2-D Gaussians were fit to the distribution and the parameters of each are given on different lines in the table.} 

\end{deluxetable*}

\begin{deluxetable*}{lccccccc}
\tablecaption{Gaussian Fits to Coronal-Line Flux Distribution \label{t_cordist}} 
\tablewidth{0pt}
\tablehead{
\colhead{Seyfert 1} & 
\colhead{Trasition} & 
\colhead{Rest \lam(\mic)} &
\colhead{FWHM\tablenotemark{a}} &
\colhead{Axis Ratio} & 
\colhead{PA (\deg)}  &
\colhead{Method\tablenotemark{b}} & 
\colhead{Extent\tablenotemark{a}} \\ 
}
\startdata

NGC 3227	&	[Ca VIII] &	 2.3213	&	0.046$\pm$0.013	&	0.58$\pm$0.23	&	97$\pm$20	&	2-D	&	0.6	\\
		&		   &			&	3.45$\pm$0.98			&						&		&		&	45	\\ 
		&	[Si VII] &	 2.4833	&	0.046$\pm$0.006	&	0.53$\pm$0.09	&	106$\pm$15	&	2-D	&	0.5	\\
		&		   &			&	3.45$\pm$0.45			&						&		&		&	37	\\
NGC 3516	&	[Ca VIII] &	 2.3213	&	0.532$\pm$0.273	&	\nodata	&	\nodata	&	2-D, C		&	0.6	\\
		&		   &			&	91.50$\pm$46.96			&						&		&		&	103	\\
NGC 4051	&	[Ca VIII] &	 2.3213	&	0.175$\pm$0.040	&	\nodata	&	\nodata	&	2-D, C	&	0.5	\\	
		&		   &			&	7.88$\pm$1.80			&						&		&		&	23	\\
		&	[Si VII]  &	 2.4833	&	0.153$\pm$0.011	&	\nodata	&	\nodata	&	2-D, C	&	0.7	\\
		&		   &			&	6.89$\pm$0.50			&						&		&		&	32	\\
NGC 4151	&	[Ca VIII] &	 2.3213	&	0.167$\pm$0.005	&	\nodata	&	\nodata	&	2-D, C	&	0.7	\\
		&		   &			&	10.86$\pm$0.33			&						&		&		&	45	\\
		&	[Si VII]  &	 2.4833	&	0.164$\pm$0.006	&	\nodata	&	\nodata	&	2-D, C	&	0.7	\\	
		&		   &			&	10.66$\pm$0.39			&						&		&		&	45	\\	
NGC 5548	&	[Ca VIII] &	 2.3213	&	0.165$\pm$0.068	&	\nodata	&	\nodata	&	2-D, C	&	0.3	\\	
		&		   &			&	55.27$\pm$22.78			&						&		&		&	100	\\
NGC 7469	&	[Ca VIII] &	 2.3213	&	0.080$\pm$0.015	&	0.76$\pm$0.20	&	94$\pm$29	&	2-D	&	0.4	\\
		&		   &			&	25.44$\pm$4.77		&						&		&		&	127	\\

\enddata
\tablenotetext{a}{The first row the FWHM and extent of the coronal-line flux distribution is given in arcseconds and in the second row it is in parsecs.  The extended emission region is defined by the average radius out to where the flux drops to 5\% of the peak value.}
\tablenotetext{b}{ Method of fitting the flux distribution: `2-D' is fitting the full 2-D map and `2-D, C' is fitting the full 2-D map with circular symmetry.} 

\end{deluxetable*}

\begin{figure*}[!h]	
\epsscale{0.3}
\plotone{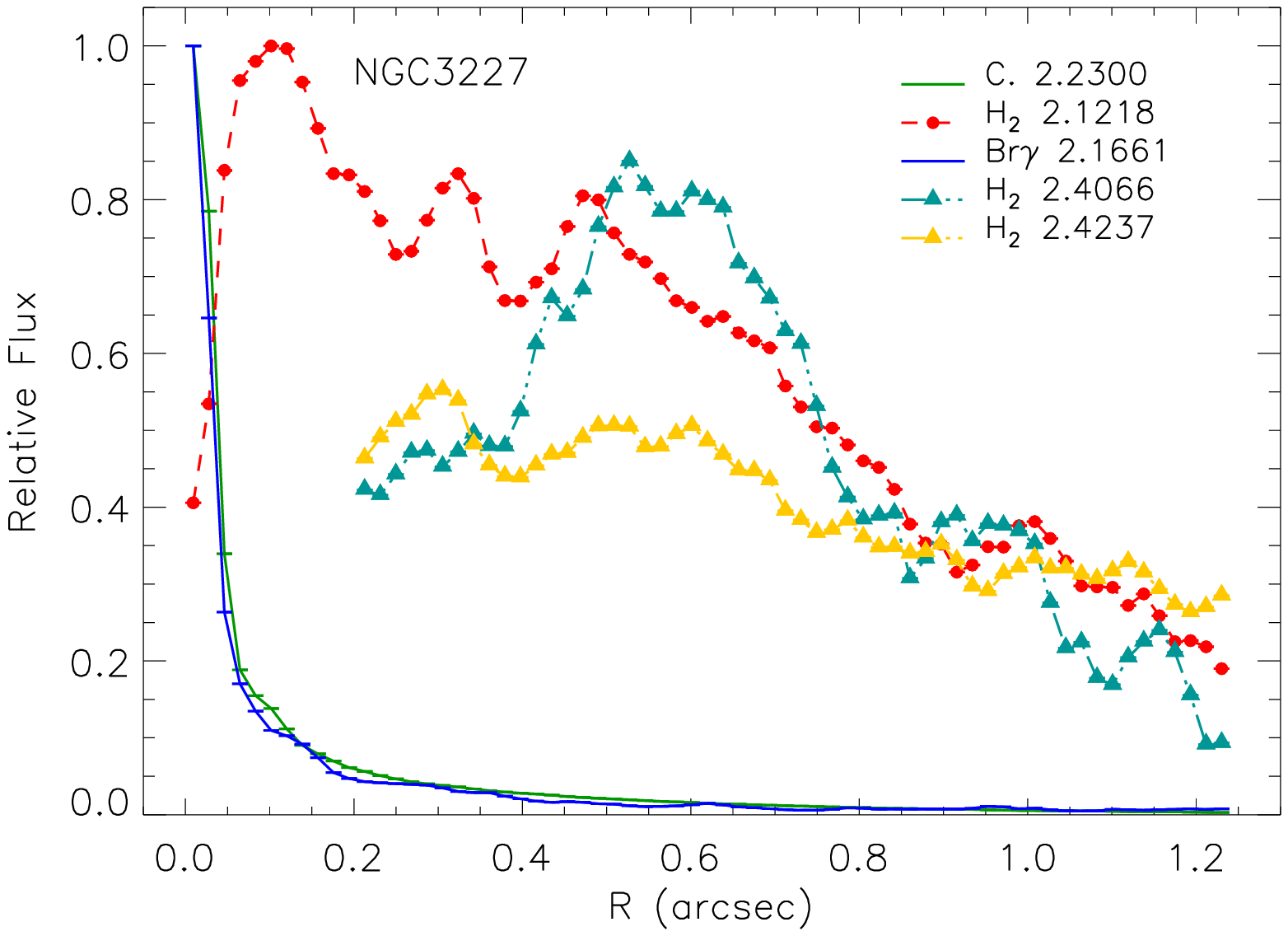}
\plotone{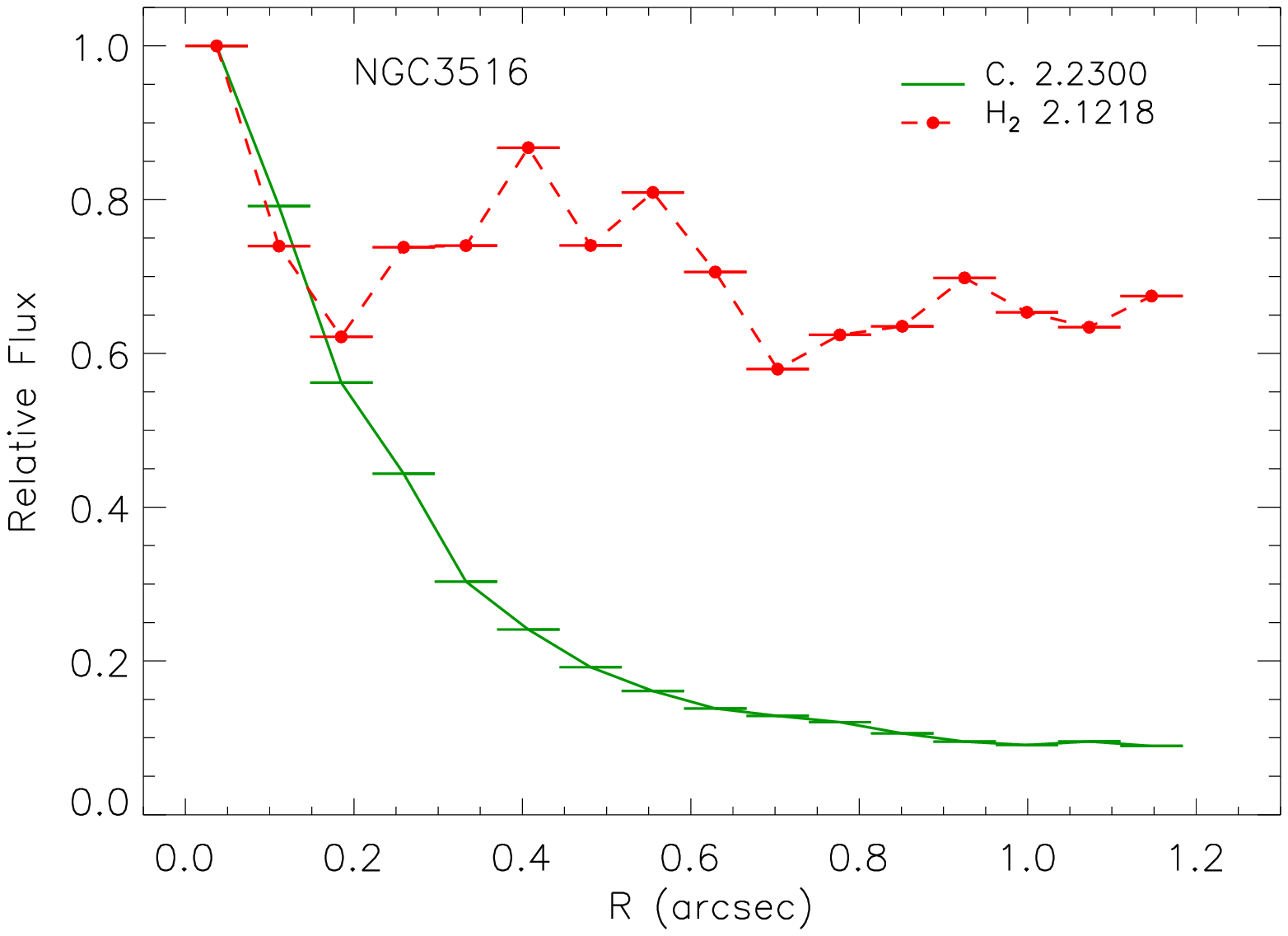}
\plotone{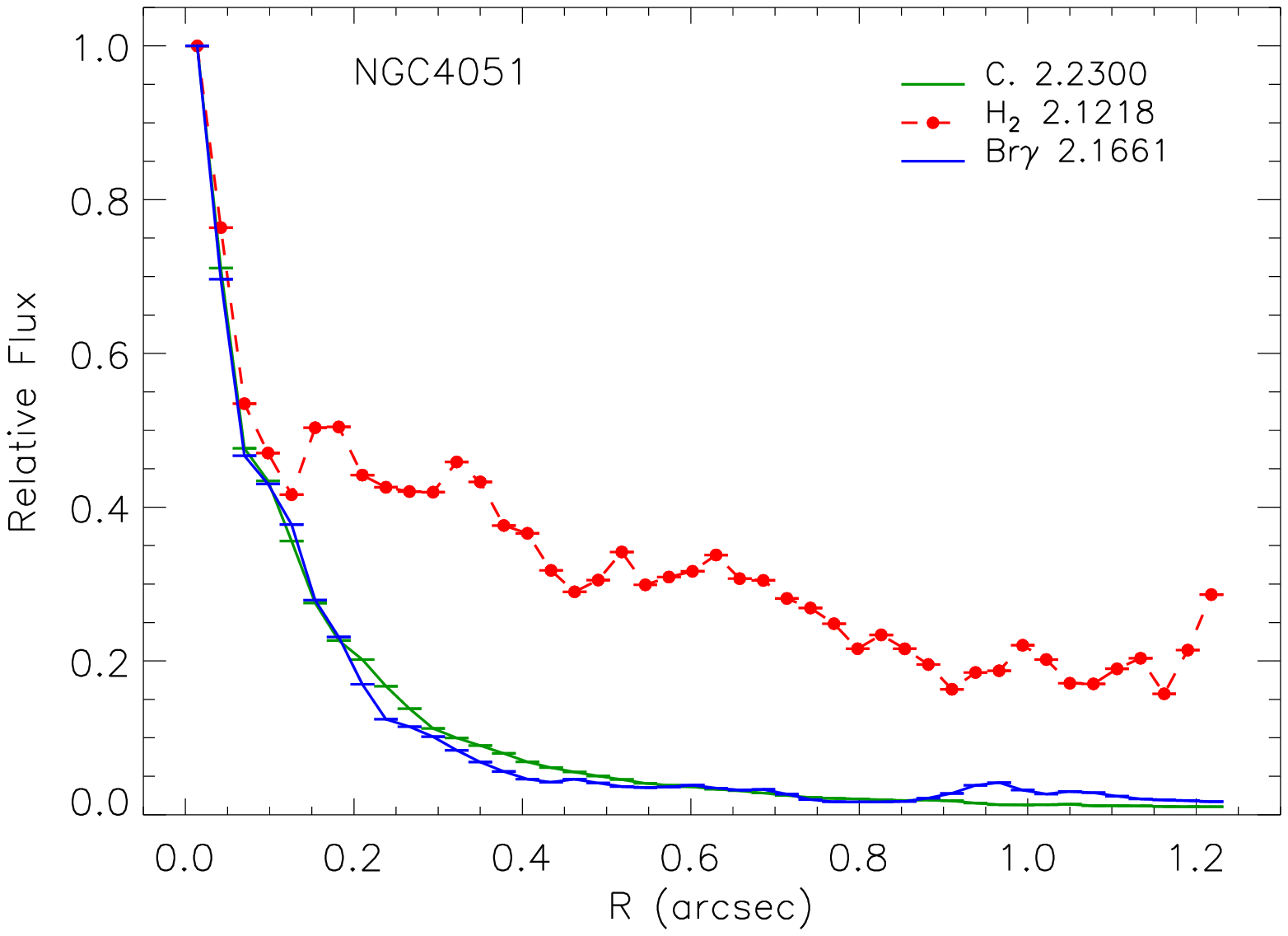}
\plotone{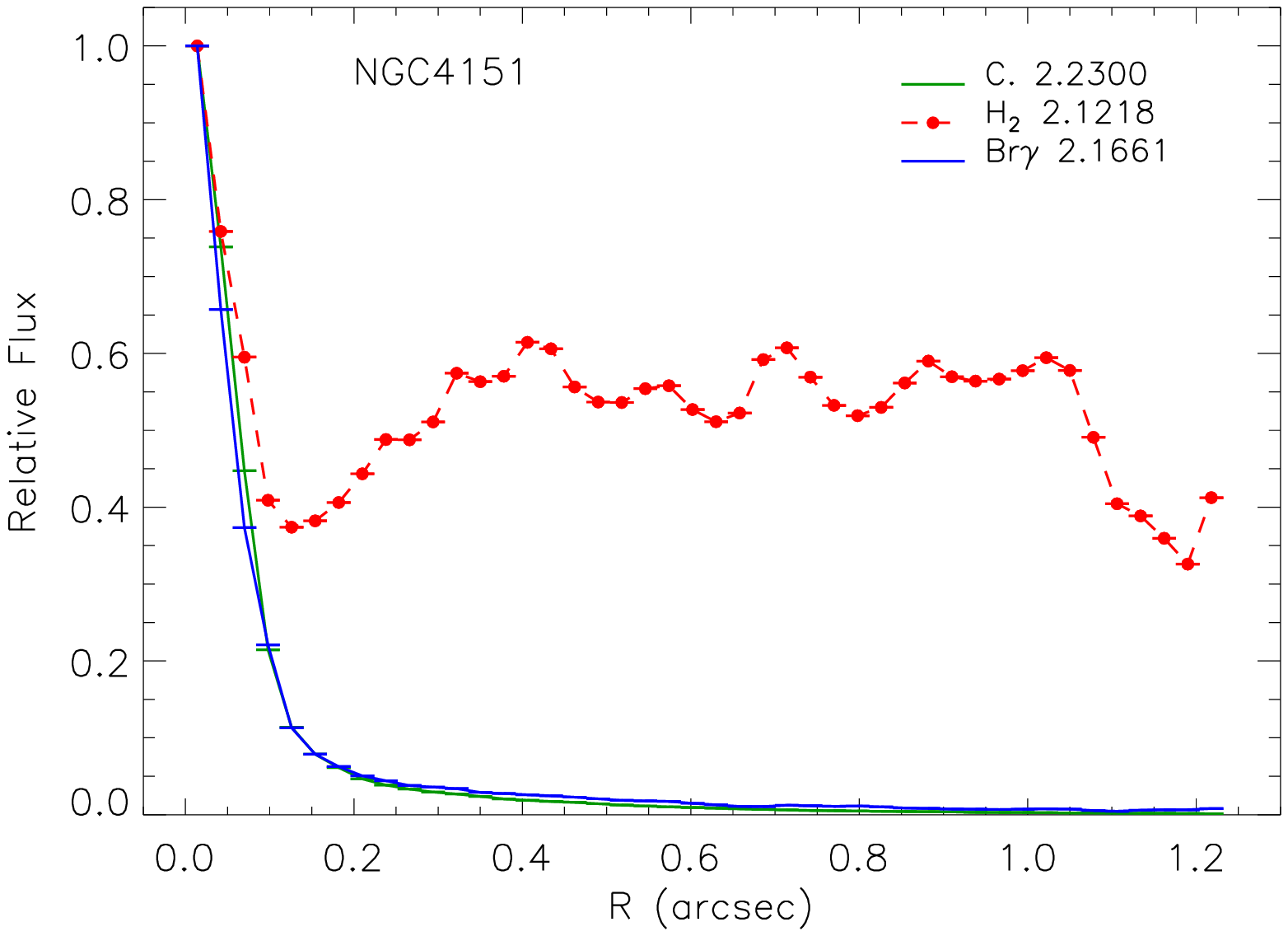}
\plotone{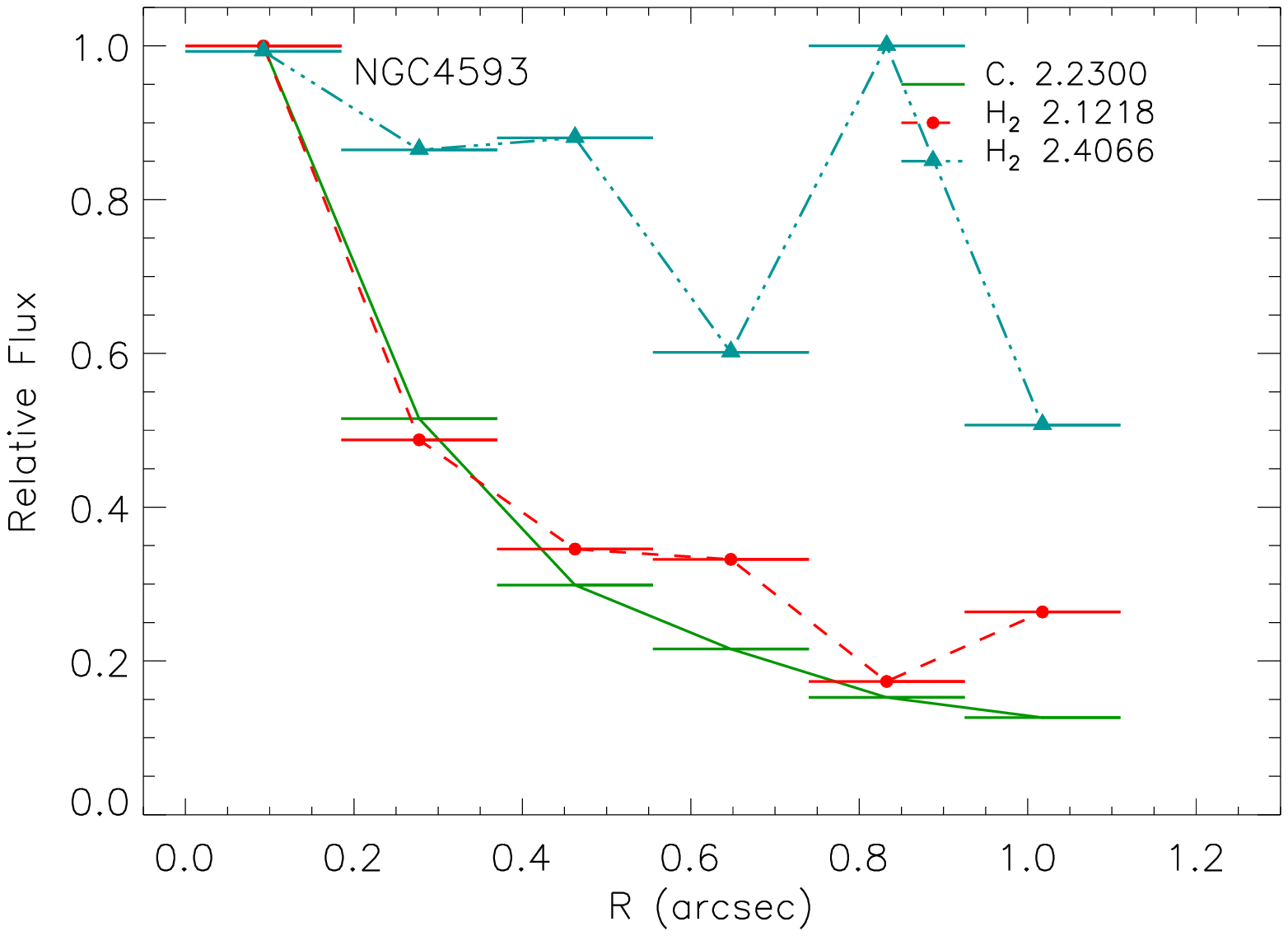}
\plotone{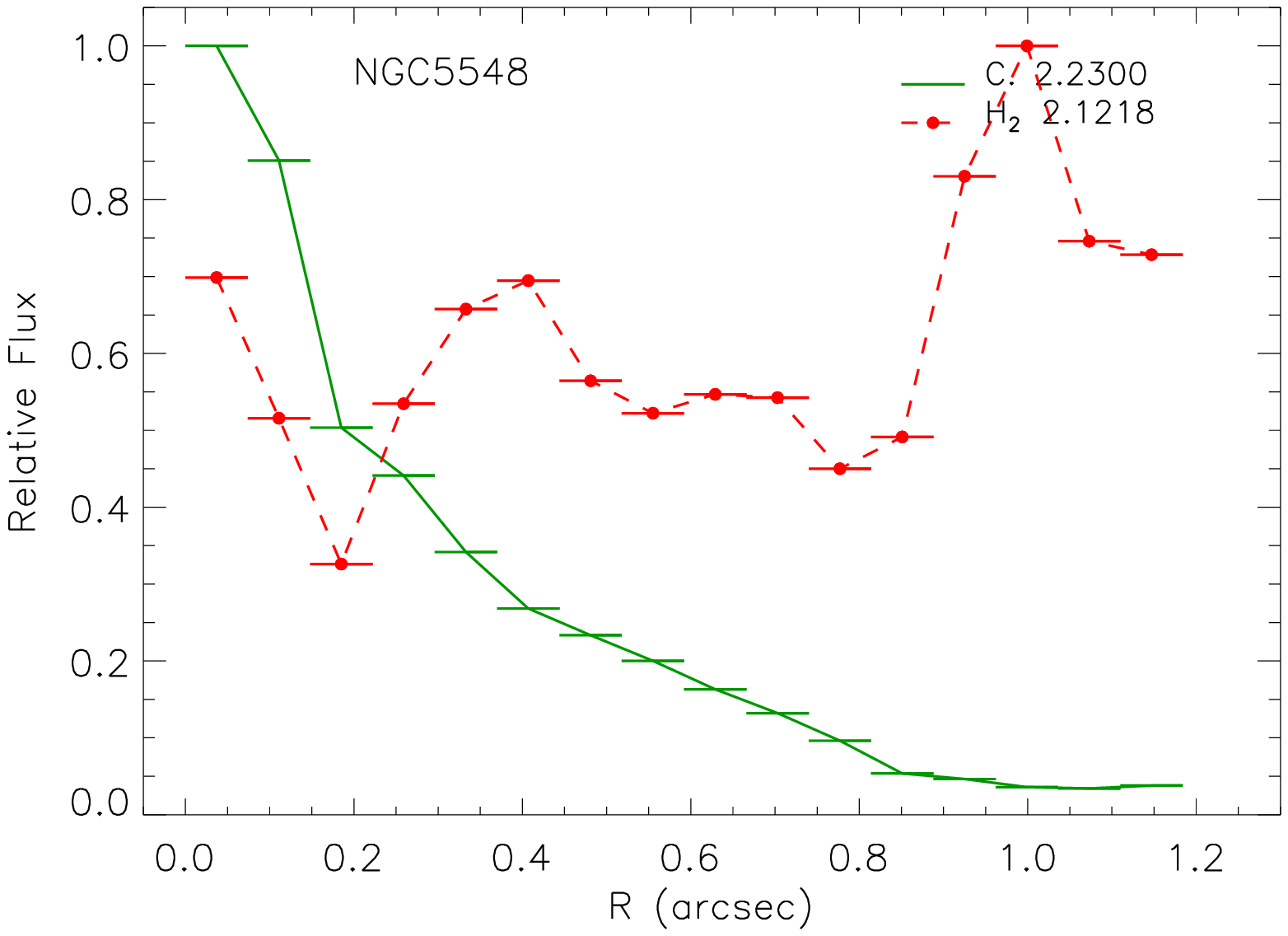}
\plotone{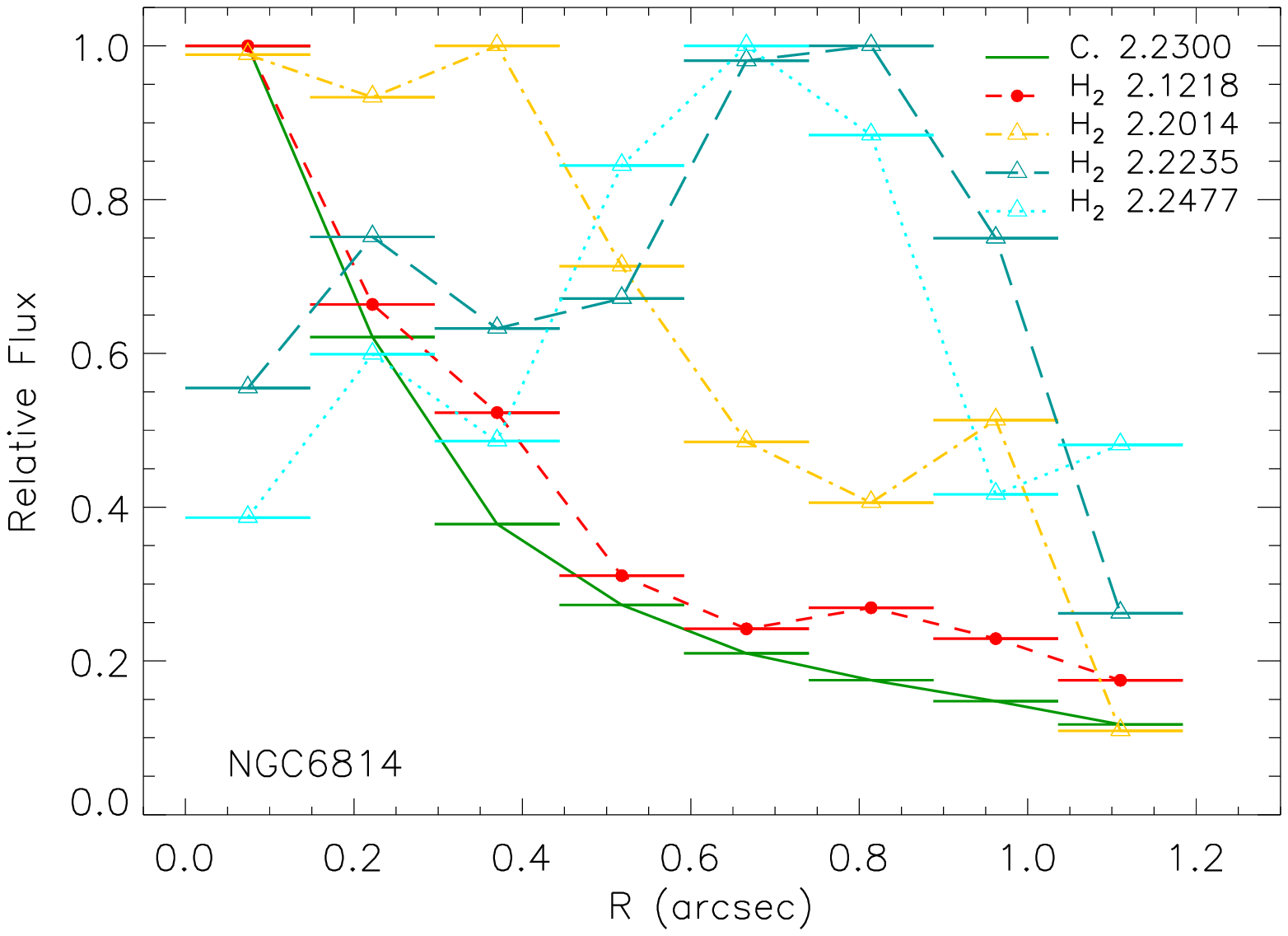}
\plotone{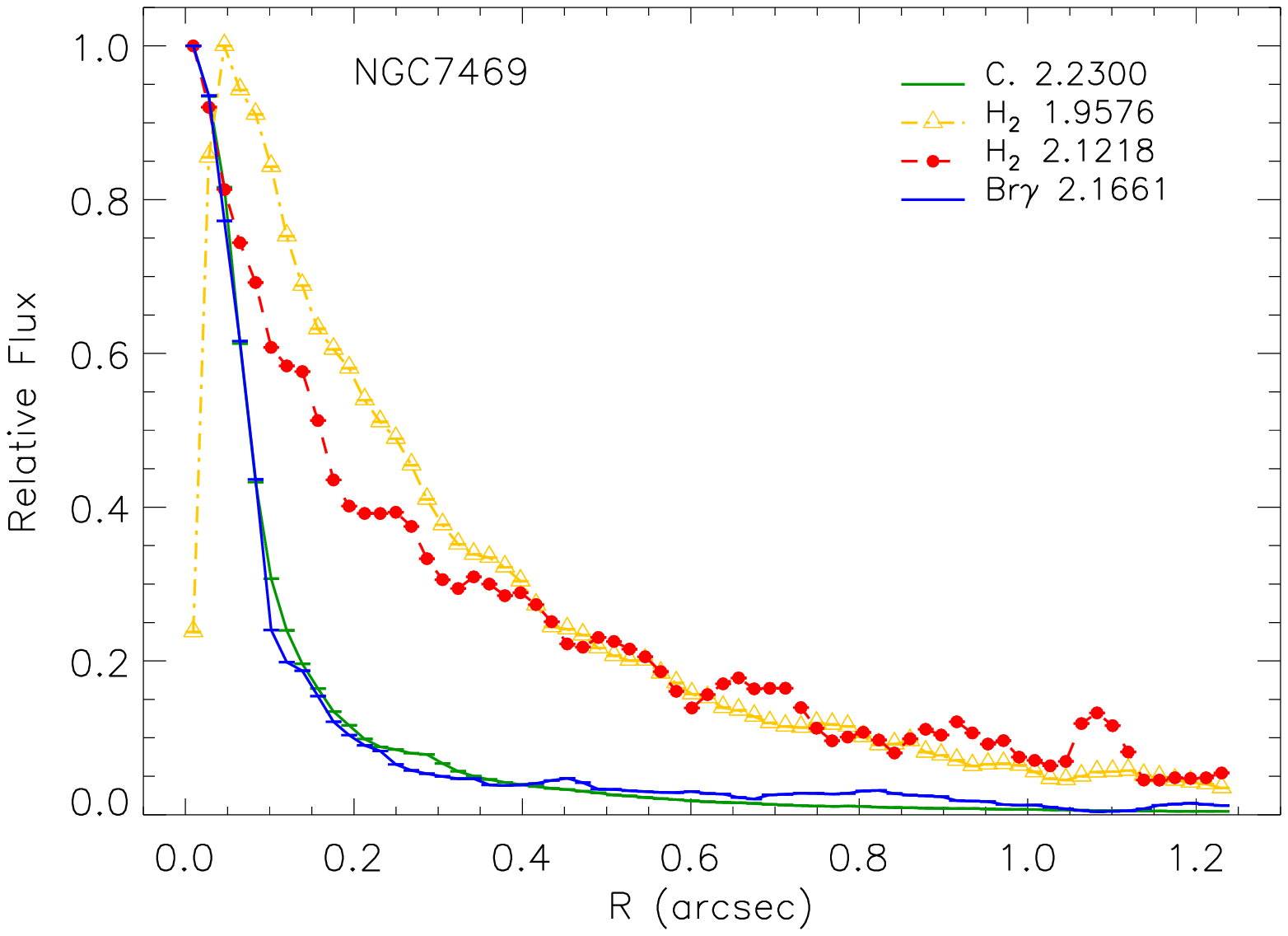}
\plotone{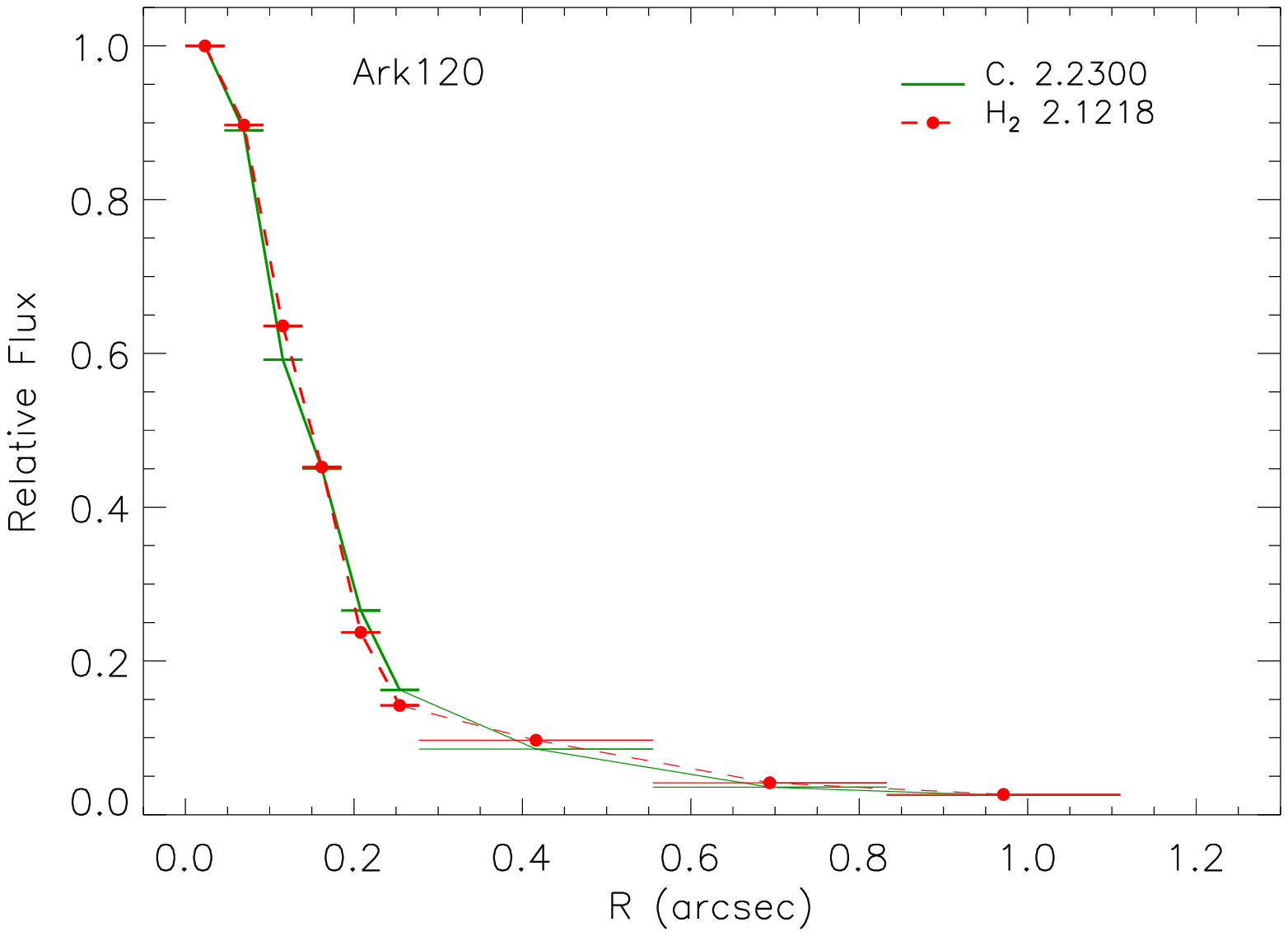}
\caption[ \htwons, \bgns, and \kb\s Radially Averaged Flux Distributions]{Radial averages of the 2-D \htwons, \bgns, and \kb\s flux distributions for each galaxy as labeled in the left of each plot.  The curves are as indicated by the legend and the continuum (C. 2.2300) is measured at 2.23-2.24 \micns.  \label{radflux}}
\end{figure*} 

\begin{figure*}[!h]	
\epsscale{0.3}
\plotone{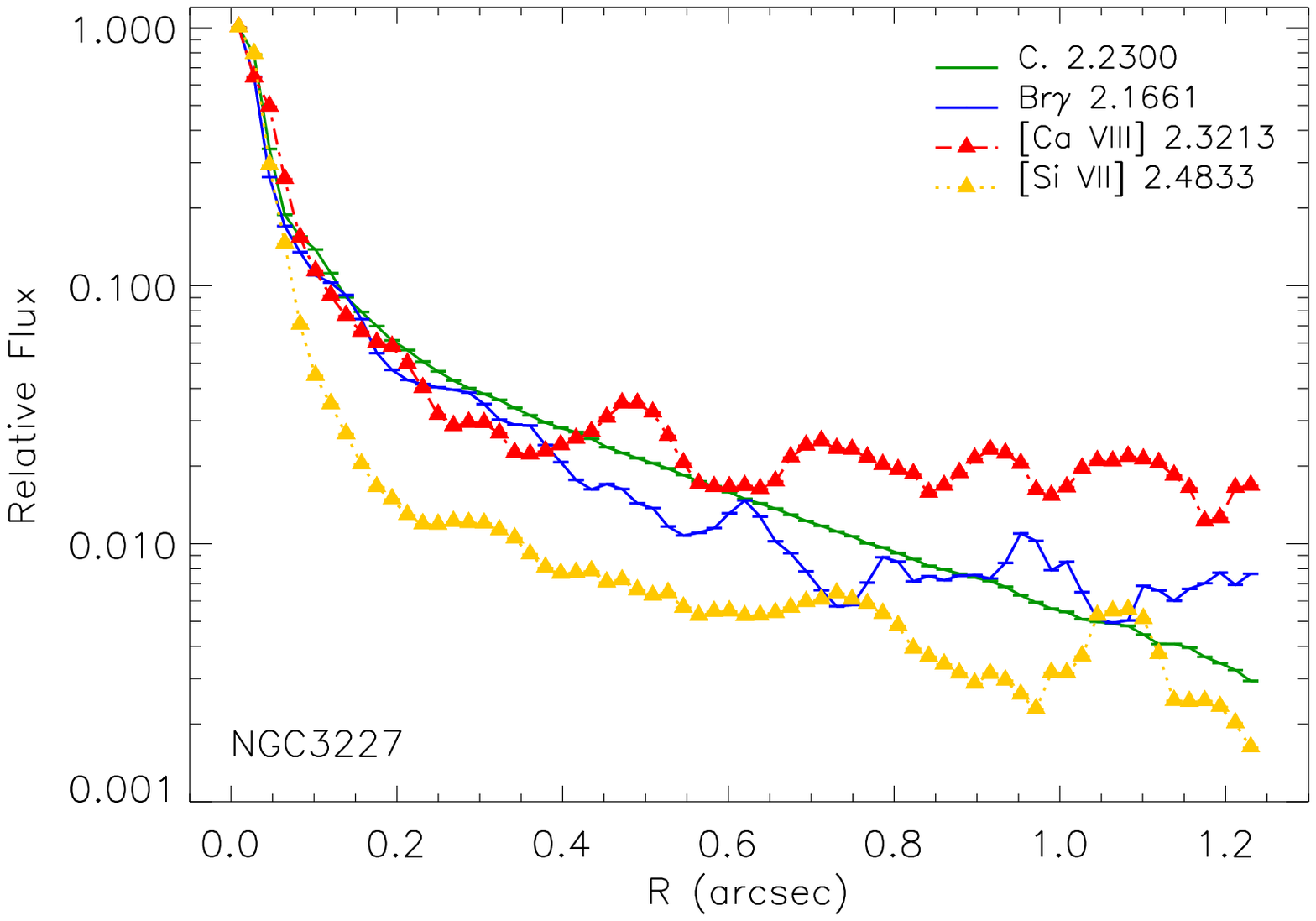}
\plotone{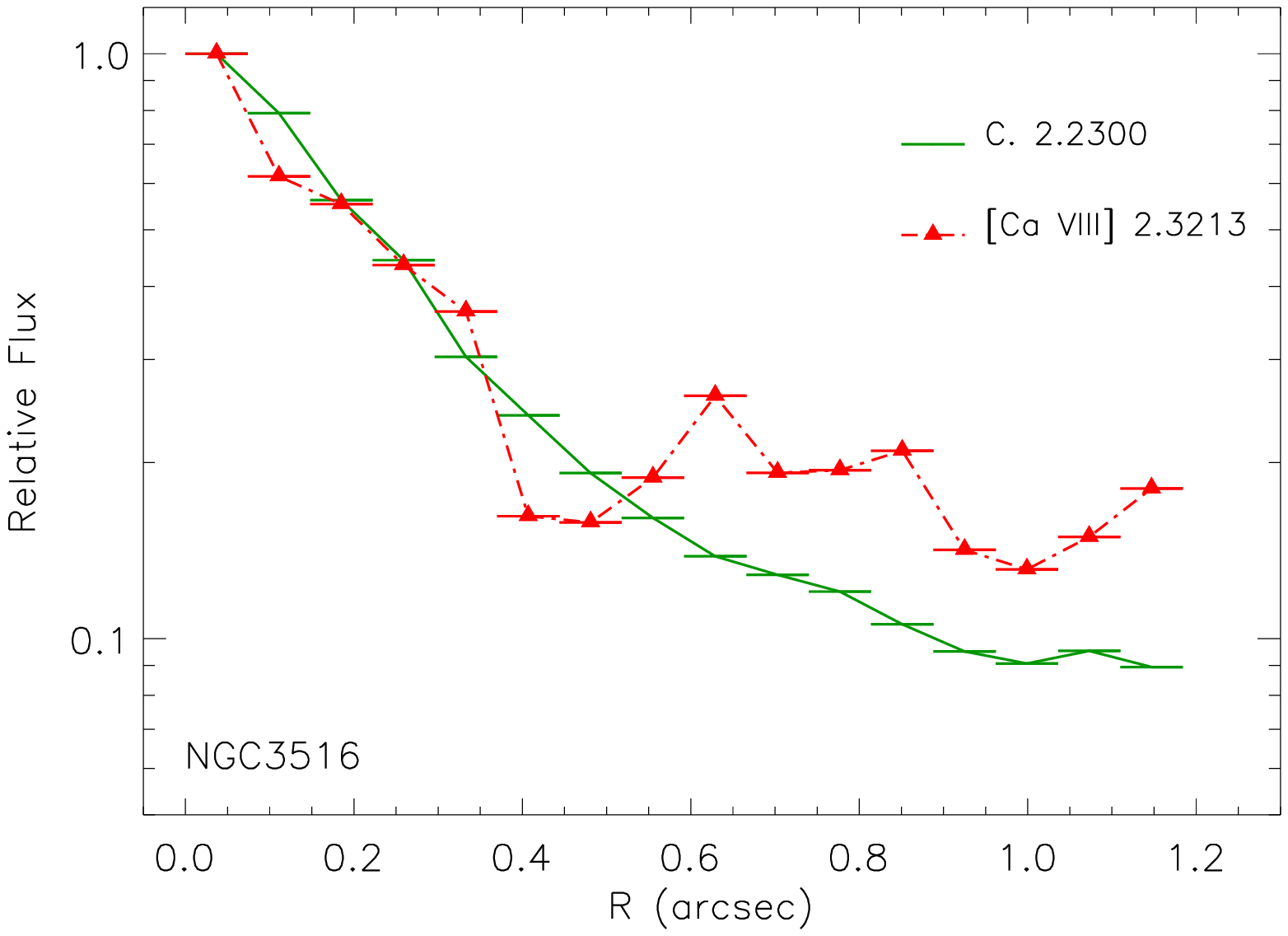}
\plotone{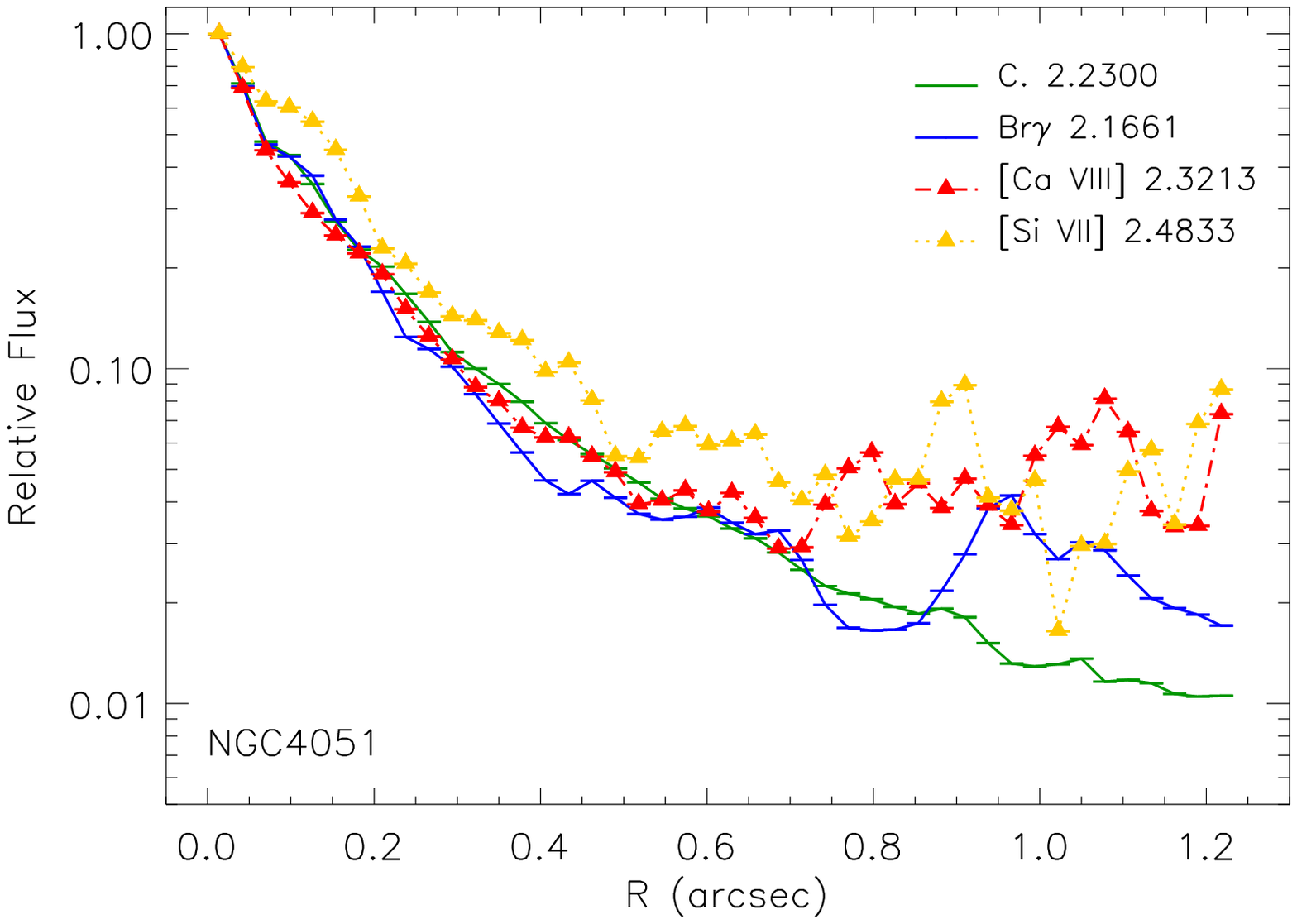}
\plotone{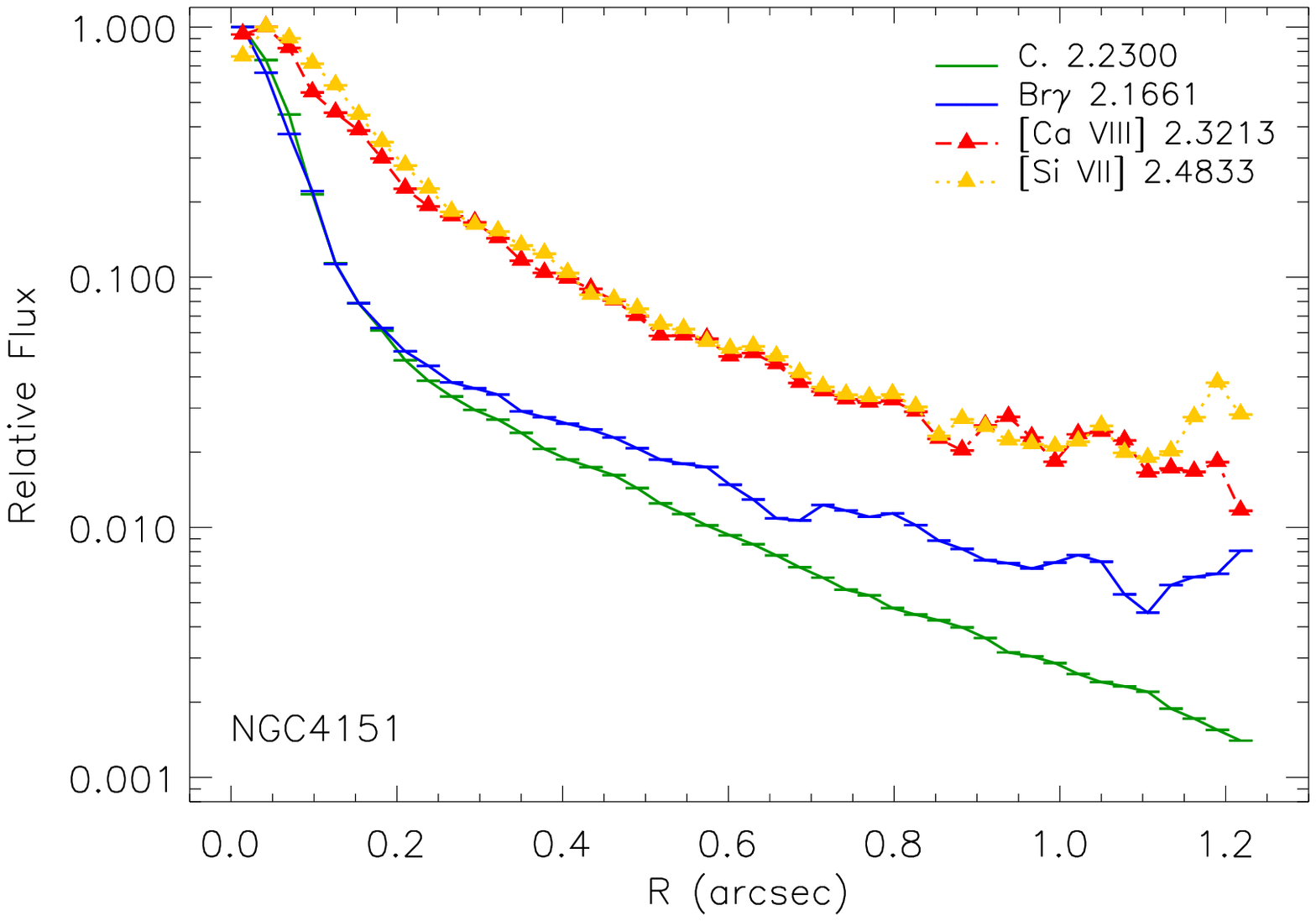}
\plotone{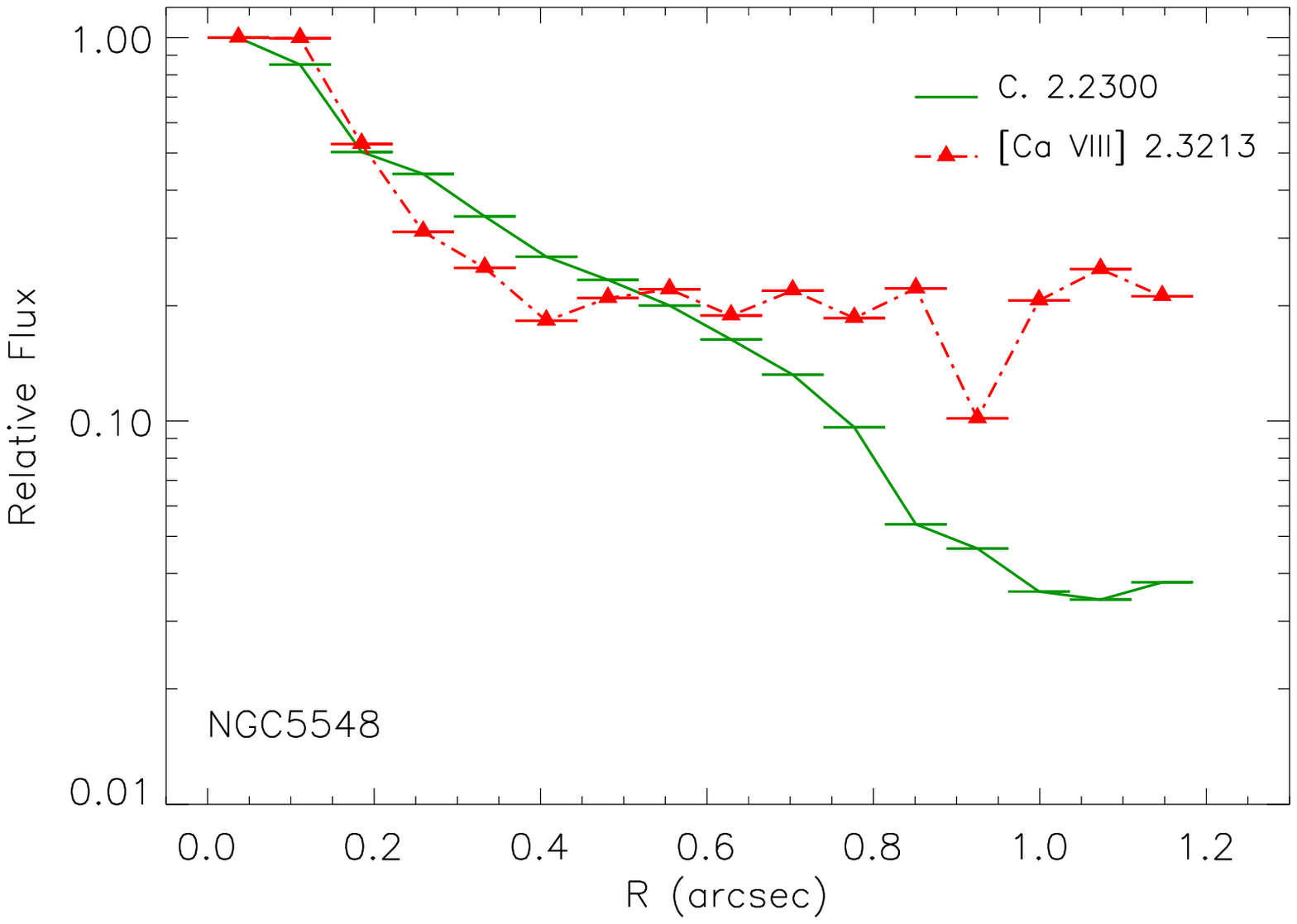}
\plotone{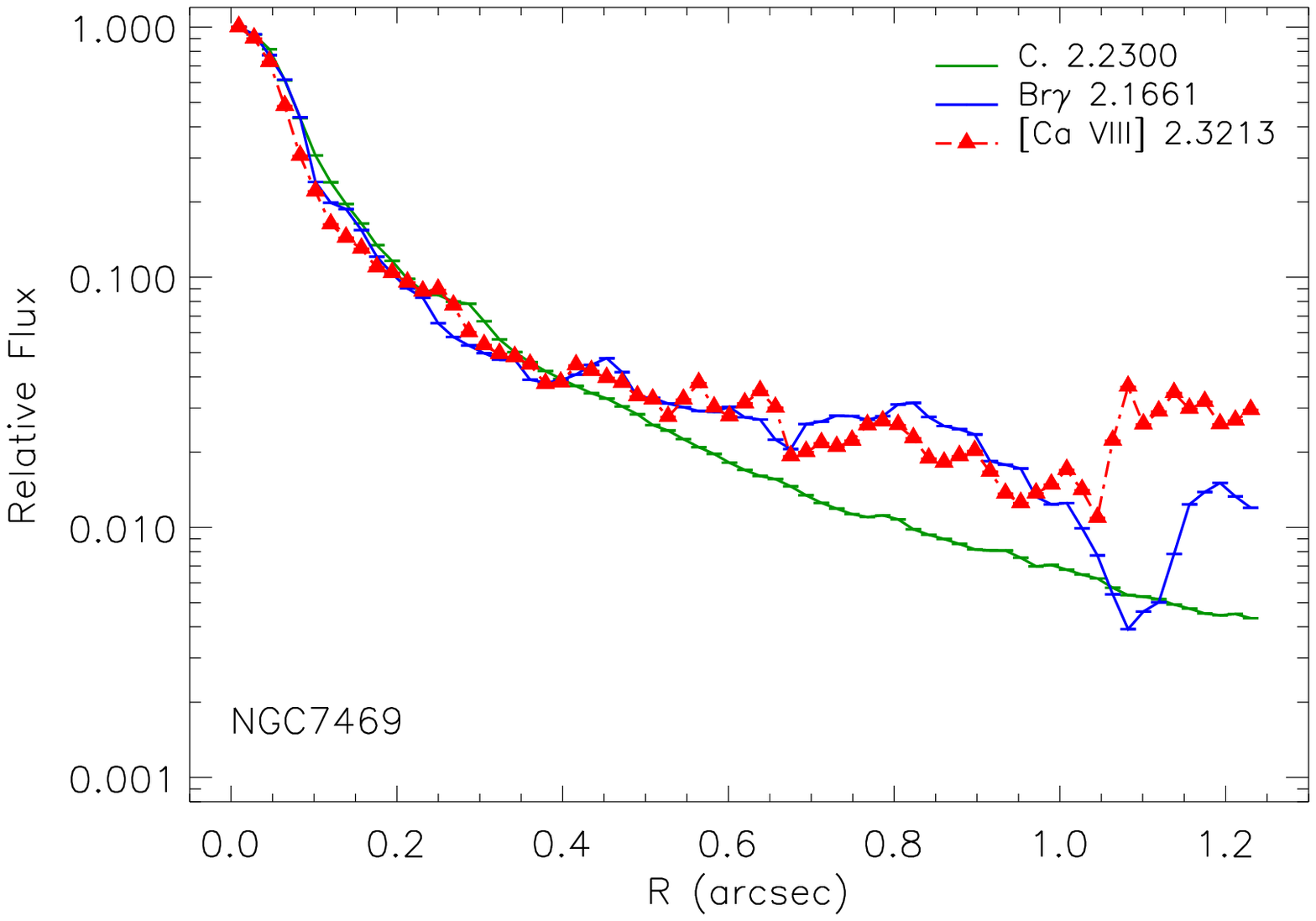}
\caption[Coronal Line Radially Averaged Flux Distributions]{Radial averages of the 2-D coronal line  flux distributions for each galaxy as labeled in the lower left corner of each plot.  The curves are as indicated by the legend and the continuum (C. 2.2300) is measured at 2.23-2.24 \micns.   \label{corflux}}
\end{figure*} 

\tabletypesize{\scriptsize}
\begin{deluxetable*}{lccccc}
\tablecaption{2D Gaussian Fits to Continuum Flux Distribution \label{t_contgfits}} 
\tablewidth{0pt}
\tablehead{
\colhead{Seyfert 1} & 
\colhead{Window \tablenotemark{a}} &
\colhead{FWHM (\as)} &
\colhead{Axis Ratio} & 
\colhead{PA (\deg)} &
\colhead{Method \tablenotemark{b}} \\ 
}
\startdata

NGC 3227	&	B	&	0.052$\pm$0.005	&	0.64$\pm$0.09	&	101$\pm$9	&	2-D	\\
		&	M 	&	0.049$\pm$0.005	&	0.64$\pm$0.09	&	102$\pm$9	&	2-D	\\
		&	R 	&	0.044$\pm$0.006	&	0.65$\pm$0.13	&	102$\pm$13	&	2-D	\\
NGC 3516	&	B	&	0.346$\pm$0.075	&	\nodata	&	\nodata	&	2-D, C	\\
		&	M 	&	0.312$\pm$0.075	&	\nodata	&	\nodata	&	2-D, C	\\
		&	R 	&	0.234$\pm$0.098	&	\nodata	&	\nodata	&	2-D, C	\\
NGC 4051	&	B	&	0.190$\pm$0.015	&	\nodata	&	\nodata	&	2-D, C	\\
		&	M 	&	0.180$\pm$0.015	&	\nodata	&	\nodata	&	2-D, C	\\
		&	R 	&	0.160$\pm$0.021	&	\nodata	&	\nodata	&	2-D, C	\\
NGC 4151	&	B	&	0.049$\pm$0.001	&	\nodata	&	\nodata	&	2-D, C	\\
		&	M 	&	0.049$\pm$0.001	&	\nodata	&	\nodata	&	2-D, C	\\
		&	R 	&	0.048$\pm$0.001	&	\nodata	&	\nodata	&	2-D, C	\\
NGC 4593\tablenotemark{c}	&	B	&	0.633$\pm$0.212\tablenotemark{c}	&	\nodata	&	\nodata	&	S	\\
		&	M 	&	0.631$\pm$0.218\tablenotemark{c}	&	\nodata	&	\nodata	&	S	\\
		&	R 	&	0.630$\pm$0.235\tablenotemark{c}	&	\nodata	&	\nodata	&	S	\\
NGC 5548	&	B	&	0.343$\pm$0.127	&	\nodata	&	\nodata	&	2-D, C	\\
		&	M 	&	0.329$\pm$0.150	&	\nodata	&	\nodata	&	2-D, C	\\
NGC 6814	&	B	&	0.395$\pm$0.233	&	\nodata	&	\nodata	&	2-D, C	\\
		&	M 	&	0.348$\pm$0.212	&	\nodata	&	\nodata	&	2-D, C	\\
		&	R	&	0.313$\pm$0.313	&	\nodata	&	\nodata	&	2-D, C	\\
NGC 7469	&	B	&	0.094$\pm$0.006	&	0.96$\pm$0.09	&	90$\pm$15	&	2-D	\\
		&	M 	&	0.090$\pm$0.007	&	0.95$\pm$0.11	&	90$\pm$17	&	2-D	\\
		&	R 	&	0.085$\pm$0.013	&	0.97$\pm$0.29	&	90$\pm$15	&	2-D	\\
Ark 120	        &	B	&	0.159$\pm$0.016	&	0.87$\pm$0.12	&	88$\pm$26	&	2-D	\\
		&	M 	&	0.153$\pm$0.873	&	0.87$\pm$0.16	&	88$\pm$34	&	2-D	\\

\enddata
\tablenotetext{a}{The three continuum windows are defined as follows: (B) `blue' window from 2.1400-2.1500 \micns, (M) `middle' window from 2.2300-2.2400 \micns, and (R) `red' window from 2.448-2.453 \micns.} 
\tablenotetext{b}{Method of fitting the flux distribution: `2-D' is fitting the full 2-D map, `2-D, C' is fitting the full 2-D map with circular symmetry, `S' is the FWHM determined from a single slit observation.}
\tablenotetext{c}{Neither of the two slits obtained for NGC 4593 go through the nucleus.  The flux distribution reported here is from a slit positioned 0\as.09 from the nucleus, and because of the exponential PSF wings, the widths given are much higher than the widths expected if measured at the nucleus.}

\end{deluxetable*}

\begin{figure*}[!ht]	
\epsscale{0.3}
\plotone{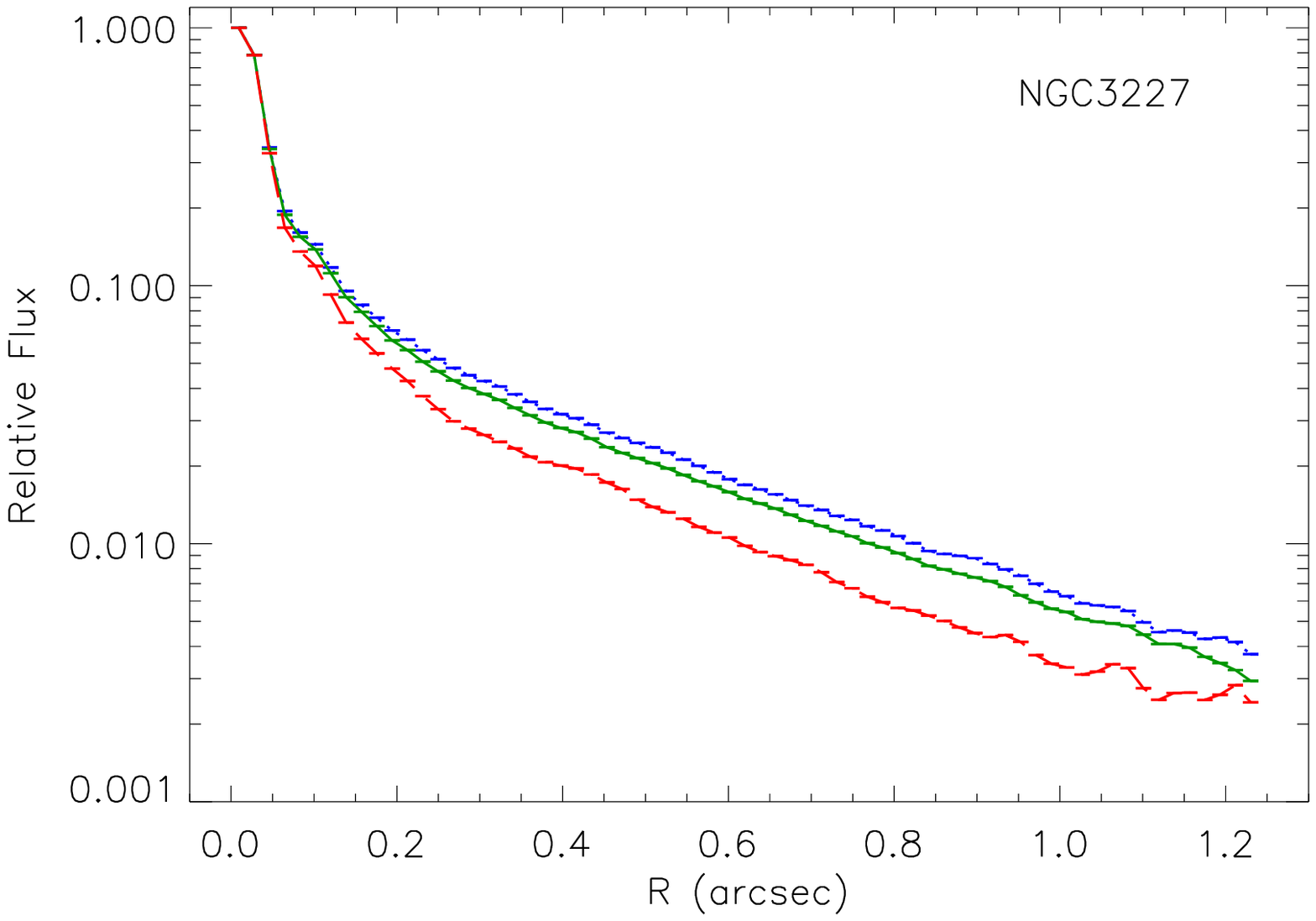}
\plotone{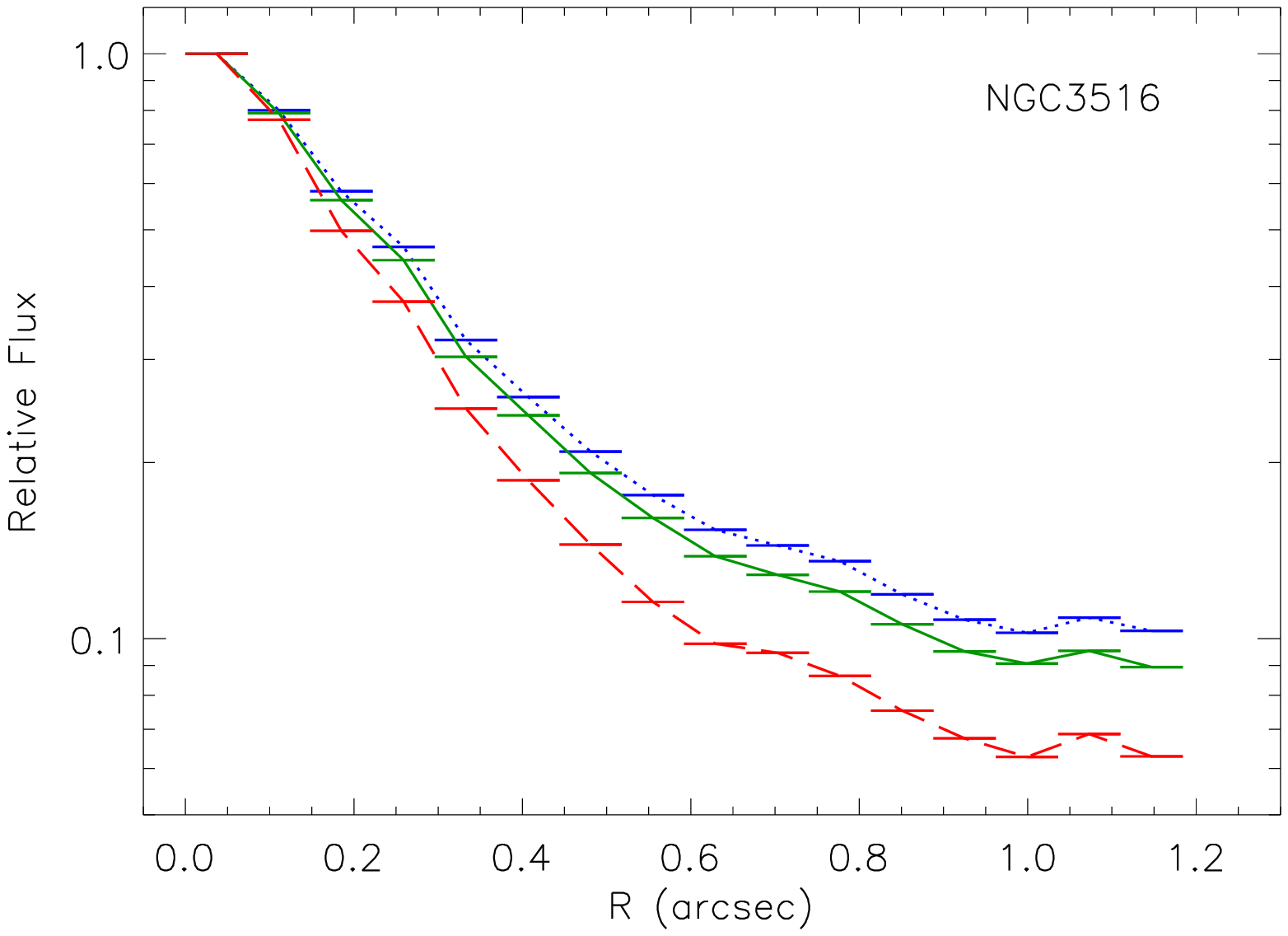}
\plotone{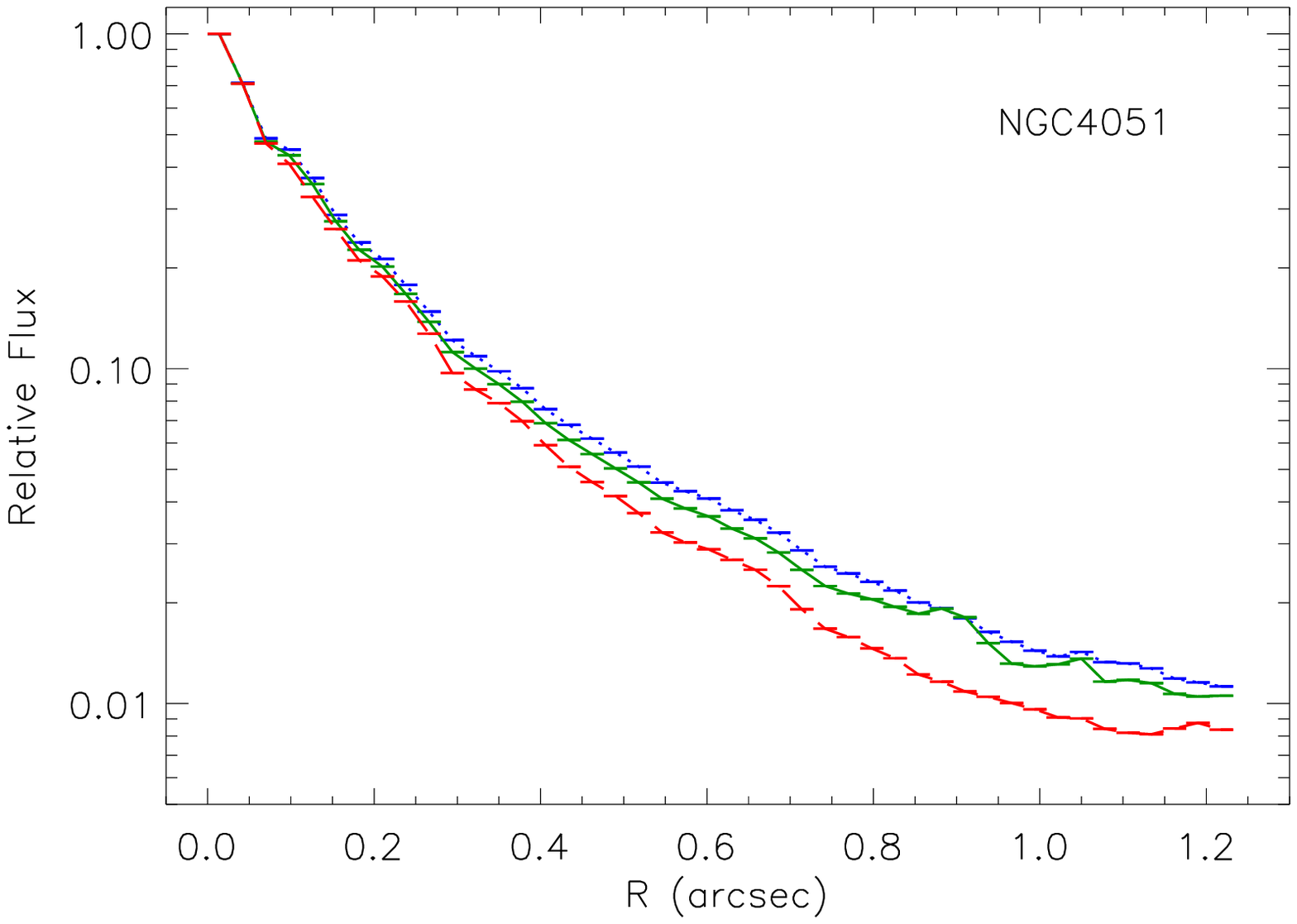}
\plotone{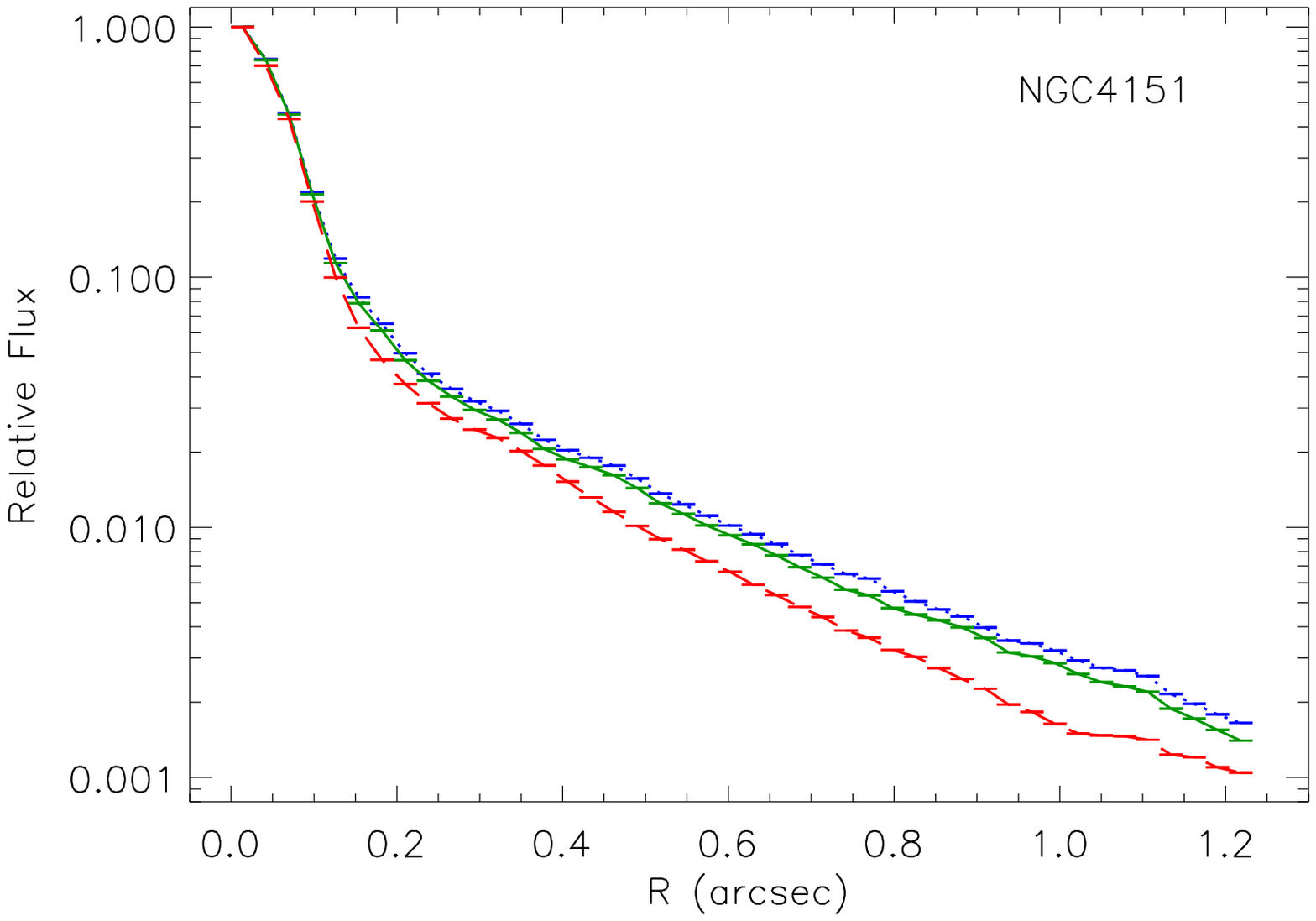}
\plotone{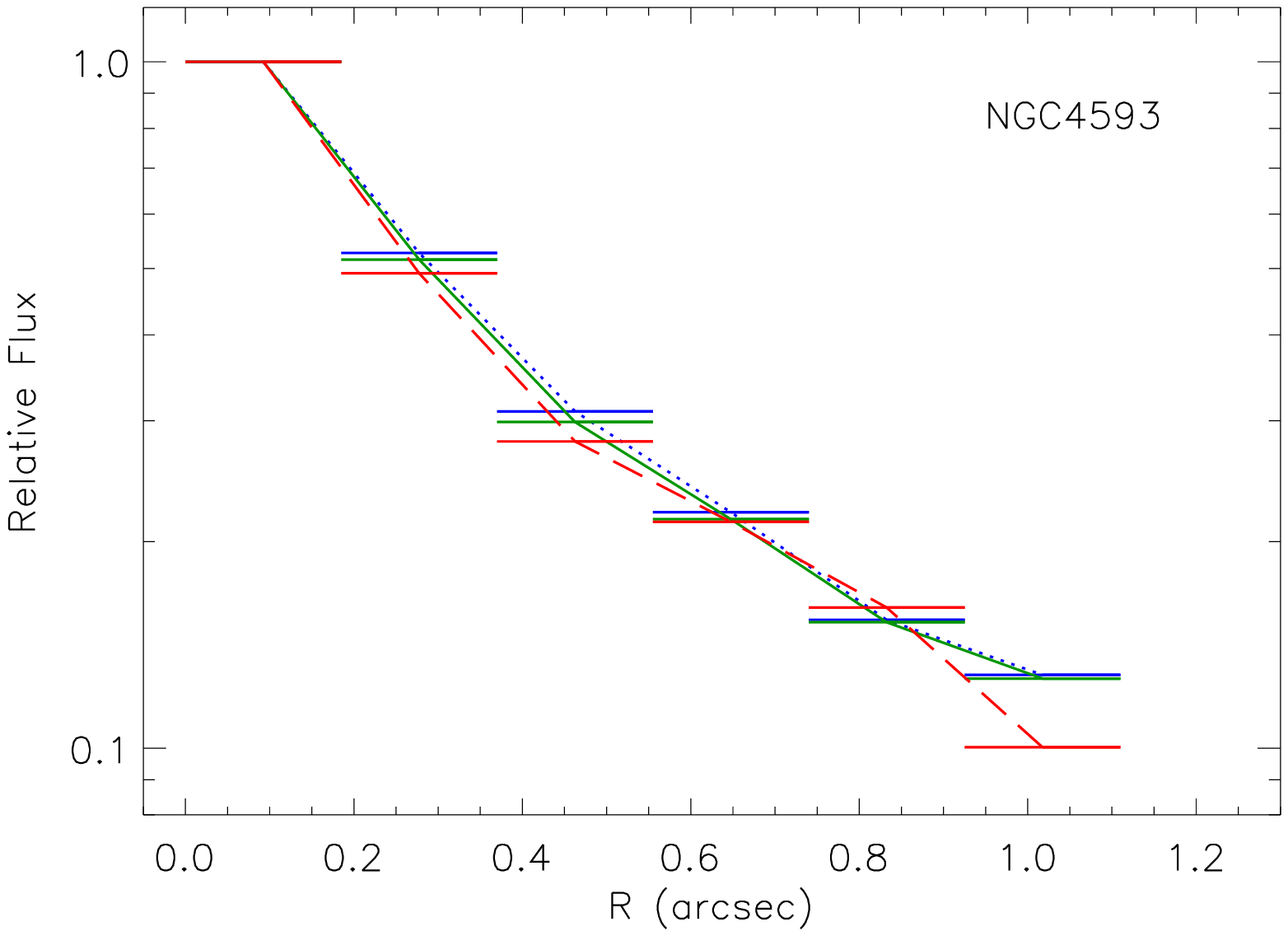}
\plotone{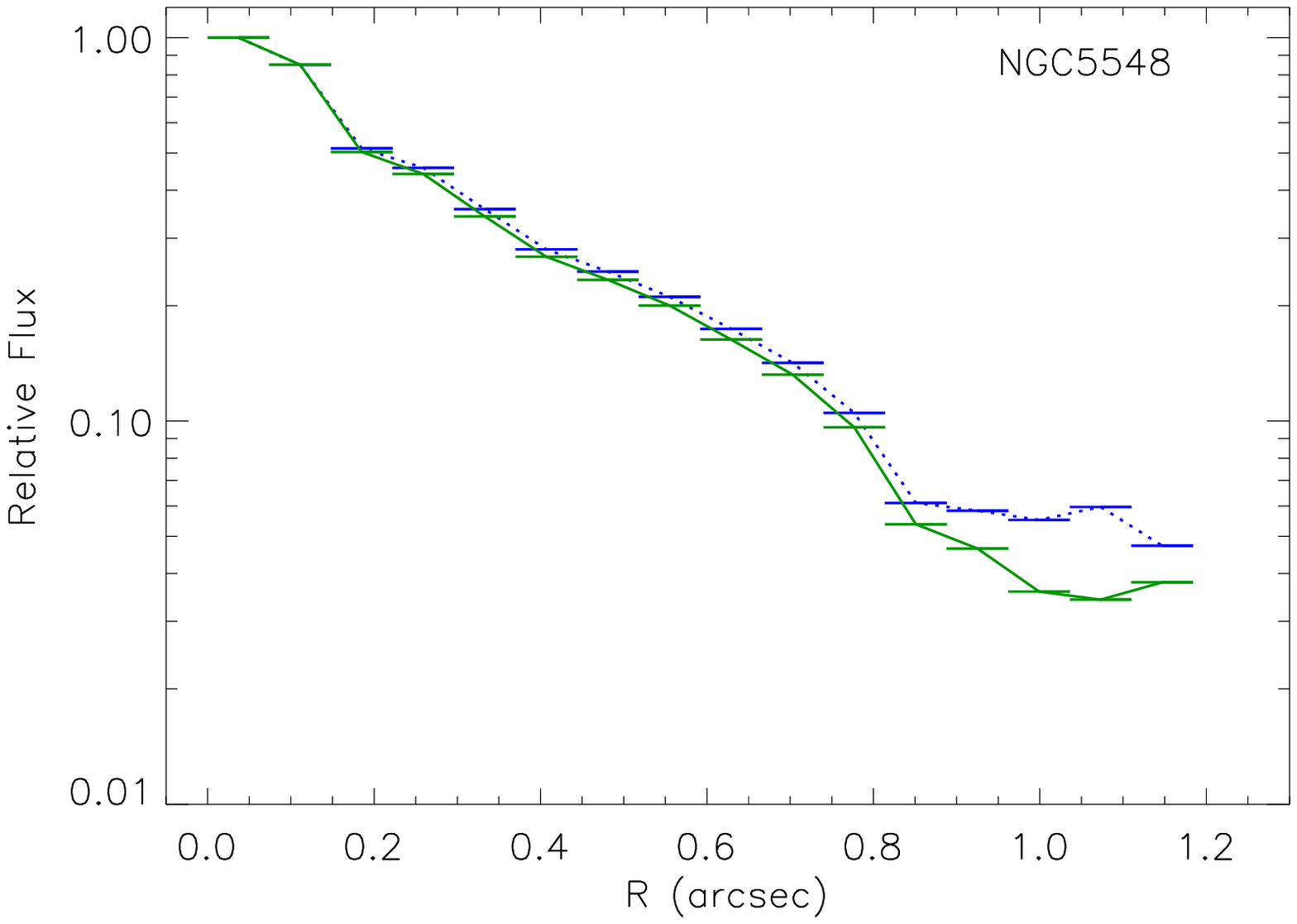}
\plotone{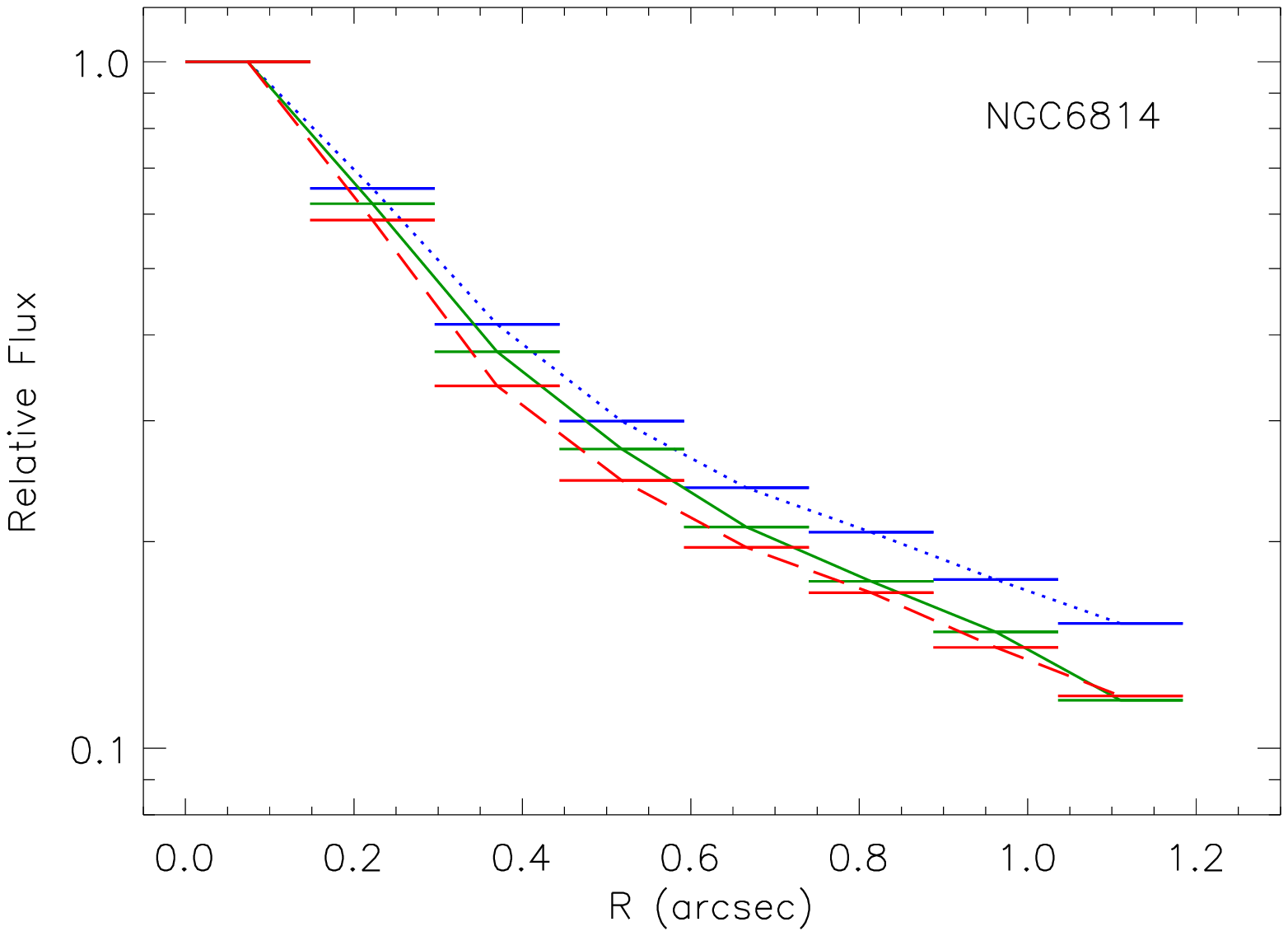}
\plotone{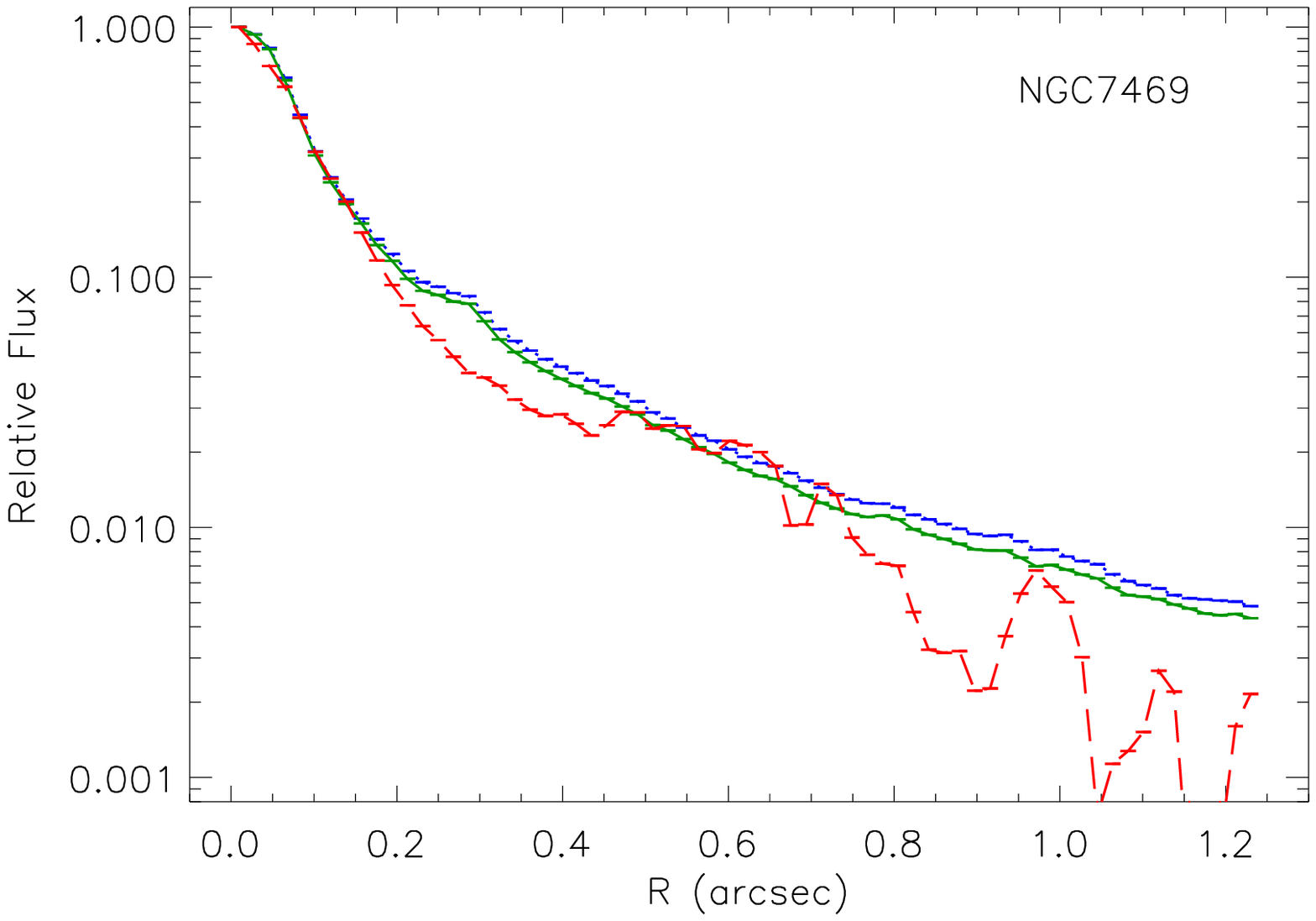}
\plotone{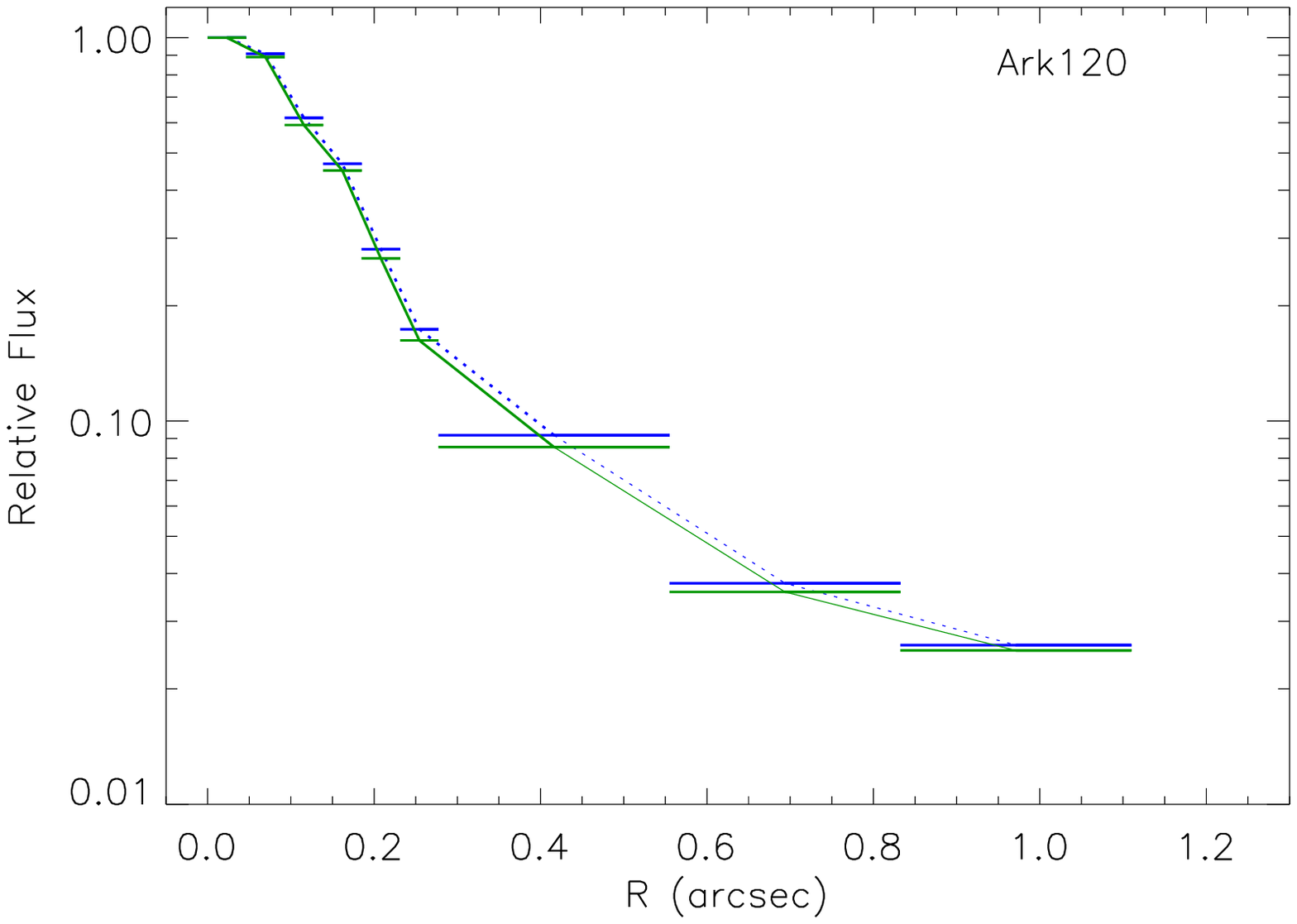}
\caption[Continuum Flux Distribution of All Seyfert 1s]{ Continuum radially averaged flux distribution in three wavelength windows: 2.14-2.15 (top dotted curve), 2.23-2.24 (middle solid curve), and 2.448-2.533 \mic (bottom dashed curve).   The distribution is shown for each of the nine Seyfert 1 galaxies, as labeled in the upper right corner.  For two of the galaxies the obtained spectra did not extent to the reddest window.  See the note on NGC 4593 in Table \ref{t_contgfits}.  \label{contflux}}
\end{figure*}

\clearpage

Fits to the flux distributions of the emission-line gas were performed using a 2-D Gaussian function, the results of which are presented in Tables \ref{t_fluxgfits} and \ref{t_cordist} for hydrogen and coronal emission, respectively.  For some galaxies, particularly those with less of the 2-D FOV measured, the 2-D Gaussian fits were done by restricting the Gaussian to be circular so the fit would not be influenced by uneven spatial coverage.  For other galaxies the flux distribution of emission, especially that of \htwons, is clumpy, and in these cases multiple 2-D Gaussians were fit to the flux distribution.  The extent of the coronal emission given in Table \ref{t_cordist} is defined as the radius at which the brightness drops to 5\% of the peak value.  For comparison to the flux distributions of the emission-line gas, the results of the same 2-D Gaussian fits of the \kb\s continuum flux distribution are given in Table \ref{t_contgfits}.  The radial average of the flux distributions of molecular hydrogen and narrow \bgns, the coronal emission, and the three continuum windows are shown in Figures \ref{radflux}, \ref{corflux}, and \ref{contflux}, respectively.  

As can be seen in Fig. \ref{contflux} and Table \ref{t_contgfits}, the continuum flux distribution in many of the galaxies becomes more compact as the wavelength increases.  Based on measurements of standard stars, when the PSF FWHM is less than 0\as.1, instrumental effects can be ruled out as the cause of this wavelength dependence.  The most likely explanation for the intrinsic narrowing of the \kb\s continuum is a change in the fraction of the continuum from the stellar population and the point-like reradiated AGN emission, and the observed wavelength dependence can then be used to estimate the relative contributions.  This is tested for the two galaxies for which the data clearly fall within the 0\as.1 PSF FWHM regime by using AGN plus stellar S\`{e}rsic fits to NICMOS {\em H}-band images (see $\S$ 5.1.2).  The {\em H}- to \kb\s conversion is taken to be 0.25, 0.30, and 0.40 magnitudes for the stellar light (\citealt{hunt99}, \citealt{tokunaga00}) and 0.77, 0.90, and 1.15 magnitudes for the AGN emission (assumes a power law flux density with a slope of unity) for the {\em H-K$_{[2.14]}$}, {\em H-K$_{[2.23]}$}, and {\em H-K$_{[2.44]}$} continuum windows, respectively.  In NGC 3227 the AGN contribution to the {\em H}-band continuum is estimated to be 20\% within 1\as.  The modeling then predicts that for NGC 3227 the blue window should be 16.2\% wider than the red window, and 5.5\% wider than the middle window, which agrees well with the measured changes of 18\% and 6\%, respectively.  In contrast, for NGC 7469 the {\em H}-band continuum is found to be dominated by AGN emission (71\% AGN, which is consistent with other independent measurements, e.g. \citealt{malkan83}), and the modeling predicts virtually no change in the flux distribution with wavelength.  This prediction also agrees with an observed widening of only 2\% in the red window and no observed change in the middle window compared to the flux distribution in the blue continuum window.  

A discussion of the morphology and kinematics of the nuclear gas in each of the Seyfert 1 galaxies follows.  The properties of the molecular hydrogen, ionized hydrogen, and coronal emission are compared to each other, as well as to measurements reported in the literature, including measurements of \oiii\s and radio emission.  The galaxies are presented in order of the significance of their detected kinematics and the coverage of the 2-D field.  See the electronic edition of the Journal for color versions of the figures, including those from this section. 

\subsubsection{NGC 3227}
\begin{figure}[!b]
\epsscale{1}	
\plotone{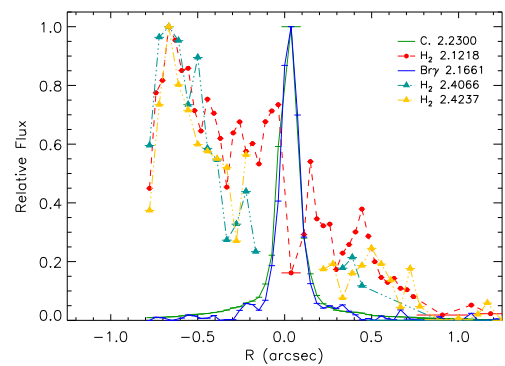}
\caption[NGC 3227 \htwons, \bgns, and \kb\s Flux Distributions]{NGC 3227 flux distributions along a single slit passing through the southeast \htwo emission knot (see Fig. \ref{allmaps}).  Negative radii are to the east.  The curves are as indicated by the legend and the continuum (C. 2.2300) is measured at 2.23-2.24 \micns.  For NGC 3227, \htwo \lam2.4066 and \lam2.4237 are not measured reliably in the inner 0\as.4. \label{3227_flux}}
\end{figure}

Molecular hydrogen is strongly detected in NGC 3227 out to the edge of the 2\as~FOV at \lam2.1218 1-0 S(1), \lam2.4066 1-0 Q(1), and \lam 2.4237 1-0 Q(3).    The flux distribution of all three \htwo lines is spatially asymmetric, with several spatially resolved knots along PA$\sim$135\deg, the brightest of which is centered 0\as.5 southeast of the nucleus.  As can be seen in Fig. \ref{3227_flux}, the \htwo flux is about five times brighter at the peak of the southeast knot compared to its flux at the nucleus.  The radial averages of the flux distributions are shown in Fig. \ref{radflux}.  It is tempting to attribute the \htwo emission structure to an edge-on disk or nuclear bar, but it is important to remember that the flux observed is merely the fraction of the gas that is locally excited, rather than resembling the true molecular gas distribution.  In fact, \htwo is observed at a significant level ($>$10 sigma detections) throughout the full NIRSPEC FOV, confirming that molecular hydrogen is present out to a radius of 2\as\s in all directions.  In addition, the kinematics do not indicate the presence of a nuclear bar.    

\begin{figure}[!t]
\epsscale{0.95}	
\plotone{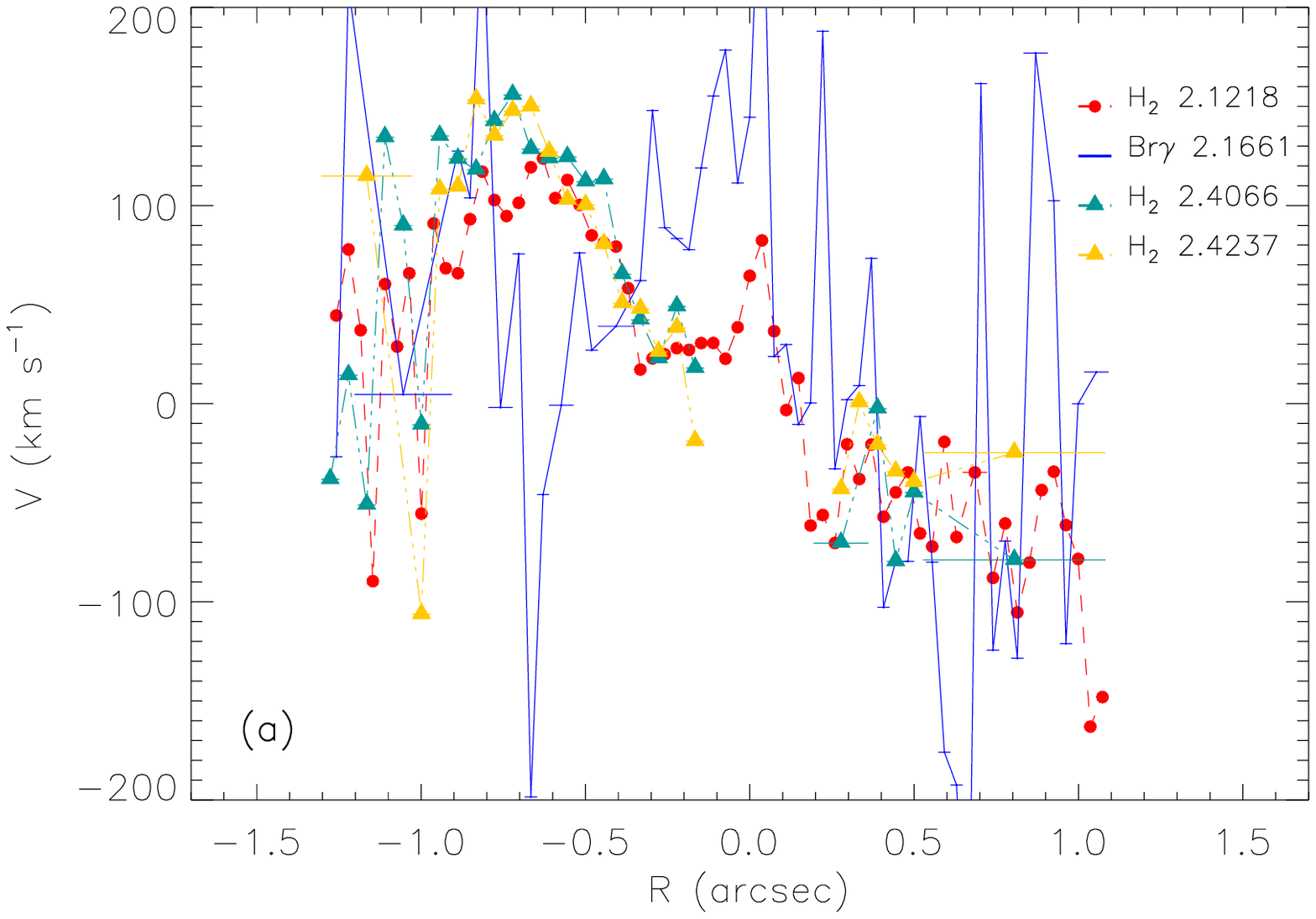}
\plotone{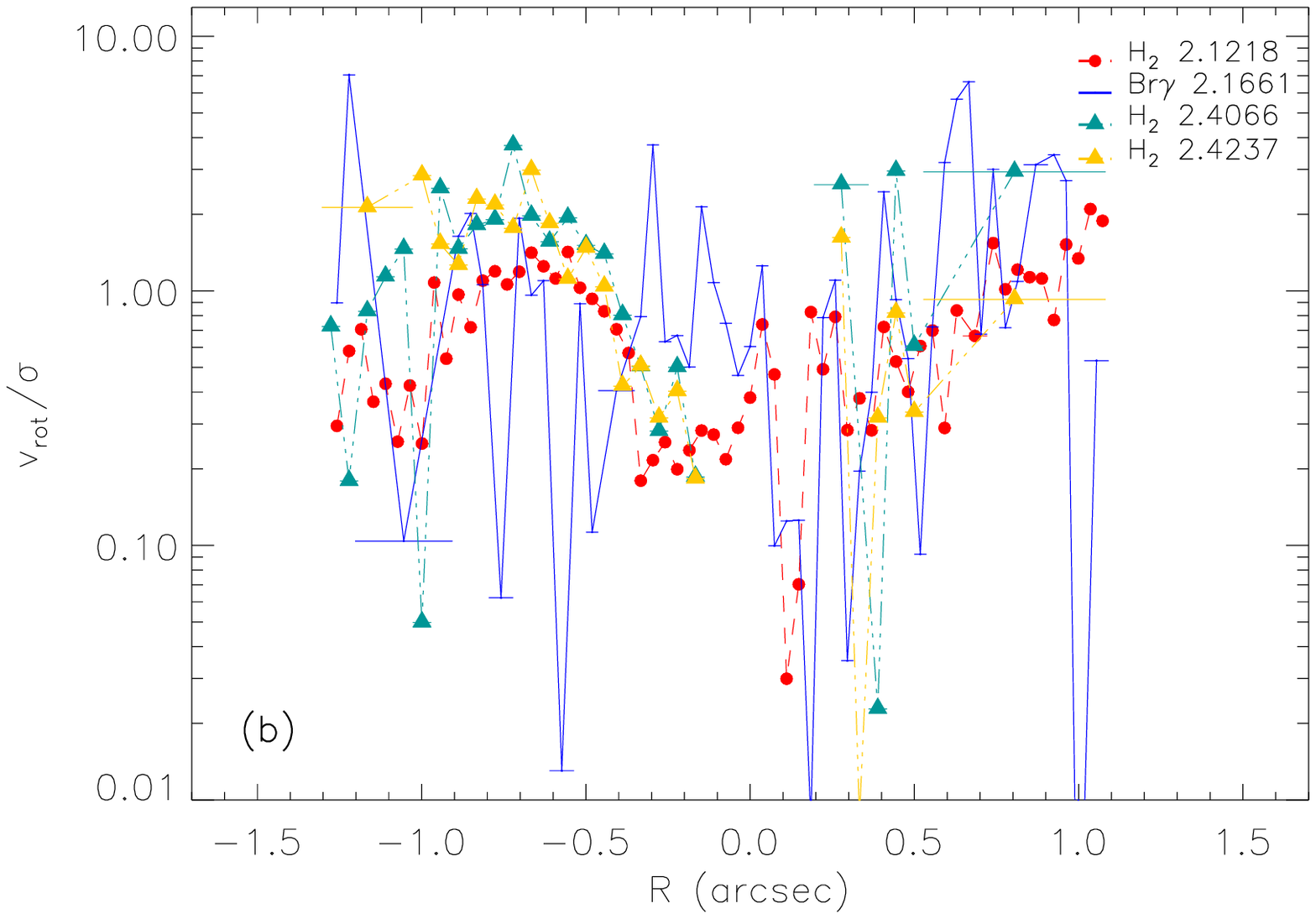}
\caption[NGC 3227 \htwo and \bg Rotation Curves]{NGC 3227 (a) rotation curves and (b)  v$_{rot}$/\sig of all three \htwo transitions and the narrow component of \bg (legend at upper right) along a single slit position offset 0\as.02 from the nucleus going through the southeast \htwo knot (negative radii are to the east).  \label{3227_vel_vs}}
\end{figure}

A consistent velocity field is measured in all three \htwo transitions (Fig. \ref{3227_vel_vs}a), and the field exhibits organized rotation with a steep gradient of 150 \kms across the central 0\as.5 along a major axis PA of about 140\deg.  The velocity dispersion of \htwo \lam2.1218 varies by more than 50 \kmsns, with the highest dispersion of $\sim$150 \kms measured in the central 0\as.5 to the west and north of the nucleus.  The longer wavelength \htwo emission lines have consistent velocity dispersions, but neither is measured reliably in the inner 0\as.4 because of residuals from the atmospheric absorption.  The relatively high velocity dispersion, which is comparable to the observed \htwo velocities, is not necessarily an indication of non-circular motions as it can be explained by a combination of a turbulent medium, such as a thick disk, and blurring of a steep velocity gradient.  The variation of v$_{rot}$/\sig with radius is shown in Fig. \ref{3227_vel_vs}b. 

Unlike the molecular gas, the flux distribution of the narrow component of \bg is centrally located and unresolved (Fig. \ref{3227_flux} and \ref{radflux}).  The velocity field of \bg shows rotation similar to that of \htwo (Fig. \ref{3227_vel_vs}a), with a gradient of $\sim$ 150 \kms across the central 0\as.5 along the same major axis (PA$\sim$140\deg).  In addition, within the central 0.\as5 \bg is blueshifted by about 30 \kms with respect to the molecular hydrogen.  The velocity dispersion increases toward the nucleus, where it peaks at 170 \kmsns.

Neither the \htwo nor \bg flux distributions follow the distribution of the highly ionized \oiii\s \lam 5007\AA\s gas.  Ground based \oiii\s imaging of NGC 3227 shows extended emission out to 7\as\s northeast of the nucleus along PA \sm30\deg\s (\citealt{mundell95}).  HST images of \oiii\s show a compact nucleus and extended emission out to 0\as.9 at a PA of 15\deg\s (\citealt{schmitt96}).  Radio maps presented by \citet{mundell95} reveal a double source separated by 0\as.4 at a PA of -10\deg.  Both \htwo and \bg have extended emission in this region, but neither shows structure at the location of the radio emission.  

Although the flux distributions of \htwo and \bg do not bear any resemblance to the emission seen in \oiii\s or at radio wavelengths, they are each consistent with the lower spatial resolution \htwo and \bg 2-D intensity maps presented by \citet{sosa01}.  The flux distributions are also consistent with that measured by \citet{davies06} at a similar spatial resolution using the integral field spectrometer SINFONI with AO.  Their 0\as.85 FOV shows extended \htwo emission toward both the southeast and the northwest (PA $\sim$ 135\deg).  The NIRSPEC data also show this structure, and, with the wider FOV, show that at least the southeast component (there is a gap in the coverage in the northwest) continues to brighten out to 0\as.5 with bright extended emission reaching out to 1\as.  

\begin{figure}[!t]	
\epsscale{0.95}
\plotone{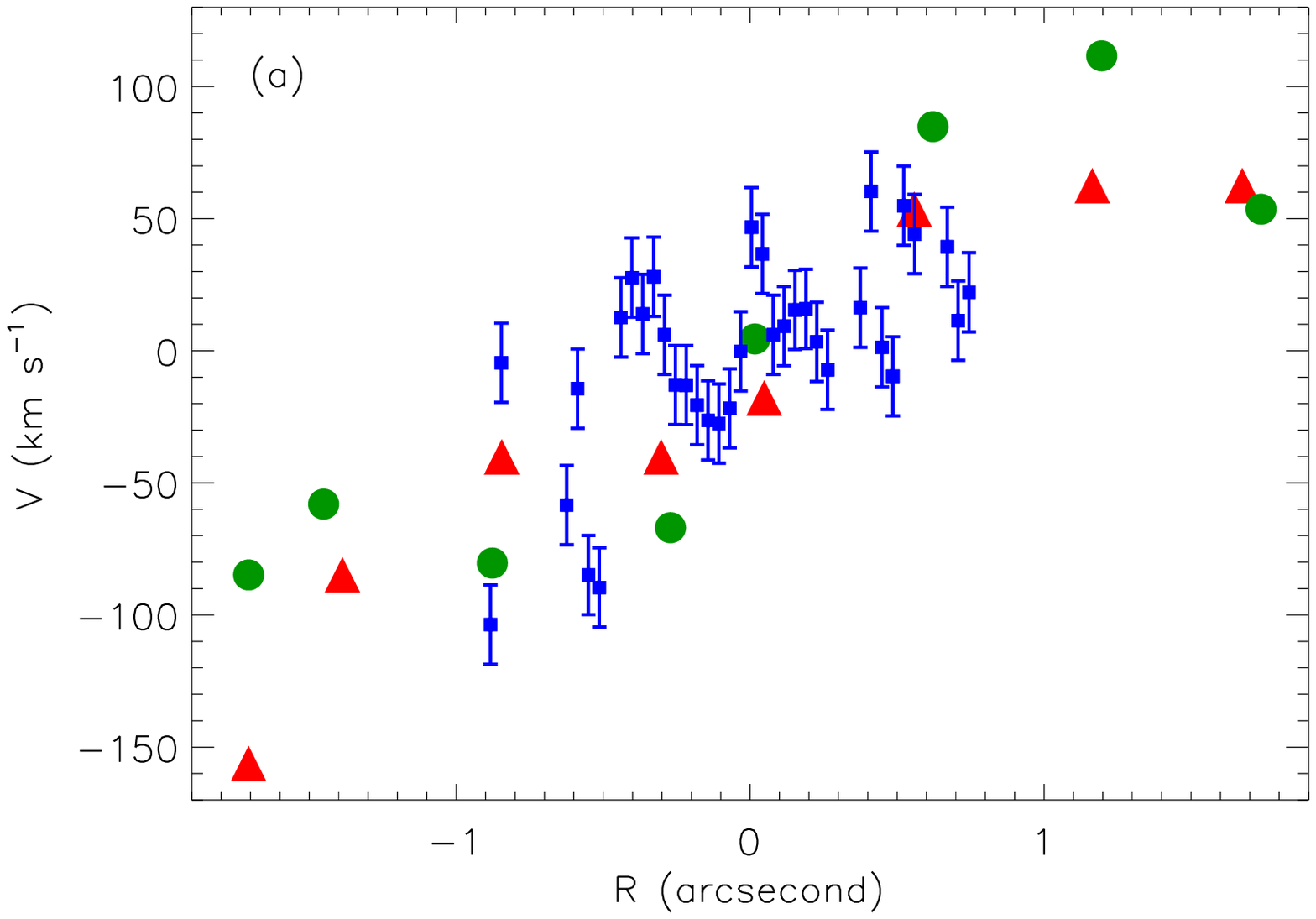}
\plotone{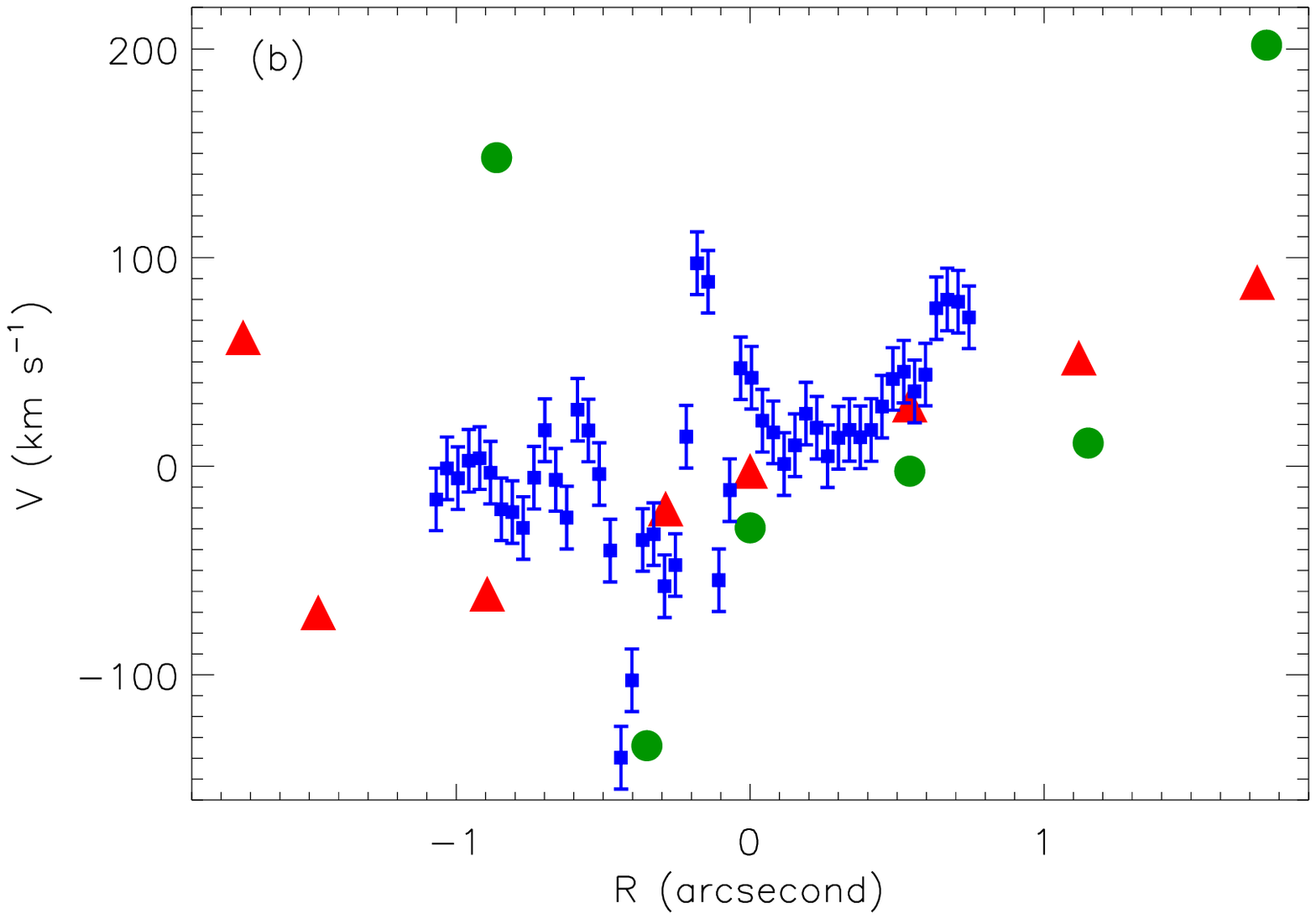}
\caption[NGC 3227 \htwo \lam 2.1218 Rotation Curve Comparison]{ NGC 3227 \htwo \lam 2.1218 rotation curves (squares) compared to those obtained by \citet{reunanen02} for \htwo \lam 2.1218 (triangles) and [Fe II] (circles) along (a) PA=15\deg\s and (b) PA=105\deg, which are parallel and perpendicular, respectively, to the \oiii\s ionization cones. \label{r3227rc}}
\end{figure}

The nuclear region of NGC 3227 has been the focus of many previous studies, which have measured the kinematics of the ionized and highly ionized gas, as well as molecular gas, at various spatial resolutions.  The nuclear kinematics measured with NIRSPEC are consistent with the large scale rotation seen in \oiii, H$\alpha$, [N II], and [S II] \lam6716\AA\s at a spatial resolution of \sm1\as\s (\citealt{gl01,arribas94}).  The \htwo velocity field is also consistent with the \htwo kinematics measured by \citet{reunanen02} on a 1\as\s scale.  Fig. \ref{r3227rc} is a comparison of cuts through our 2-D velocity field to the data along their two slit positions

\begin{figure}[!t] 	
\epsscale{1}	
\plotone{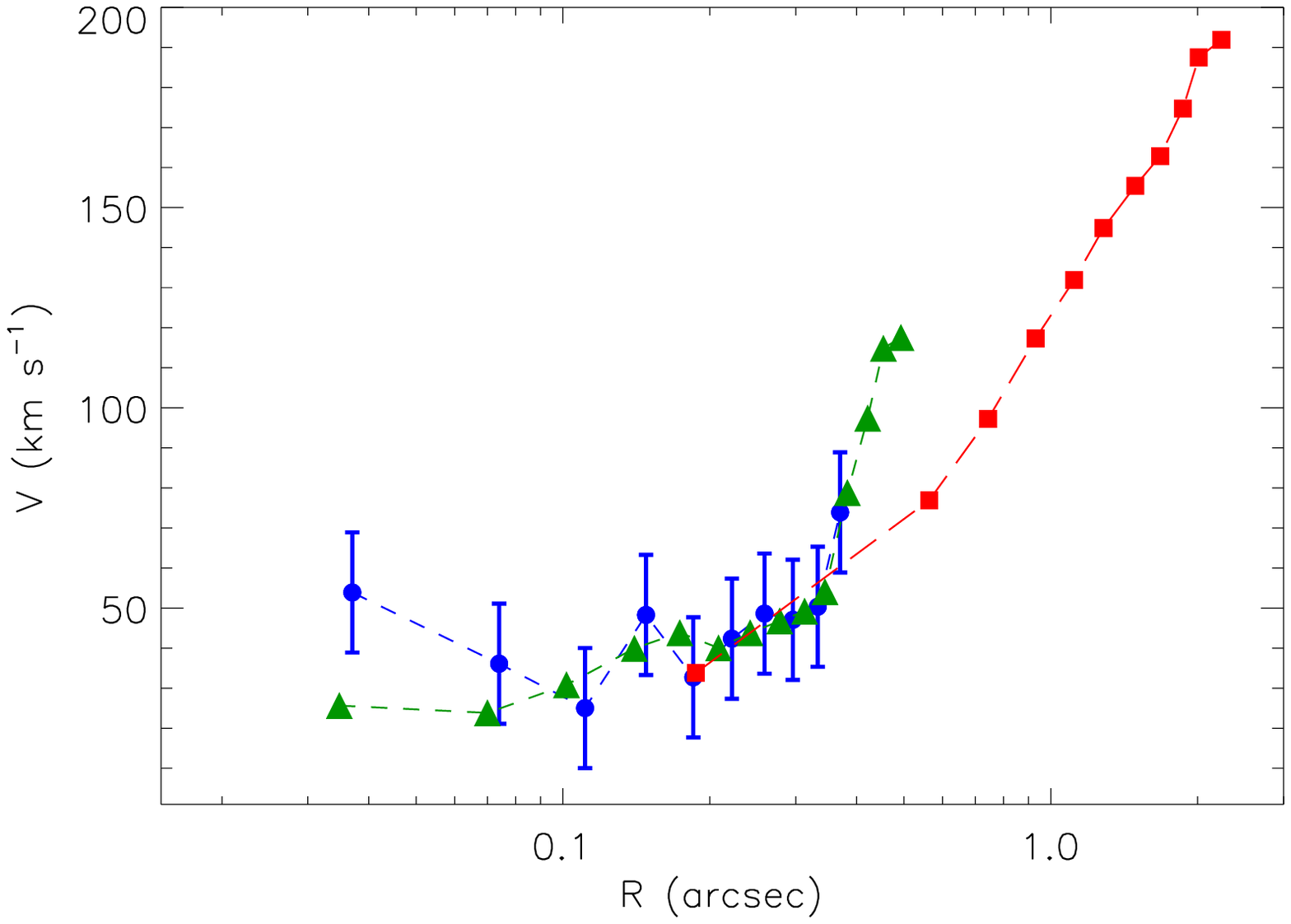}
\caption[NGC 3227 \htwo \lam 2.1218 Rotation Curve Comparison]{ NGC 3227 \htwo \lam 2.1218 rotation curve, along PA=140\deg\s (very close to the major axis of rotation), with the two sides averaged (circles), compared to that measured by \citet{davies06} (triangles) and the rotation of CO detected in emission by \citet{schinnerer00} (squares).  The discrepancy between the rotation of \htwo and CO at R$>$0.\as3 is partially, if not completely, explained by averaging of the velocity field by the larger beam used for the CO observations. \label{d3227rc}}
\end{figure}

Our 2-D \htwo \lam 2.1218 and \bg velocity fields are also consistent with the 2-D maps from the \citet{davies06} study mentioned above (Fig. \ref{d3227rc}).  In addition, the 2-D maps of the \htwo \lam 2.1218 velocity dispersion are also consistent, showing peaks in dispersion \sm0\as.25 west and north of the nucleus.  \citet{davies06} also measured the stellar kinematics in the nuclear region of NGC 3227 (the longer integration times afforded by the integral field spectrometer provided higher quality CO bandhead absorption compared to what was possible with our relatively short exposure single slit spectra).  They find that the stellar kinematics follow a very similar pattern to that seen with \htwons.  Finally, the kinematics of the cold molecular gas, measured in CO and HCN emission (\citealt{schinnerer00}), are consistent with the kinematics seen in the hotter \htwo molecular gas, as is shown in Fig. \ref{d3227rc}.

Also detected in the spectra of NGC 3227 are both [Ca VIII] and [Si VII].  The 2-D Gaussian fits to the flux distributions of both coronal lines are consistent with the distributions found for \bg and the \kb\s continuum (Fig. \ref{corflux}).  Similarly, the measured extended coronal line emission is within the same region as the extended \bg emission, which extends out to 40-50 pc (around 0\as.55; measured at 5\% of the peak flux).  The extensions of \bg emission toward the south and west are also seen in [Ca VIII] and [Si VII].  Also like \bgns, as well as \htwons, neither coronal line has a flux distribution similar to that of \oiii\s or the radio emission, which are reported to be extended along position angles of 30\deg\s and 10\deg, respectively (\citealt{mundell95}, \citealt{schmitt96}).

[Ca VIII] is found to be blueshifted by about 100 \kms compared to the systemic velocity.  The velocity dispersion is around 80 \kms throughout most of the nuclear region, but is closer to 100 \kms in the central 0\as.2, as well as to the northeast and southwest.  The kinematics of [Si VII] is generally consistent with that measured for [Ca VIII], but not as strongly blueshifted.  However, as discussed, measurements of this emission line are uncertain due to the strong atmospheric absorption.  In addition, for NGC 3227 the [Si VII] emission line is at the edge of acceptably measured region of the spectra, and therefore measurements of the line are even more uncertain.  The velocity field measured for [Si VII] shows a region of redshifted gas in the south and a region of blueshifted gas to the north.  Similar to [Ca VIII], the velocity dispersion is not uniform, with the southeast having a dispersion of around 60 \kms and the northwest at 60-100 \kmsns.  The general trend for lower velocity dispersion in the southeast is also seen in \htwo \lam2.1218, however the coronal lines have dispersions several tens of \kms lower than \htwons, and over 50 \kms lower than the \bg velocity dispersion.  

In addition to the emission lines detected in NGC 3227, all three CO bandheads ($^{12}$CO(2,0) \lam 2.2935, $^{12}$CO(3,1) \lam 2.3226, and $^{13}$CO(2,0) \lam 2.3448) are also present in the composite 1\as.4 aperture spectrum.  The velocity dispersion measured in these stellar absorption features is consistent with that measured in the shorter wavelength Ca II triplet feature (\lam\lam 8498, 8542, and 8662 \AA; \citealt{onken04}).  This implies that despite the spectral features being produced by stars of different spectral classifications, and potentially different ages, the stellar populations have similar kinematics.  This conclusion is further supported by a similar agreement between the velocity dispersion of the 2.29 \mic CO bandhead and the Ca II triplet feature in the composite spectra of both NGC 7469 and NGC 4051.

In summary, the flux distributions of \htwo and \bg in NGC 3227 are consistent with the distributions reported by other authors, both at lower and similar spatial resolutions, but neither shows the same structure as that seen in \oiii\s or radio emission.  However, the \htwo and \bg kinematics are consistent with what is seen at larger scales in ionized and highly ionized gas, as well as with the stellar dynamics.  The coronal emission in NGC 3227 is spatially consistent with the distribution of the \kb\s continuum and the \bg emission and the kinematics of the coronal emission is consistent with the blueshift seen in \bgns.  Furthermore, the coronal emission does not show any emission structures coincident with the reported \oiii\s or radio emission, as has been suggested to be the preferential distribution for coronal emission in Seyfert galaxies (\citealt{reunanen03}; \citealt{prieto06}), suggesting that coronal emission is excited by something other than shocks associated with a nuclear jet.  Results from modeling of the 2-D molecular hydrogen velocity field with a simple coplanar disk are presented in $\S$ 6.1.

\subsubsection{NGC 7469}
NGC 7469 is well known for its circumnuclear ring 1\as.5-2\as.5 from the Seyfert 1 nucleus.  This ring of recent star formation has been detected at many wavelengths, including radio (\citealt{wilson91,condon91,colina01}), optical (\citealt{mauder94}), and mid-infrared (\citealt{miles94,soifer03}), as well as in the near-infrared (\citealt{genzel95,scoville00}).  Within the NIRSPEC  2\as\s FOV, which falls mostly inside the ring, molecular, ionized, and highly ionized emission line gas are detected.

Molecular hydrogen is detected in both \htwo \lam2.1218 and \htwo \lam1.9576 and the centrally concentrated flux distribution is spatially resolved with a FWHM of 0\as.45 (Fig. \ref{radflux}).  The \htwo 1-0 S(1) velocity field is well organized with a gradient of 150 \kms across the central 0\as.25 with a major axis PA of around 130\deg.  The \htwo \lam2.1218 velocity dispersion is 75-100 \kms at 0\as.5-1\as.0 from the nucleus and decreases to \sm60 \kms in the inner 0\as.5.  In contrast, the \htwo \lam1.9576 velocity dispersion is relatively constant across the FOV at 115 \kmsns

\begin{figure}[!t] 	
\epsscale{1}	
\plotone{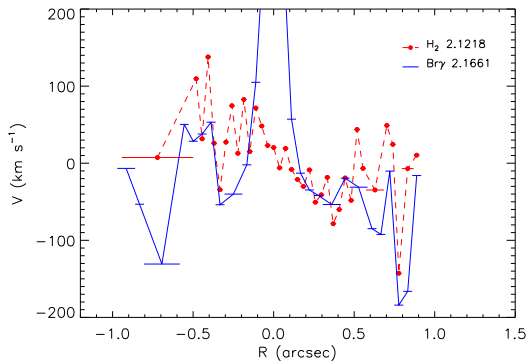}
\caption[NGC 7469 \htwo and \bg Rotation Curves]{NGC 7469 rotation curves of \htwo and the narrow component of \bg (legend at upper right) along a single slit position going through the nucleus at PA=90\deg.  Negative radii are to the west.  \label{7469_vel}}
\end{figure} 

The narrow component of \bg is also measured in NGC 7469 and is found to have an unresolved flux distribution consistent with that of the \kb\s continuum (Fig. \ref{radflux}).  The velocity field is consistent with that measured in \htwo (Fig. \ref{7469_vel}), except in the inner 0\as.2, where it is redshifted with respect to \htwo kinematics. Also, the \bg velocity dispersion rises steeply in the central 0\as.5 to above 150 \kmsns. 

The flux distribution of \htwo \lam2.1218 is consistent with that observed by \citet{davies04}, who obtained two slit positions using NIRSPEC and AO in a similar configuration.  In addition, our 2-D flux distributions for \htwo and \bg are consistent with the 2-D images of Genzel et al. (1995; 0\as.4 spatial resolution) and Sosa-Brito et al. (2001; 0\as.5 spatial resolution).  In addition to the diffuse radio emission associated with the nuclear starburst ring, an east-west extended nuclear source is also detected (\citealt{thean01}), which, at high spatial resolution, separates into five sources along an east-west direction within 0\as.1 of the nucleus (\citealt{lonsdale03}).  There is no indication in the \htwo or \bg flux distribution of a counterpart to this radio emission.  

\begin{figure}[!t] 	
\epsscale{1}	
\plotone{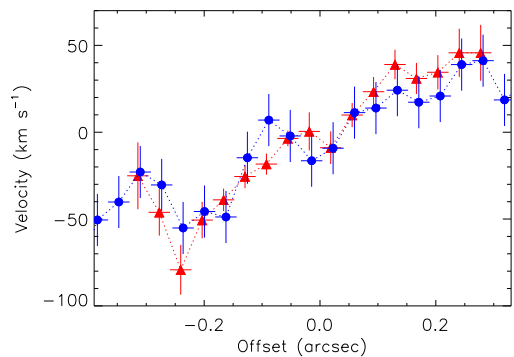}
\caption[NGC 7469 \htwo \lam 2.1218 Rotation Curve Comparison]{NGC 7469 \htwo \lam 2.1218 rotation curve created from a cut through the 2-D velocity map (circles) compared to the rotation curve measured by \citet{davies04} at PA=100\deg\s (triangles).  Negative radii are to the east.  \label{davies7469rc}}
\end{figure}

The kinematics of the nuclear region of NGC 7469 has previously been measured by \citet{davies04} with IRAM interferometric CO 2-1 and HCN 1-0 observations at \sm1\as\s resolution. The \htwo kinematics measured are consistent with this larger scale rotation, with a major axis at a PA of near 130\deg.  In addition, as mentioned, the same authors obtained near infrared spectroscopy at two slit positions also using NIRSPEC.  Although we did not get data at identical slit positions, we can take a cut through our constructed 2-D field to compare the rotation curves measured.  As shown in Fig. \ref{davies7469rc}, the same small scale wiggles in the rotation curves are seen in both sets of data, indicating that these are real changes in the velocity field that are superimposed on the general organized rotation.

The [Ca VIII] coronal emission distribution in NGC 7469 is found to be consistent with the \kb\s continuum and \bg emission, with an unresolved core and extended emission out to 0\as.4 (Fig. \ref{corflux}).  The distribution does not show any extension along the east-west direction, as is seen in the radio emission (\citealt{lonsdale03}).  The [Ca VIII] velocity field shows a continuous increase in velocity toward the center from 30 to a redshift of 120 \kmsns, similar to the redshift seen in \bgns.  The velocity dispersion, which is low compared to that of \bgns, is around 35 \kms throughout the nuclear region.

The smooth flux distribution of both \htwo and \bgns, as well as their regular rotation and relatively low velocity dispersions, suggests that the nuclear kinematics of NGC 7469 are less complex than that of some other Seyfert 1 galaxies (e.g. NGC 4151).  Modeling of the \htwo \lam2.1218 nuclear gas kinematics is presented in $\S$ 6.2. 

\subsubsection{NGC 4151}
Both \htwo \lam2.1218 and \bg are strongly detected in NGC 4151.  In addition, NGC 4151 has the strongest coronal lines, [Ca VIII] and [Si VII], of all Seyfert 1s in the sample.  Despite the complexity seen in the flux distributions of \htwo and \bgns, the molecular hydrogen velocity field indicates there is a dominant component in rotation. 

\begin{figure}[!t]	
\epsscale{1}
\plotone{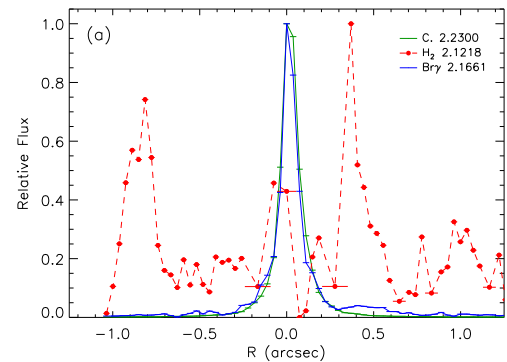}
\plotone{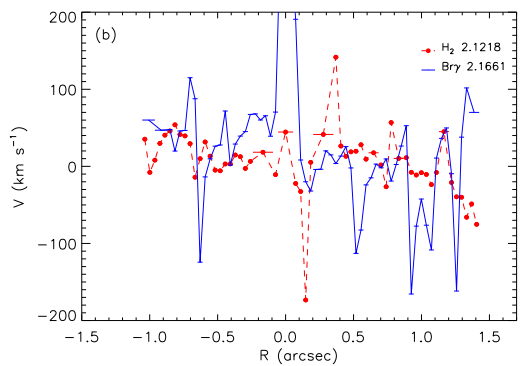}
\caption[NGC 4151 Flux Distributions \& Rotation Curve]{NGC 4151 (a) flux distributions and (b) rotation curves along a single slit position located close to the nucleus at PA=90\deg\s (negative radii are to the east).  Curves are as given by the legends in the upper right. \label{4151_flux_rc}} 
\end{figure} 

\bg has an unresolved core coincident with the \kb\s continuum peak (Fig. \ref{radflux} and \ref{4151_flux_rc}), as well as extended emission out to 1\as.  The \htwo gas, on the other hand, has only weak emission in the central 0\as.5, and a patchy distribution beyond this radius.  This is most evident in the data obtained under the best seeing conditions (21 Apr 03), in which several knots along a PA of 90\deg\s are detected (Fig. \ref{allmaps}).  Several of the \htwo knots can be seen in the flux distribution along a single slit position, see Fig. \ref{4151_flux_rc} for an example.  

\begin{figure}[!b]
\epsscale{1}	
\plotone{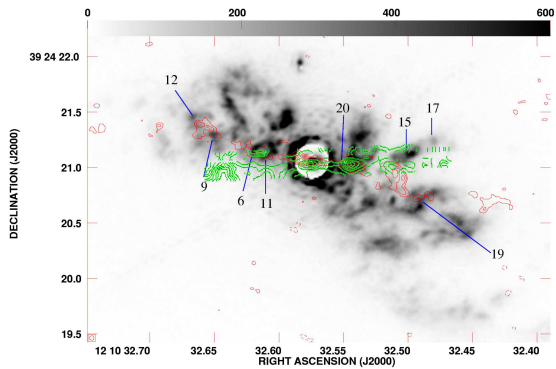}
\caption[NGC 4151 \htwo \lam 2.1218, \oiii, and Radio Emission]{ NGC 4151 \htwo \lam 2.1218 (darker contours), radio (lighter contours; \citealt{mundell03}), and \oiii\s emission (gray scale; \citealt{winge97}).  The \htwo flux distribution measured only during the best seeing (22 Apr 2003, PA=90\deg) is shown.  Contour levels are 0.01, 0.05, 0.1, 0.2, 0.3, 0.4, 0.5, and 0.8 times the maximum flux.  The \oiii\s regions of high velocity dispersion are labeled following (\citealt{hutchings98} and \citealt{kaiser00}). \label{radio4151}}
\end{figure}

At radio wavelengths a complex structure of knots has been identified in NGC 4151 within the central few arcseconds along a PA of 77\deg\s (e.g. \citealt{mundell03}).  \oiii\s also has a clumpy flux distribution in the inner arcseconds that is distributed along a PA of 50\deg\s (\citealt{mediavilla95,schmitt96}).  As can be seen in Fig. \ref{radio4151}, several of the \htwo knots are aligned with knots seen in the radio, and possibly with those in \oiii.  In addition, \citet{mediavilla95} find that H$\alpha$ emission is elongated along the axis of the \oiii\s emission.  Although \htwo has no extended emission along this axis, \bg does have emission along a PA of 50\deg.  

[Ca VIII] and [Si VII] are both spatially resolved with FWHMs over three times greater than those found for the \kb\s continuum and \bg (Fig. \ref{corflux}).  The coronal emission is detected out to a radius of 0\as.7 and the same extended structures are seen in both coronal lines.  In addition, the knot of \bg emission 0\as.4 to the west is also detected in both coronal lines.  This can be seen in a comparison of the 2-D maps (Fig. \ref{allmaps}). As discussed above, this knot is spatially coincident with a knot of \htwo emission as well as one of the components of the nuclear radio jet (\citealt{mundell03}; component C).  

The kinematics of \htwo and \bg are consistent (Fig. \ref{4151_flux_rc}) and show organized rotation along a major axis of PA about 10\deg.  However, \bg differs from \htwo in the central 0\as.2, where it is measured to be redshifted by over 120 \kmsns.  The nuclear kinematics of \htwo and \bg follow the same pattern seen at larger scales with H$\alpha$ and \oiii\s (\citealt{mediavilla95}, \citealt{kaiser00}), as well as with Pa$\beta$ and [Fe II] (\citealt{knop96}).  Complex emission line profiles of lines such as \oiii\s reveal that at least in the highly ionized gas there are multiple velocity components within the nuclear region, including a possible outflow in the northeast-southwest direction (e.g. \citealt{mediavilla95}, \citealt{winge99}).  However, a careful analysis of the \oiii\s line profile uncovers a dominant rotating component, which can still be separated from the complex kinematics in the inner 1\as\s (\citealt{winge99}) and is consistent with the rotation seen in \htwons.

Both [Ca VIII] and [Si VII] are redshifted by more than 100 \kms in the central 1\as\s with respect to the systemic velocity, which is consistent with the redshift seen in \bgns.  The velocity dispersion of the coronal lines is around 100 \kmsns, with a range of 65-150 \kmsns.  This dispersion is generally higher than is measured for \htwons, which varies from 50-100 \kmsns, and lower than the narrow \bg dispersion, which is at least 150 \kmsns.

Since rotation can be traced down to the central 1\as\s in the \oiii\s ionized gas, albeit with careful separations of the different kinematic components, it is not unexpected that the molecular gas would also be dominated by general rotation.  Modeling of the 2-D \htwo kinematics is presenting in $\S$ 6.3.  Despite being dominated by rotation, the complex kinematics of the nuclear region seen in NGC 4151 is confirmed by the irregularities in \htwo and \bg velocity seen along a PA similar to that of the radio jet.

\subsubsection{NGC 4051}
\htwo \lam2.1218 and \bg are detected in the nuclear region of NGC 4051, as well as both coronal lines, [Ca VIII] and [Si VII].  \htwo has a patchy flux distribution mainly concentrated in the central 0\as.5, while the distribution of \bg is the same as that of the \kb\s continuum, both of which are spatially resolved with a FHWM of 0.2\as (Fig. \ref{radflux}).  Both emission lines have a gradient in the velocity field of at least 60 \kms along the PA of about 135\deg.  However, the central 1\as\s of the \bg velocity field is blueshifted to -125 \kms with respect to \htwo and has a velocity dispersion over 200 \kmsns.  In contrast, \htwo has a velocity dispersion of 60-90 \kms throughout the FOV.

At radio wavelengths the core of NGC 4051 separates into two components separated by 0\as.4 in roughly an east-west (PA\sm90\deg) orientation (\citealt{ulvestad84}).  With a coarser spatial resolution of 1\as, \citet{ho01} find that the extended emission is `two-sided' at PA=41\deg.  This emission extends out to several arcseconds, with the northeast component much weaker than the southwest one.  The emission is perpendicular to the optical major axis of the host galaxy.  NICMOS \oiii\s images show a similar flux distribution (\citealt{sosa01}), with an unresolved nucleus and a knot of emission 0\as.5 offset from the center along a PA of 100\deg, as well as extended emission out to 1\as.2.  The \htwo flux distribution is consistent with the radio and \oiii, in that it also has knots of emission along a PA of 100\deg, one 0\as.3 to the east and another 0\as.15 to the west, but no extended emission is seen toward the northeast.   With similar spatial resolution as the \bg apertures used here, \citet{sosa01} detect extended \bg emission in their 3120 second exposure that is not detected in our 600 second exposures.  

The blue shifted \bg in the central 1\as of NGC 4051 supports previous studies that have found evidence of a blue shifted velocity component or blue wing in the \oiii\s line profile.  For example, \citet{veroncetty01} report a second, very broad, component in the line profile of \oiii, which is blue shifted by 175 \kms with respect to the narrower component.  On the other hand, the \htwo velocity field does not show any indication of this blue shift, and instead is consistent with organized rotation down to the smallest radii measured.

The 2-D Gaussian fits to the extended coronal line emission of both [Ca VIII] and [Si VII] are consistent with distributions of the \kb\s continuum and \bg emission (Fig. \ref{corflux}).  Both coronal lines have a tendency toward blueshifted velocities, ranging from -90 to +30 \kms within the central 1\as.  In comparison, \bg is found to have a blueshift of -125 \kms within this same region.  The velocity dispersion of [Ca VIII] is around 75 \kmsns, while [Si VII] has a slightly higher dispersion of around 110 \kmsns, although this difference is not significant given the uncertainty of the measurements.  Neither of the coronal lines have a flux distribution similar to that of \oiii\s or radio emission, both of which have knots of emission within the inner 1\as\s (\citealt{ulvestad84}, \citealt{sosa01}).

\subsubsection{NGC 3516}
All emission lines detected in the spectra of NGC 3516 are weak (with the exception of the broad \bgns) and 2-D maps are constructed for only \htwo \lam2.1218 and [Ca VIII].  The \htwo \lam 2.1218 emission is spatially resolved (FHWM of 0\as.2) with an emission peak that is offset by 0\as.1 to the south compared to the peak in the \kb\s continuum (Fig. \ref{allmaps} and Fig. \ref{radflux}).  A velocity gradient of 75 \kms is detected in the 20 Feb 2001 data, which is at a PA of 28\deg, and the velocity dispersion is $\sim$60 \kms throughout the 1\as.2 FOV.  The velocity gradient is not confirmed in the 20 Feb 2003 data set, but the poor AO correction (PSF FWHM=0\as.17), which smooths any intrinsic rotation, is most likely responsible for this.  Although a narrow \bg component was too weak to construct meaningful 2-D maps, the 1\as.4 aperture spectra indicate the velocity dispersion is $\sim$100 \kmsns, which is marginally consistent with the \htwo velocity dispersion.

\begin{figure}[!b]
\epsscale{1}	
\plotone{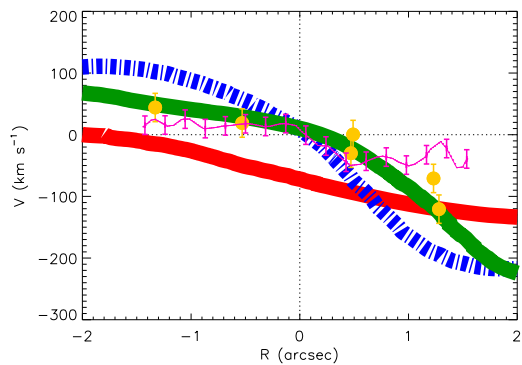}
\caption[NGC 3516 \htwo \lam2.1218 Rotation Curve Comparison]{NGC 3516 \htwo \lam 2.1218 averaged rotation curve (curve with error bars) along a PA of 28\deg.  The \htwo velocity field is quite noisy due to the weakness of the emission, so all data from the 20 Feb 2001 run have been averaged (10 slit positions all within 0\as.15 of the nucleus) and smoothed by the PSF FWHM, which was 0\as.12.  For comparison, rotation curves measured by \citet{arribas97}, all along PA=37\deg, are also shown.  The \htwo rotation is similar to that of \oiii\s and [S II], represented by the upper solid and dashed curves, respectively.  The lower curve is the stellar rotation, as measured using Mg I {\em b}.  The measurements of H$\alpha$ along a PA of 30\deg\s by \citet{mulchaey92} are represented by the large circles.  \label{3516rc}}
\end{figure} 

The [Ca VIII] flux distribution in NGC 3516 is about two times wider than the \kb\s continuum based on the 2-D Gaussian fits, but they are consistent within the errors of the measurements (Fig. \ref{corflux}).  The [Ca VIII] emission is more extended toward the south, which is also seen in the \htwo emission.  [Ca VIII] is redshifted in the nuclear region with a velocity dispersion of 50-90 \kmsns.

Using an integral field spectrometer with a spatial resolution of 0\as.9, Arribas et al. (1997; also see \citealt{gl01}) found that both ionized and highly ionized gas (e.g. H$\alpha$, \oiii, and [S II]) have a flux distribution with a PA \sm 20\deg\s elongation and a 'Z'-shape morphology in their 5\as\s FOV.  Within the central 2\as, the optical emission is elongated at PA \sm10-15\deg.  In the region of the 2-D field where data were obtained (39\% of the field within 1\as), there is no similarity between the optical emission and the flux distribution of \htwons.  Nor is there any similarity between the distributions of the optical emission and the coronal gas ([Ca VIII]), with the exception of the extended emission toward the south.

The \htwo velocity gradient is consistent with the larger scale \oiii\s and [S II] ionized gas kinematics detected by \citet{arribas97}.  This is shown in Fig. \ref{3516rc}, which compares an average of all the 20 Feb 2001 data (10 slit positions all within 0\as.15 of the center) to the rotation of the ionized gas.  
\citet{arribas97} conclude the ionized gas is in an outflow, rather than in rotation about the center of mass, based on the complex emission line profiles and the alignment of the extended emission line flux with the line of nodes of the ionized gas velocity field, as well as with the extended radio emission (\citealt{nagar99,ho01}).  In contrast, the \htwo line profiles, at least at our spectral resolution, are well represented by a single Gaussian.  \citet{arribas97} also measured the stellar kinematics (using Mg I {\em b} \lam\lam5184, 5173, 5167 \AA), and found that within the central 2\as\s the gas velocities are blue shifted with respect to the stellar kinematics, which is also true of the \htwo velocities (Fig. \ref{3516rc}).  In addition, they find that the stellar kinematics are less disturbed compared to the ionized gas, and appear to be in regular rotation.

Although the data presented do not provide clear evidence that the nuclear molecular gas is in rotation, there is some support for this in the best data with a major axis of rotation along a PA of around 30\deg.  In addition, the velocity gradient is consistent with the regular rotation of the stellar kinematics measured by \citet{arribas97}.  It is possible that longer integration times and higher spatial resolutions will yield a more definitive measurement of the nuclear kinematics in NGC 3516 and that rotation about the center of mass will be confirmed.  However, with the current data, no modeling of the 2-D velocity field is performed.

\subsubsection{NGC 5548}
Only weak emission features were detected in the spectra of NGC 5548, the strongest of which is \htwo \lam2.1218.  Both the \htwo and the \kb\s continuum flux distributions are resolved.  The 2-D Gaussian fit to the continuum gives a FHWM of 0\as.33 and \htwo is detected throughout the FOV (see Fig. \ref{radflux}).  No velocity gradient is seen along either of the two perpendicular position angles and the velocity dispersion is relatively high at \sm100 \kmsns.  The narrow \bg velocity dispersion measured in the 1\as.4 aperture spectra ($\sim$140 \kmsns) is higher than the \htwo dispersion, but still marginally consistent given the measurement errors.

NGC 5548 has a triple radio source aligned at PA 168\deg, which is composed of a compact core and two oppositely directed lobes that peak in flux several arcseconds from the core (\citealt{nagar99,ho01}).  \citet{wilson89} find that, on a 1\as\s scale, the nuclear \oiii\s emission is aligned with the radio emission but that the H$\alpha$ emission is not, and instead is extended northeast of the nucleus.  The \htwo flux distribution detected might hint at an elongation aligned with the radio and \oiii\s emission.  \citet{wilson89} also find that the velocity field of \oiii\s is dominated by rotation with PA 140\deg\s major axis.  No such rotation is detected in \htwo at the two PA measured, which are each 45\deg\s off of the \oiii\s major axis.  In addition,  \citet{wilson89} report that at the nucleus  \oiii\s is blueshifted with respect to the systemic velocity found from the rotation.  No blueshift is detected in \htwons.  

Since rotation is detected in ionized gas, it is likely that the kinematics of the molecular gas is also dominated by rotation, although this may break down in the region of blueshifted \oiii.  Given the weakness of the \htwo emission measured, organized rotation of the gas cannot be ruled out and it is still feasible that rotation could be detected with longer integration times.

[Ca VIII] is detected in NGC 5548 and is found to have flux distribution that is comparable to both the \kb\s continuum and \htwo emission (Fig. \ref{corflux}).  In the fraction of the 2-D field measured, the coronal emission is extended toward the west, which is consistent with the nuclear radio and \oiii\s emission (\citealt{nagar99}, \citealt{ho01}).  The coronal emission has no consistent structure in the velocity field, with measurements ranging from -70 to +70 \kmsns, but the velocity dispersion steadily rises in the central 0\as.4 to 130 \kmsns, consistent with the narrow \bg dispersion.

\subsubsection{Ark 120}
The only emission detected in the Ark 120 spectra is weak \htwo \lam 2.1218 and \bg (neither of the coronal lines are available) and 2-D maps are only constructed for \htwons.  The unresolved peak of the \htwo flux distribution is coincident with that of the unresolved \kb\s continuum (Fig. \ref{radflux}).  No rotation is detected in the \htwo velocity field, and instead the central apertures (0\as.37) indicate that the gas is redshifted to greater than 125 \kms with respect to the systemic velocity.  In this same region the velocity dispersion is above 150 \kmsns, confirming that the bulk of the gas is not in simple rotation about the center of mass.  With higher signal-to-noise data it might be possible to distinguish the redshifted component from one that traces the gravitational potential, but with the current data no modeling of nuclear gas rotation is possible.

Ark 120 has a radio quiet nucleus with a slight extension of emission toward the northeast out to 30\as\s (\citealt{condon98}).  At a spatial resolution of 1\as.3, the flux distributions of both \oiii\s and H$\alpha$ emission are symmetric out to several arcseconds (\citealt{mulchaey96}), consistent with the distribution seen in \htwons.  At a spatial resolution of 0\as.5 (compared to the 0\as.37 apertures used here), \citet{sosa01} find a slight elongated at PA \sm135\deg\s in Br$\gamma$ out to 1\as.5, which there is no sign of in the \htwo distribution.  

\subsubsection{NGC 6814}
The spectra from the slit positions obtained (PA of 35\deg) for NGC 6814 have several strong emission lines, including four molecular hydrogen emission lines and broad \bg emission, but no [Ca VIII] is detected ([SI VII] is not available).  \htwo is measurable throughout the 2\as\s FOV and its peak is coincident with that of the \kb\s continuum (Fig. \ref{radflux}).  The \kb\s continuum is resolved with a FHWM of 0\as.35.  No rotation is detected in any of the \htwo emission lines. However, organized rotation of the molecular gas cannot be ruled out since data are only available at a few slit positions at a single PA, leaving open the possibility that the PA observed is the minor axis of rotation.  The \htwo velocity dispersion is generally $\sim$75 \kmsns.  For comparison, the narrow \bg detected in the 1\as.4 aperture spectra has a higher velocity dispersion of 170 \kmsns.

Both the \oiii\s (\citealt{schmitt96}) and radio emission (\citealt{ulvestad84}) show an extension along PA \sm150\deg\s within the inner 2\as. There is also a western extension in the radio at about 1\as.5 for which there is no counter part in seen in \oiii.  The \htwo distribution in the four transitions measured does not show any similarity to these flux distributions, but the lack of 2-D coverage of the nuclear region of NGC 6814 limits any detailed comparison.       

No modeling of the nuclear \htwo kinematics is possible with the NIRSPEC data presented; however, NGC 6814 shows potential for a detailed study of the nuclear region of a Seyfert 1.  If the \htwo \lam2.1218 emission is distributed throughout the nuclear region (as is suggested by the NIRSPEC data), then the relatively low velocity dispersion implies that it is likely a reliable tracer of the gravitational potential.  

\subsubsection{NGC 4593}
Two molecular hydrogen lines, \htwo \lam2.1218 and \lam2.4066, are detected in the spectra of NGC 4593 from the two slit positions obtained at a PA of 90\deg.  Both \htwo transitions have a flux distribution similar to that of the \kb\s continuum (Fig. \ref{radflux}), which is resolved with a FHWM of 0.13\as.  No rotation is detected in the \htwo velocity field, and therefore no modeling of the field is carried out, but, as with NGC 6814, rotation cannot be ruled out based on only a couple slit positions along a signal PA.  The velocity dispersion is 60 \kms in most of the data, but in a few apertures the dispersion is above 100 \kms.  However, given the low signal-to-noise ratio of the \htwo emission, this difference in velocity dispersion is unlikely to be significant.  NGC 4593 does not have any known extended radio emission, but has a weak, unresolved radio core less than 0\as.2 in size (\citealt{ulvestad84}).  Only [Ca VIII] could potentially have been measured in the spectra for NGC 4593 and no emission is detected.

\section{2-D Dynamical Modeling Technique}

\subsection{Modeling of the 2-D Velocity Field}
Modeling of the observed emission line gas kinematics consists of two main steps.  First a model velocity field is created assuming a co-planar thin disk undergoing circular rotation in a gravitational field created by both the stellar population and a central point mass, presumably a supermassive BH.  There are four free parameters in the model: the disk inclination and position angle of its major axis, both of which can be constrained based on previous studies, \mbh, and the mass-to-light ratio (\ml) used to determine the stellar mass distribution.  The model velocity field is then synthetically ``observed'', which simulates the same configuration used for the actual observations.  This step takes instrumental effects into account, which smooth or blur the intrinsic velocity field.  These effects include convolution with the point spread function (PSF) and the intensity weighted averaging over apertures determined by the slit width and the number of pixels integrated along the slit.  This technique of modeling a gas velocity field is similar to methods used by \citet{macchetto97} and \citet{barth01}, as well as many other authors.  The parameters of the best fit model to the observed velocity field is determined using a chi-squared (\ch) quality of fit statistic, which will be discussed further in $\S$ 5.4. 

In addition to the observed velocity field, there are three other constituents of the model that must be acquired: the PSF, the stellar mass distribution, and the emission line flux distribution.  Each of these components is discussed in detail below.  

\begin{figure}[!t] 
\epsscale{1}	
\plotone{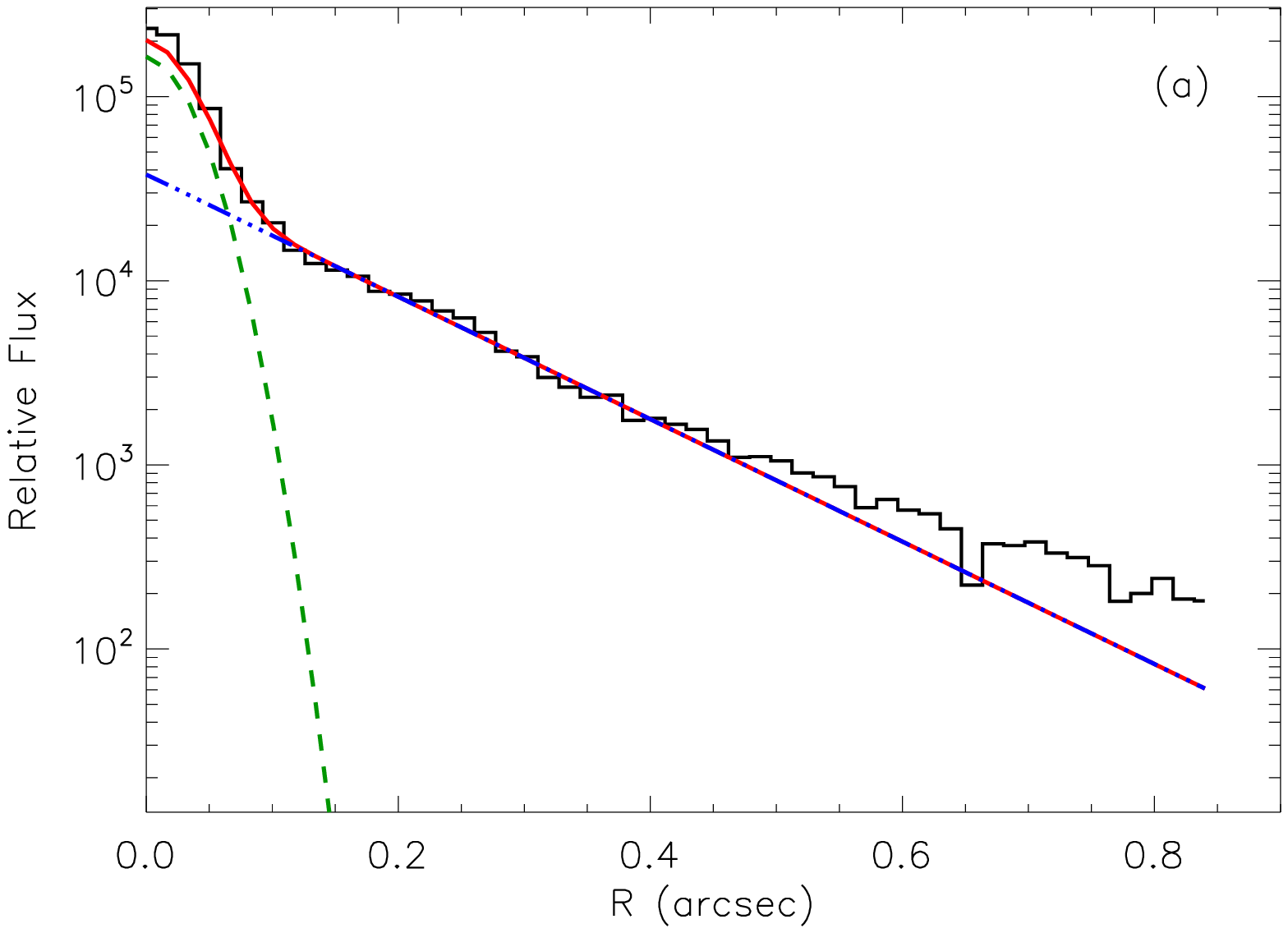}
\plotone{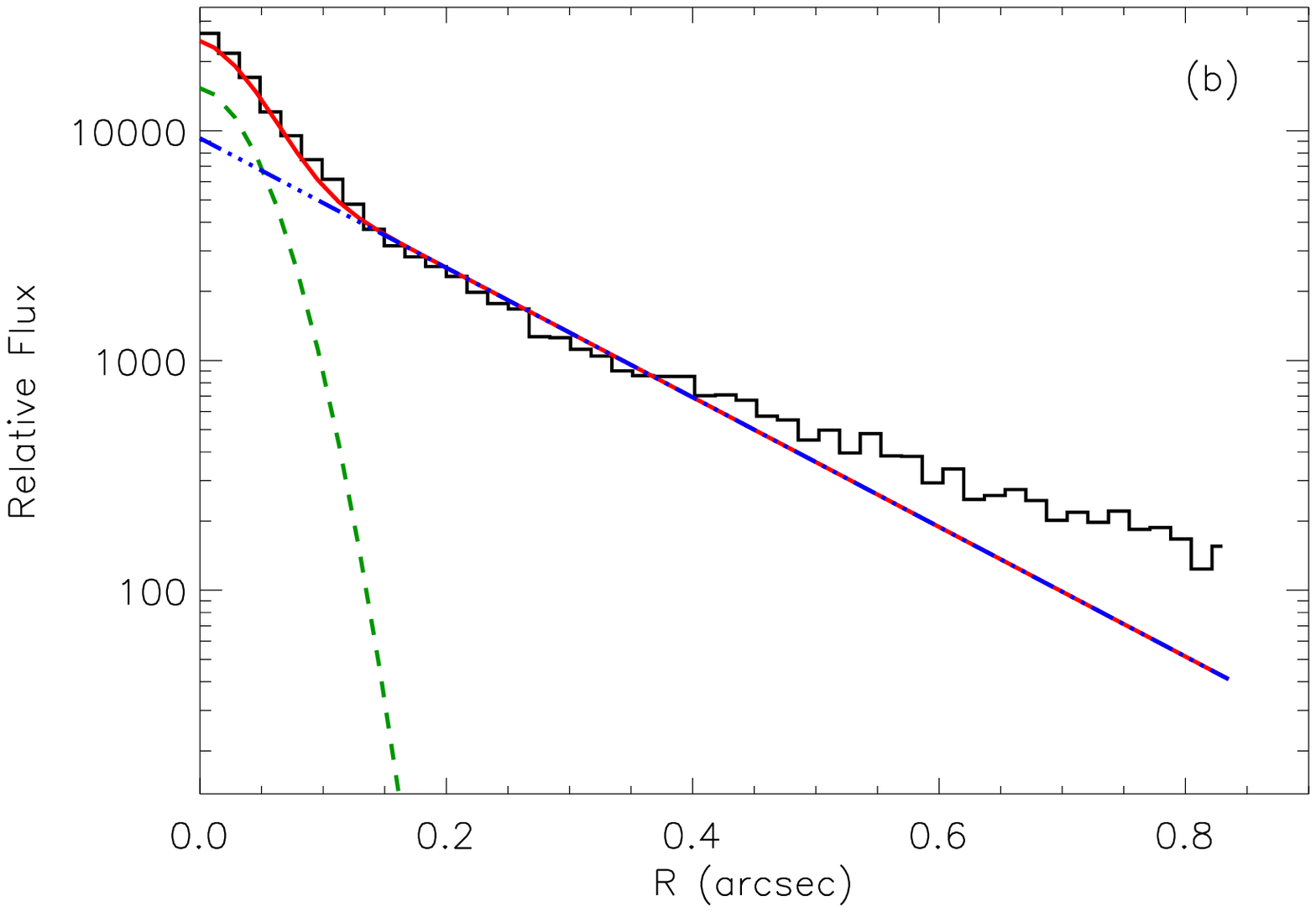}
\caption[Example Best Fit Two-Function PSF and Predicted PSF]{(a) Example of a best fit two function PSF.  The histogram is the stellar light profile, and the solid curve is the PSF function fit to the data, which is made up of a Gaussian core (dashed curve) and exponential wings (dash-dotted curve).  (b) A typical predicted PSF from an SCAM image of a Seyfert 1 galaxy (in this case NGC 3227).  A cut through the 2-D galaxy light profile is shown by the histogram, the predicted PSF (solid curve), is composed of a Gaussian core (dashed curve) and exponential wings (dash-dotted curve). \label{psf_func_g}}
\end{figure}

\subsubsection{Characterization of the PSF}
The variable AO PSF is measured throughout each spectroscopic exposure using the slit viewing camera SCAM, enabling the seeing and AO correction quality for each slit position to be accounted for in the model.  Since the SCAM images are relatively short (typically only ten seconds), it is not expected that a significant amount of star light will be detected, and the images will therefore be dominated by AGN light (i.e. a point source).  However, to ensure the accuracy of the PSF estimation, it is not assumed that these galaxy images represent the PSF.  Instead, a method is developed to predict the PSF from a core weighted single Gaussian fit to the central 0\as.08 of the galaxy profile, which can be safely assumed to be dominated by the point source AGN.

\begin{figure}[!t] 
\epsscale{1}	
\plotone{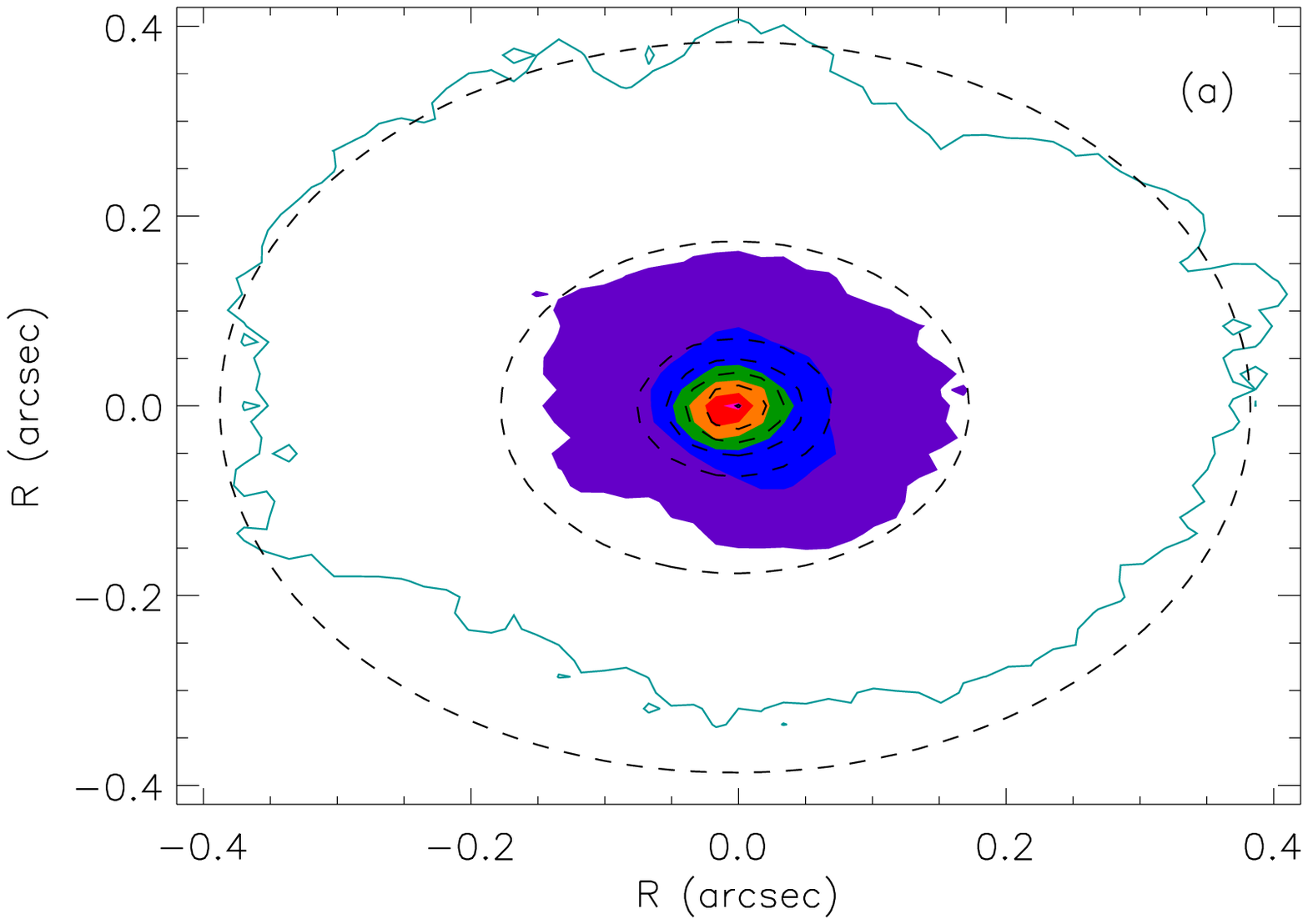}
\plotone{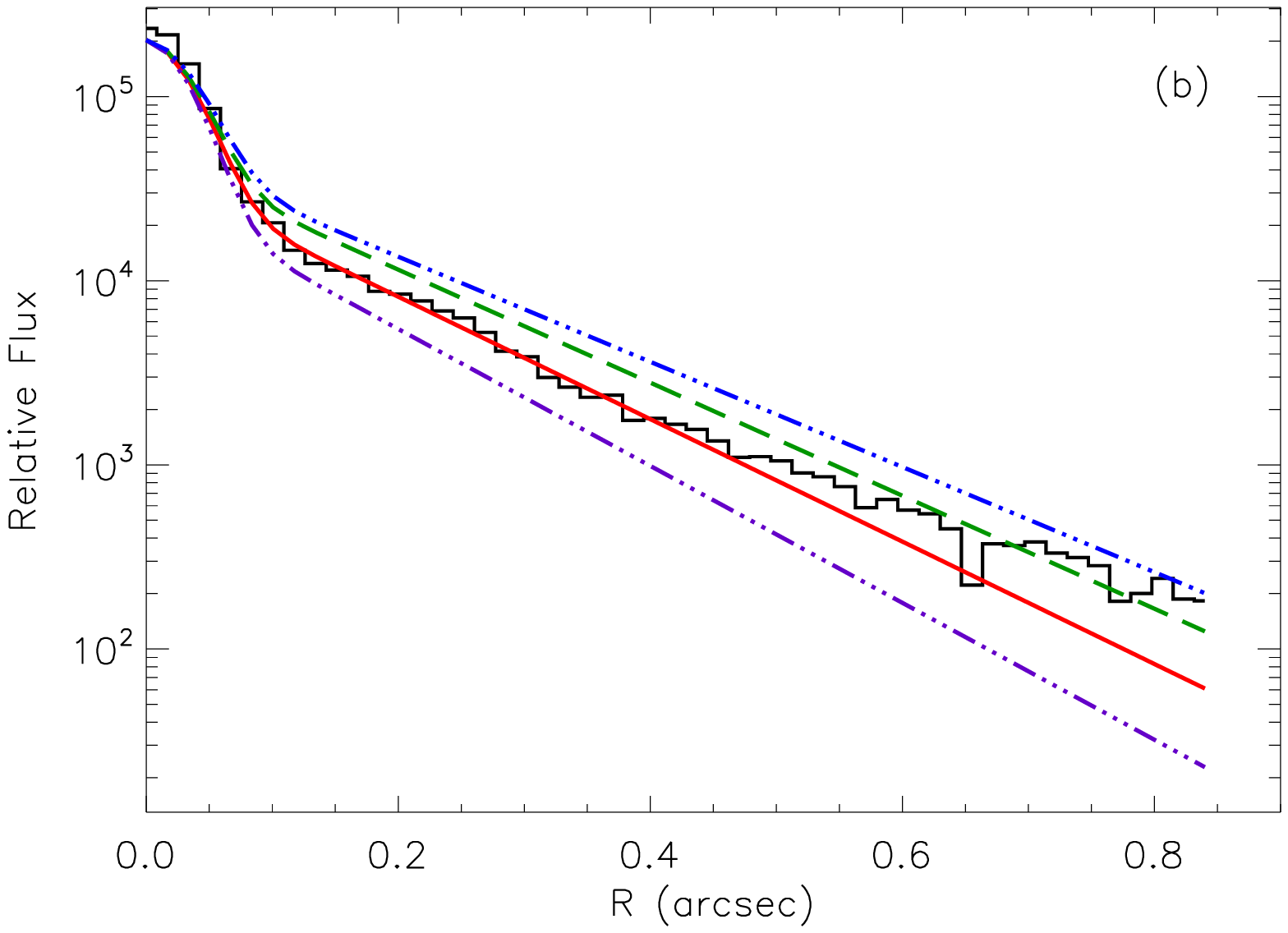}
\caption[Predicted versus Two-Function Model PSF]{ Model versus a predicted PSF for a point source.  (a) The stellar light profile (solid contours) compared to the model (dashed contours). (b) Cuts through the 2-D stellar light profile (histogram) compared to the best fit 2-D PSF (solid curve) and the predicted PSF (dashed curve) and the error of the predicted PSF (dash-dotted curves). \label{spsf}}
\end{figure}

The PSF, as measured in SCAM images of known point sources (stars) positioned offset from the slit, is found to be well characterized by a Gaussian core and exponential wings (Fig. \ref{psf_func_g}a).  A relationship between a core weighted single Gaussian fit to the central 0\as.08 and a Gaussian plus exponential PSF was established using these stellar observations (see Appendix B).  All fits were done in two dimensions assuming circular symmetry and the galaxy light profiles were corrected across the slit location as described in $\S$ 3.2.2.  An example of a PSF predicted from a Seyfert 1 SCAM image is shown in Fig. \ref{psf_func_g}b.  In most cases, very little star light remains after accounting for the AGN flux using the predicted PSF.

\begin{figure}[!htt] 
\epsscale{0.95}
\plotone{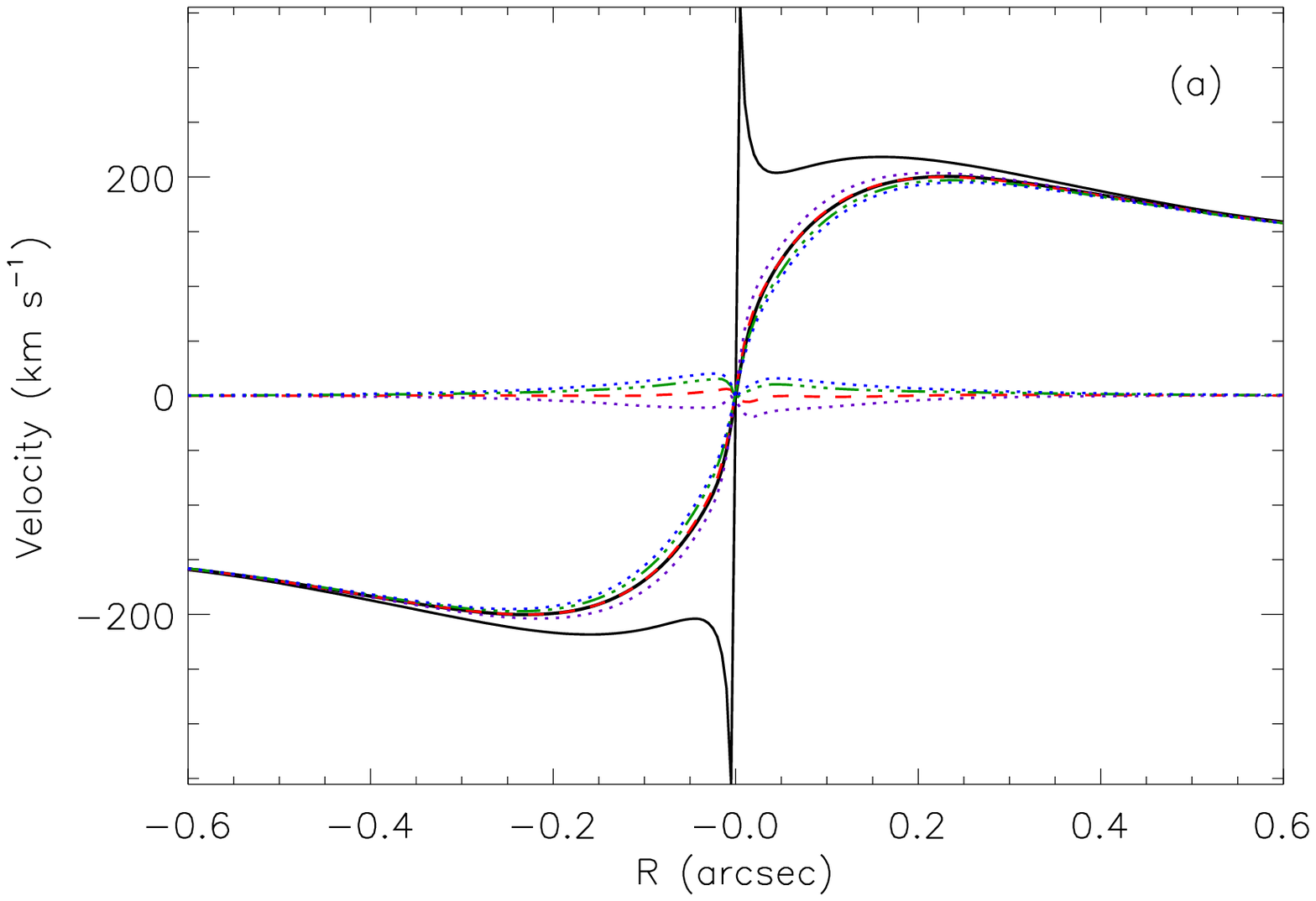}
\plotone{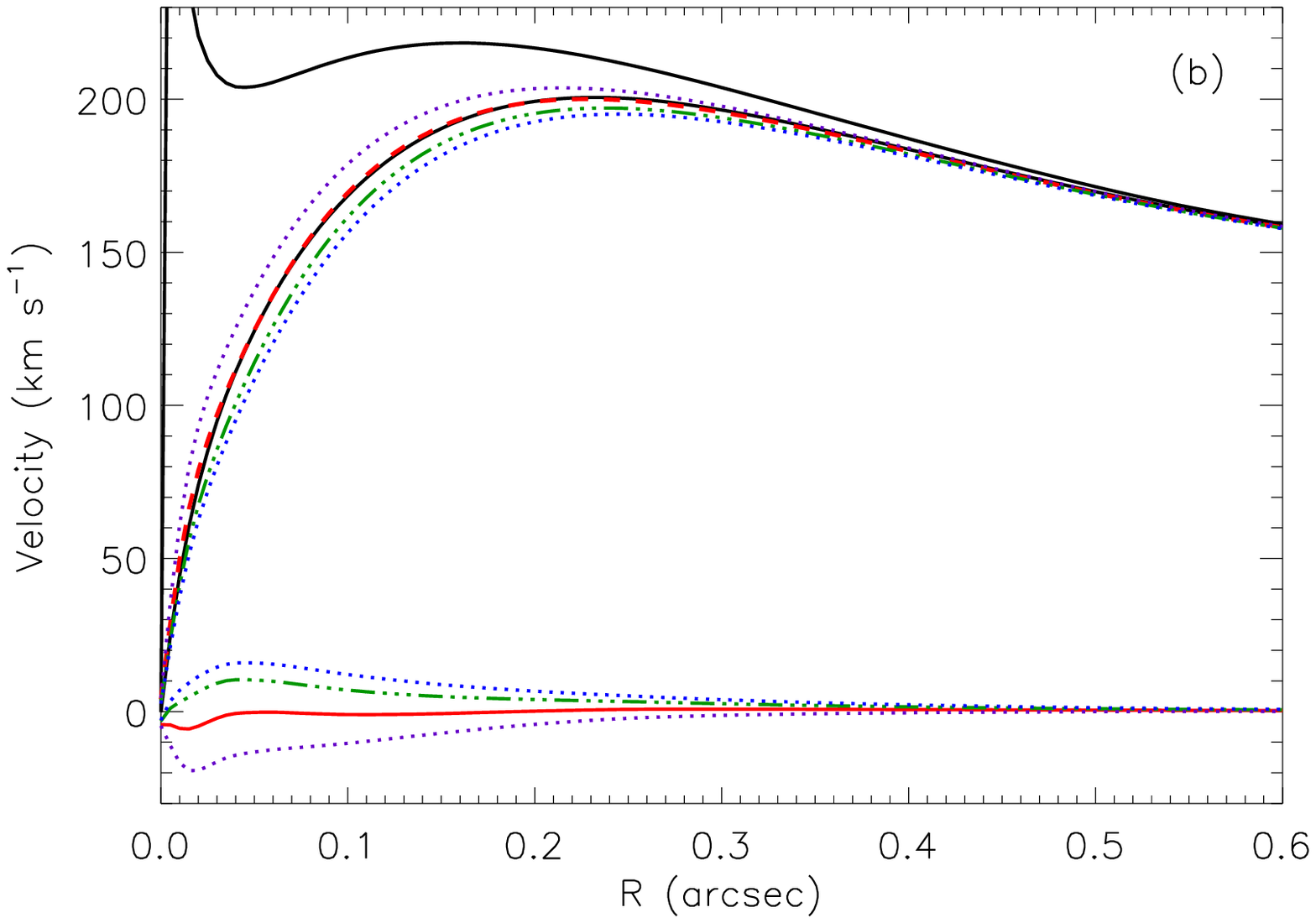}
\caption[PSFs convolved with an Example Velocity Field]{ (a) An example stellar plus BH velocity field (thick solid curve at top and bottom) convolved with the actual PSF (thinner solid curve), the best fit two function model PSF (dashed curve), the predicted PSF (dash-dotted curve), and range due to the error of the predicted PSF (dotted curves).  The residuals of the convolved rotation curves with the various PSFs compared to the actual PSF are shown with curves of the same line style.  (b) A blow up of half of the rotation curves.  \label{psfconv}}
\end{figure}

The accuracy of the predicted PSFs is such that the resulting velocity fields are in error by typically 5-10 \kmsns, and at most 20 \kmsns, with the largest error at small radii.  Fig \ref{spsf} is a comparison of the actual PSF, the model PSF, and the predicted PSF and in Fig. \ref{psfconv} these PSFs are convolved with an example velocity field.  Most often the prediction overestimates the PSF in a sense that the resulting model velocity fields are over smoothed.  This has the potential to make detection of a BH more difficult, but is safer than under-smoothing the velocities, which would lead to erroneously small error bars in the \mbh\s estimates. 

\subsubsection{Determination of the Stellar Mass Distribution}
The Seyfert 1 galaxies in this sample have not been included in previous studies of the nuclear structure in AGN because of the difficulty of subtracting the bright non-stellar AGN light.  Although there is uncertainty in the characterization of the PSF, and thus the flux attributed to the point source AGN, an estimate of the stellar light distribution in the central 1\as, at an accuracy sufficient for this study, is possible using archival HST NICMOS F160W images (similar to the {\em H}-band).  At the small radii of importance for this study, the galaxies are dominated by light from bulge stars with no significant contribution from disk stars (\citealt{nelson04}), and thus no fit to the disk is performed.

Separation of the point source from the star light is done by fitting a point source represented by a Tiny Tim PSF (\citealt{krist98}) plus a generalized exponential or S\`{e}rsic function (\citealt{sersic68}, which is given by
\begin{equation}
I(r) \, = \, I_{e} \, exp \left \{-\alpha_{n} \, \left [ \left (\frac{r}{r_{e}} \right )^{1/n} -1 \right ] \right \}
\end{equation}
\noindent where {\em I}$_{e}$ ($\mu_{e}$ in magnitude units) and {\em r}$_{e}$ are the effective (or half-light) surface brightness and radius, respectively, and $\alpha_{n}$ is a constant relating the effective brightness and radius to the exponential values (\citealt{moriondo98a}).  The bulge index \n\s = 4 corresponds to the standard R$^{1/4}$ law.

First a S\`{e}rsic function is fit to the light profile from a radius of 1\as.5 out to 5\as.0 using a minimization of chi-squared technique.  The point source and S\`{e}rsic function are then both fit to the entire light profile by constraining the S\`{e}rsic function to the parameters found based on the outer radii.  The effective radius is allowed to change by no more than 0\as.4 and the surface brightness at this radius by no more than 30\%.  S\`{e}rsic functions with n values of 1, 2, 3, and 4 are fit, and all fits are done in 2-D with circular symmetry assumed.

Deprojection of the best-fit 2-D light profile is done using the strip brightness method (\citealt{moriondo98b}).  A constant mass-to-light ratio in the range of \ml=0.3-1.1 {\em H}-band units (\mlu) is then used to convert the light distribution to a stellar mass distribution (the spatially distributed dark matter is also included in \ml).  Observations and theory show that, in the {\em H}-band, \ml\s can vary from 0.05 to as high as 4.0 (see e.g. \citealt{oliva99}, \citealt{bell01}), but different types of galaxies lie within a much smaller range of \ml.  Seyfert 1 galaxies, for example, are observed to have \ml=0.3-1.1 \mlu, while galaxies with older populations, such as elliptical galaxies, have \ml=0.3-2.8 \mlu, and starburst galaxies are observed to have \ml=0.1-1.6 \mlu\s (\citealt{oliva99}).  For the gas dynamical modeling the accepted \ml\s values are assumed to be those observed for Seyfert 1 galaxies, although values as low as 0.1 \mlu\s and as high as 2.0 will be considered as extreme cases.

The range of acceptable fits to the light profiles were determined by eye rather than by a statistical method (see $\S$ 6 for the fits to each modeled galaxy).  This is due to the difficulty in matching the model PSF to the NICMOS image, which can cause statistical routines to find a best fit that by eye is clearly not acceptable.  Notably, compared to most non-active spiral galaxies (\citealt{scarlata04}, \citealt{seigar02}), all three galaxies that were fit appear to have higher stellar surface densities in the inner two arcseconds.  Of the 44 non-active galaxies in the \citet{scarlata04} sample, NGC 3227 and NGC 4151 have higher surface densities than all but one galaxy, and only six galaxies have a higher density than NGC 7469. 

As a check of the above method, a scaled Tiny Tim PSF, representing the non-stellar AGN continuum emission, was subtracted from the NICMOS image to determine the minimum and maximum plausible stellar light.  Using this method it is possible to detect any asymmetries in the light distribution or features such as nuclear rings, which cannot be represented by a smooth S\`{e}rsic function.  The AGN-subtracted light distribution is also used as a check of the S\`{e}rsic  fits, which should match the AGN-subtracted light distribution outside of the radius effected by the NICMOS PSF (r$\ge$ 1\as).  For only one Seyfert 1 galaxy, NGC 7469, it was necessary to use the AGN-subtracted light distribution rather than the S\`{e}rsic fit because of an excess of stellar light in the inner 1\as.  A radial average of the AGN-subtracted light distribution is used and the deprojection is carried out as described above.  For the remaining galaxies, the AGN-subtracted stellar light distributions are consistent with those estimated from the best S\`{e}rsic fits.  A comparison of the AGN-subtracted light distribution and the S\`{e}rsic fits for the individual galaxies is discussed in $\S$ 6.

\subsubsection{Intrinsic Emission Line Flux Distribution}
The last input for the model velocity field is the flux distribution of the line emitting gas.  This is considered because the velocity observed from a given aperture is defined by the average of the gas velocities within the spatial scale of the aperture weighted by the flux distribution.  Since no narrow band images at sufficiently high spatial resolution (a resolution higher than, or matching, that of the spectroscopy) are available for the galaxies in this sample, the flux distribution is estimated from the 2-D maps constructed from the NIRSPEC/AO single slit observations.  The intrinsic flux distribution is modeled with multiple Gaussians and convolved with the PSF (see $\S$ 5.1.2) to find the best fit to the observed distribution.  This method provides a suitable fit to the observed distribution as well as an interpolation of the flux distribution across gaps in the 2-D coverage.  The change in best fit model parameters due to taking the flux distribution into account is relatively small compared to other uncertainties, as will be discussed in more detail for each modeled galaxy in $\S$ 6.

In cases of an irregular flux distribution, such as \htwo in NGC 3227, including the effects of this distribution in the modeling will account for the small scale wiggles in the observed velocity field and thus improves the quality of the fit (i.e. lowers \ch).  However, using a relatively smooth flux distribution, such as the multi-Gaussian fits used here, merely raises the \ch\s value and has little effect on the best fit model parameters (e.g. \mbh; \citealt{barth01}).  In contrast, as \citet{marconi06} show, modeling of the velocity dispersion (which is not attempted here) can be highly dependent on the intrinsic emission line flux distribution, and they conclude that the gas velocity dispersion is not a reliable indicator of \mbh.

\subsubsection{Summary of the Model Accuracy}

Considering the error associated with each of the three components discussed above, the model velocity fields are typically accurate to 20-25 \kmsns, with the most uncertain models accurate to at least 35 \kmsns.  The uncertainty in the PSF contributes an error of typically 5-10 \kmsns, up to at most 20 \kmsns.  The model error due to the range of acceptable fits to the stellar light distribution can be substantial (20-30 \kmsns) in the innermost region of the galaxies, but, as discussed in $\S$ 6, this error is found to not significantly alter the resulting \mbh\s estimates.  The modeled flux distributions contribute an error of less than a few \kms to the model (see discussion for each galaxy in $\S$ 6).  The total error for the model velocity field as a result of these three components is comparable to the error in the velocity field measurements.

\subsection{\mbh\s Dependence on Disk Inclination and Mass-to-Light Ratio}

As expected, the best fit \mbh\s is dependent on both the inclination angle of the gas disk (\inc=0\deg\s is face on, \inc=90\deg\s is edge on) and the stellar mass-to-light ratio, \ml, with a significant increase in either parameter resulting in a lower \mbh\s estimate.  However, as discussed in $\S$ 5.1.2, upper and lower limits can be placed on value of \ml, and this in turn constrains the range of acceptable \inc values.  For example, if an upper limit is placed on \ml, then the stellar component cannot exceed a given mass and the additional observed velocity (outside the influence of the \mbh) must be accounted for by inclination effects, placing a lower limit on \inc.  Furthermore, with sufficient spatial resolution, the steep central velocity gradient due to the BH point mass cannot be fit by arbitrarily increasing either \inc\s or \ml.  As a result, for each of the three galaxies modeled, the best fit \mbh\s is largely independent of both \ml\s and \inc.  In each case, an adjustment in one parameter is compensated for by the other parameter such that the best fit \mbh\s remains unchanged over a reasonably small range of \ml\s and \inc\s values (see $\S$ 6). 

\subsection{Data Quality Considerations and Fitting Radius} 
Another consideration in determining the dependence of \mbh\s on the modeling technique is what data should be included in the fit.  For example, a data quality cut can be made by using the Pearson correlation coefficient (PCC) of the Gaussian fits to the emission lines.  Low PCC values indicate a poor fit (possible explanation of which include poor sky subtraction, non-Gaussian line profiles, or simply weak emission), and therefore the velocity determined from this fit has more uncertainty.  The quality of the Gaussian fit is taken into account in determining the error in the particular velocity measurement; however the quality of the best fit model is often improved by excluding the worst data.  

The fit of the models can also be improved by restricting the fit to a given radius.  This could, for example, be used to avoid potential regions of non-circular velocities, or to see if the best fit model parameters are dependent on the radius fit (r$_{fit}$).  Such a situation might indicate that a simple co-planar gas disk is a poor fit to the data and something like a warped disk might provide a better fit.  The best PCC value for the data quality cut and r$_{fit}$ are determined for each galaxy individually and is discussed in more detail in $\S$ 6. 

\subsection{Best Fit Model Parameters: the Significance of \ch}
The velocity fields of the nuclear gas observed in both quiescent and active galaxies are in practice much more complicated than the simple models used to describe their kinematics.  A consequence of applying these overly simplified models is that a comparison of the model to the data does not formally yield a statistically acceptable fit.  The best fit model to an observed velocity field often has a reduced \ch\s (\ch\s divided by the number of degrees of freedom) that is much greater than one.   Although the model fails to represent the small scale complexities of the data, the statistic can serve to indicate which model parameter(s) provide the best fit to a generally rotating velocity field.  A confidence interval is then used to determine the range of parameter values that provide a reasonable fit.  However, the definition of this confidence interval is problematical when the reduced \ch\s is much larger than one.

Two methods have been used in the literature to deal with the high reduced \ch\s values.  The first method is to rescale the error in the velocity measurements so that the irregularities are essentially treated as random error in the velocity (e.g. \citealt{barth01}, \citealt{marconi06}).  The additional error, which is constant throughout the FOV, is chosen such that when added in quadrature with the formal velocity uncertainties \rch$\simeq$1.0.  In this case the confidence intervals (e.g. 1\signs, 2\signs, etc.) are defined by the \ch\s probability distribution for a given number of parameters (\citealt{press92}).  The second method uses a relative likelihood statistic, which gives the best fit parameters and the interval of parameter values that yield an equally good (or bad) fit (e.g. \citealt{vandermarel98}, \citealt{gebhardt00}).  The parameter that gives the lowest \ch\s value, \ch$_{min}$, is considered to be the best fit value and the confidence intervals are determined by a range of \ch\s values greater than \ch$_{min}$, defined as $\Delta$\ch = \ch-\ch$_{min}$.  If it is assumed that the observational errors are normally distributed, then $\Delta$\ch\s also follows a \ch\s probability distribution with the number of degrees of freedom equal to the number of model parameters (\citealt{lampton76}, \citealt{cash76}).  Although both of these techniques seem dubious given the inherent poor fits to the data, the field has continued to use these methods for lack of a better way to handle the statistics.

The modeling results presented here are based on the relative likelihood or $\Delta$\ch\s method.  In addition, the error in the best fit model parameters have also been estimated using the half-sample bootstrap method (e.g. \citealt{babu96}).  This method uses different realizations of the data set to directly determine the statistical distribution of the model parameters.  For each realization half of the data set is randomly selected and the best fit model parameter values are determined.  The errors in the best fit parameters are then determined by the distribution of the values for 1000 realizations.  This method is more robust in that the probability of a given parameter value being the best fit to the data is directly determined, rather than depending on the definition of a confidence interval.  In general, the distributions of the parameter values from the half-sample bootstrap method are consistent with the relative likelihood statistics. 

The sensitivity of the reduced \ch\s value to small irregularities in the observed rotation causes the reduced \ch\s to increase as more data are systematically included in the fit.  For example, the reduced \ch\s increases as r$_{fit}$ is increased or the PCC limit is lowered, both of which result in systematically fitting more data.  Therefore, even though a fit to all available data may give a more reasonable fit, the reduced \ch\s will be higher compared to the reduced \ch\s of a fit to a systematically chosen subset of the data.  In contrast, a randomly selected subset of the data, such as is done in the half-sample bootstrap method, does not give a lower reduced \ch\s value. 
 
In addition to assessing the parameter errors with the above methods, the true reduced \ch\s value will be given so the reader can make their own assessment of the quality of the fit to the data. It should be keep in mind that the confidence intervals used only give formal limits, and are perhaps not truly representative of the quality of the model parameter estimates.  Throughout this paper the confidence intervals used for 1\signs, 2\signs, and 3\sig confidence limits (corresponding to 68.3\%, 95.4\%, and 99.7\% limits, respectively) are taken to be, following \citet{press92} (also see \citealt{lampton76}), $\Delta$\ch=1.0, 4.0, 9.0 for one interesting parameter, $\Delta$\ch=2.3, 6.2, 11.8 for two interesting parameters, and $\Delta$\ch=3.5, 8.0, 14.2 for three interesting parameters.

\section{\mbh\s Estimates from Gas Dynamical Modeling} 

\subsection{Modeling Results for NGC 3227}
Of the Seyfert 1 galaxies in the sample, NGC 3227 has the most well constrained \mbh\s from the gas dynamical modeling.  This is in part due to how well the \htwo velocity field is measured, but is also because the other parameters and input to the dynamical model are reasonably constrained.  The following sections discuss the best fit model parameters, as well as the dependencies of these parameters on each other and on the model input.  

\begin{figure}[!t] 
\epsscale{0.9}
\plotone{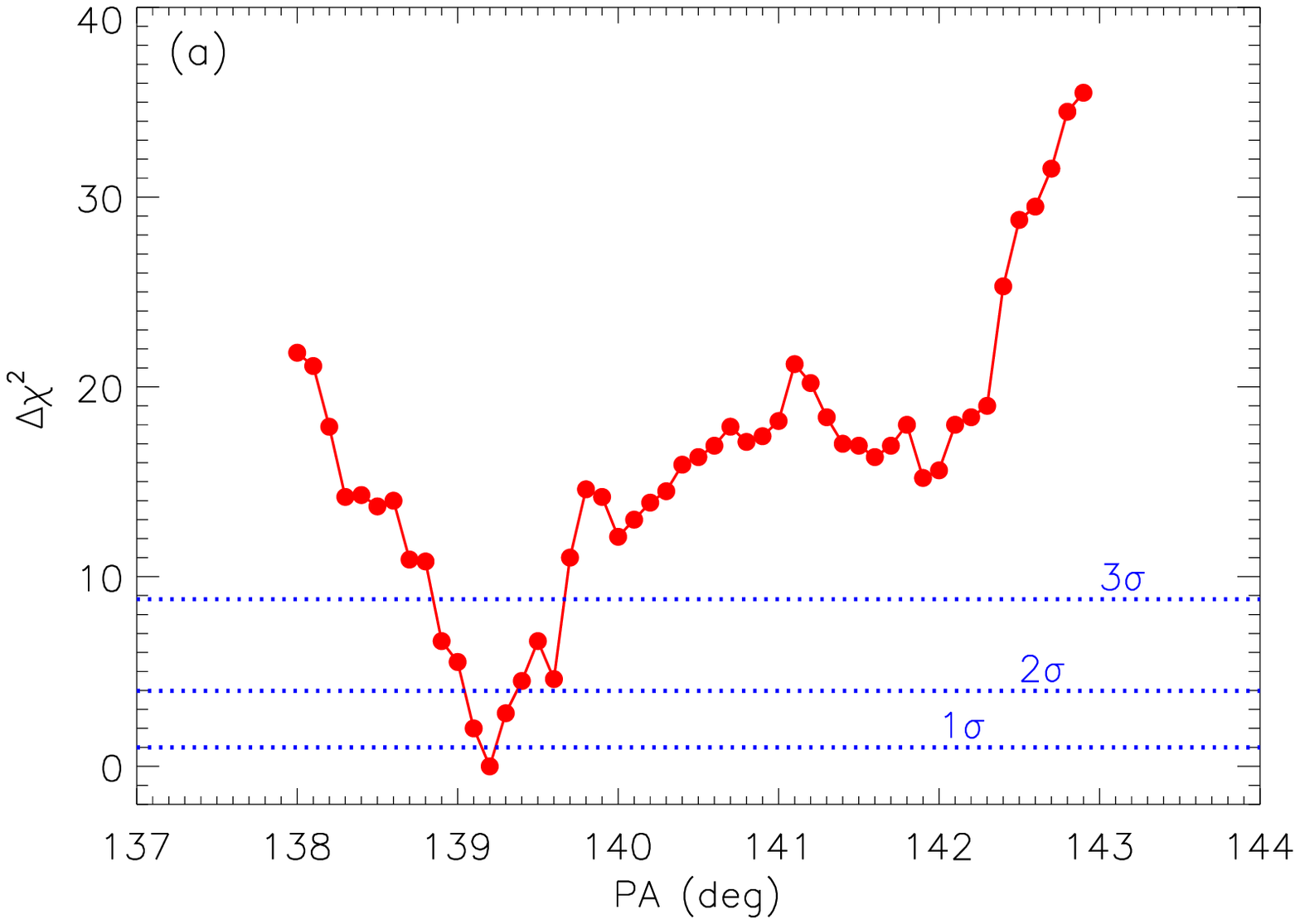}
\plotone{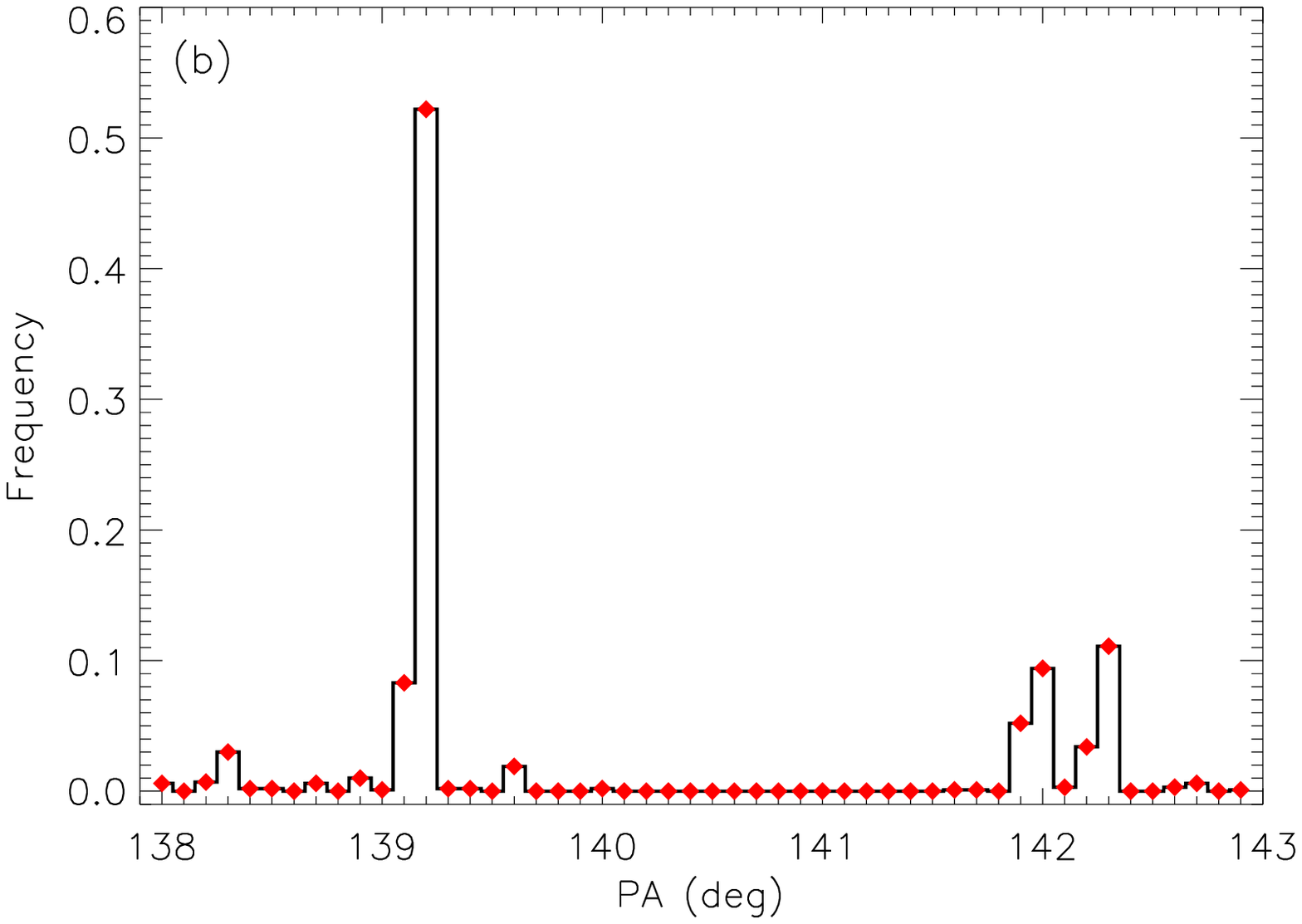}
\caption[NGC 3227: Best Fit Position Angle]{Best fit position angle of the gas disk major axis in NGC 3227. (a) Dependence of \ch\s on the position angle, \t, where the best fit model (minimum \ch) for each \t\s is found by varying \inc, \ml, and \mbh.  The horizontal lines show the 1\signs, 2\signs, and 3\sig confidence intervals for one parameter ($\Delta$\ch\s = 1, 4, 9).  Circles indicate parameters for which models were run and \ch\s calculated. (b) The distribution of the best fit position angle for 1000 realizations of the half-sample bootstrap method (see $\S$ 5.1.2).  The specific model parameters fit to the data are indicated by the diamonds.  \label{3227_chi_bs_pa}}
\end{figure}

\subsubsection{Major Axis Position Angle}
The line of nodes of the \htwo velocity field is found to be \t=139.2$\pm$0.1\deg\s at the 1\sig confidence level (\t=139.2$\pm$0.5\deg\s 3\sig limit).  This result is especially robust since \t\s is not dependent on any of the other model parameters (\mbh, \inc, and \ml).  The minimized \ch, found by varying the other three model parameters for a range of \t\s values, is shown in Fig. \ref{3227_chi_bs_pa}a.  The bootstrap method implies that the probability \t=139.2$\pm$0.1\deg\s is at least 60\% (see Fig. \ref{3227_chi_bs_pa}b), which is reasonably consistent with the relative likelihood statistics.  This kinematic major axis is consistent with the major axis position angle of the \htwo flux distribution, as well as with that of the larger scale optical emission. 

\begin{figure}[!htt]
\epsscale{1}
\plotone{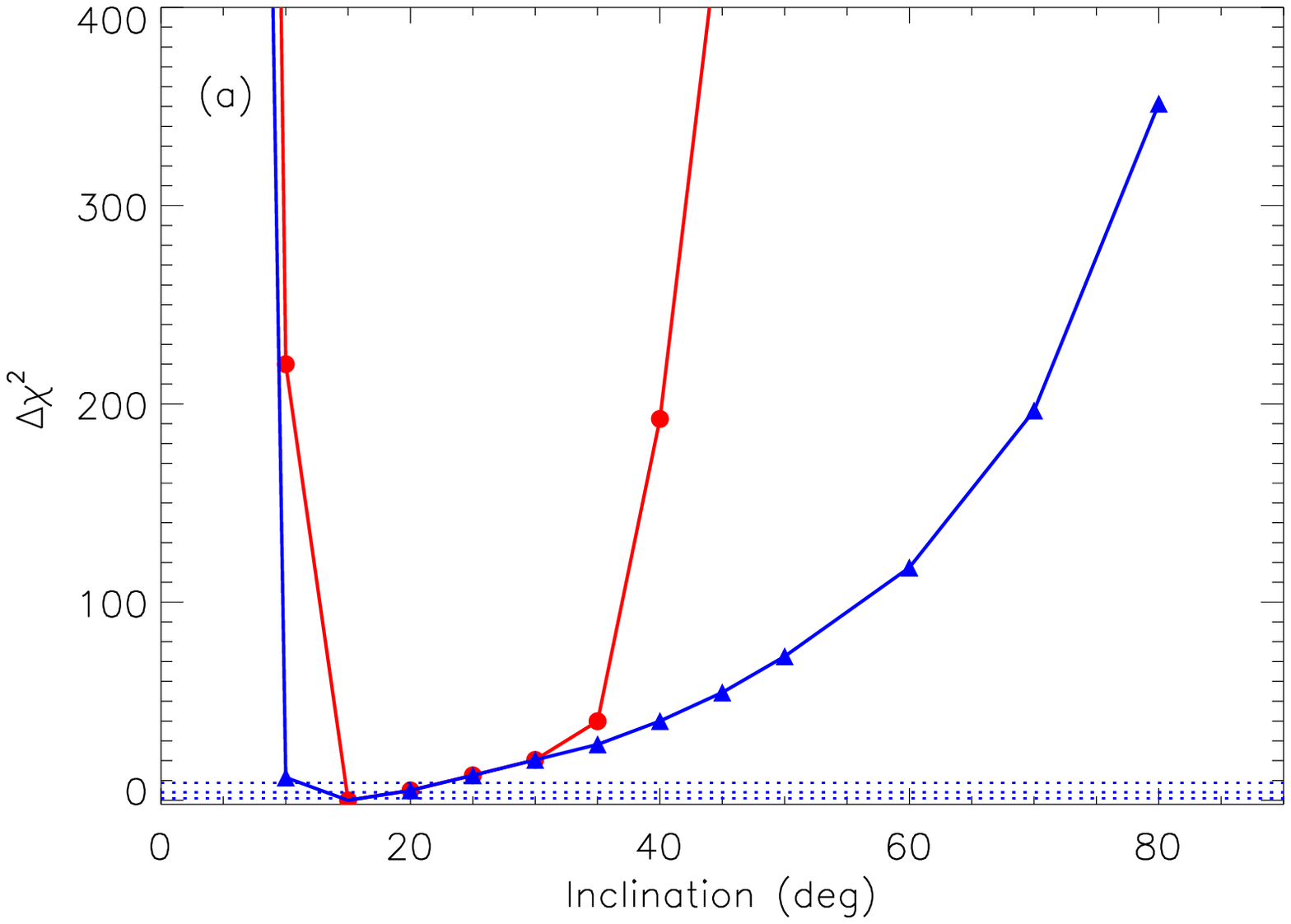}
\plotone{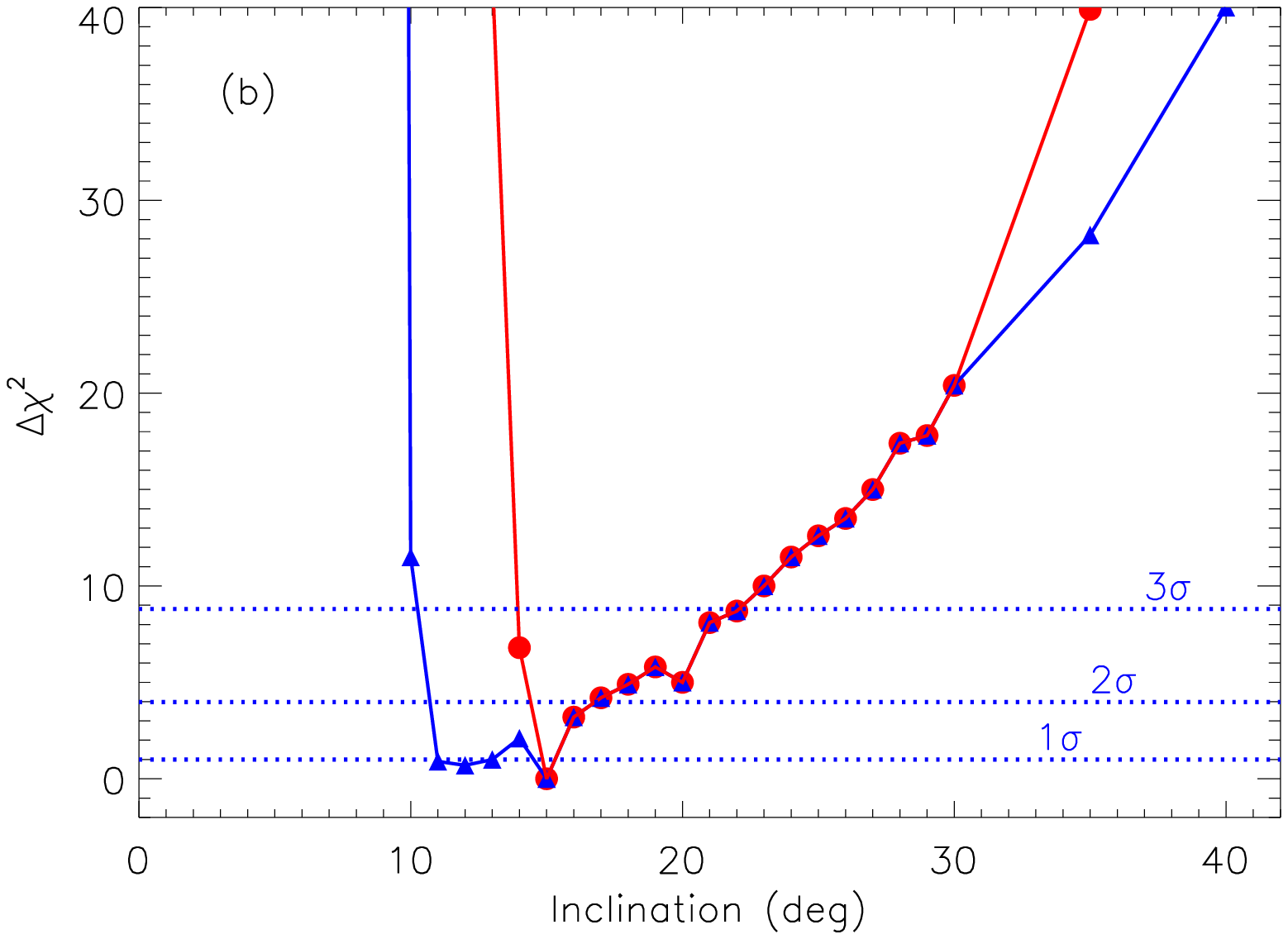}
\plotone{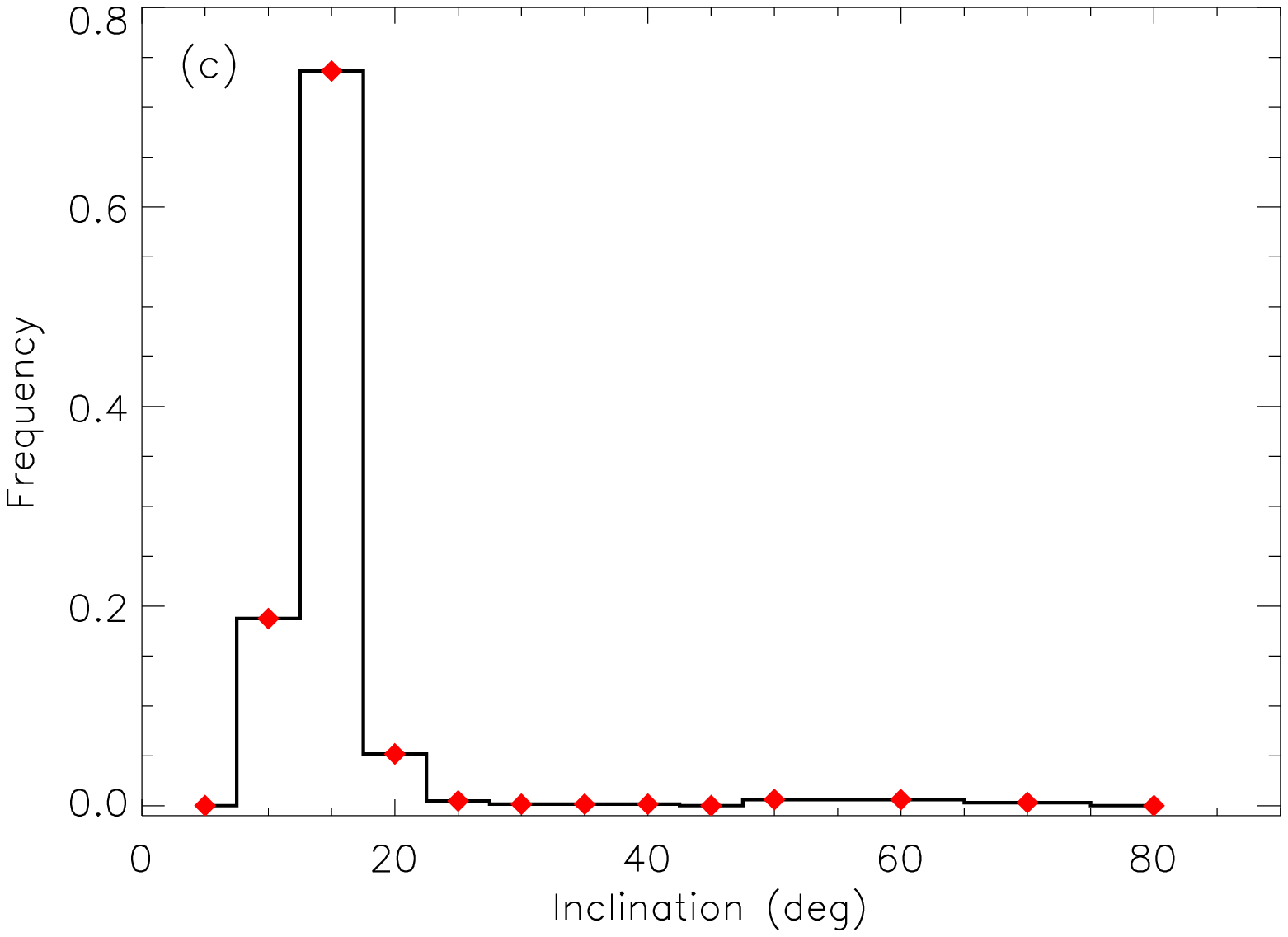}
\caption[NGC 3227: Best Fit Inclination Angle]{Best fit gas disk inclination angle, \inc, for NGC 3227. (a) Dependence of \ch\s on \inc, where the best fit model for each \inc\s is found by varying \mbh, \ml, and \t.  The circles are for \ml=0.3-1.1 \mlu, and the triangles are for \ml=0.1-2.0 \mlu.  The symbols indicate models that were run and the horizontal lines show the 1\signs, 2\signs, and 3\sig confidence intervals for one parameter.  (b) A detailed view of the lower \i\s values.  (c) The distribution of the best fit \inc\s for 1000 realizations of the half-sample bootstrap method with \ml=0.1-2.0 \mlu.  The specific model parameters fit to the data are indicated by the diamonds. \label{3227_chi_bs_i}}
\end{figure}

\subsubsection{Disk Inclination and Mass-to-Light Ratio}
Models with lower values of \inc\s are preferred, with an angle below 22\deg\s favored at the 3\sig confidence level (see Fig. \ref{3227_chi_bs_i}).  The range of best fit \inc\s based on the acceptable range of \ml\s (0.3-1.1 \mlu, see $\S$ 5.1.2), is \inc=14-22\deg\s at the 3\sig level.  If the wider range of \ml=0.1-2.0 \mlu\s is allowed, then \inc\s can go down to \inc=10-22\deg.  The bootstrap method confirms the preference of low inclination angles with greater than 70\% of the best fits finding \inc=15$\pm$2.5\deg, and greater than 97\% finding \inc$\le$20\deg\s (Fig. \ref{3227_chi_bs_i}c) when \ml=0.1-2.0 \mlu.  These low inclinations are in contrast to \inc=56\deg\s found at greater radii (\citealt{mundell95}).

\begin{figure}[!htt] 
\epsscale{1}
\plotone{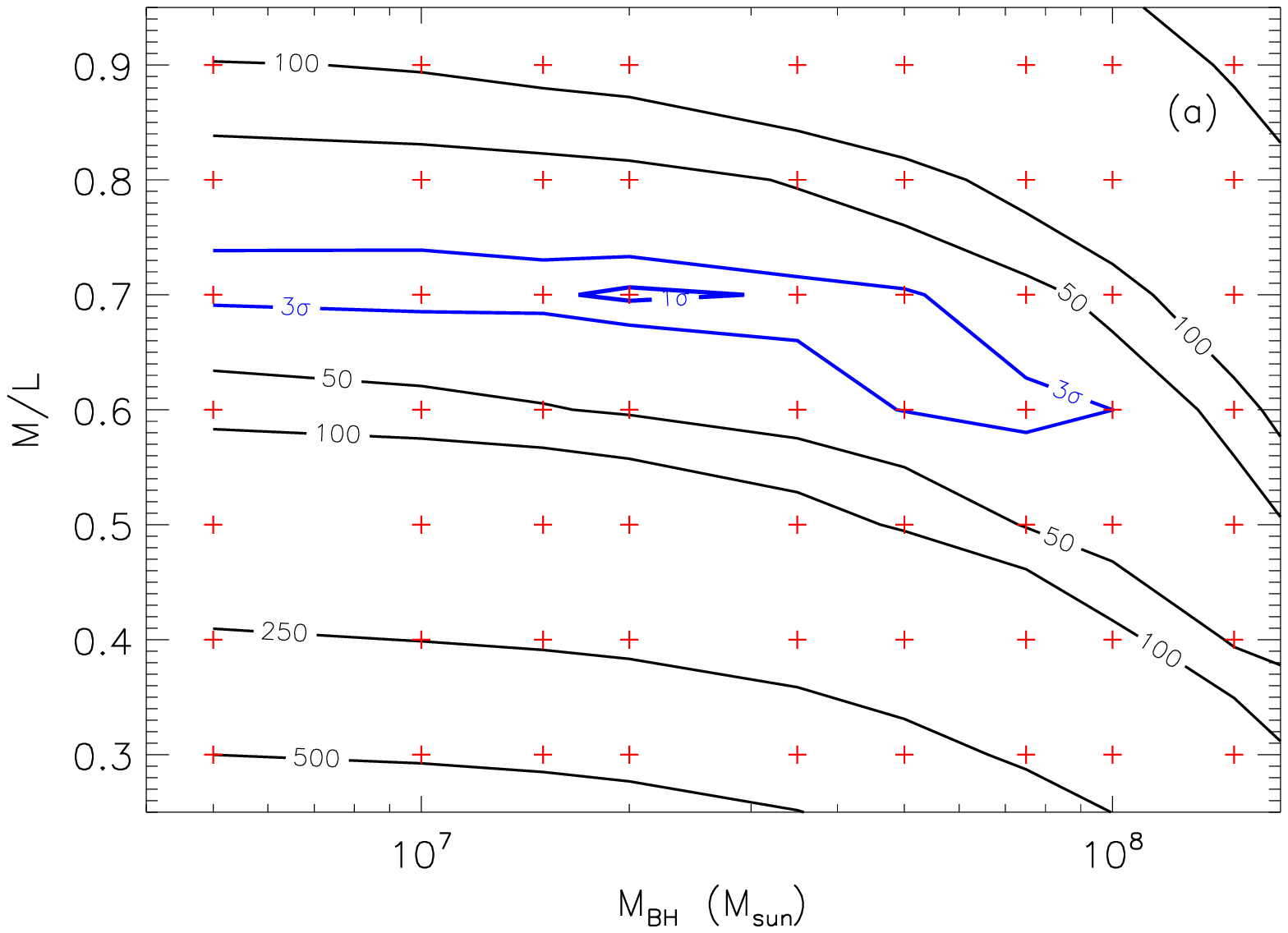}
\plotone{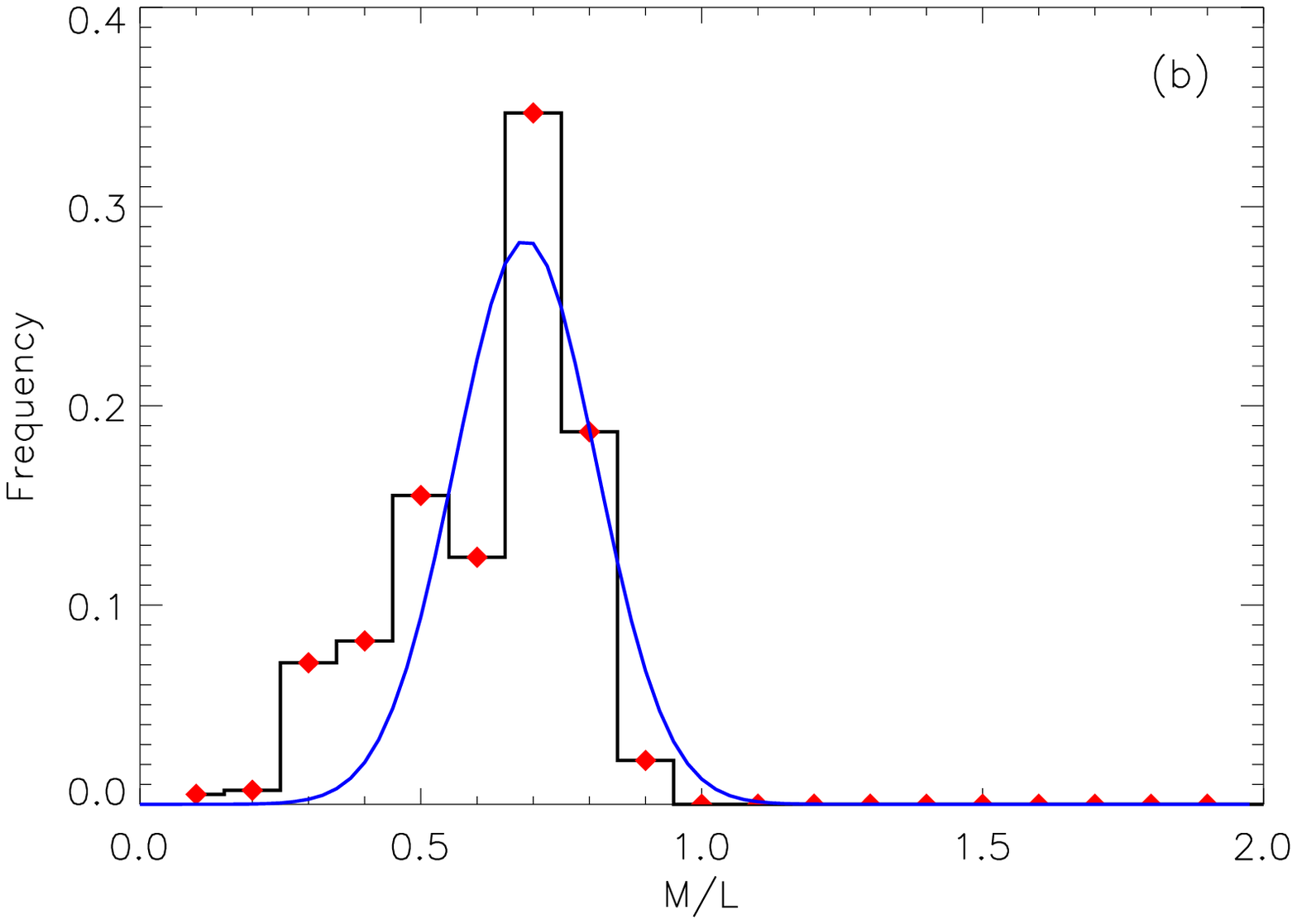}
\plotone{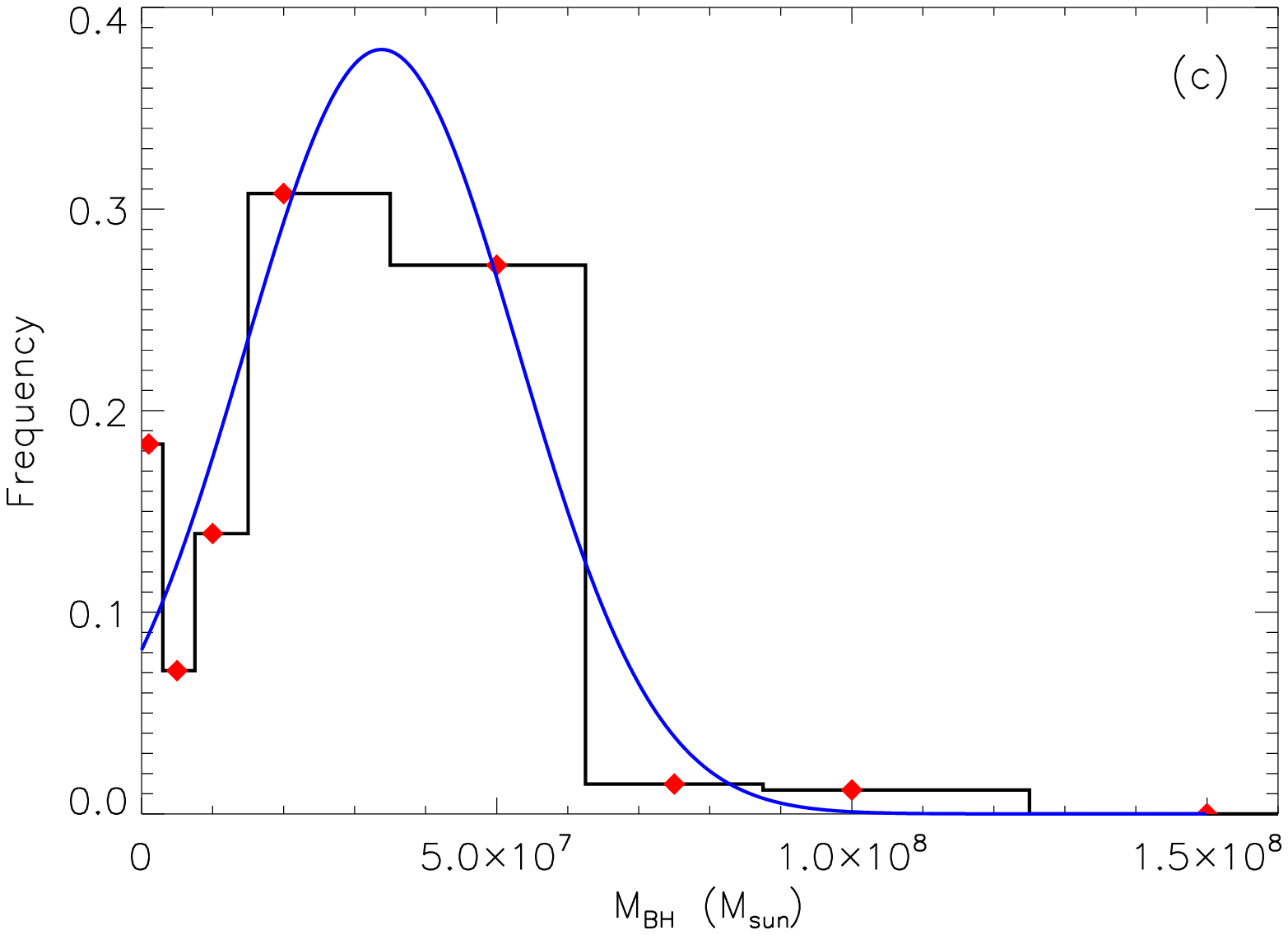}
\caption[NGC 3227: Best Fit \mbh\s and \ml]{Best fit \mbh\s and \ml\s for NGC 3227.  (a) Contours of constant \ch\s with \inc=20\deg\s and PA=139.2\deg, where \ch\s has been rescaled as discussed in the text.  The 1\sig and 3\sig confidence intervals (for two parameters) are labeled.   For comparison the distributions of the best fit (b) mass-to-light ratio and (c) \mbh\s for 1000 realizations of the half-sample bootstrap method are also shown.  The specific model parameters fit to the data are indicated by the diamonds and Gaussian fits to the distributions are also shown.  The best fit \mbh\s found with the bootstrap method assumes \ml=0.7 \mlu, \inc=20\deg, and \t=139.2\deg.  \label{3227_M_MLR}}
\end{figure}

\tabletypesize{\small}
\begin{deluxetable*}{ccccccl}
\tablecaption{Comparison of Different Models for NGC 3227 
\label{t_3227mbhi}} 
\tablewidth{0pt}
\tablehead{
\colhead{Model \tablenotemark{a}} &
\colhead{\inc} & 
\colhead{\mbh} &
\colhead{\ml} &
\colhead{Reduced \ch} & 
\colhead{DOF} &
\colhead{Comments \tablenotemark{a}} \\
\colhead{} &
\colhead{(\deg)} &
\colhead{(10$^{7}$ \Msun)} &
\colhead{(\mlu)} &
\colhead{}  &
\colhead{} &
\colhead{} \\
}
\startdata

A	&	10	&	50.0	&	1.90	&	14.598	&	1794	&  \n=3 \\
A	&	15	&	10.0	&	1.10	&	14.592	&	&		\\
A	&	20	&	2.0	&	0.70	&	14.595	&	&		\\
A	&	25	&	2.0	&	0.45	&	14.599	&	&		\\
A	&	30	&	3.5	&	0.30	&	14.603	&	&		\\
A	&	35	&	1.5 	&	0.25	&	14.608	&	&		\\
A	&	40	&	2.0	&	0.20	&	14.614	&	&		\\
A	&	45	&	1.5	&	0.20	&	14.622	&	&		\\
A	&	50	&	1.5	&	0.15	&	14.632	&	&		\\
A	&	60	&	1.0	&	0.15	&	14.657	&	&		\\
A	&	70	&	0.0	&	0.20	&	14.702	&	&		\\
A	&	80	&	0.0	&	0.35	&	14.788	&	&		\\
\cline{1-7}
B	&	20	&	10.0	&	0.60	&	14.621	&	1794	& \n=2\\
C	&	20	&	2.0	&	0.70	&	14.601	&	1794	& w/o flux \\
\cline{1-7} 
D	&	20	&	2.0	&	0.60	&	26.034	&	730	& \htwo 2.4066 \\
E	&	20	&	1.5	&	0.70	&	29.772	&	933	& \htwo 2.4237 \\

\enddata
\tablenotetext{a}{All models have \t=139.2\deg and have a stellar light distribution and emission line flux distribution as follows.  For \htwo \lam2.1218, Model A - S\`{e}rsic \n=3 and the two component Gaussian function, Model B - S\`{e}rsic \n=2 and the two component Gaussian function, and Model C - S\`{e}rsic \n=3 and a constant distribution.  For \htwo \lam2.4066, Model D - S\`{e}rsic \n=3 and the two component Gaussian function.  For \htwo \lam2.4237, Model E - S\`{e}rsic \n=3 and the two component Gaussian function.  The degrees of freedom of each model is given in the DOF column.}

\end{deluxetable*}

For \inc=20\deg\s and \t=139.2\deg, the best fit for the remaining parameters is \mbh= 2.0$^{+1.0}_{-0.4}\times10^{7}$ \Msun\s and \ml=0.70$\pm$0.05 \mlu\s at the 1\sig confidence level (using two interesting parameters; Fig. \ref{3227_M_MLR}a).  The distributions of the best fit \ml\s and \mbh\s values found with the bootstrap method are also shown in Fig. \ref{3227_M_MLR} and are given in Table \ref{t_results}.  The most frequently best fit \ml\s value is 0.7 \mlu, confirming the value given by the relative likelihood method, and the best fit \ml\s is found to be between 0.5-0.8 \mlu\s 65\% of the time.  Although the \ml\s distribution is not necessarily Gaussian, by fitting a Gaussian to the distribution a formal 1\sig error for the best fit value is estimated.  In this case a Gaussian fit implies a best fit of \ml=0.69$\pm$0.13 \mlu.  The distribution of best fit \mbh\s values is also consistent with the relative likelihood statistics, giving a best fit of \mbh=2$\times$10$^{7}$ \Msun\s over a third of the time with \ml=0.7 \mlu, and the distribution gives \mbh=1-5$\times$10$^{7}$ \Msun\s 72\% of the time.  A Gaussian fit to the \mbh\s distribution indicates that the best fit \mbh=3.4$\pm$1.9$\times$10$^{7}$ \Msun.  Also of note is the fact that no black hole is the best fit in 18\% of the cases, meaning the presence of a black hole is significant at about the 2\sig level. 

\begin{figure}[!b] 
\epsscale{1}
\plotone{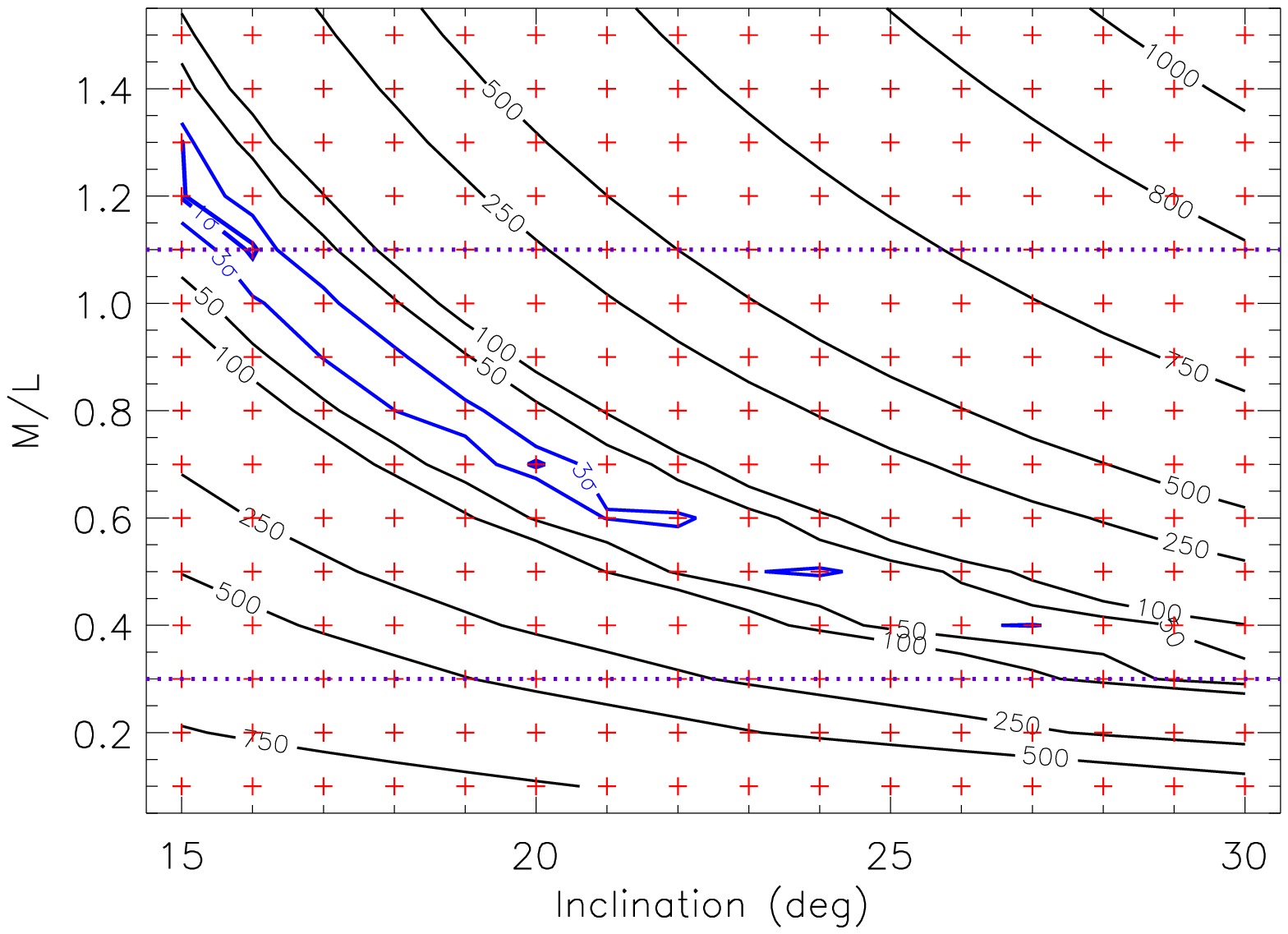}
\caption[NGC 3227: \inc\s versus \ml]{Contours of constant \ch\s for NGC 3227 with \mbh=2.0$\times10^{7}$ \Msun\s and \t=139.2\deg, where \ch\s has been rescaled as discussed in the text.  The 1\sig and 3\sig confidence intervals (for two parameters) are labeled.  The limits of the accepted range in \ml\s are indicated by the dashed lines. \label{3227_INCvMLR}}
\end{figure}

The best fit \mbh\s is largely independent of both \inc\s and \ml, indicating that the inner velocity gradient cannot be matched without including a point-like mass distribution, such as a BH.  As discussed in $\S$ 5.2, the combination of \ml\s and \inc\s is well constrained, and the two parameters compensate for each other such that the best fit \mbh\s remains unchanged (Fig. \ref{3227_INCvMLR}).  In NGC 3227, over a reasonable range of \inc\s from 20-35\deg\s the best fit \mbh\s is constant at 2.0$\times10^{7}$ \Msun\s with a \ml\s varying from 0.70 to 0.45 \mlu.  Table \ref{t_3227mbhi} lists values of \mbh\s for a wide range of \inc\s values.

\subsubsection{Modeling of \htwo \lam2.4066, and \lam2.4237}
\begin{figure}[!ht] 
\epsscale{1}
\plotone{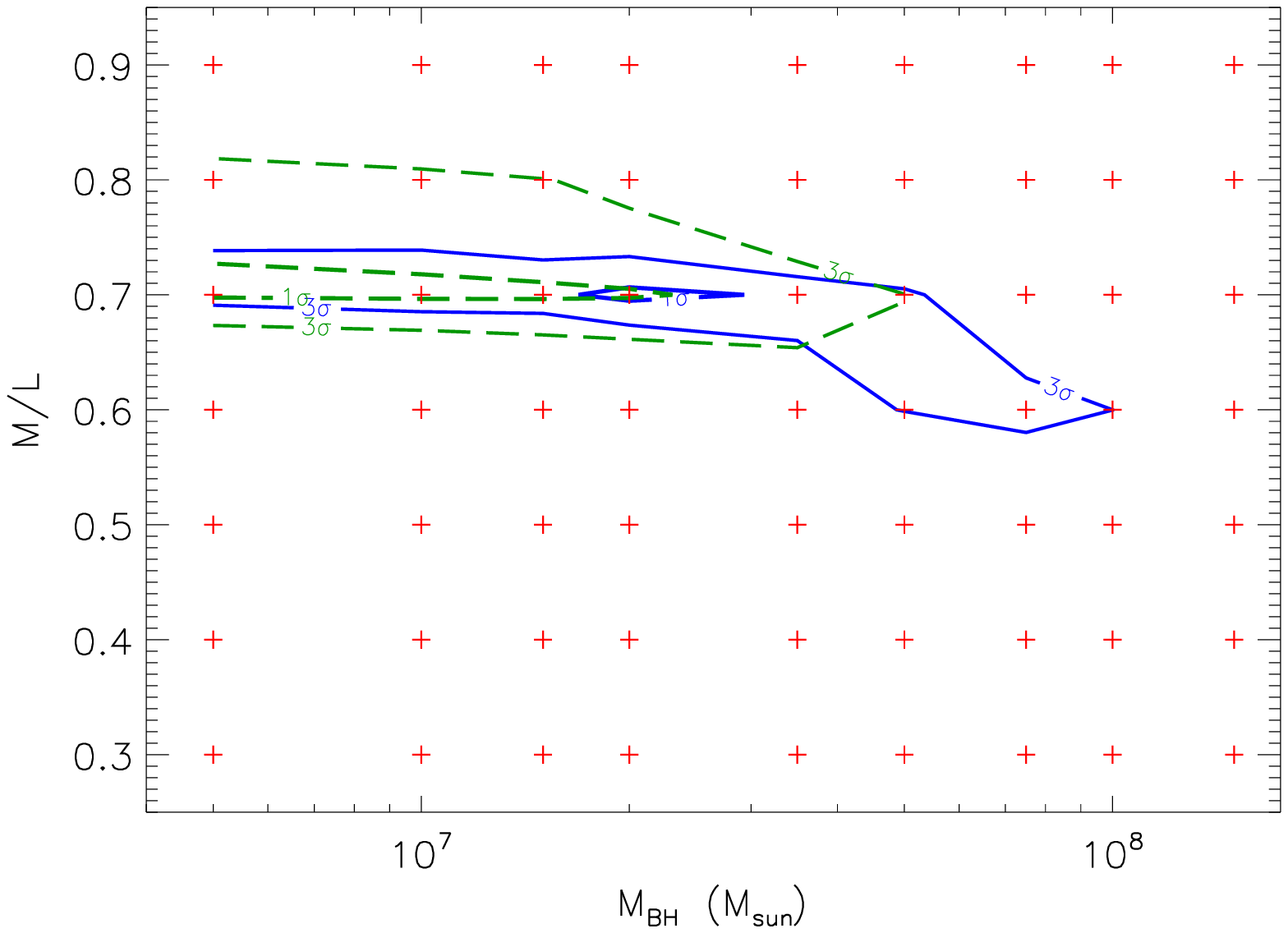}
\caption[NGC 3227: Best Fit \mbh\s for Different \htwo lines]{Contours of constant \ch\s for \htwo \lam2.1218 and \lam2.4237 in NGC 3227 with \inc=20\deg\s and PA=139.2\deg, where \ch\s has been rescaled as discussed in the text.  The 1\sig and 3\sig confidence intervals (for two parameters) for \htwo \lam2.1218 (solid curves) and \lam2.4237 (dashed curves) are shown.  \label{3227_MvsMLR_all}}
\end{figure}

The best fit model parameters found for \htwo \lam 2.1218 are confirmed with modeling of the other two \htwo emission lines present in the spectra.  Both \htwo \lam2.4066 and \lam2.4237 have relatively strong emission, but their longer wavelengths place them in a region of strong atmospheric absorption, and the emission line profiles of each are affected by an absorption line just to the red of their expected wavelength.  The effects of this are most prominent in the central 0\as.2, where the residuals of the atmospheric absorption are greatest and significantly bias the Gaussian fits to the emission lines.  The fits of these lines are thus more uncertain then those of \htwo \lam2.1218, and the resulting velocity fields are considerably noisier.  Consequently, only upper limits on \mbh\s can be determined from these lines, but both lines nevertheless confirm the best fit \mbh, \inc, \ml, and \t\s found with the more reliable \htwo \lam 2.1218 line described above.  Fig. \ref{3227_MvsMLR_all} is a comparison of the best fit \mbh\s and \ml\s from modeling of \htwo \lam2.4237 and \lam 2.1218. 

\subsubsection{Stellar Light Distribution}

\begin{figure}[!t] 
\epsscale{1}
\plotone{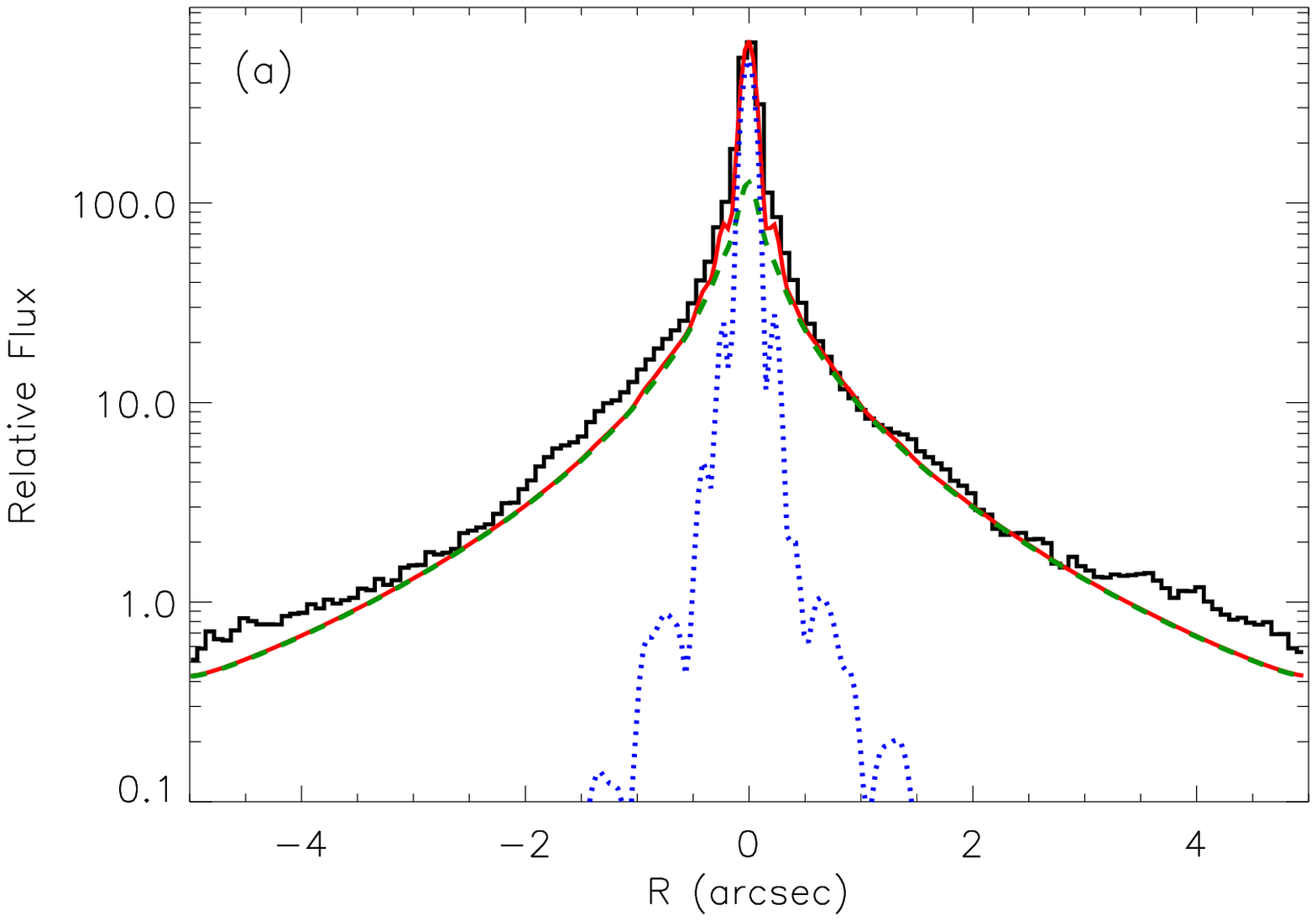}
\plotone{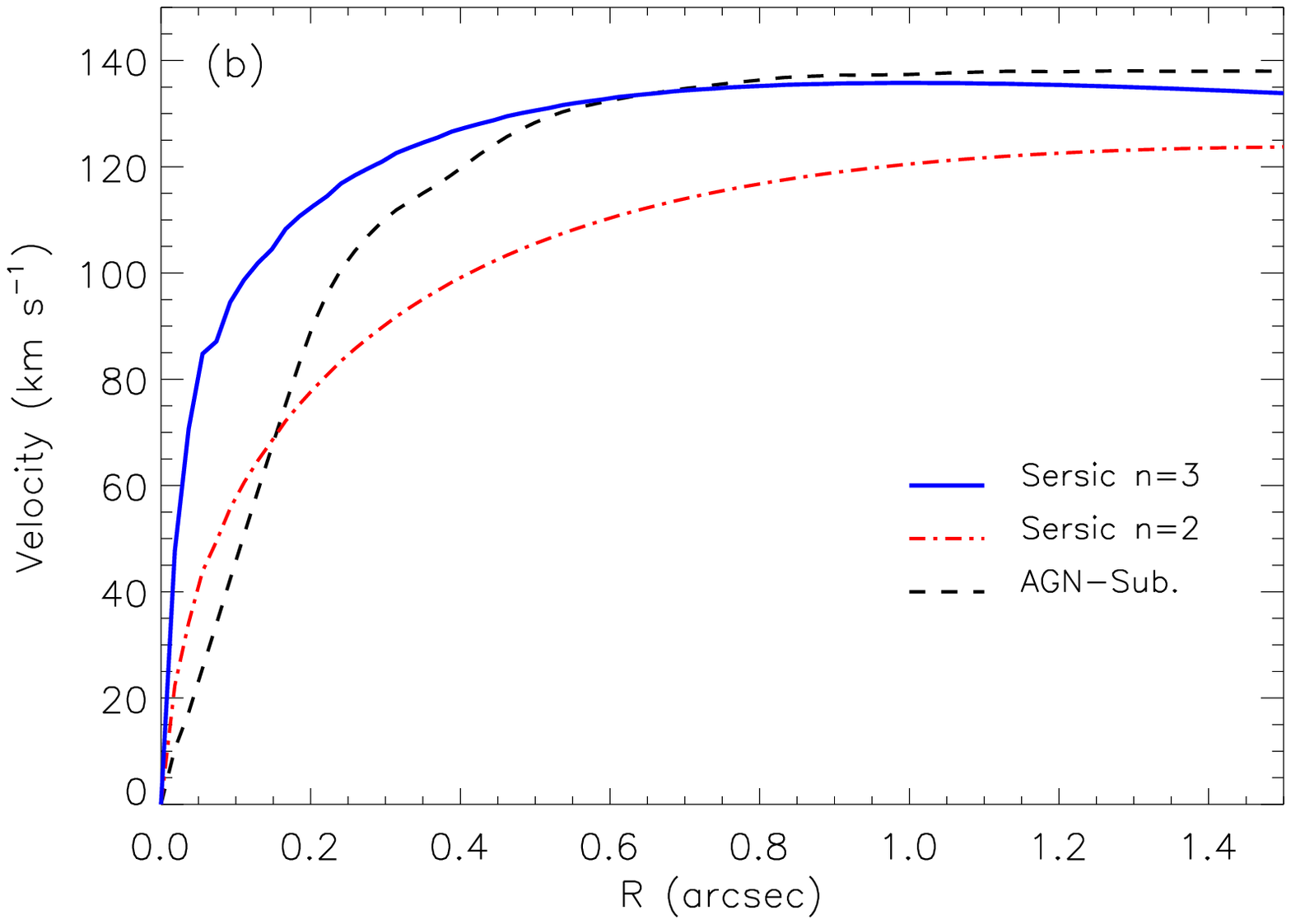}
\plotone{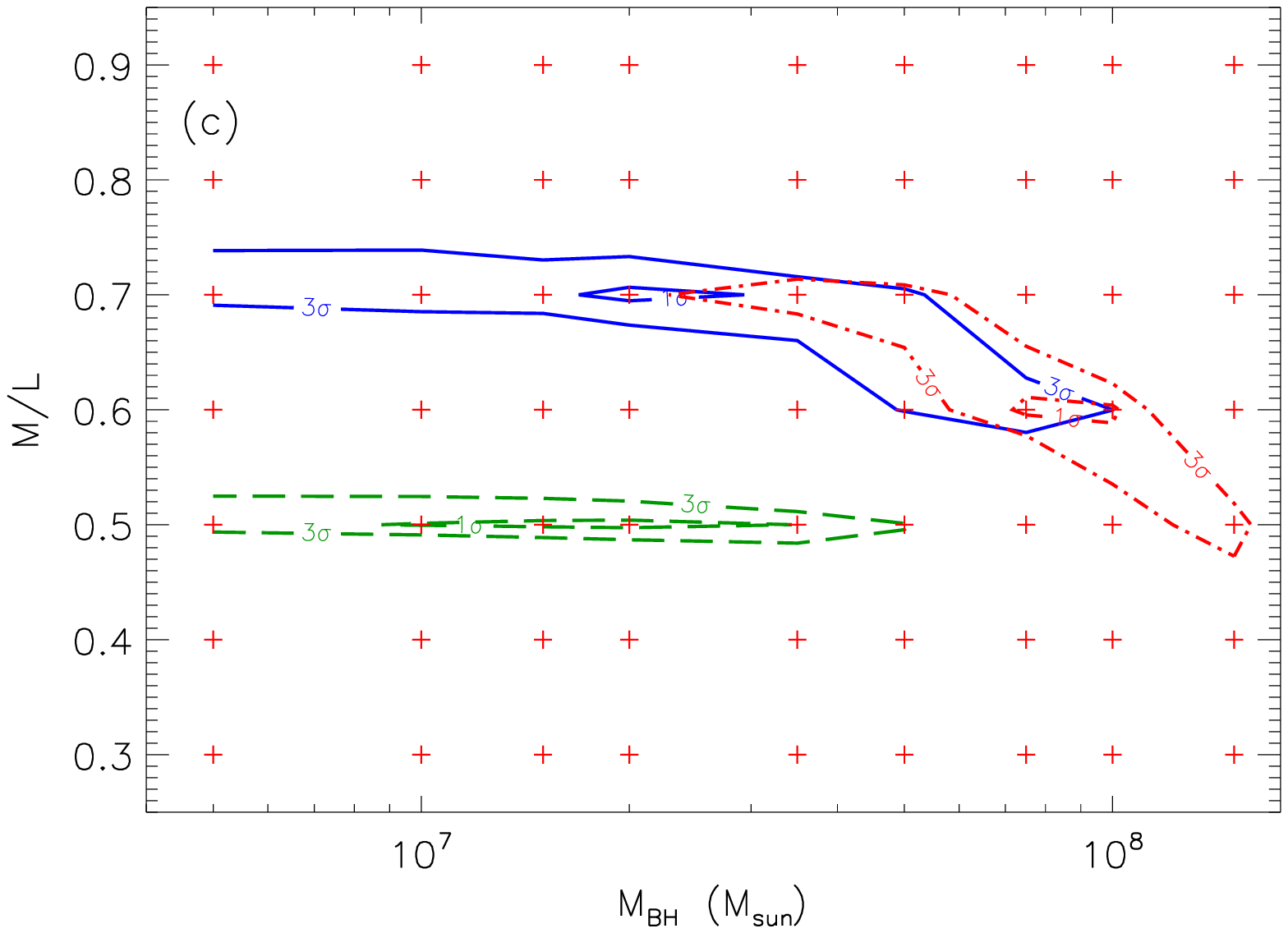}
\caption[NGC 3227: Stellar Light Distribution]{The effect of the stellar light distribution in NGC 3227 on the best fit model.  (a)  The best fit to the stellar light distribution (solid curve) in the NICMOS F160W image (histogram) was found with a S\`{e}rsic \n=3 function (dashed curve) plus a point source represented by a Tiny Tim PSF (dotted curve). (b) The resulting rotation curves in the plane of the gas disk are shown for \n=3 and \n=2 S\`{e}rsic functions and the AGN-subtracted light distribution (as labeled in the legend) assuming an inclination angle of 56\deg\s and \ml=0.5 \mlu. (c) Contours of constant \ch\s at the 1\sig and 3\sig confidence intervals, for two parameters, are shown for the range of the S\`{e}rsic \n=3 fit (solid for the minimum, and dashed for maximum stellar light) and for the S\`{e}rsic \n=2 fit (dash-dotted), where \ch\s has been rescaled as discussed in the text.  \label{3227_sersic}}
\end{figure}

The best fit \mbh\s for NGC 3227 is, as expected, dependent on the choice of stellar light distributions.  The best S\`{e}rsic plus point source fit to the NICMOS F160W light distribution is with a \n=3 S\`{e}rsic function, with reduced \ch=0.7 (see Fig. \ref{3227_sersic}a).  The best fit S\`{e}rsic parameters are given in Table \ref{t_sersic} and the resulting rotation due to the corresponding stellar mass distribution is shown in Fig. \ref{3227_sersic}b.  The light distribution remaining after subtracting the non-stellar AGN continuum emission (see $\S$ 5.1.2) is found to be relatively flat at r$<$0\as.2.  It is more likely, however, that the stellar density continues to increase toward the center.  For this reason a S\`{e}rsic function is believed to provide a more realistic stellar light distribution for NGC 3227.  Nonetheless, the AGN-subtracted light distribution provides a guide for what the distribution should be outside of this radius, and even more definitively outside of the radius affected by the PSF (r$>$ 1\as).  The S\`{e}rsic function most similar to the AGN-subtracted light distribution at these radii is also the \n=3 function.  The S\`{e}rsic \n=3 fit is therefore used as the best representation of the stellar light distribution because it provides the best fit to the F160W image and the AGN-subtracted light distribution, especially at radii free of significant complications from the PSF.    

The best fit model found using a \n=3 S\`{e}rsic function stellar light distribution gives, as stated before, \mbh=2.0$^{+1.0}_{-0.4}\times10^{7}$ \Msun\s and \ml=0.70$\pm$0.05 \mlu\s at the 1\sig confidence level for \inc=20\deg.  The uncertainty in the \n=3 S\`{e}rsic fit allows a change in \ml\s down to 0.5 \mlu, but does not change the best fit \mbh\s estimate, which can be seen in Fig. \ref{3227_sersic}c.

If instead the \n=2 S\`{e}rsic function is used, then the rotation due to the stellar component in the inner radii is not as steep (Fig. \ref{3227_sersic}b), resulting in the need for a higher \mbh\s to generate the additional velocity.  In this case the best fit is found to be \mbh=1.0$^{+0.5}_{-2.5}\times10^{8}$ \Msun\s and \ml=0.60$\pm$0.05 \mlu\s with \inc=20\deg.  However, this fit is ruled out at greater than the 99.99\% level compared to the best fit with the \n=3 function (see Table \ref{t_3227mbhi}).  The \n=1 and \n=4 S\`{e}rsic functions are not considered because they do not fit the F160W light distribution nearly as well (reduced \ch$\sim$1.5 compared to \ch=0.7 with \n=3).  The \n=1 function misses a significant amount of the light in the inner 1\as, as well as beyond r=3\as.  The \n=4 function rises so steeply in the middle that only a very weak AGN component is needed and this fit is at the expense of missing light beyond r$>$1\as.

\begin{deluxetable*}{lcccccc}
\tablecaption{Point Source Plus S\`{e}rsic Fits to the Stellar Light Distributions \label{t_sersic}} 
\tablewidth{0pt}
\tablehead{ 
\colhead{} &
\multicolumn{3}{c}{S\`{e}rsic Function} &
\colhead{} &
\colhead{} \\
\cline{2-4} \\
\colhead{} & 
\colhead{} &
\colhead{$\mu_{e}$} &
\colhead{R$_{e}$} &
\colhead{$M_{AGN}$} &
\colhead{F$_{S}$} &
\colhead{F$_{AGN}$}	\\
\colhead{Galaxy} & 
\colhead{\n} &
\colhead{(H mags)} &
\colhead{(\as)} &
\colhead{(H mags)} &
\colhead{(1\as)} &
\colhead{(1\as)}	\\
}
\startdata

NGC 3227	&	3	&	14.40	&	1.55	&	-25.28	&	0.80	&	0.20	\\
NGC 4151	&	3	&	14.78	&	2.31	&	-27.56	&	0.29	&	0.71	\\
NGC 7469\tablenotemark{a}	&	2	&	14.96	&	1.81	&	-30.03	&	0.32	&	0.68	\\				

\enddata
\tablenotetext{a}{This fit is not used in the modeling; see the text for further details.}


\end{deluxetable*}

\begin{figure}[!t] 
\epsscale{1}
\plotone{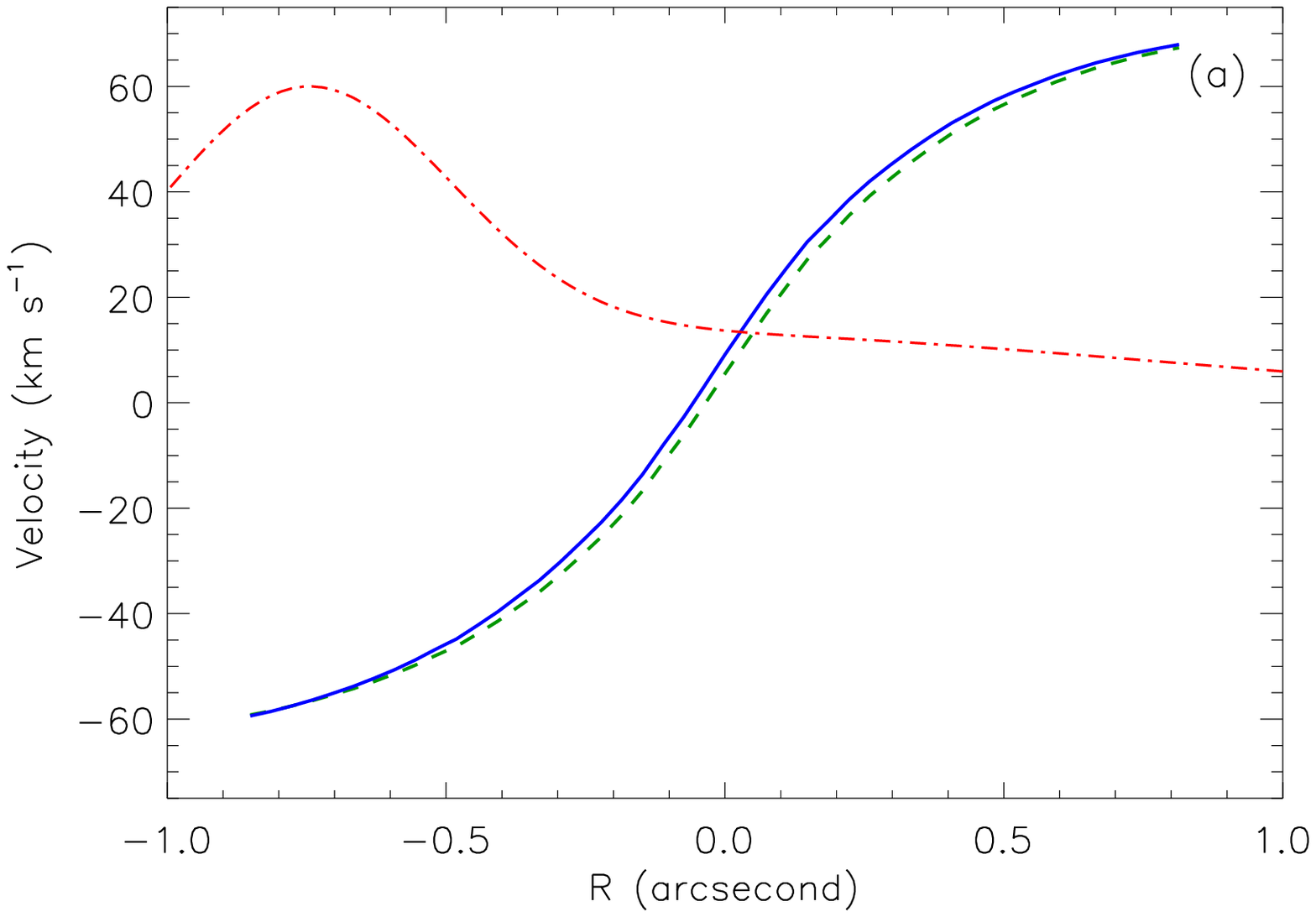}
\plotone{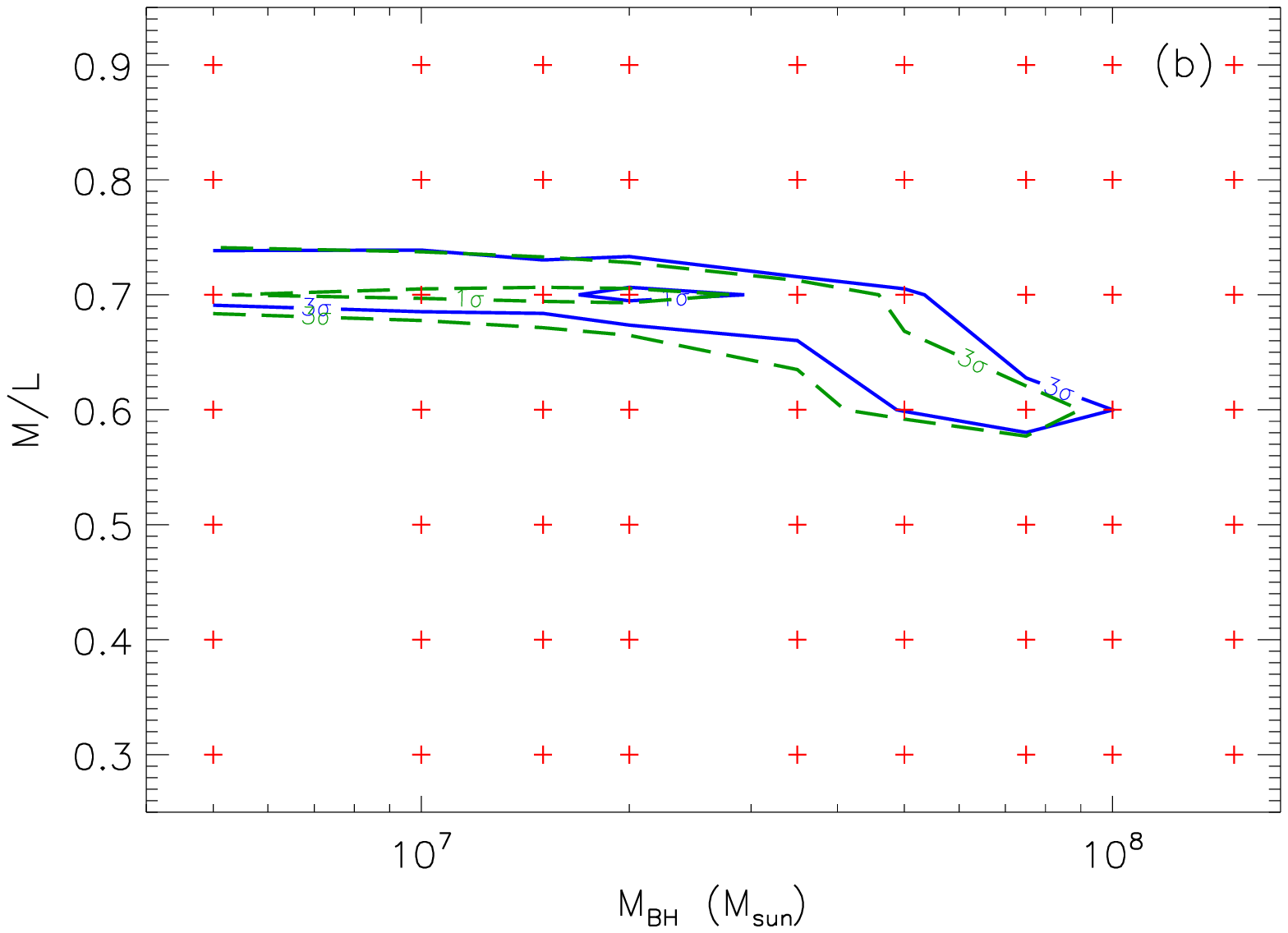}
\caption[NGC 3227: Different Flux Distributions]{The effect of the \htwo flux distribution in NGC 3227 on the best fit model.  (a) Comparison of the rotation curve expected along a particular slit position (PA=163\deg going through the nucleus) with the parameterized intrinsic flux distribution (solid curve; see text for details) and a constant flux distribution (dashed curve). The relative flux distribution along the slit is represented by the dash-dotted curve. (b) Contours of constant \ch\s when taking into account the asymmetric flux distribution as discussed in the text (solid curves), and when a constant flux distribution is used (dashed curves).  The 1\sig and 3\sig confidence intervals, for two parameters, are shown. \label{3227_rc_chi_flux}}
\end{figure}

\subsubsection{Emission Line Flux Distribution}
The Gaussian parameters that produce the best fit to the observed \htwo flux distribution after convolution with the PSF are given in Table \ref{t_gflux}.  For NGC 3227 two components are used: a component offset 0\as.5 to the southeast as well as a central component 4.7 times brighter.  An example of the change in the observed velocities due to including the intrinsic flux distribution rather than a constant distribution is shown in Fig. \ref{3227_rc_chi_flux}a.  The effect of including the asymmetric distribution of the \htwo emission is to decrease the \ch of the fit, but it does not change the best fit \mbh\s (see Fig. \ref{3227_rc_chi_flux}b and Table \ref{t_3227mbhi}).

\subsubsection{Region Fit and Data Quality Considerations}
In NGC 3227, the \htwo emission is strong throughout the FOV and eliminating the worst data (e.g. all data with PCC $<$ 0.5) has little effect on the best fit model parameters, although it does decrease \ch.  All results presented for NGC 3227 have a data quality cut of PCC $>$ 0.5 for \htwo \lam 2.1218 and PCC $>$ 0.3 for \htwo \lam 2.4066 and \lam 2.4237.  The lower PCC constraint is used for the longer wavelength \htwo lines in order to include the poorly measured central 0\as.2 (due to residuals from strong atmospheric absorption lines).  Furthermore, the best fit model parameters for all three \htwo transitions are not dependent on r$_{fit}$, and for all results presented the full 2\as\s FOV was included in the model fits.

\subsubsection{Summary of \mbh\s Estimate for NGC 3227}
\begin{figure*}[!htt]
\epsscale{0.5}
\plotone{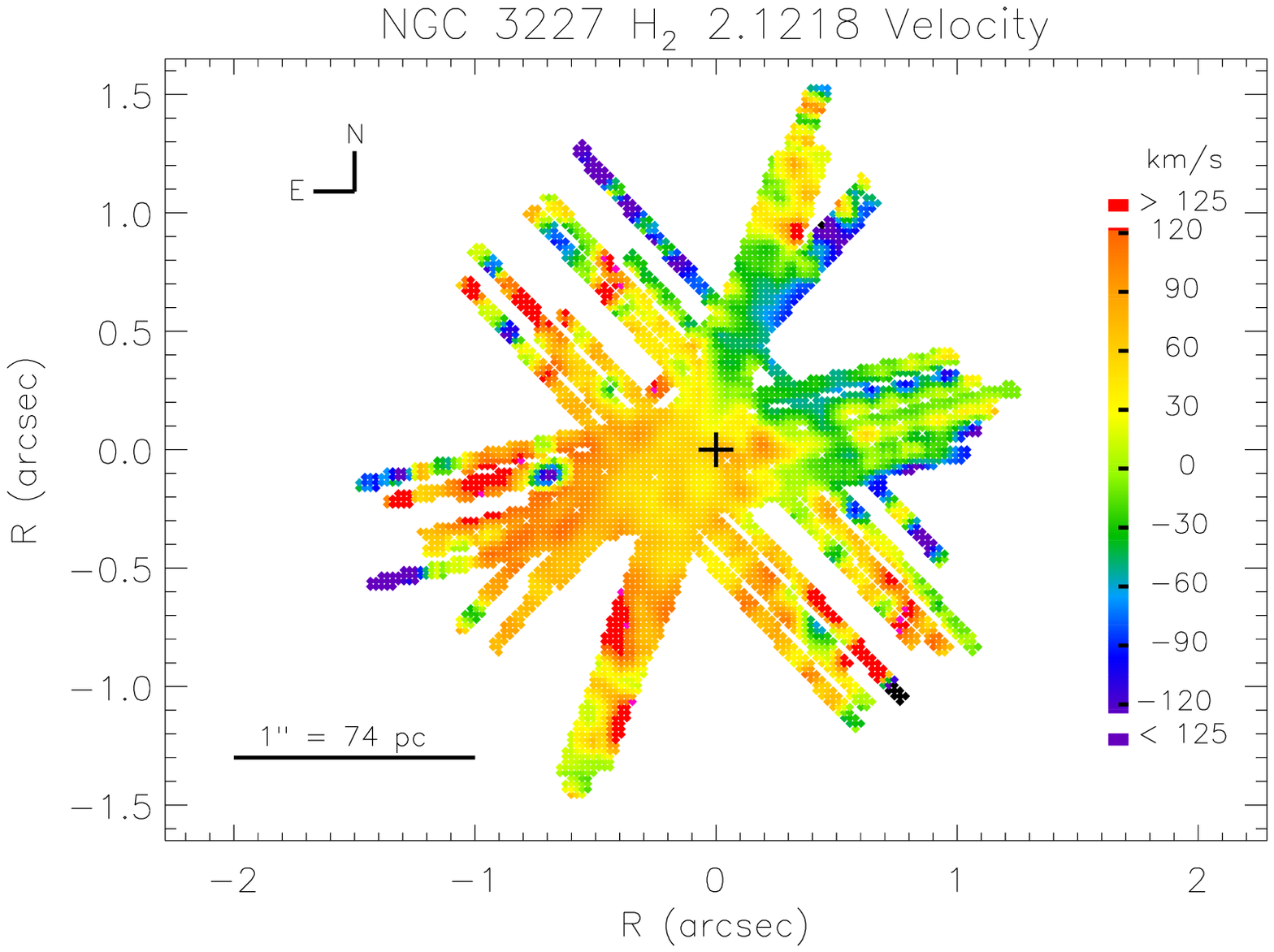}
\plotone{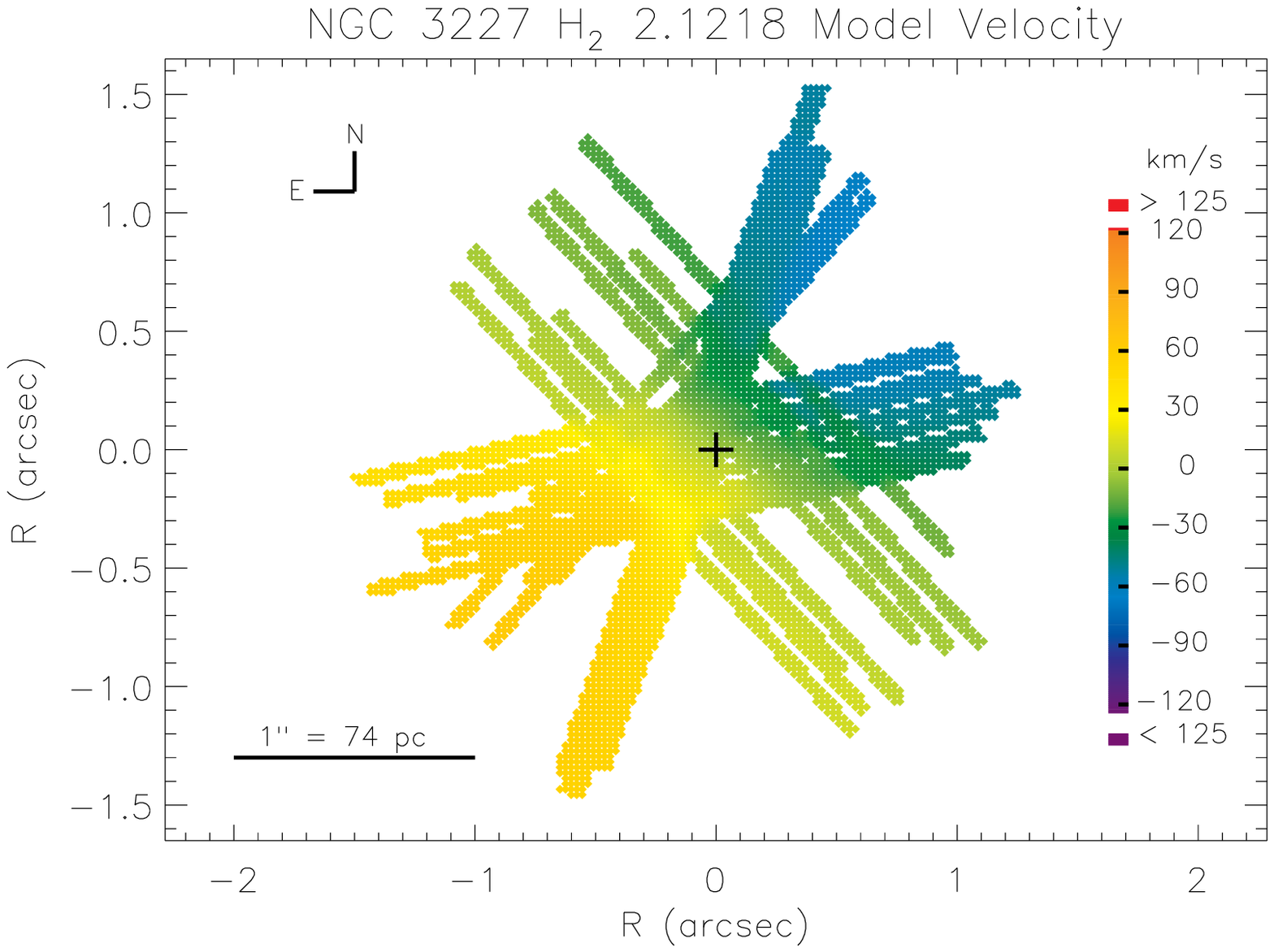}
\plotone{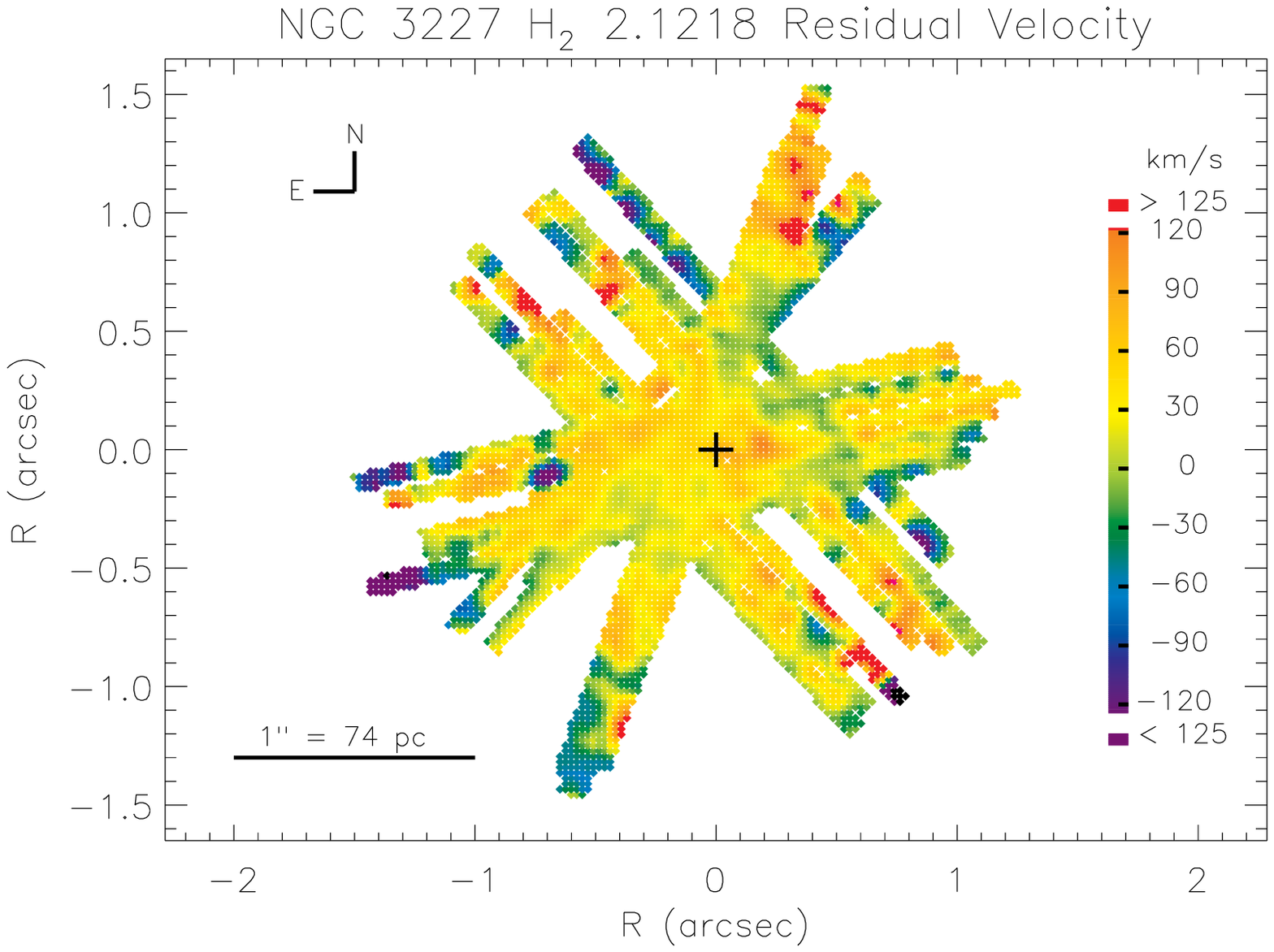}
\caption[NGC 3227: 2-D Map of Model Velocity Field and Residuals]{2-D map of the NGC 3227 \htwo 2.1218 measured velocity field, model velocity field, and the residuals of the model minus the data, as indicated by the labels at the top of each plot.  The model has \mbh=2.0$\times10^{7}$ \Msun\s with \inc=20\deg, \ml=0.70 \mlu, \t=139.2\deg\s and uses the \n=3 S\`{e}rsic stellar light distribution. \label{3227_maps}}
\end{figure*}

In summary, the best fit \mbh\s in NGC 3227 is found to be \mbh=2.0$^{+1.0}_{-0.4}\times10^{7}$ \Msun\s with \inc=20\deg, \ml=0.70$\pm$0.05 \mlu, \t=139.2$\pm$0.5\deg\s and with the \n=3 S\`{e}rsic stellar light distribution.  These results are confirmed with the half-sample bootstrap method, which indicates the best fit \mbh=3.4$\pm$1.9$\times10^{7}$ \Msun\s with \ml=0.7 \mlu.  A comparison of the estimates for all four model parameters based on the different statistical methods is given in Table \ref{t_results}.  2-D maps of the measured and model velocity fields, as well as the residuals, are shown in Fig. \ref{3227_maps}.  The best fit \mbh\s is not dependent on \t, and \ml\s and \inc\s compensate for each other to keep the best fit \mbh\s relatively constant.  In addition, the best fit \mbh\s is not dependent on the emission line flux distribution.  Being generous with the stellar light distribution and considering the \n=2 S\`{e}rsic stellar light distribution, the best fit model changes to \mbh=1.0$^{+0.5}_{-2.5}\times10^{8}$ \Msun, but this fit is ruled out at greater than the 99.99\% confidence level compared to the best fit model using the \n=3 distribution. 

The estimated \mbh\s of 2.0$^{+1.0}_{-0.4}\times10^{7}$ \Msun\s agrees well with other estimates in the literature for NGC 3227.  Using CO and HCN emission at millimeter wavelength to trace the gas dynamics, \citet{schinnerer00} found \mbh$> 2 \times 10^{7}$ \Msun.  In addition, \citet{davies06} recently measured the stellar dynamics using an integral field spectrometer, and found \mbh=0.7-2$\times 10^{7}$ \Msun.  In the case of NGC 3227, it appears that both gas and stellar dynamics result in consistent \mbh\s estimates.  A comparison of directly measured \mbh\s estimates and those from the commonly used indirect measurement techniques, such as reverberation mapping, will be discussed in $\S$ 6.5.

\tabletypesize{\small}
\begin{deluxetable*}{lccccc}
\tablecaption{Intrinsic Emission Line Flux Distribution Gaussian Components \label{t_gflux}} 
\tablewidth{0pt}
\tablehead{
\colhead{Galaxy} & 
\colhead{Component} &
\colhead{Peak Offset} &
\colhead{Peak Scaling} &
\colhead{FWHM (\as)} &
\colhead{Ellipticity} \\ 
\colhead{} & \colhead{} & \colhead{$\Delta$$\alpha$(\as),$\Delta$$\delta$(\as)} & \colhead{} & \colhead{} & \colhead{}
}
\startdata

NGC 3227	&	1	&	-0.13,-0.90	&	0.18	&	0.93	& 0.4 \\
		&	2	&	-0.58,-0.24	&	0.82	&	0.52	& 1.0 \\
NGC 4151	&	1	&	0.0,0.0	&	1.0	&	1.85	& 1.0 \\
NGC 7469	&	1	&	0.0,-0.37	&	0.5	&	0.19	& 1.0	\\
		&	2	&	0.0,-0.37	&	0.5	&	0.37	& 1.0	\\				

\enddata

\end{deluxetable*}

\subsection{Modeling Results for NGC 7469}
The rotation detected in the \htwo \lam2.1218 velocity field of NGC 7469 is well measured; however, uncertainty in the stellar gravitational field and the greater distance of the galaxy make detection of a BH more challenging in this Seyfert 1 galaxy.  The \mbh\s estimate based on the gas dynamics, and its dependence on the other model parameters, the model input, and the data selection are discussed in detail below.

\begin{figure}[!t] 
\epsscale{1}
\plotone{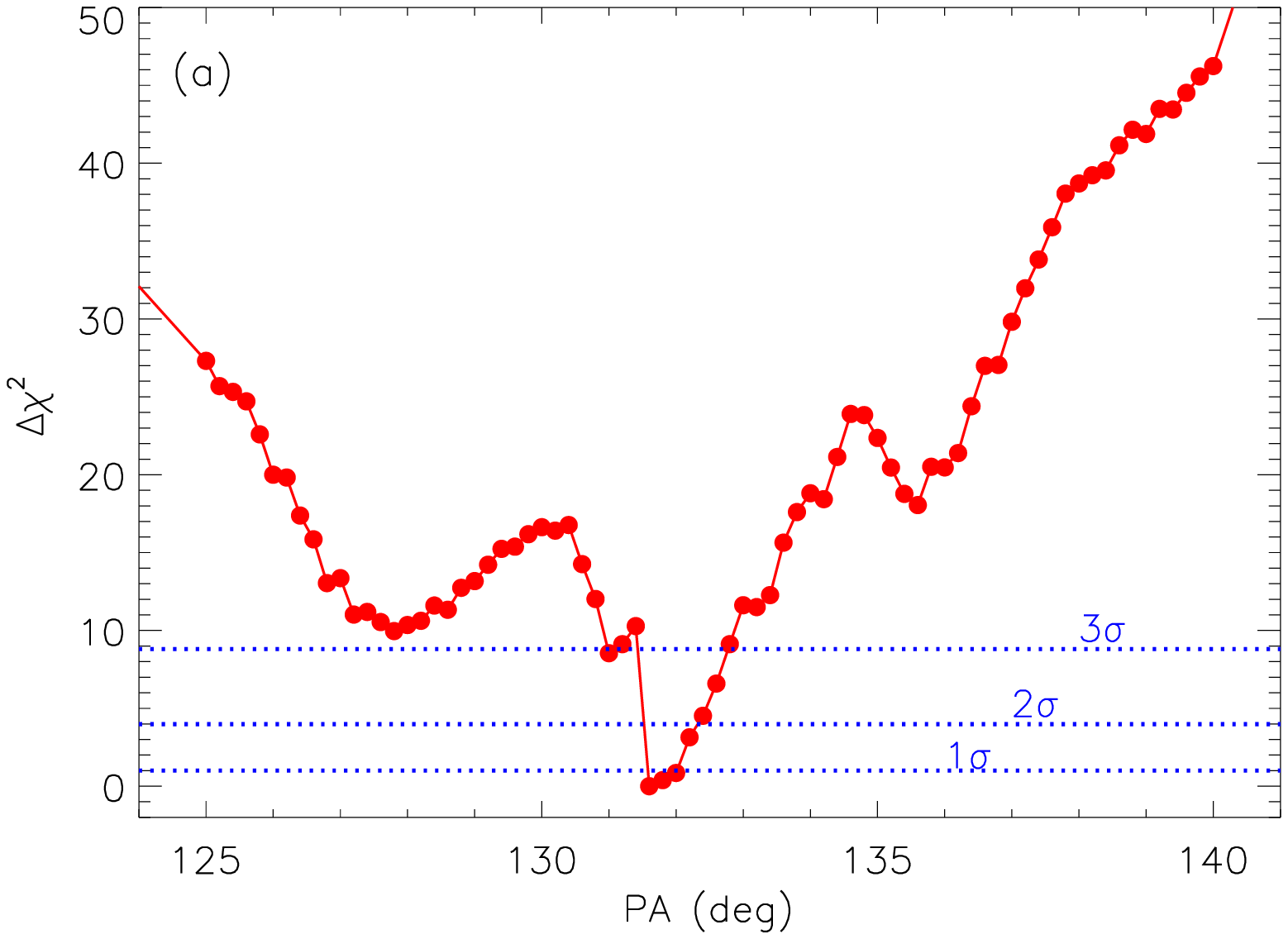}
\plotone{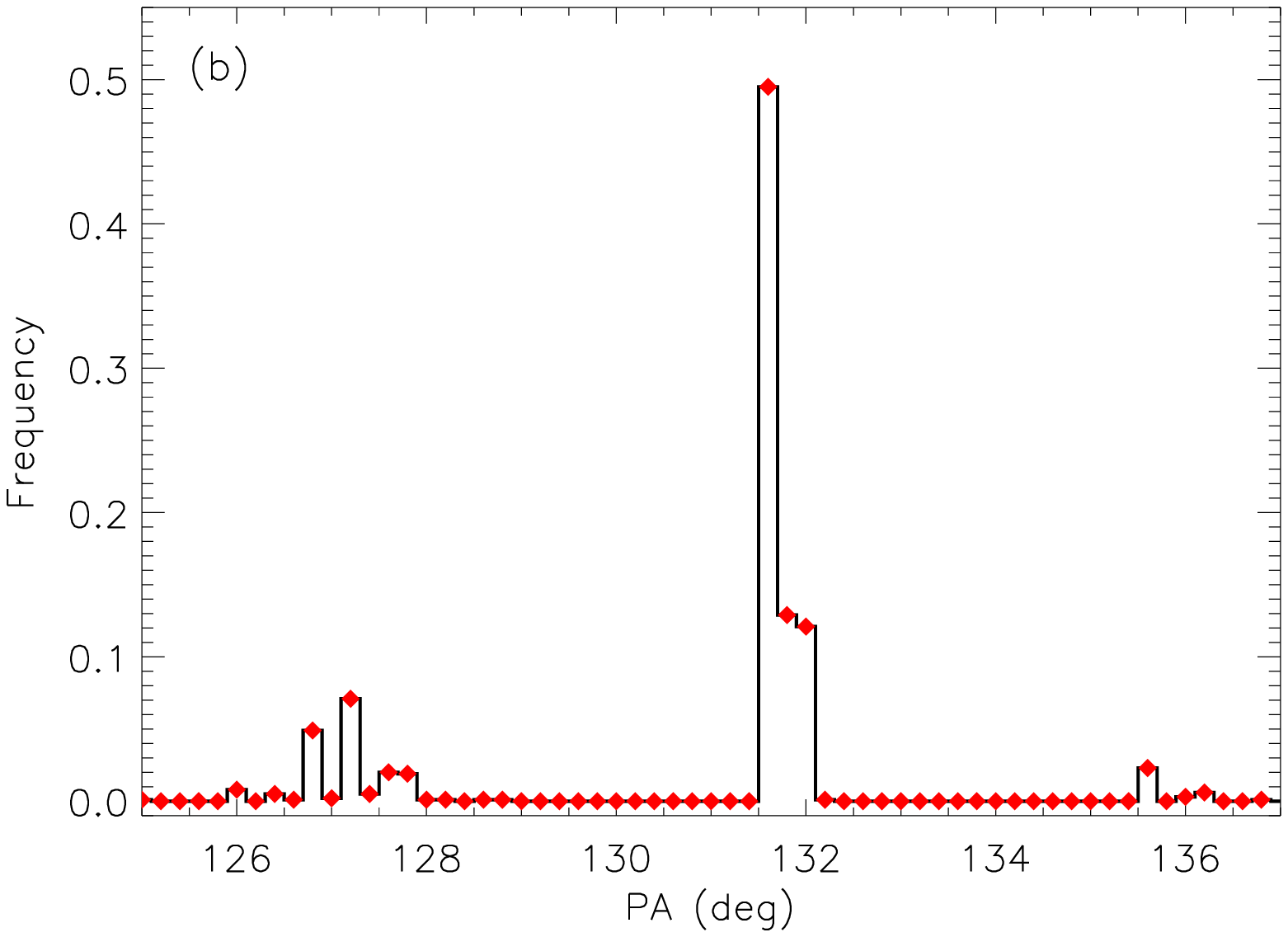}
\caption[NGC 7469: Best Fit Position Angle]{Best fit position angle of the gas disk major axis in NGC 7469.  (a) Dependence of \ch\s on the position angle, \t, where the best fit model (minimum \ch) for each \t\s is found by varying \inc, \ml, and \mbh.  The horizontal lines are the 1\signs, 2\signs, and 3\sig confidence intervals for one parameter.  Circles indicate parameters for which models were run and \ch\s calculated. (b) The distribution of the best fit position angle for 1000 realizations of the half-sample bootstrap method.  The specific model parameters fit to the data are indicated by the diamonds. \label{7469_chi_bs_pa}}
\end{figure}

\subsubsection{Major Axis Position Angle}
The position angle of the major axis of rotation of the \htwo gas in NGC 7469 is constrained to \t=131.6$^{+0.4}_{-0.1}$\deg\s at the 1\sig confidence level and 131.6$^{+1.2}_{-0.6}$\deg\s at the 3\sig level (Fig. \ref{7469_chi_bs_pa}a).  The position angle is the only parameter on which \mbh\s does not depend, and \t\s is not found to depend on any of the other parameters or model input.  As can be seen in Fig. \ref{7469_chi_bs_pa}b, the half-sample bootstrap method (see $\S$ 5.4) indicates the best fit \t=131.6$\pm$0.05\deg, with a 75\% probability of \t=131.6-132.0\deg, which confirms the best fit based on the relative likelihood method.  This \t\s is consistent with the major axis of 128$\pm$10\deg\s observed in the rotation of CO and HCN gas observed at a resolution of 0\as.7 (\citealt{davies04}), as well as with that of the \htwo flux distribution.

\subsubsection{Disk Inclination and Mass-to-Light Ratio}
\begin{figure}[!t] 
\epsscale{1}
\plotone{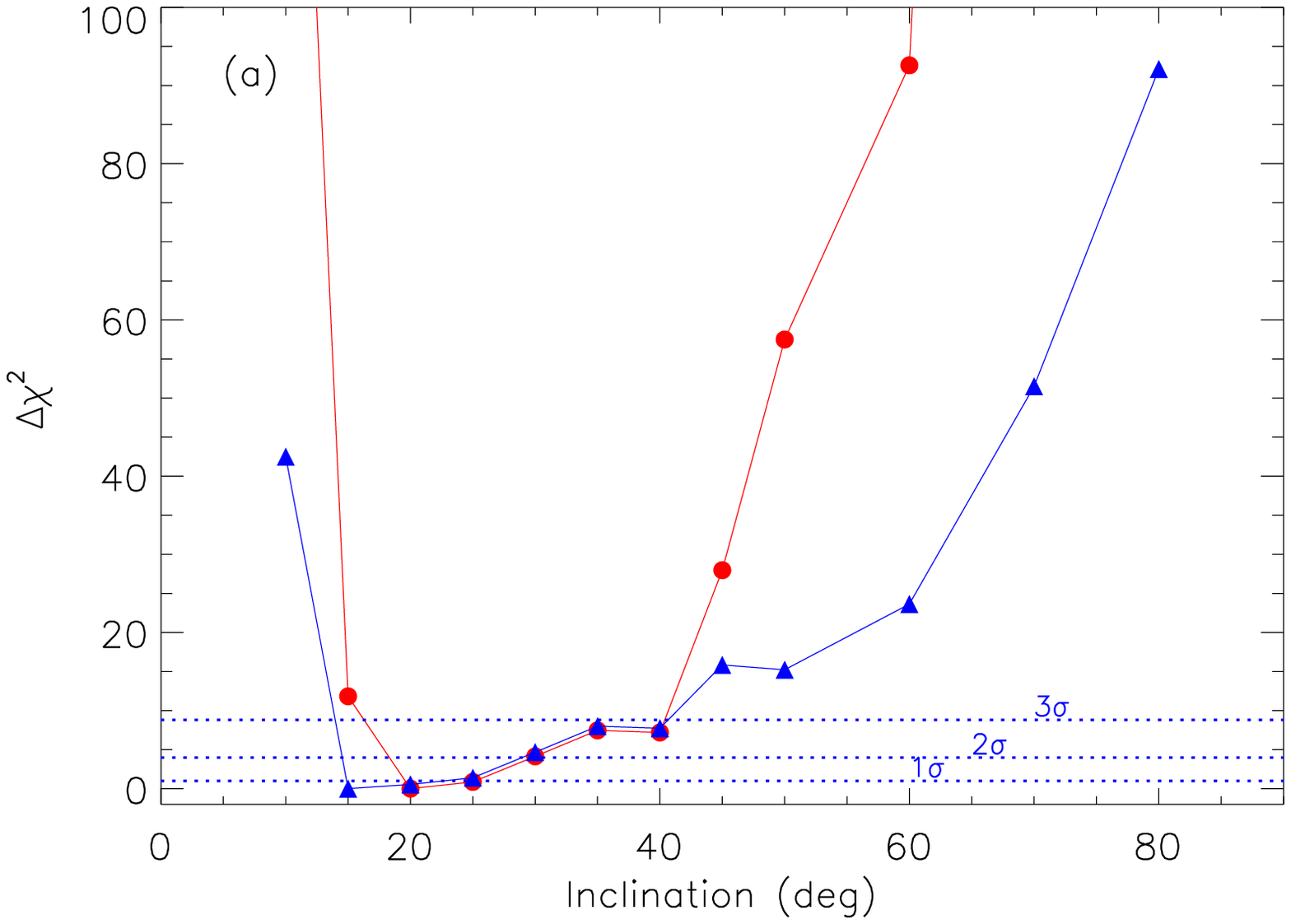}
\plotone{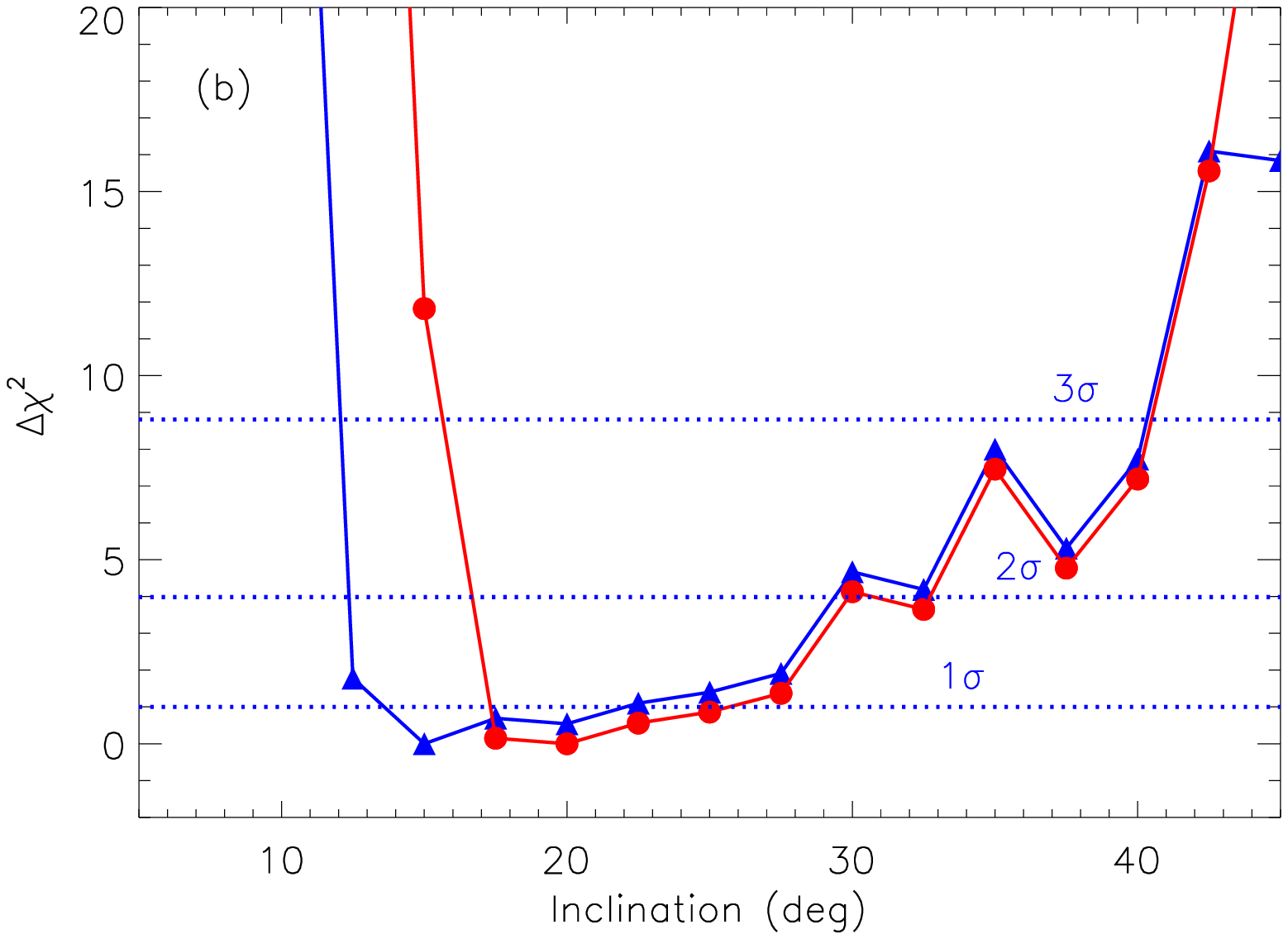}
\caption[NGC 7469: Best Fit Position Angle]{Dependence of \ch\s on the inclination of the gas disk in NGC 7469.  The best fit model for each \inc\s (i.e. the minimum \ch\s was found by varying \mbh, \ml, and \t) is given (a) as well as a more detailed plot at lower \inc\s (b).  The circles are for \ml=0.3-1.1 \mlu, and the triangles are for \ml=0.1-2.0 \mlu.  The symbols indicate model parameters that were run and the horizontal lines are the 1\signs, 2\signs, and 3\sig confidence intervals for one parameter. \label{7469_chi_inc}}
\end{figure}

\begin{figure}[!htt] 
\epsscale{1}
\plotone{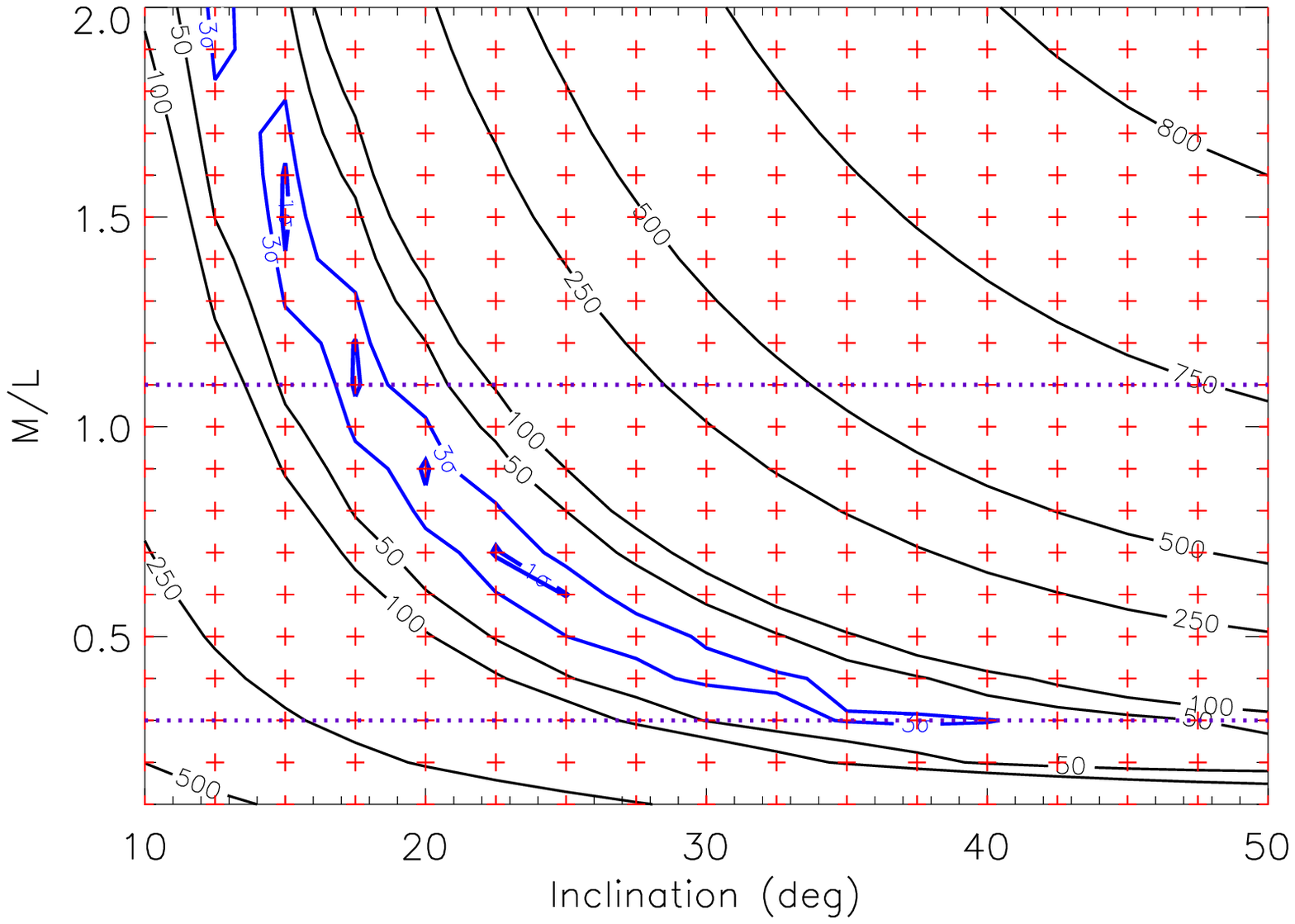}
\caption[NGC 7469: \inc\s versus \ml]{Contours of constant \ch\s for NGC 7469 with \mbh=1.0$\times10^{7}$ \Msun\s and \t=131.6\deg, where \ch\s has been rescaled as discussed in the text.  The 1\sig and 3\sig confidence intervals (for two parameters) are labeled.  The limits of the accepted range in \ml\s are indicated by the dashed lines. \label{7469_INCvMLR}}
\end{figure}

As in NGC 3227, the best fit inclination angle places the gas disk nearly face-on.  With the range of acceptable \ml\s (0.3-1.1 \mlu; see $\S$ 5.1.2), the inclination is found to be \inc=17-25\deg\s at the 1\sig level and \inc=15-40\deg\s at the 3\sig level (Fig. \ref{7469_chi_inc}).  Increasing \ml\s to 0.1-2.0 \mlu\s changes the range of suitable inclinations at the 1\sig level to \inc=12-22\deg.  These results are confirmed with the half-sample bootstrap method, which finds a best fit of \inc=15$\pm$2.5\deg at a frequency of 89\% with \ml=0.1-2.0 \mlu.  As can be seen in Fig. \ref{7469_INCvMLR}, at \inc=17\deg\s the best fit to the observed velocity field hits the upper limit of acceptable \ml\s values.  At the other extreme, the lower limit on \ml\s is hit at \inc=40\deg.  For comparison, the inclination inferred from the outer stellar disk is 45\deg\s (\citealt{marquez94}), and is ruled out in the nuclear region at a significance level greater than 99.99\% based on the gas dynamics.  

\begin{figure}[!htt] 
\epsscale{1}
\plotone{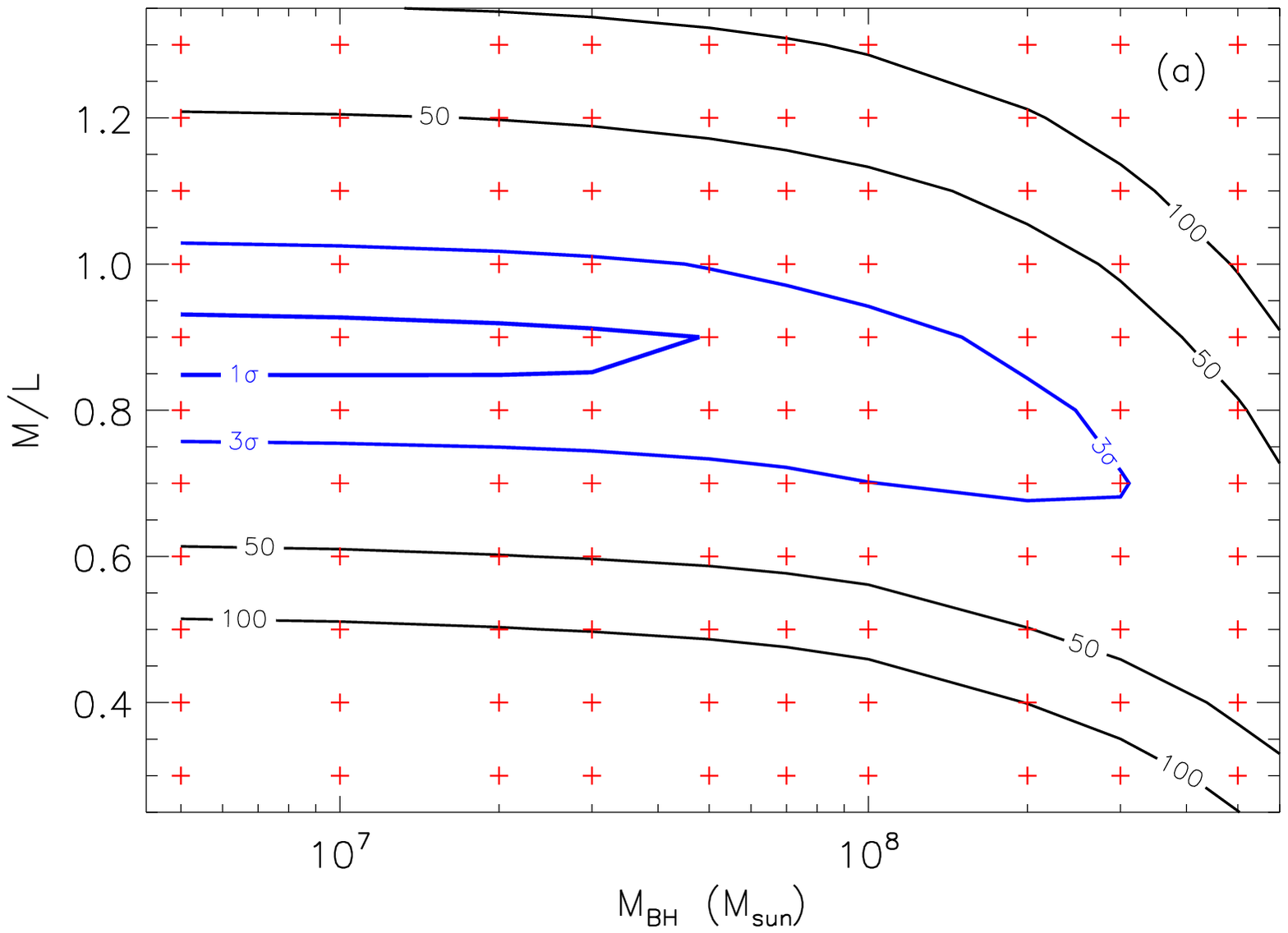}
\plotone{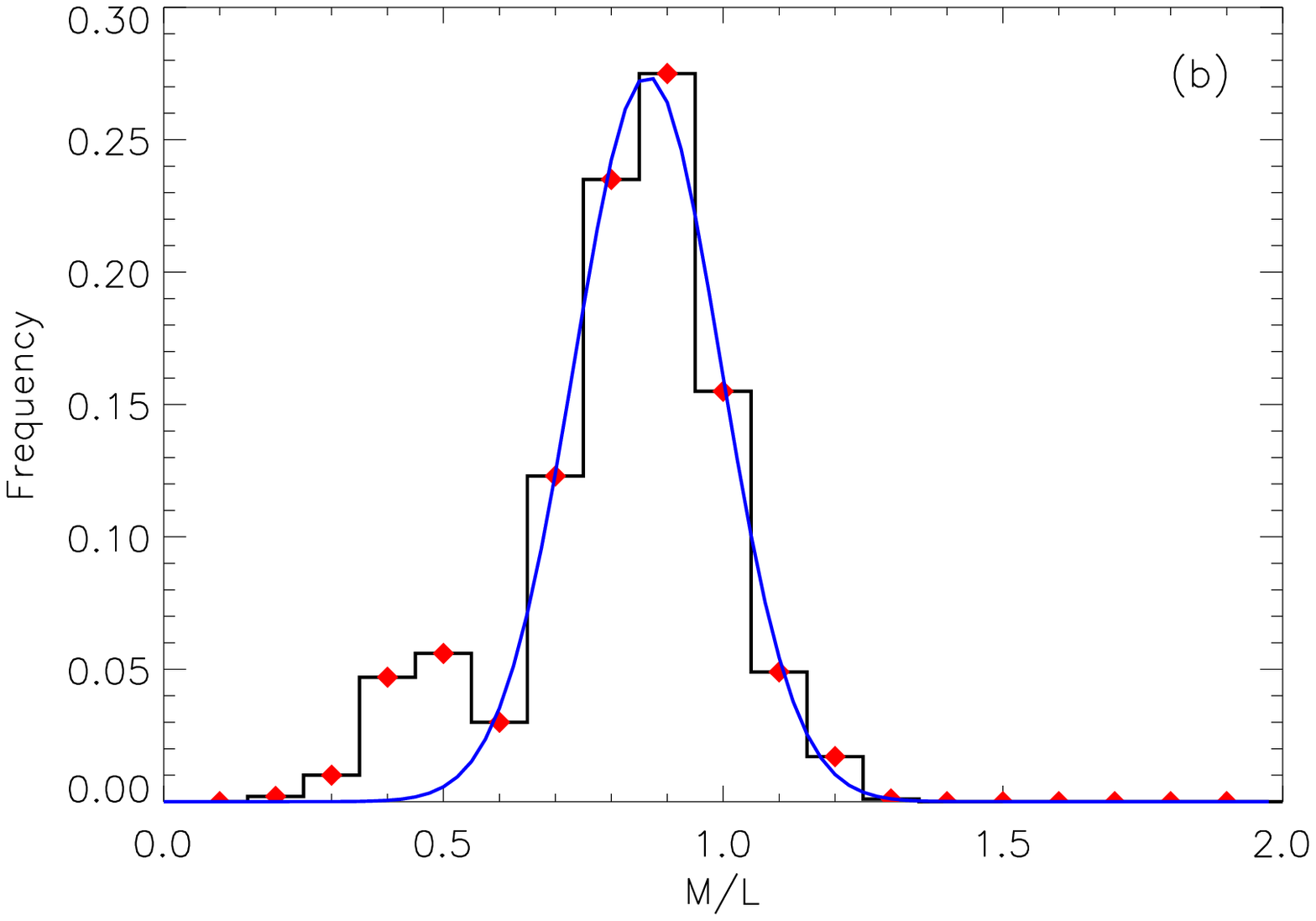}
\plotone{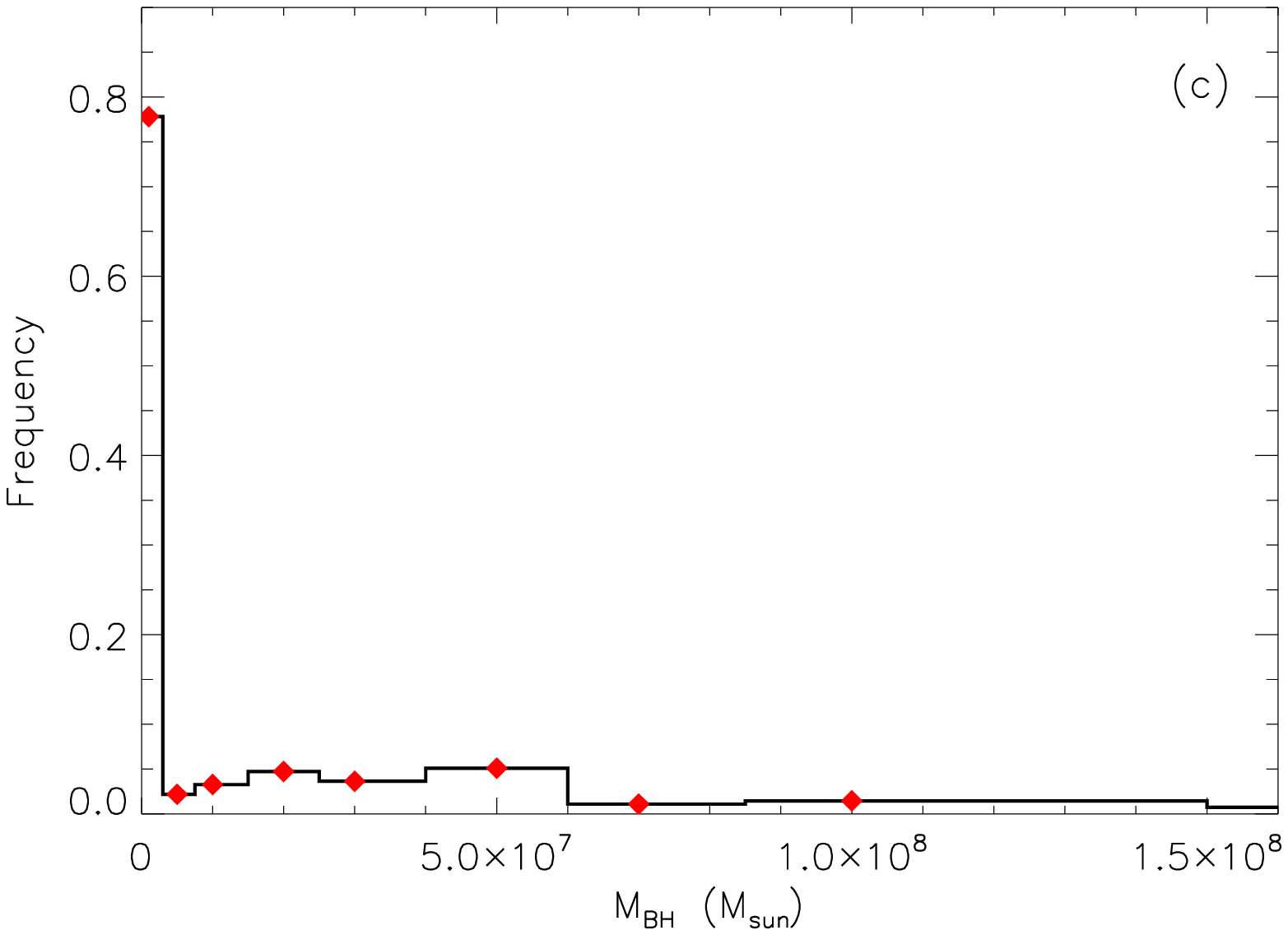}
\caption[NGC 7469: Best fit \mbh\s and \ml]{Best fit \mbh\s and \ml\s for NGC 7469.  (a) Contours of constant \ch\s with \inc=20\deg\s and PA=132\deg, where \ch\s has been rescaled as discussed in the text.  The 1\sig and 3\sig confidence intervals (for two parameters) are labeled.  For comparison the distributions of the best fit (b) mass-to-light ratio and (c) \mbh\s for 1000 realizations of the half-sample bootstrap method are also shown.  The specific model parameters fit to the data are indicated by the diamonds and a Gaussian fit to the \ml\s distribution is also shown.  The best fit \mbh\s found with the bootstrap method assumes \ml=0.9 \mlu, \inc=20\deg, and \t=131.6\deg.  \label{7469_M_MLR}}
\end{figure}

\begin{figure}[!htt] 
\epsscale{1}
\plotone{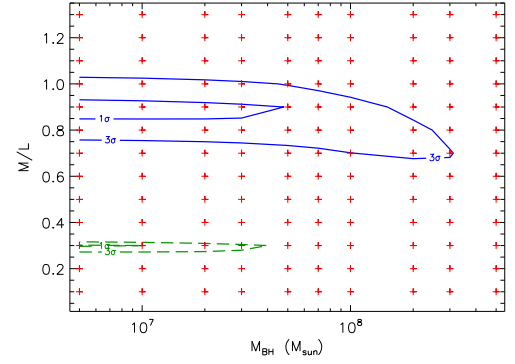}
\caption[NGC 7469: Best Fit \mbh\s for different \inc]{Contours of constant \ch\s for NGC 7469 with different values of \inc, where \ch\s has been rescaled as discussed in the text.  The 1\sig and 3\sig confidence intervals, for two parameters, are shown for \inc=20\deg\s (solid curves) and \inc=40\deg\s (dashed curves). \smallskip \smallskip \label{7469_MvMLR_i}}
\end{figure}

With a fixed \inc=20\deg, the best fit model is \mbh$<5.0\times10^{7}$ \Msun, with \ml=0.9 \mlu\s (1\sig confidence limit; see Fig. \ref{7469_M_MLR})a, assuming a stellar light distribution determined directly for the NICMOS image (the assumed stellar light distribution will be discussed further in a moment).  The values of these parameters are confirmed by the half-sample bootstrap method (see Table \ref{t_results}), which finds the best fit \ml=0.9 \mlu, with 80\% of the distribution having \ml=0.7-1.0 \mlu\s (Fig. \ref{7469_M_MLR}b).  The bootstrap method also confirms the \mbh\s upper limit, with 80\% of the distribution giving \mbh$<$1.0$\times10^{7}$ \Msun\s and 93\% of the distribution giving \mbh$<5.0\times10^{7}$ \Msun\s (Fig. \ref{7469_M_MLR}c).  If the inclination angle is increased to \inc=40\deg, then the best fit model is \mbh$<1.0\times10^{7}$ \Msun\s with \ml=0.3 \mlu.  The best fit \mbh\s and \ml\s for these two inclinations are shown in Fig. \ref{7469_MvMLR_i}, and the best fit model parameters for additional values of \inc\s are listed in Table \ref{t_7469mbhi}.

\subsubsection{Stellar Light Distribution}
\begin{figure}[!ht] 
\epsscale{1}
\plotone{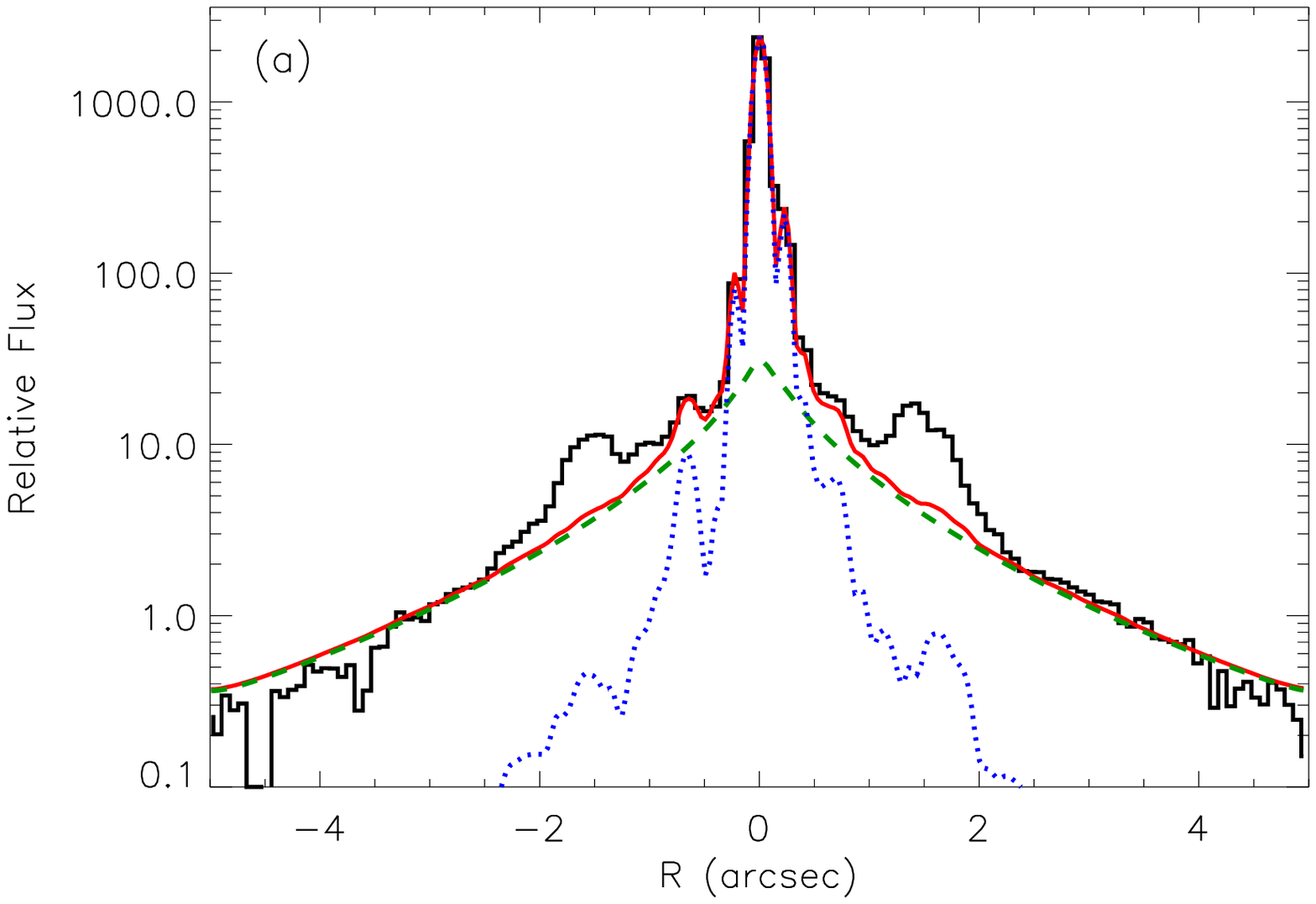}
\plotone{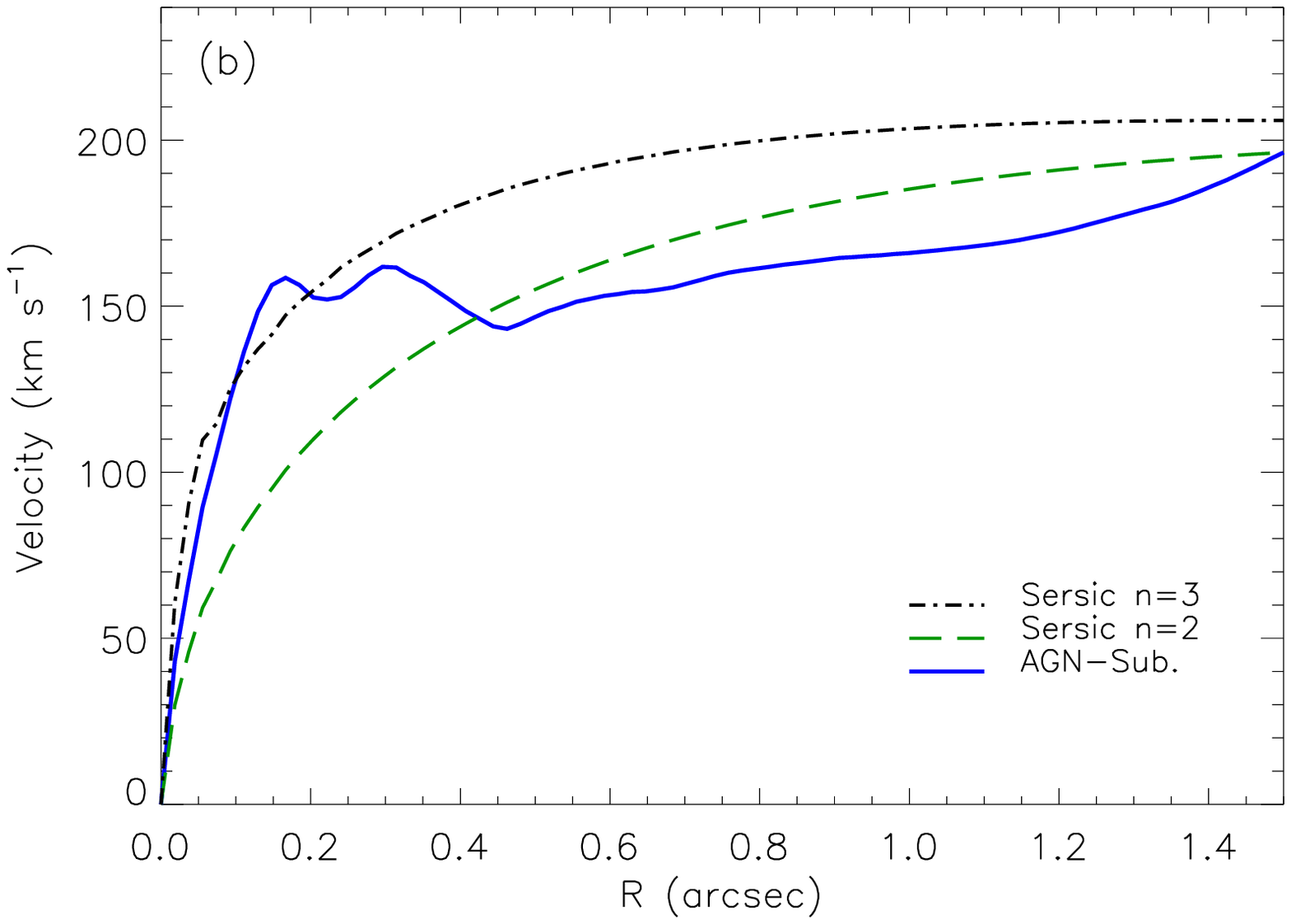}
\plotone{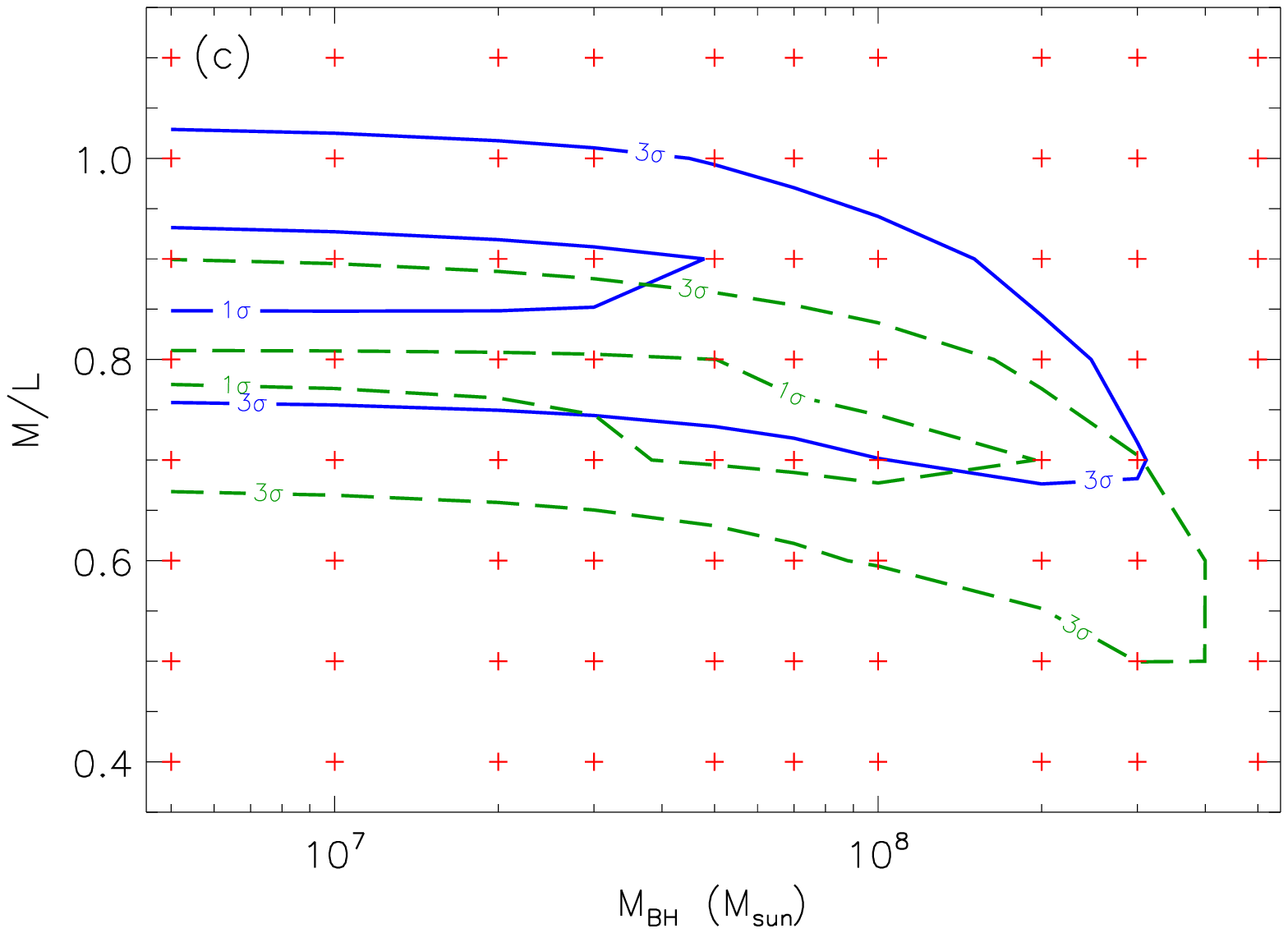}
\caption[NGC 7469: Stellar Light Distribution]{The effect of the stellar light distribution in NGC 7469 on the best fit model. (a) The best fit to the stellar light distribution (solid curve) in the NICMOS F160W image (histogram) is found with a S\`{e}rsic \n=2 function (dashed curve) plus a point source represented by a Tiny Tim PSF (dotted curve).  (b) The rotation curves in the plane of the gas disk are shown for \n=3 and \n=2 S\`{e}rsic functions and the AGN-subtracted light distribution (as labeled in the legend) assuming an inclination angle of 45\deg\s and \ml=0.5 \mlu.  (c) Contours of constant \ch\s at the 1\sig and 3\sig confidence intervals, for two parameters, are shown for the AGN-subtracted light distribution (solid curves) and the \n=2 S\`{e}rsic function (dashed curves), where \ch\s has been rescaled as discussed in the text. \label{7469_sersic}}
\end{figure}

The greatest uncertainty in the gas dynamical modeling of the velocity field in NGC 7469 comes from the stellar mass distribution.  The stellar light remaining after subtraction of the non-stellar AGN emission (see discussion in $\S$ 5.1.2) reveals an excess of stellar light in the inner 1\as\s that cannot be represented by a S\`{e}rsic function, and fits to the stellar light give reduced \ch $>$ 3.2 for all \n\s values.  The S\`{e}rsic function that best fit is \n=2, but as can be seen in Fig. \ref{7469_sersic}a, the rotation curve produced by this light distribution has velocities that are less than is predicted from the AGN-subtracted light distribution at r$<$0\as.5 (Fig. \ref{7469_sersic}b).  

Another point of comparison is the mass distribution determined by \citet{davies04} from millimeter CO emission as well as \htwo emission.  They find a fit to the data with a three component mass distribution: a disk component, a ring at r=2\as.3, and a nuclear component, which they suggest is in a ring structure at r=0\as.2 with a FWHM=0\as.6.  Their mass distribution, and the resulting rotation in the plane of the disk, is very similar to that determined from the AGN-subtracted distribution.  Although subtracting the AGN emission from a NICMOS image by using a scaled PSF is tricky, especially with a model PSF, the fact that an independent method yields a similar mass distribution lends some confidence that the stellar light distribution is reasonable.  

For the reasons stated above, the AGN-subtracted light distribution is thought to provide the most accurate description of the stellar mass distribution and is used in the modeling of the velocity field in NGC 7469.  Using this light distribution gives best fit model parameters assuming \inc=20\deg\s of \mbh$<5.0\times10^{7}$\Msun\s and \ml=0.9 \mlu, which includes the error contributed by the uncertainty in the AGN-subtracted light distribution

If the best fit S\`{e}rsic function (\n=2) is used instead of the AGN-subtracted stellar light distribution, then the parameters of the best fit model to the \htwo velocity field change to \mbh$<2.0\times10^{8}$\Msun\s and \ml=0.7-0.8 \mlu\s with \inc=20\deg\s (\t\s remains unchanged).  Fig. \ref{7469_sersic}c is a comparison of the best fit model parameters using both the AGN-subtracted and \n=2 S\`{e}rsic light distributions.  Since the S\`{e}rsic function cannot reproduce the stellar light in the inner 1\as, it instead attributes the velocities produced by this mass to the BH, and as a result fits a model with a much higher \mbh.  However, the best fit with the \n=2 S\`{e}rsic stellar light distribution has a higher reduced \ch, which rules the S\`{e}rsic \n=2 model out at the 99.9\% confidence level over the AGN-subtracted model.  

\subsubsection{Emission Line Flux Distribution}
The \htwo surface brightness in NGC 7469 is well represented by a combination of two Gaussians, the parameters of which are given in Table \ref{t_gflux}.  Since the peak of the flux is located at, or very near the kinematic center, the flux distribution only affects velocities measured at the innermost radii.  The best fit model parameters with a constant flux distribution are not significantly changed compared to those with the parameterized flux distribution.

\subsubsection{Region Fit and Data Quality Considerations}
To avoid complications from the nuclear ring located at r$\sim$1\as.5-2\as.5, only model-data comparisons with \rfit$<$1\as.0 are considered.  Increasing \rfit\s results in a significant increase in the minimum \ch\s and the best fit models are at the limit of acceptable values for \ml.  The lowest \ch\s values are found with \rfit=0\as.5 and PCC$>$0.6.  Using a PCC cut off of PCC$>$0.6 only eliminates a small percentage of data while significantly decreasing the minimum \ch.  In addition, the best fit model parameters are not dependent on the \rfit\s or PCC cut off values used.

\subsubsection{Summary of \mbh\s Estimate for NGC 7469}
\begin{figure*}[!ht]
\epsscale{0.45}
\plotone{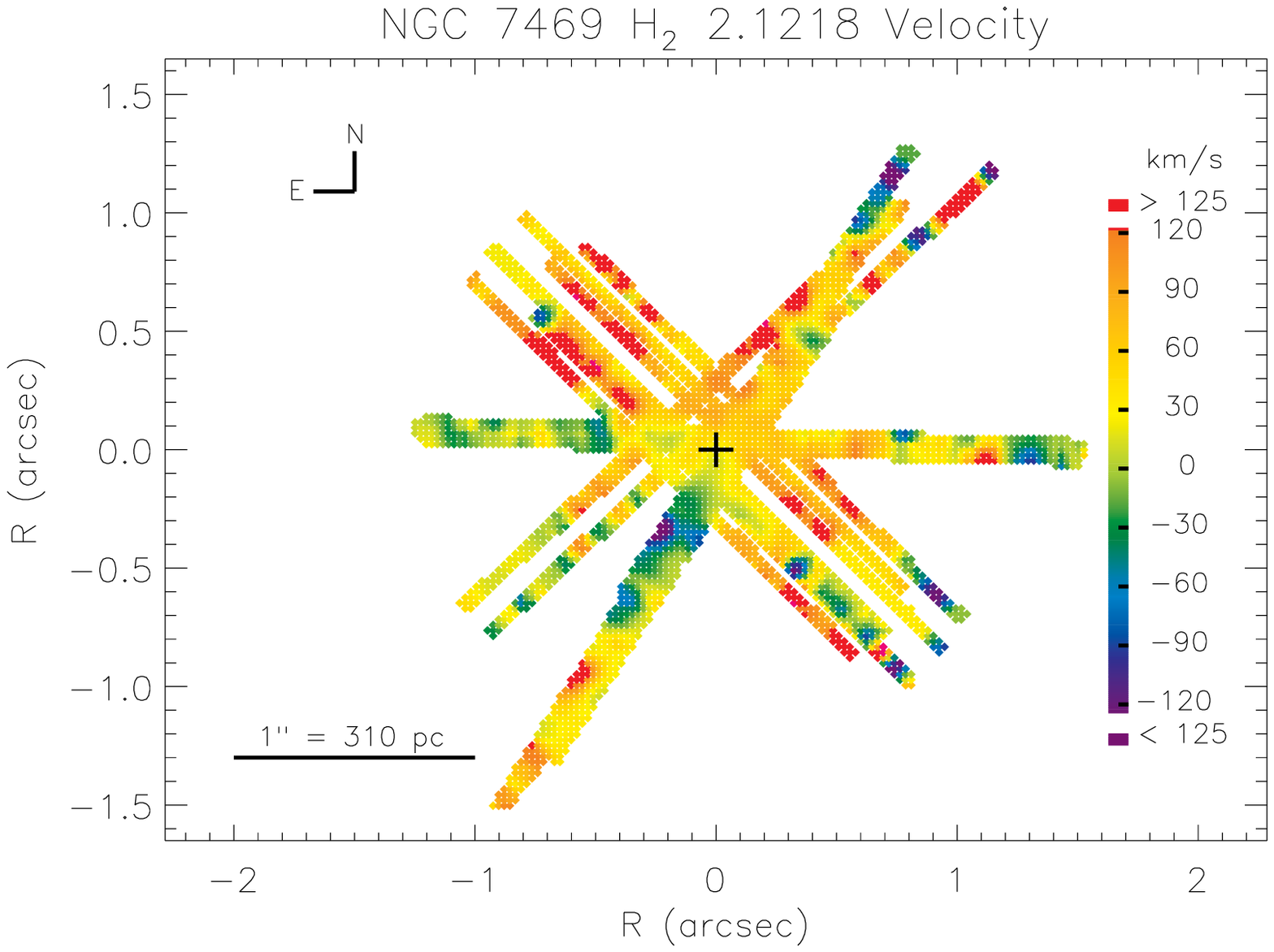}
\plotone{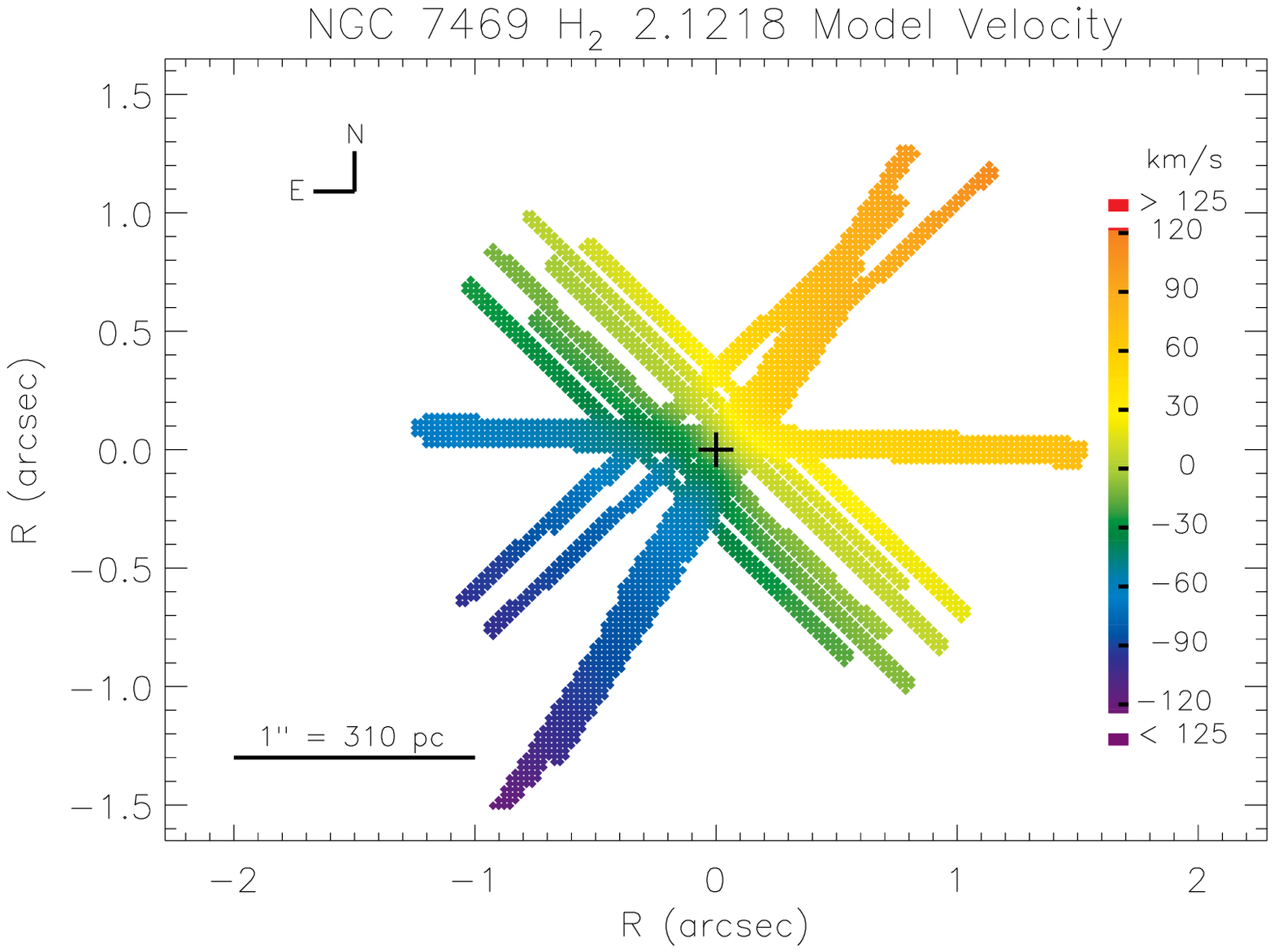}
\plotone{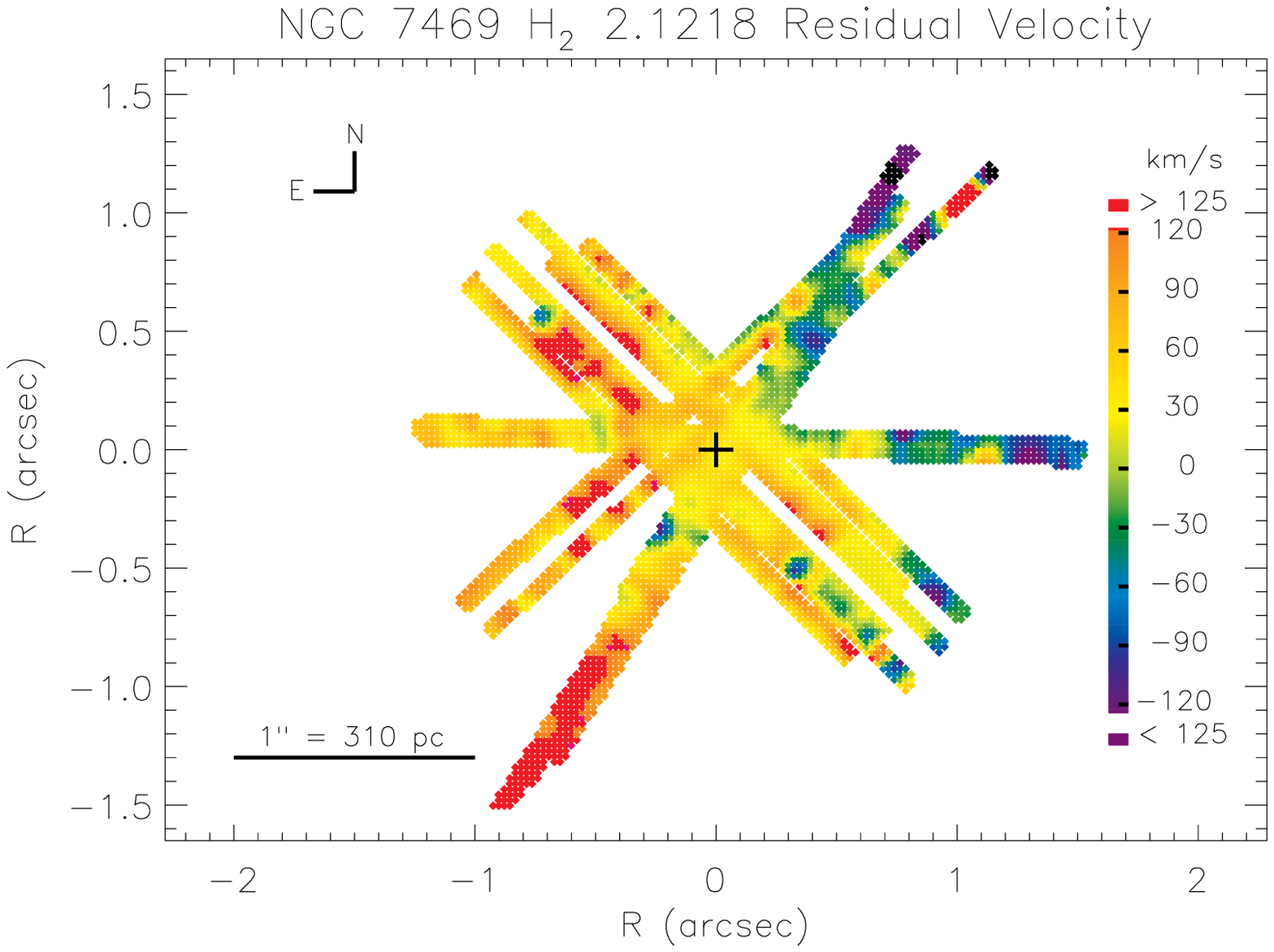}
\caption[NGC 7469: 2-D Map of Model Velocity Field and Residuals]{2-D map of the NGC 7469 \htwo 2.1218 measured velocity field, model velocity field, and the residuals of the model minus the data, as indicated by the labels at the top of each plot.  The model has \mbh=$3.0\times10^{7}$ \Msun\s with \inc=20\deg, \ml=0.9 \mlu, \t=131.6\deg\s and uses the AGN-subtracted stellar light distribution. \label{7469_maps}}
\end{figure*}

In summary, the best fit model parameters to the \htwo velocity field in NGC 7469 are found to be \mbh$<5.0\times10^{7}$ \Msun\s with \inc=20\deg, \ml=0.9 \mlu, and \t=131.6$^{+0.4}_{-0.1}$\deg, and are confirmed with the half-sample bootstrap method.  See Table \ref{t_results} for a comparison of the parameter estimates for the two statistical methods.  2-D maps of the measured and model velocity fields, as well as the residuals, are shown in Fig. \ref{7469_maps}.  The stellar light distribution has been estimated from the AGN-subtracted image to account for the additional light in the inner 1\as.  If instead a S\`{e}rsic \n=2 function is used for the stellar light distribution, then the best fit \mbh\s changes to \mbh$<2.0\times10^{8}$ \Msun.  

A previous attempt to measure \mbh\s in NGC 7469 by \citet{davies04} using radio CO emission, as well as \htwo emission along two position angles, gives \mbh$\le$5$\times10^{7}$ \Msun, which is the same as the gas dynamical measurement from the 2-D \htwo velocity field.  To date, there are no stellar dynamical measurements of \mbh\s in NGC 7469.

\begin{deluxetable*}{lcccccl}
\tablecaption{Comparison of Different Models for NGC 7469
\label{t_7469mbhi}} 
\tablewidth{0pt}
\tablehead{
\colhead{Model \tablenotemark{a}} &
\colhead{\inc} & 
\colhead{\mbh} &
\colhead{\ml} &
\colhead{Reduced \ch}  &
\colhead{Comments \tablenotemark{a}} \\
\colhead{} &
\colhead{(\deg)} &
\colhead{(10$^{7}$ \Msun)} &
\colhead{(\mlu)} &
\colhead{}  &
\colhead{} \\
}
\startdata

A	&	10	&	70.0	&	2.0	&	5.292		&	AGN-sub.	\\
A	&	15	&	0.0	&	1.5	&	5.189		&	\\
A	&	20	&	0.0	&	0.9	&	5.190		&		\\
A	&	25	&	0.0	&	0.6	&	5.192		&	\\
A	&	30	&	2.0	&	0.4	&	5.200		&		\\
A	&	35	&	3.0	&	0.3	&	5.208		&		\\
A	&	40	&	0.0	&	0.3	&	5.208		&		\\
A	&	45	&	5.0	&	0.2	&	5.227		&		\\
A	&	50	&	2.0	&	0.2	&	5.226		&		\\
A	&	60	&	0.0	&	0.2	&	5.246		&		\\
A	&	70	&	5.0	&	0.2	&	5.314		&		\\
A	&	80	&	1.0	&	0.4	&	5.413		&		\\
\cline{1-6}
B	&	20	&	0.0	&	0.9	&	5.215		&	\n=2	\\
C	&	20	&	10.0	&	0.7	&	5.211		&	w/o flux	\\

\enddata
\tablenotetext{a}{All models have \t=132\deg and a stellar light distribution and emission line flux distribution as follows.  For \htwo \lam2.1218, Model A - AGN-subtracted and the two component Gaussian function, Model B - S\`{e}rsic \n=2 and the two component Gaussian function, and Model C - AGN-subtracted and a constant flux distribution.  All models have 411 degrees of freedom.}

\end{deluxetable*}

\subsection{Modeling Results for NGC 4151}
The \htwo \lam2.1218 velocity field in NGC 4151 has a generally organized rotation pattern with a possible additional velocity component superimposed on the organized rotation.  The \htwo flux distribution results in a dependence of the model parameters on \rfit, and the details of this dependence, as well as the dependence of \mbh\s on other model parameters, are discussed below. 

\subsubsection{Major Axis Position Angle}
\begin{figure}[!h] 
\epsscale{1}
\plotone{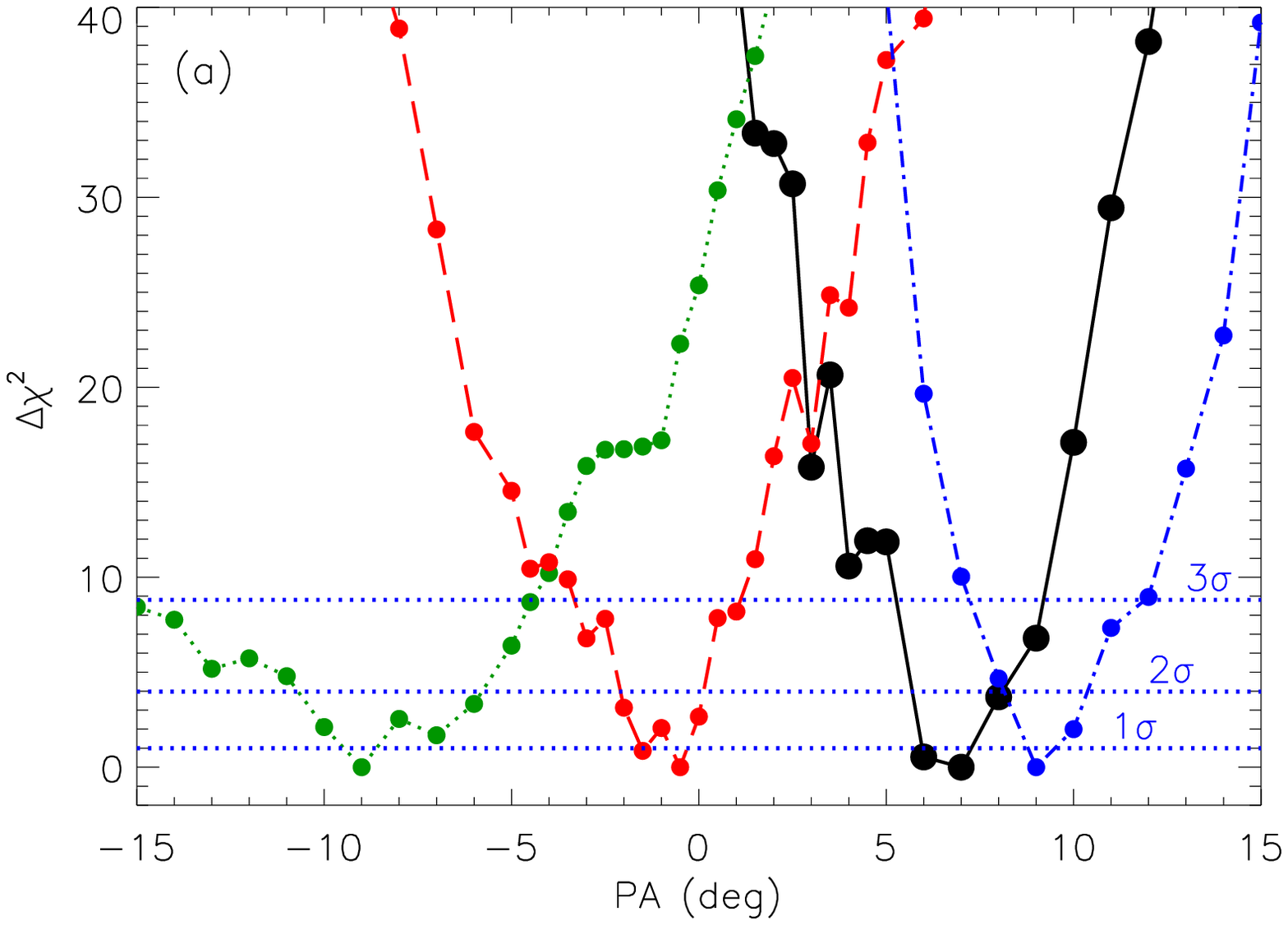}
\plotone{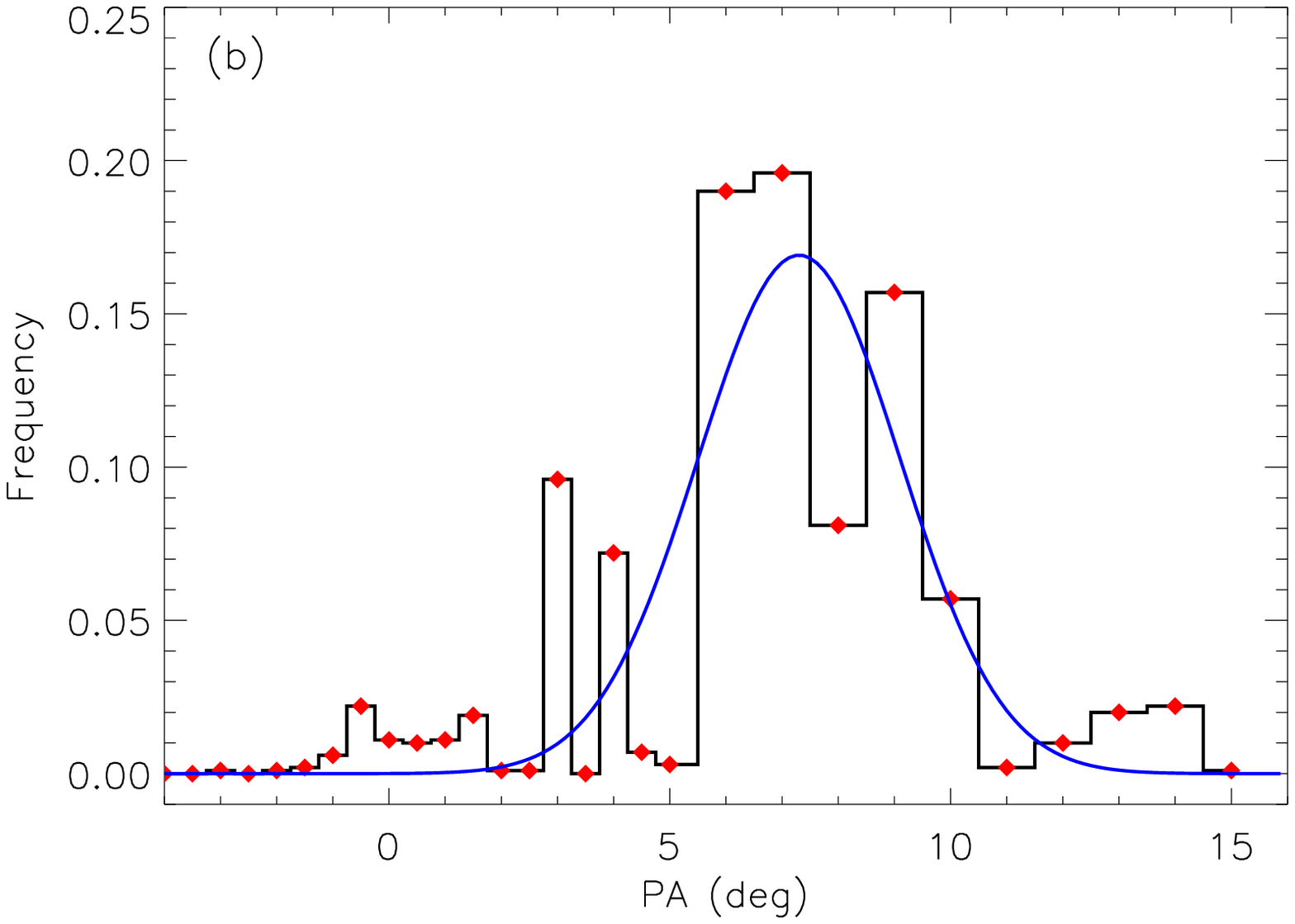}
\caption[NGC 4151: Best Fit Position Angle]{Best fit position angle of the gas disk major axis in NGC 4151. (a) Dependence of \ch\s on the position angle, \t, where the best fit model (minimum \ch) for each \t\s is found by varying \inc, \ml, and \mbh.  The best fit \t\s with \rfit=0\as.5, 0\as.75, 1\as.0, and 1\as.25 are indicated by the dotted, dashed, solid, and dash-dotted curves, respectively.  The horizontal lines are the 1\signs, 2\signs, and 3\sig confidence intervals for one parameter.  Circles indicate parameters for which models were run and \ch\s calculated. (b) The distribution of the best fit position angle for 1000 realizations of the half-sample bootstrap method with \rfit=1\as.  The specific model parameters fit to the data are indicated by the diamonds and a Gaussian fit to the distribution is also shown. \label{4151_chi_bs_pa}}
\end{figure}

In NGC 4151 the major axis of the \htwo \lam2.1218 rotation is constrained to \t= 6.5$\pm$0.5\deg\s at the 1\sig confidence level and \t= 6.5$\pm$2.0\deg\s at the 3\sig level with \rfit=1\as.0 (Fig. \ref{4151_chi_bs_pa}a).  This position angle is consistent with the major axis of the general \htwo flux distribution, and is confirmed with the half-sample bootstrap method (see $\S$ 5.4), which, as can be seen in Fig. \ref{4151_chi_bs_pa}b, gives a best fit of \t=6-7\deg.  The distribution of the best fit position angle implies \t=6-10\deg\s at about the 1\sig level, which is a greater distribution implied by the relative likelihood statistics.  In addition, fitting a Gaussian to the distribution gives a best fit \t=7.3$\pm$1.8\deg.  Unlike for NGC 3227 and NGC 7469, \t\s is dependent on how far out the data are fit, i.e. the value of \rfit.  Fig. \ref{4151_chi_bs_pa}a shows the best fit \t\s (letting the other three model parameters vary) for different values of \rfit.  The major axis varies from \t=-11\deg\s to 9\deg\s for \rfit=0\as.5 to 1\as.5.  For a given fitting radius, \t\s does not depend on any of the other model parameters or on the model input (stellar light distribution, etc.).  

\begin{figure}[!h] 
\epsscale{1}
\plotone{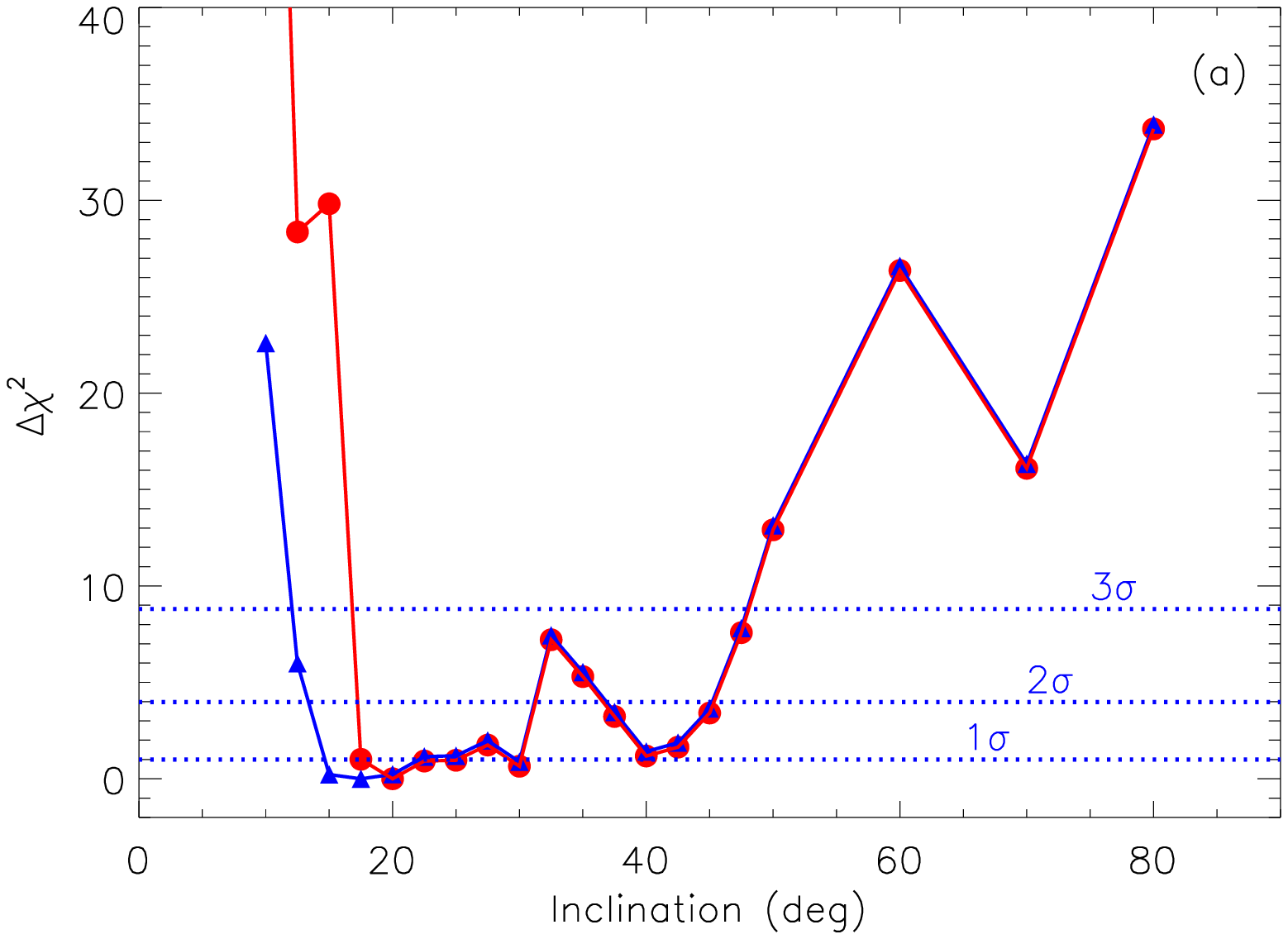}
\plotone{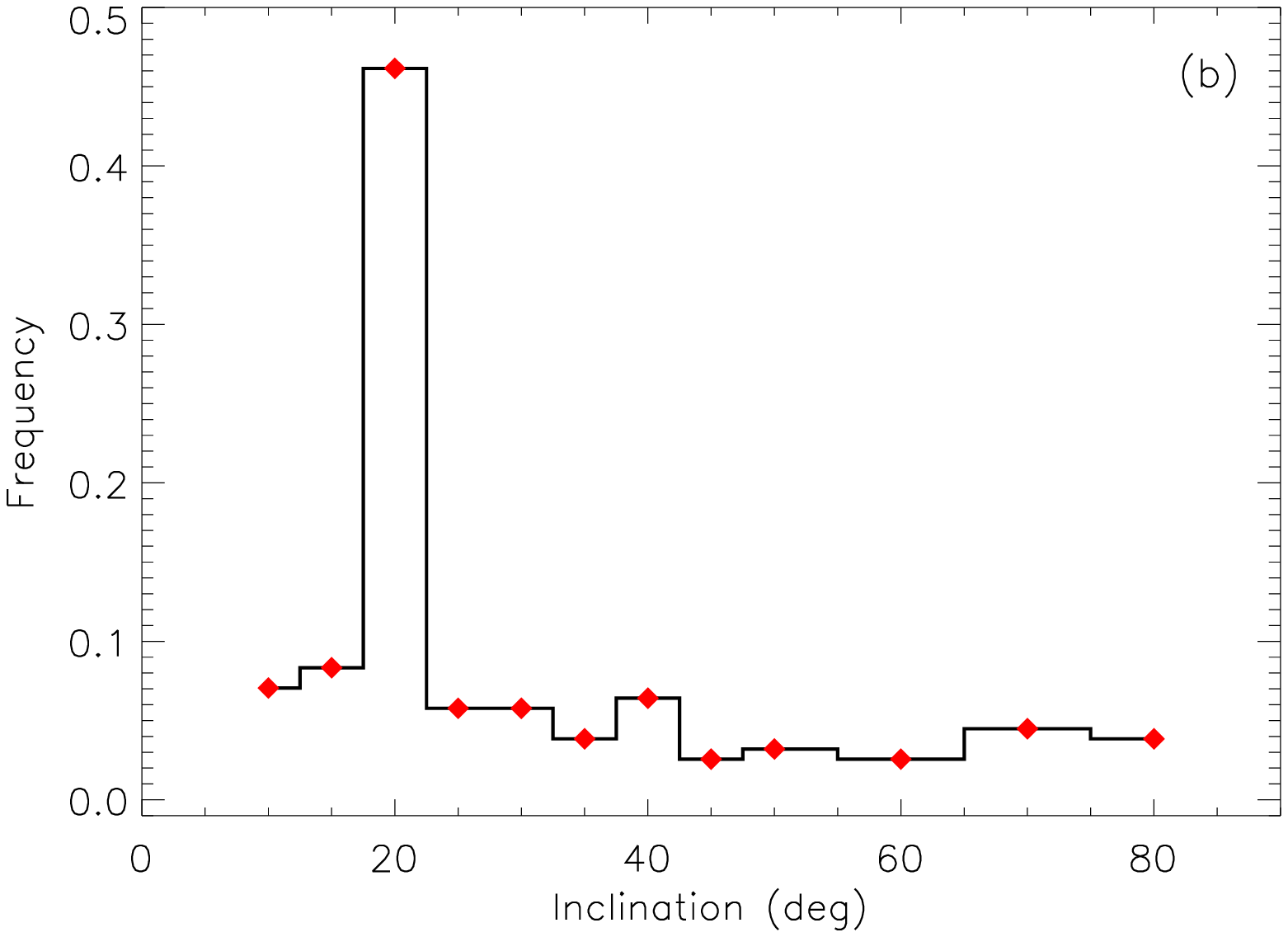}
\caption[NGC 4151: Best Fit Inclination Angle]{Best fit gas disk inclination angle, \inc, for NGC 4151. (a) Dependence of \ch\s on \inc, where the best fit model for each \inc\s is found by varying \mbh, \ml, and \t.  The circles are for \ml=0.3-1.1 \mlu, and the triangles are for \ml=0.1-2.0 \mlu.  The symbols indicate models that were run and the horizontal lines are the 1\signs, 2\signs, and 3\sig confidence intervals for one parameter.  (b) The distribution of the best fit \inc\s for 1000 realizations of the half-sample bootstrap method with \ml=0.1-2.0 \mlu.  The specific model parameters fit to the data are indicated by the diamonds. \label{4151_chi_bs_i}}
\end{figure}

\begin{figure}[!t] 
\epsscale{1}
\plotone{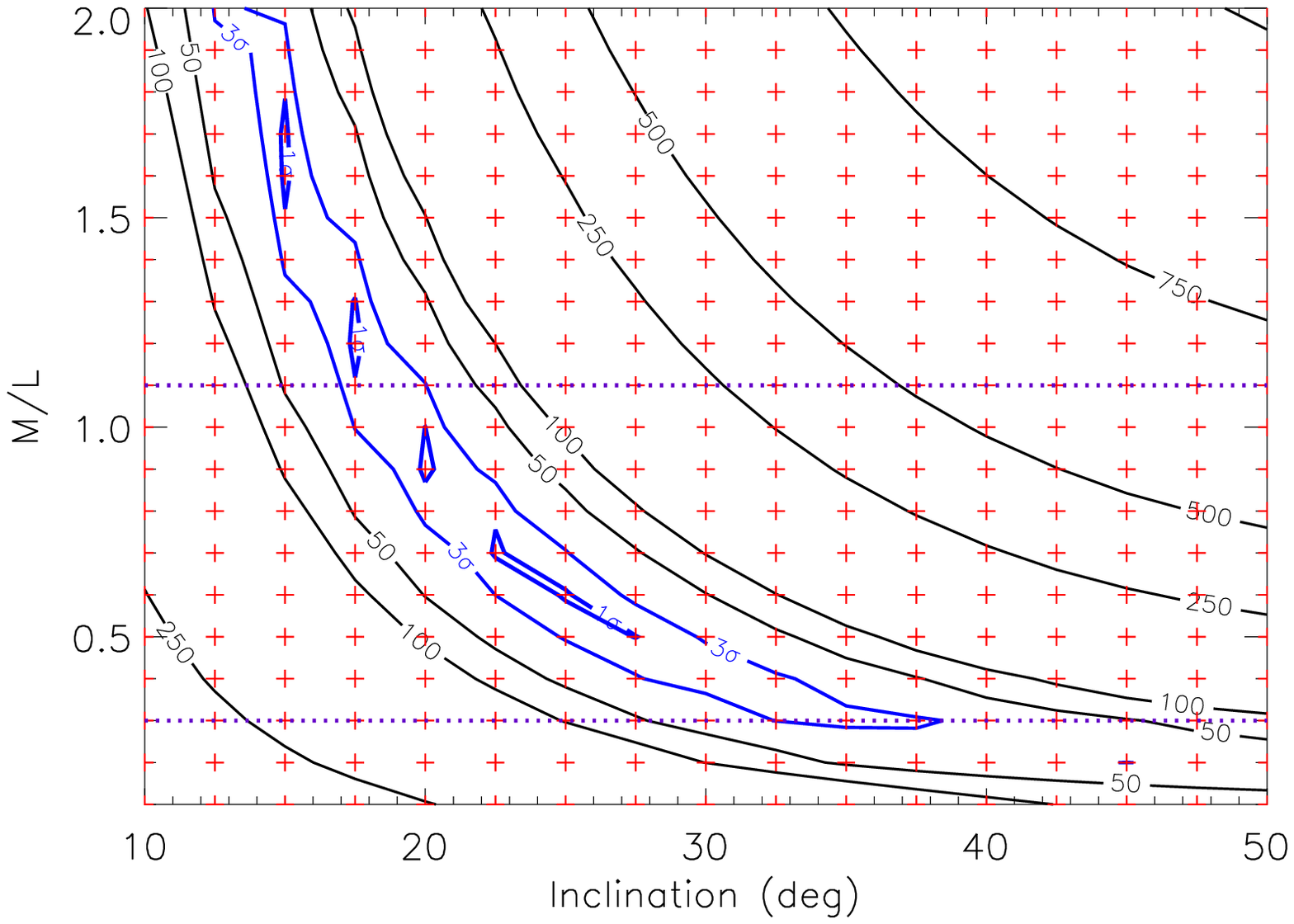}
\caption[NGC 4151: \inc\s versus \ml]{Contours of constant \ch\s for NGC 4151 with \mbh=3.0$\times10^{7}$ \Msun\s and \t=6.5\deg, where \ch\s has been rescaled as discussed in the text.  The 1\sig and 3\sig confidence intervals (for two parameters) are labeled. \label{4151_INCvMLR}}
\end{figure}

The \htwo major axis of rotation is in good agreement with the photometric major axis in the {\em I}-band (a wavelength region free of any contribution from emission lines), which, at 1\as, is found to be very close to circular with \t$\simeq$10\deg\s (\citealt{mediavilla95}).  Also of note is that the major axis of the rotating components of H$\alpha$ and \oiii\s are reported by \citet{mediavilla95} to be 30-34\deg\s and 39-43\deg, respectively, at a radius of 1\as.0-2.5\as.  The discrepancy between the major axis of rotation for H$\alpha$ and \oiii\s (\t$\sim$30-40\deg) and that of \htwo (\t$\sim$7\deg) raises the question of whether the ionized gas is rotating with the molecular gas or if they are kinematically distinct.  Even though the ionized gas appears to be dominated by rotation at r$>$1\as, rather than being influenced by the nuclear radio jet (see e.g. \citealt{mediavilla95}, \citealt{winge99}), it is possible that it would still be affected at these radii in a way that the molecular gas would not.

\subsubsection{Disk Inclination and Mass-to-Light Ratio}
As Fig. \ref{4151_chi_bs_i}a shows, the inclination angle is constrained to be \inc=17-47\deg\s (3\sig confidence limit; \inc=20$\pm$2\deg\s 1\signs) with \ml\s restricted to 0.3-1.1 \mlu\s (see $\S$ 5.1.2).  A range of \ml=0.1-2.0 \mlu\s extends the 3\sig limits to \inc=12-47\deg.  The preference toward low disk inclination angles is confirmed with the half-sample bootstrap method, which has a best fit of \inc=10-25 for 69\% of the distribution (Fig. \ref{4151_chi_bs_i}b) with \ml=0.1-2.0 \mlu.  As shown in Fig. \ref{4151_INCvMLR}, the disk inclination angle cannot be greater than 40\deg\s without a \ml\s less than 0.3 \mlu.  This close to face-on inclination is consistent with the inclination reported by \cite{winge99} of i$\sim$21\deg\s for the rotating component of \oiii\s on a scale of tens of arcseconds.  In addition, the low inclination angle is consistent with that of the stellar disk at even larger spatial scales, which has \inc$\sim$26\deg\s (\citealt{bosma77}).

\begin{figure}[!t] 
\epsscale{1.05}
\plotone{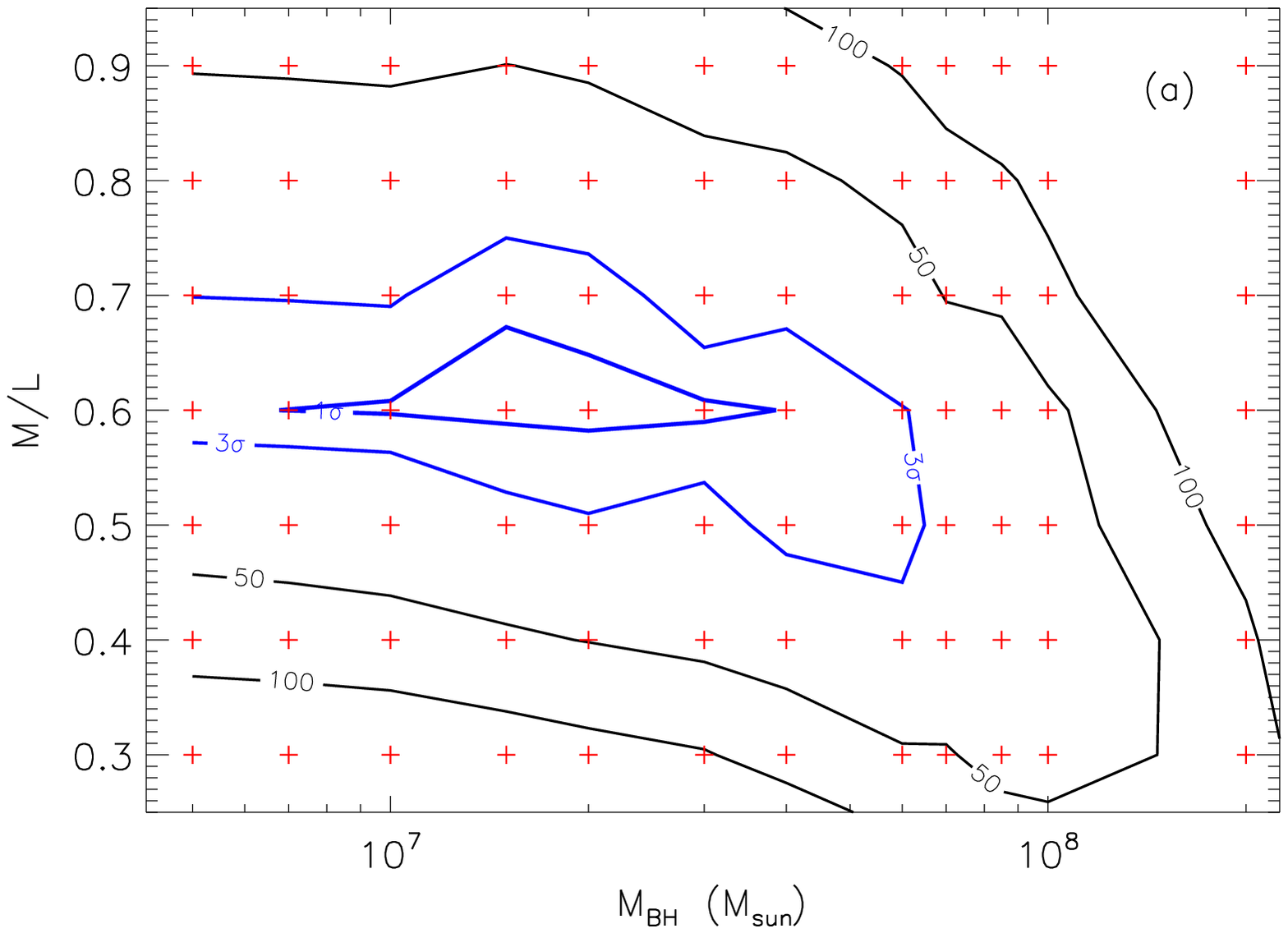}
\plotone{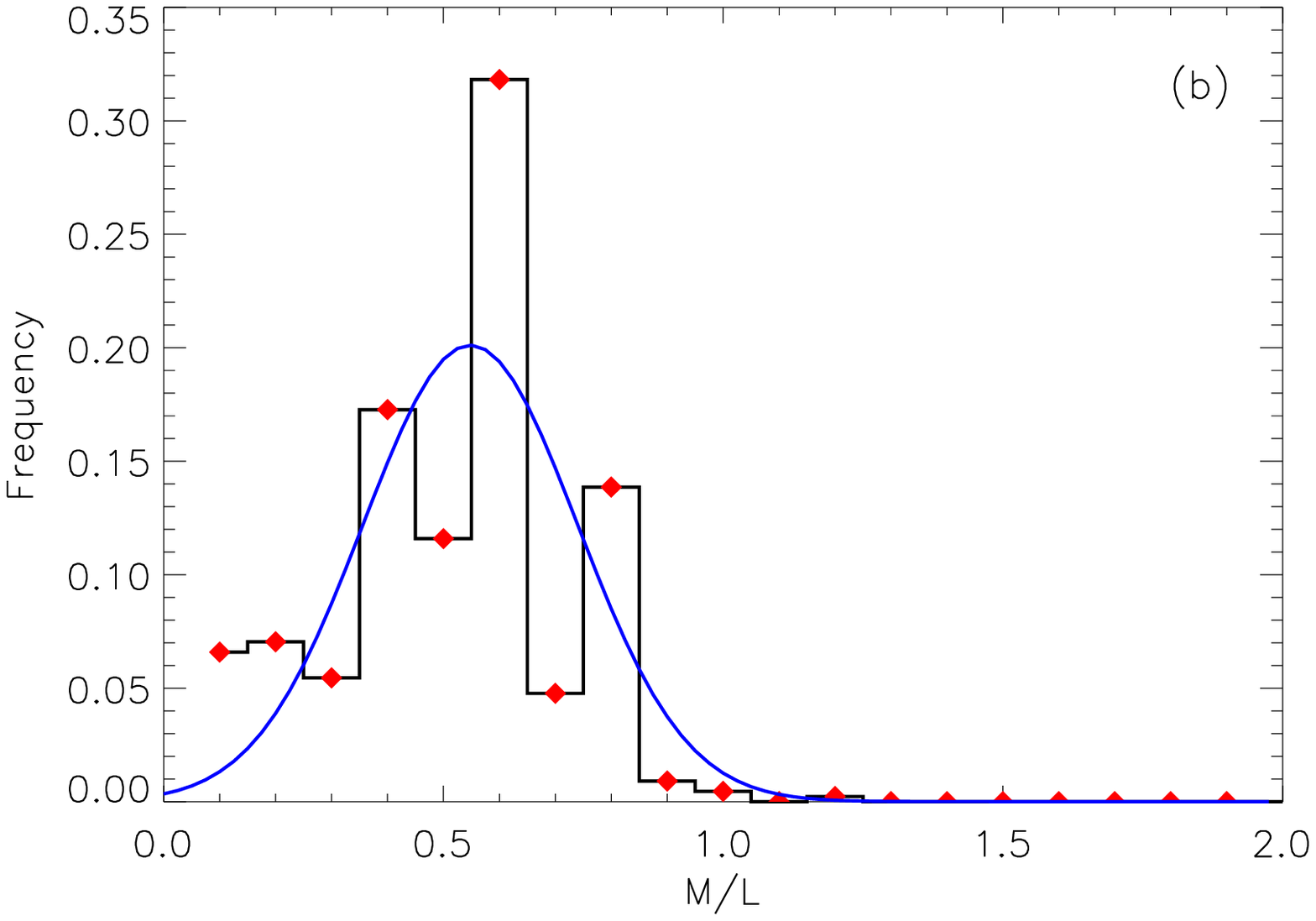}
\plotone{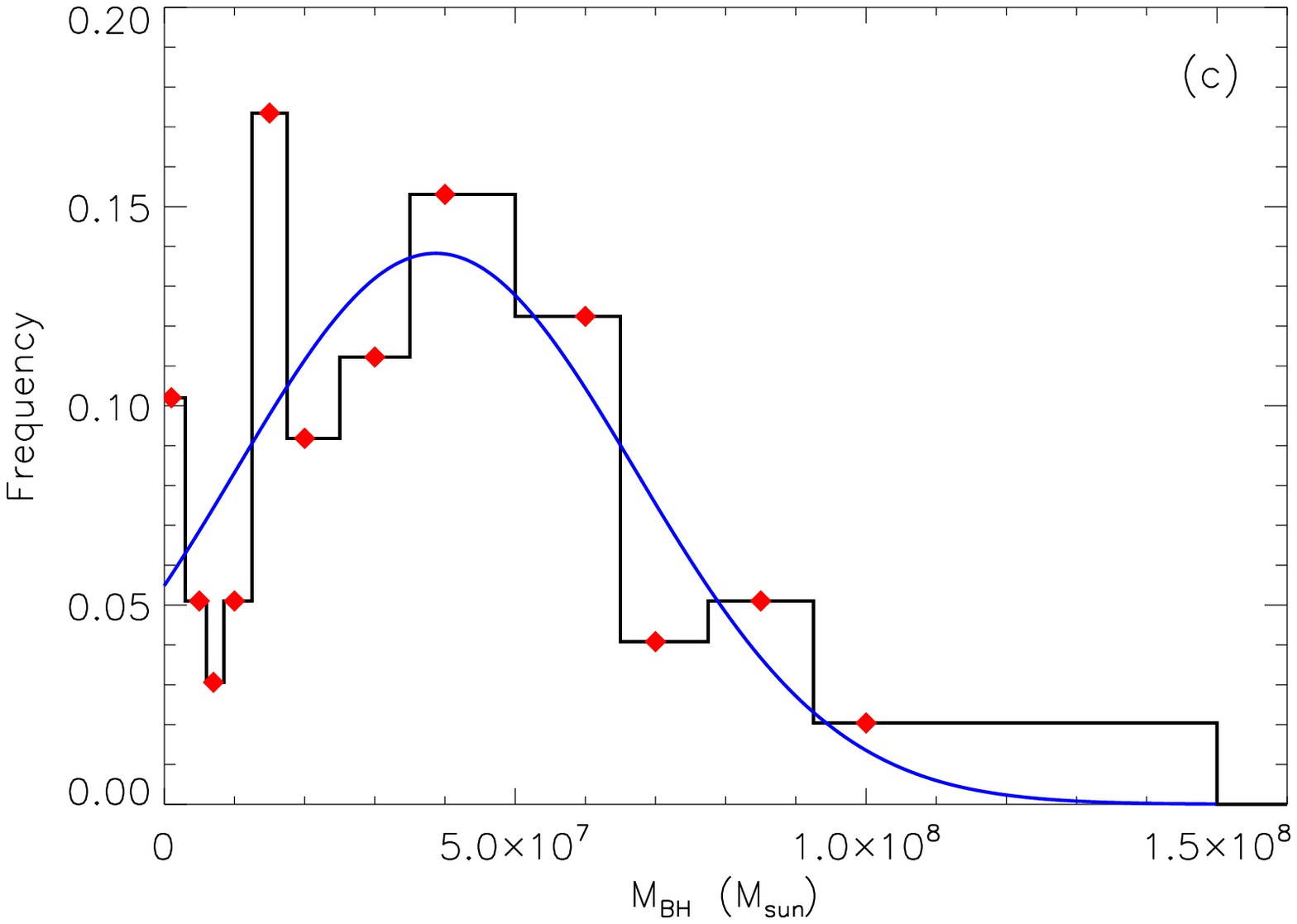}
\caption[NGC 4151: Best Fit \mbh and \ml]{Best fit \mbh\s and \ml\s for NGC 4151.  (a) Contours of constant \ch\s with \inc=25\deg\s and PA=6.5\deg, where \ch\s has been rescaled as discussed in the text.  The 1\sig and 3\sig confidence intervals (for two parameters) are labeled.  For comparison the distributions of the best fit (a) mass-to-light ratio and (c) \mbh\s for 1000 realizations of the half-sample bootstrap method are shown.  The specific model parameters fit to the data are indicated by the diamonds and Gaussian fits to the distributions are also shown.  The best fit \mbh\s found with the bootstrap method assumes \ml=0.6 \mlu, \inc=25\deg, and \t=6.5\deg.  \label{4151_M_MLR}}
\end{figure}

\begin{figure}[!t] 
\epsscale{1}
\plotone{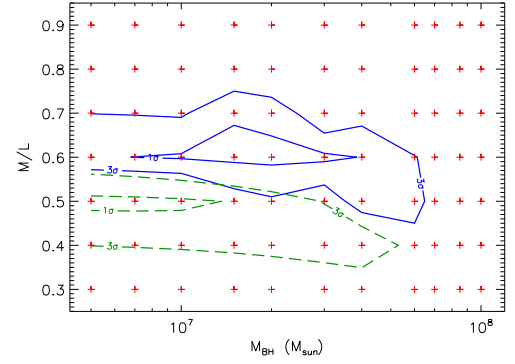}
\caption[NGC 4151: Best Fit \mbh\s for different \inc]{Contours of constant \ch\s for NGC 4151 with different values of \inc, where \ch\s has been rescaled as discussed in the text.  The 1\sig and 3\sig confidence intervals (for two parameters) are shown for \inc=25\deg\s (solid curves) and \inc=30\deg\s (dashed curves). \label{4151_MvMLR_i}}
\end{figure}

For a fixed \inc=25\deg, the best fit model finds \mbh=3.0$^{+0.75}_{-2.2}\times10^{7}$ \Msun\s with \ml=0.6 \mlu\s (1\sig limits; Fig. \ref{4151_M_MLR}a).  Fig. \ref{4151_M_MLR}b shows the distribution of best fit \ml\s values from the half-sample bootstrap method.  The best fit is \ml=0.6$\pm$0.05 \mlu\s over a third of the time, and a Gaussian fit to the distribution gives a best fit \ml=0.55$\pm$0.19 \mlu.  The distribution of the best fit \mbh\s is shown in Fig. \ref{4151_M_MLR}c.  The distribution is relatively wide, with 63\% of the values falling within 1-7$\times10^{7}$ \Msun.  A Gaussian fit to the distribution gives a best fit \mbh=3.9$\pm$2.9$\times10^{7}$ \Msun\s with \ml=0.6 \mlu.  The \ml\s and \mbh\s values given by the relative likelihood method are consistent with those given by the half-sample bootstrap method (see Table \ref{t_results} for a comparison).  Increasing \inc\s to 30\deg\s gives a best fit of \mbh$<$1.5$\times10^{7}$ \Msun\s with \ml=0.5 \mlu.  Fig. \ref{4151_MvMLR_i}\, shows the best fit \mbh\s and \ml\s for a range of \inc\s and the best fit model parameters for several different values of \inc\s are given in Table \ref{t_4151mbhi}.

\subsubsection{Stellar Light Distribution}
\begin{figure*}[!t] 
\epsscale{0.45}
\plotone{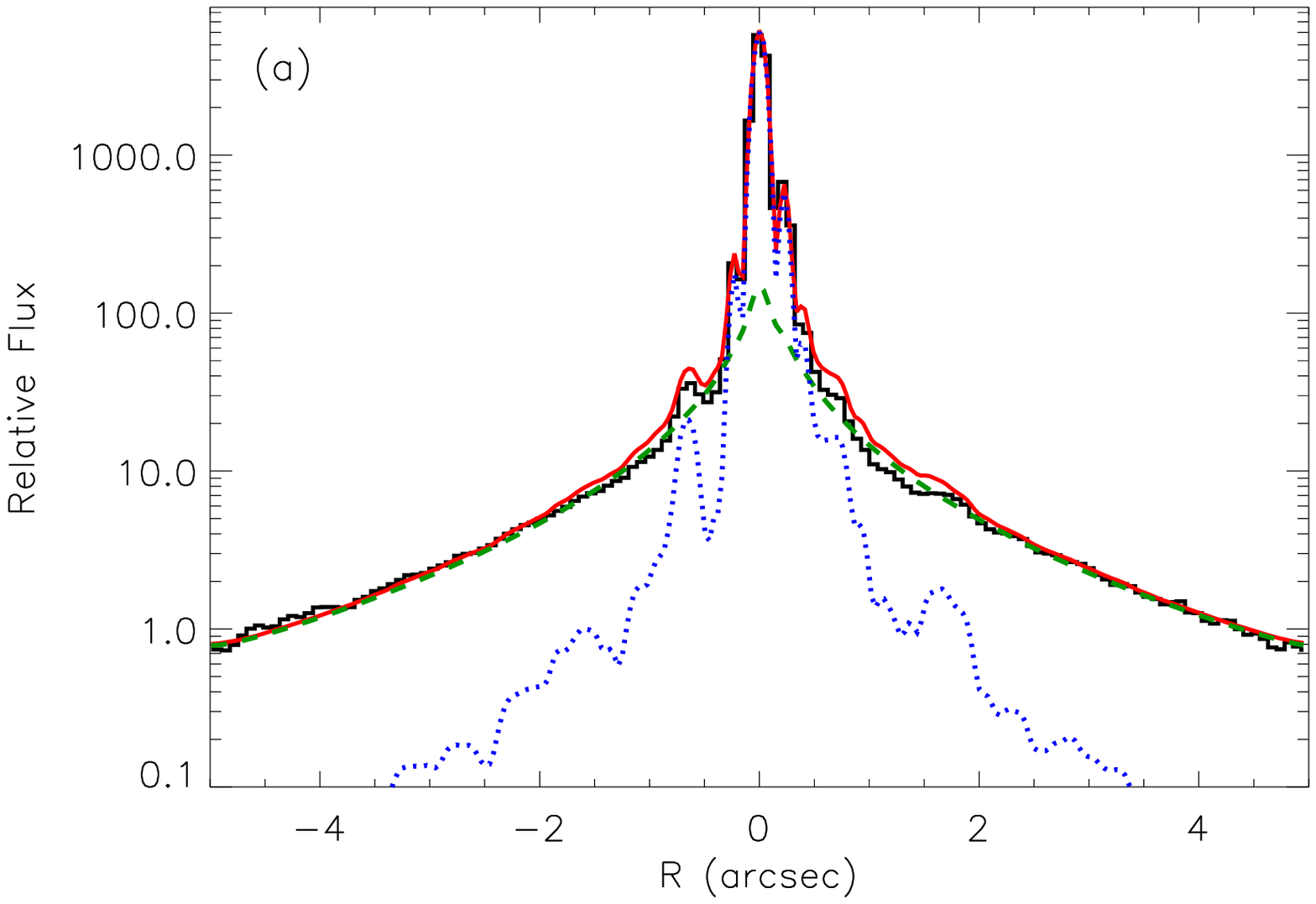}
\plotone{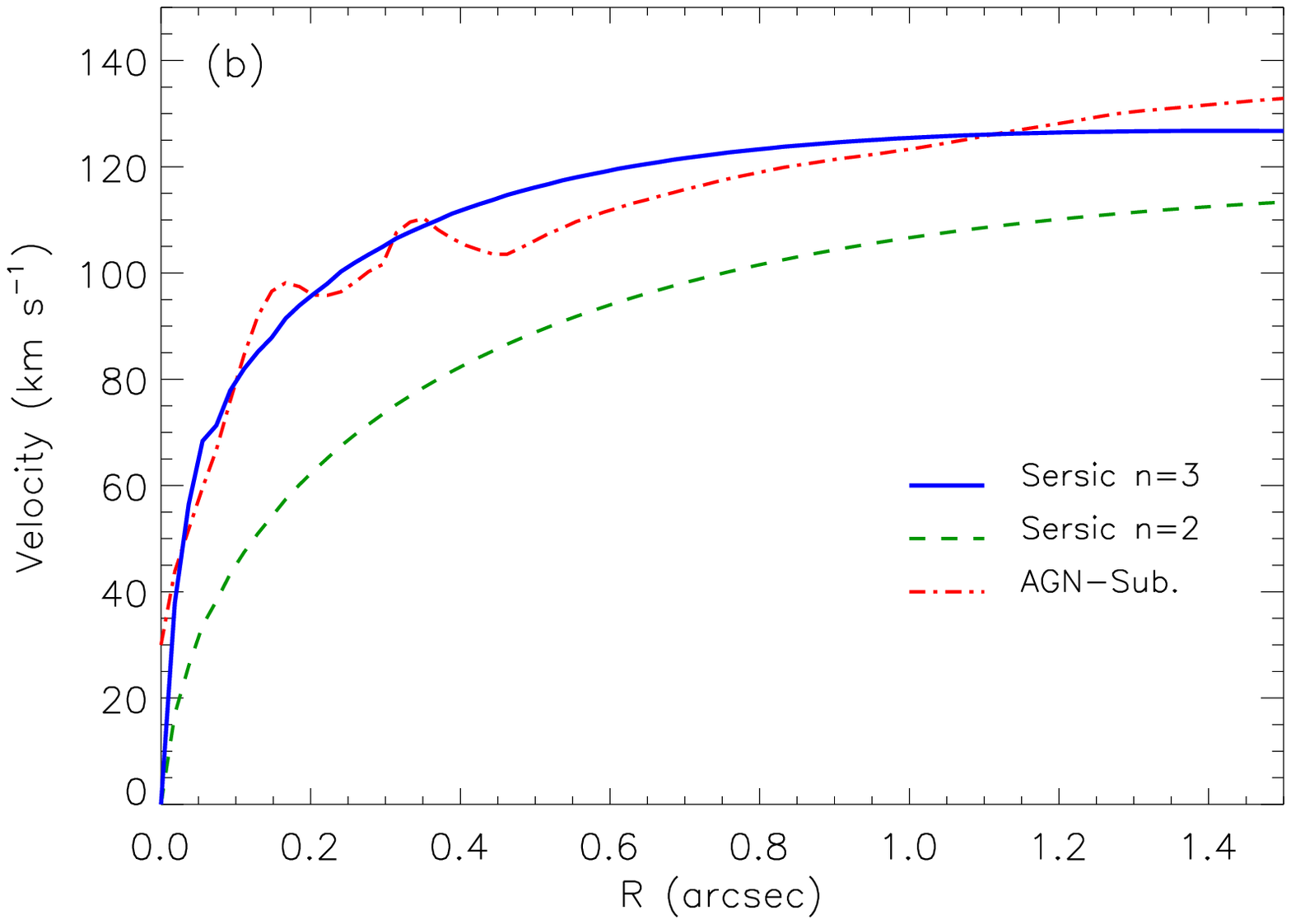}
\plotone{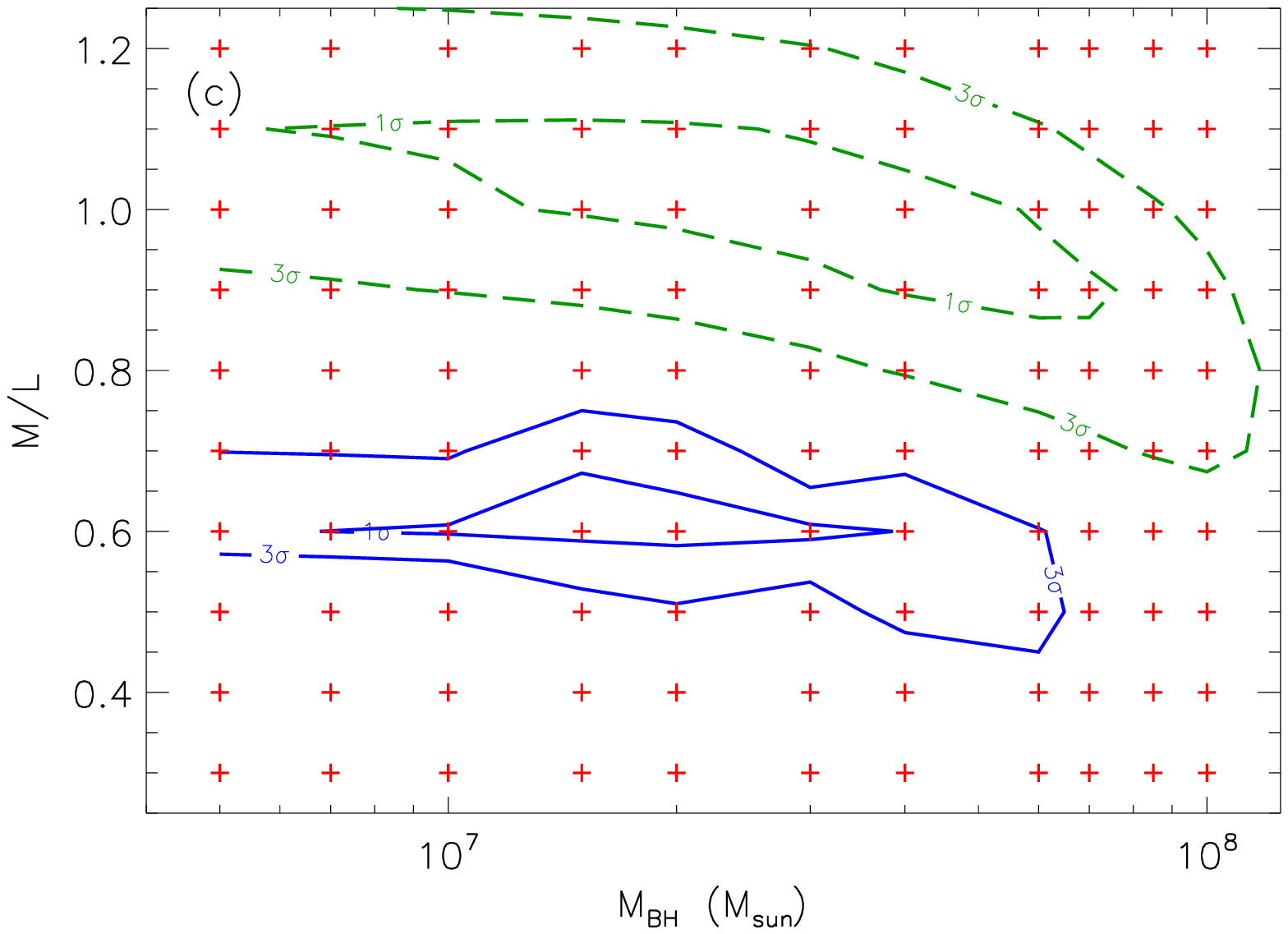}
\caption[NGC 4151: Stellar Light Distribution]{The effect of the stellar light distribution in NGC 4151 on the best fit model.  (a) The best fit to the stellar light distribution (solid curve) in the NICMOS F160W image (histogram) was found with a S\`{e}rsic \n=3 function (dashed curve) plus a point source represented by a Tiny Tim PSF (dotted curve).  (b) The rotation curves in the plane of the gas disk are shown for \n=3 and \n=2 S\`{e}rsic functions and the AGN-subtracted light distribution (as labeled in the legend) assuming an inclination angle of 40\deg\s and \ml=0.5 \mlu.  (c) Contours of constant \ch\s at the 1\sig and 3\sig confidence intervals, for two parameters, are shown for the S\`{e}rsic \n=3 fit (solid curves) and \n=2 fit (dashed curves), where \ch\s has been rescaled as discussed in the text. \label{4151_sersic}}
\end{figure*}

The best \mbh\s is not significantly dependent on the stellar light distribution used in the modeling.  The best fit S\`{e}rsic plus point source (scaled Tiny Tim PSF) fit to the NICMOS F160W image is a \n=3 function, with a reduced \ch=1.5 (Fig. \ref{4151_sersic}a).  The \n=3 S\`{e}rsic function is also the best match to the AGN-subtracted stellar light distribution, with the \n=1 and 2 distribution underestimating the central light gradient and \n=4 overestimating it.  A comparison of the velocity in the plane of the disk as a result of the \n=3 and \n=2 S\`{e}rsic fits and the AGN-subtracted light distribution is shown in Fig. \ref{4151_sersic}b. 

The best fit model with the \n=3 light distribution is, as was given earlier, 
\mbh= 3.0$^{+0.75}_{-2.2}\times10^{7}$ \Msun\s with \ml=0.6 \mlu\s at \inc=25\deg.  Alternatively, if the \n=2 S\`{e}rsic function is used for the light distribution, then the uncertainty of the best fit \mbh\s increases, but the best fit value is unchanged.  In this case the best fit model parameters are \mbh=3.0$^{+4.0}_{-2.2}\times10^{7}$ \Msun\s with \ml=0.6 \mlu\s at \inc=25\deg.  However, the \ch\s of the \n=2 fit is higher than is found with the \n=3 light distribution, ruling the \n=2 fit out at the 99.9\% confidence level compared to the \n=3 fit.  A comparison of the best fit for models with each light distribution is shown in Fig. \ref{4151_sersic}c.

\subsubsection{Emission Line Flux Distribution}

The \htwo \lam2.1218 flux distribution in NGC 4151 is very patchy, especially in the data taken under the best seeing conditions along a PA=90\deg\s (see Table \ref{t_obs}).  However, as will be discussed in $\S$ 6.3.5, these bright patches are most likely not emitted from gas associated with the general rotation, but are instead associated with a radio jet.  A single, very broad Gaussian is used to represent the underlying flux distribution of the rotating \htwo gas, the parameters of which are given in Table \ref{t_gflux}.  As in NGC 7469, the peak of the parameterized flux distribution is at the kinematic center and therefore only velocities measured at the inner most radii are affected.  There is no change in the best fit model parameters if a constant flux distribution is used instead of the single Gaussian distribution.

\subsubsection{Region Fit and Data Quality Considerations}
The minimum reduced \ch\s of the best fit models to the \htwo \lam2.1218 velocity field are not dependent on \rfit, with the minimum reduced \ch\s changing by less than one for \rfit=0.5 to 1.5.  The exception is \rfit$<$0\as.5, which gives a reduced \ch\s almost twice that found with other \rfit\s values.  This differs from the general trend for reduced \ch, which is to increase with an increase in the number of data points measured.  As discussed in $\S$ 5.4, this expected increase is due to the sensitivity of the reduced \ch\s to data that slightly departs from circular motion, and as \rfit\s is increased more of these slightly discrepant regions are included.  NGC 4151 does not follow this general trend because of the structure of its \htwo flux distribution, which is weak in the inner 0\as.6 in comparison to the patches of intense emission further out.  For this galaxy, when \rfit\s is increased, regions of higher \htwo flux, and thus regions with less velocity uncertainty (also higher PCC values) are included, which decreases \ch.  This counters the expected increase in the reduced \ch\s to the point of keeping it relatively constant.

\begin{deluxetable*}{cccccl}
\tablecaption{Comparison of Different Models for NGC 4151 
\label{t_4151mbhi}} 
\tablewidth{0pt}
\tablehead{
\colhead{Model\tablenotemark{a}} &
\colhead{\inc} & 
\colhead{\mbh} &
\colhead{\ml} &
\colhead{Reduced \ch} &
\colhead{Comments} \\
\colhead{} &
\colhead{(\deg)} &
\colhead{(10$^{7}$ \Msun)} &
\colhead{(\mlu)} &
\colhead{} &
\colhead{} \\
}
\startdata

A	&	10	&	50.0	&	1.9	&	12.339	&	\n=3	\\
A	&	15	&	3.0	&	1.6	&	12.303	&	\\
A	&	20	&	1.0	&	1.0	&	12.303	&	\\
A	&	25	&	3.0	&	0.6	&	12.304	&	\\
A	&	30	&	0.5	&	0.5	&	12.308	&	\\
A	&	35	&	3.0	&	0.3	&	12.311	&	\\
A	&	40	&	0.7	&	0.3	&	12.305	&	\\
A	&	45	&	0.0	&	0.3	&	12.308	&	\\
A	&	50	&	0.0	&	0.3	&	12.324	&	\\
A	&	60	&	0.0	&	0.3	&	12.345	&	\\
A	&	70	&	0.0	&	0.3	&	12.329	&	\\
A	&	80	&	0.0	&	0.5	&	12.357	&	\\
\cline{1-6}
B	&	25	&	3.0	&	0.6	&	12.320	&	\n=2 \\
C	&	25	&	3.0	&	0.6	&	12.305	&	w/o flux \\


\enddata
\tablenotetext{a}{These models have \t=6\deg\s and a stellar light distribution and emission line flux distribution as follows.  For \htwo \lam2.1218, Model A - S\`{e}rsic \n=3 and the two component Gaussian function, Model B - S\`{e}rsic \n=2 and the two component Gaussian function, and Model C - S\`{e}rsic \n=3 and a constant distribution.  All models have 617 degrees of freedom. }

\end{deluxetable*}

Despite the consistency of the reduced \ch\s with \rfit, the best fit model parameters are not independent of \rfit.  As mentioned, \t\s depends on \rfit, giving a best fit \t=-11\deg\s with \rfit=0\as.5 up to \t=9\deg\s with \rfit=1\as.5.  This apparent twist of \t\s with radius could be due to a warped disk or non-circular motion (e.g. radial flow), in which case the simple co-planar rotating disk model will not accurately estimate \mbh.  However, it is also possible that the twist in \t\s is simply due to the patchiness of the \htwo velocity field.  The field in general has higher velocities in the north, but, as can be seen most clearly in the 22 Apr 2003 data (for which the seeing was particularly good), there are regions where the velocity field deviates from the smooth velocity gradient.  The best fit \t\s found for a particular \rfit\s is essentially an average \t\s of the data, and as these regions of discrepant velocity are included in the fit, the averaged \t\s varies.  In addition, these regions often coincide with areas of higher flux, and thus smaller error is associated with the regions, giving them more weight in the \ch\s calculation.  The change in \t\s to the east (greater values of \t) as \rfit\s is increased is consistent with what is seen in the NGC 4151 \htwo velocity map (Fig. \ref{allmaps}).  Regions of higher velocity along a PA of $\sim$-20\deg\s at r$\sim$0\as.5 pull \t\s to the west at smaller \rfit, but when larger radii are included, \t\s is pulled to the east by a patch of high velocity at PA$\sim$90\deg\s at r$\sim$1\as.  The variation of \t\s with \rfit\s is therefore likely explained by a few regions of gas in non-circular motion, rather than a true change in the underlying generally rotating gas disk.  This uncertainty in \t\s does not affect the best fit values of the other model parameters, as they are not dependent on \t. 

Although the \htwo gas in NGC 4151 does not have a purely smooth velocity gradient, it is possible that the few discrepant regions are from another velocity component that is superimposed on the general rotation.  In this case a simple coplanar model is a valid interpretation of the general velocity field, but not necessarily of the second component.  This picture is supported by studies of the \oiii\s velocity field, which has been measured by many authors (e.g. \citealt{winge99}, \citealt{nelson00}, \citealt{das05}).  They find that there are at least two components to the \oiii\s velocity field, one of which is consistent with rotation of a disk.  The other components are associated with radial flow connected with the radio jet oriented along PA=77\deg.  As is shown in Fig. \ref{radio4151}, many of the bright knots of \htwo emission are aligned with this radio jet.  Since these same bright patches have velocities that are not in agreement with the general rotation, it is suggested that they are evidence of an additional velocity component.  Unfortunately, separation of these two \htwo velocity components is not possible at the NIRSPEC spectral resolution.  

Additional support for the presence of a rotating \htwo disk comes from the fact that the best fit \mbh\s (and \ml\s and \inc) is relatively constant with \rfit (Fig. 43).  For \rfit=0\as.25-1\as.0, the best fit \mbh\s has a range of \mbh=2-4$\times 10^{7}$ \Msun, which gives a range in \mbh\s just a bit wider than the error of the \rfit=1\as.0 mass estimate already discussed.  For \rfit$>$1\as.0, the fit becomes dominated by data at larger radii and as \rfit\s is increased the fit becomes less sensitive to the inner velocity gradient, and thus insensitive to the BH mass.  A balance between the need to find as tight a constraint as possible on the model parameters and having a large enough area to not be dominated by individual regions of discrepant velocity is found by using \rfit=1\as.0. 

\begin{figure}[!t] 
\epsscale{1}
\plotone{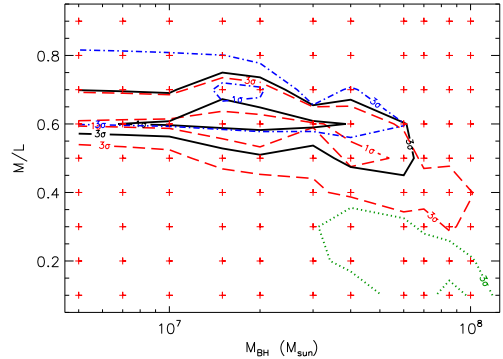}
\caption[NGC 4151: Best Fit \mbh\s for different \rfit]{Contours of constant \ch\s for NGC 4151 with different values of \rfit, where \ch\s has been rescaled as discussed in the text.  The 1\sig and 3\sig confidence intervals, for two parameters, are shown for \rfit=0\as.5 (dash-dotted curves), 0\as.75 (dashed curves), 1\as.0 (solid curves), and 2\as.0 (dotted curves). \label{415_MvMLR_rfit}}
\end{figure}

\begin{deluxetable*}{llcccc}
\tablecaption{2-D Modeling Results with Different Statistical Methods
\label{t_results}} 
\tablewidth{0pt}
\tablehead{
\colhead{Sefyert 1} & 
\colhead{Method} &
\colhead{\mbh} &
\colhead{\ml} & 
\colhead{\t} &
\colhead{\inc} \\
\colhead{} &
\colhead{} &
\colhead{(10$^{7}$ \Msun)} &
\colhead{(\mlu)} &
\colhead{(\deg)} &
\colhead{(\deg)} \\
}
\startdata

NGC 3227	&	Relative Likelihood $\Delta$\ch	&	2.0$^{+1.0}_{-0.4}$	&	0.7			&	139.2$\pm$0.1	&	15$\pm$1	\\
		&	Bootstrap Distribution			&	1-5 (72\%)			&	0.5-0.8 (65\%)	&	139.1-139.2 (60\%)	&	15 (74\%)	\\	
		&	Bootstrap Gaussian Fit			&	3.4$\pm$1.9			&	0.69$\pm$0.13	&	139.18$\pm$0.04	&	14.2$\pm$2.5	\\
\cline{1-6}
NGC4151	&	Relative Likelihood $\Delta$\ch	&	3.0$^{+0.75}_{-2.2}$	&	0.6			&	6.5$\pm$0.5		&	20$\pm$2	\\
		&	Bootstrap Distribution			&	1-7 (63\%)			&	0.4-0.7 (67\%)	&	6-10 (63\%)		&	10-25 (69\%)	\\
		&	Bootstrap Gaussian Fit			&	3.9$\pm$2.9			&	0.55$\pm$0.19	&	7.30$\pm$1.80	&	19.8$\pm$2.9	\\
\cline{1-6}

NGC 7469	&	Relative Likelihood $\Delta$\ch	&	$<$5.0				&	0.9			&	131.6$^{+0.5}_{-0.1}$	&	17-25	\\
		&	Bootstrap Distribution			&	$<$1 (80\%)			&	0.7-1.0 (80\%)	&	131.6-132.0 (75\%)	&    15 (89\%) \\
		& 	Bootstrap Gaussian Fit			&	$<$0.2				&	0.86$\pm$0.13	&	131.66$\pm$0.08	&    15.0$\pm$1.8 \\
\enddata
\tablecomments{The estimates given for the relative likelihood $\Delta$\ch\s and bootstrap Gaussian fit methods are 1\sig estimates.  For the half-sample bootstrap distribution estimates the number in parentheses is the percentage of the distribution within the range of values given (chosen to be as close to 1\signs, or 68\%, as possible).  See $\S$ 5.4 for a description of the different statistical methods used to determine the range of best fit parameter values. }

\end{deluxetable*}

\begin{figure*}[!htt]
\epsscale{0.5}
\plotone{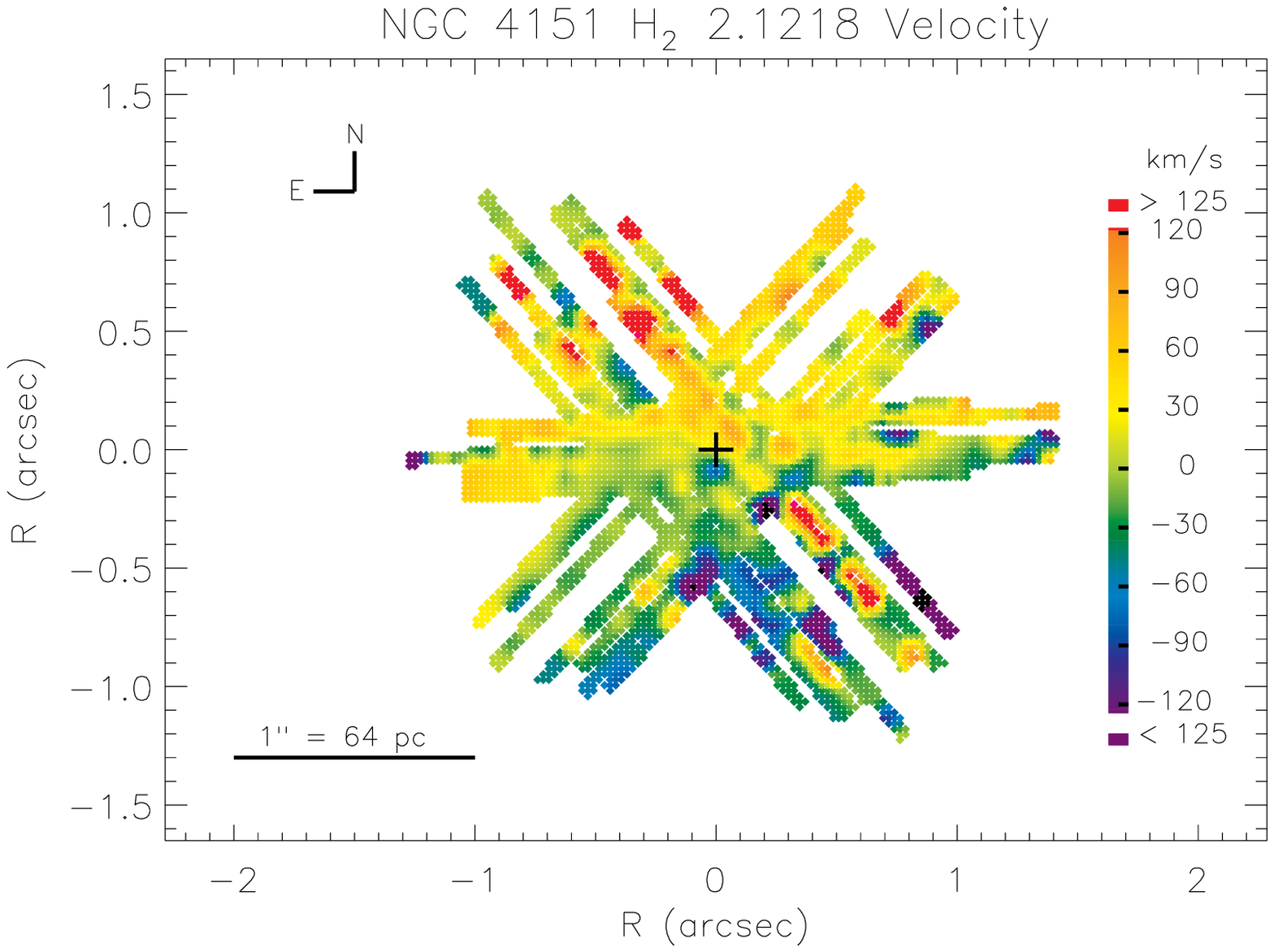}
\plotone{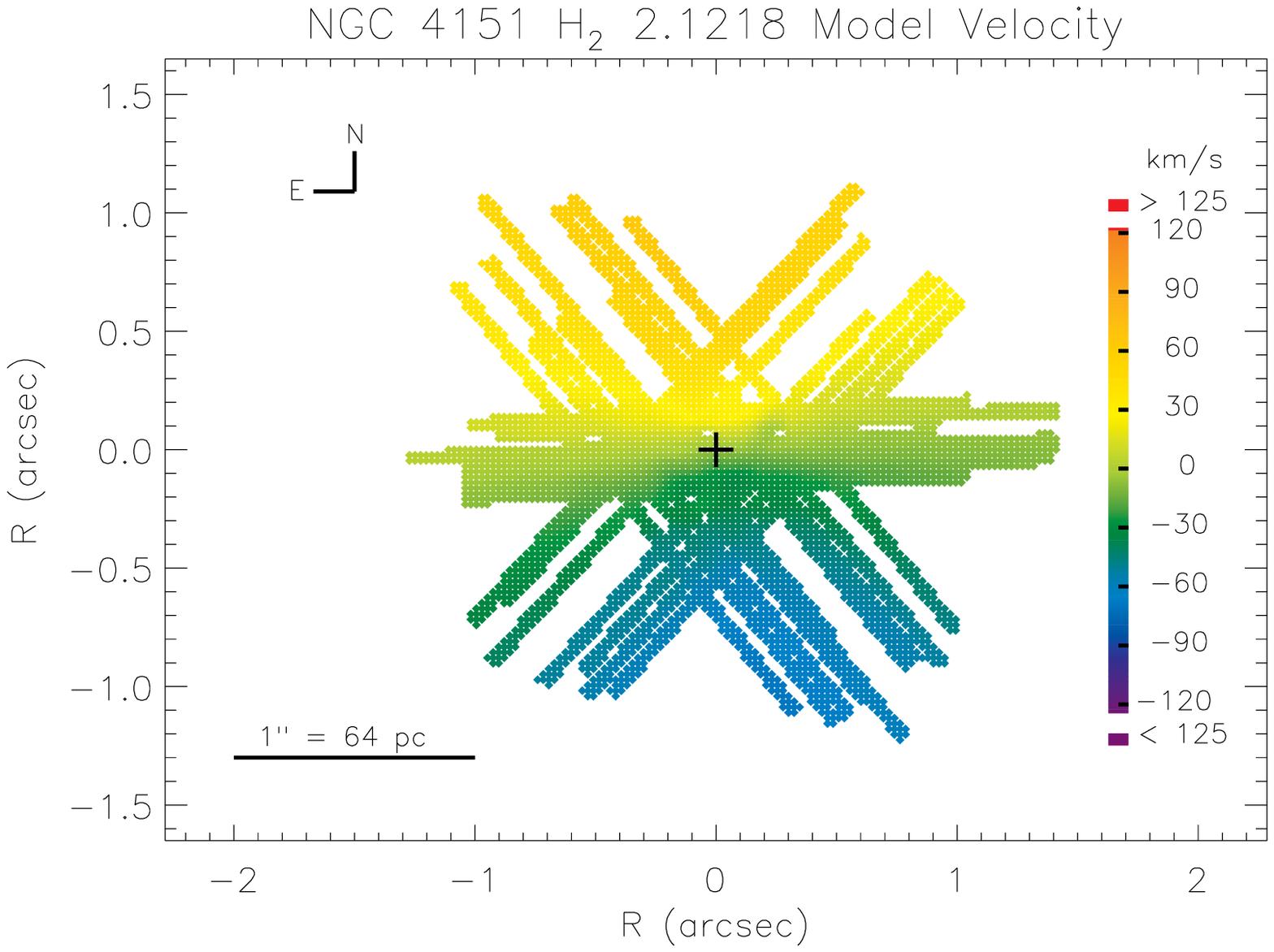}
\plotone{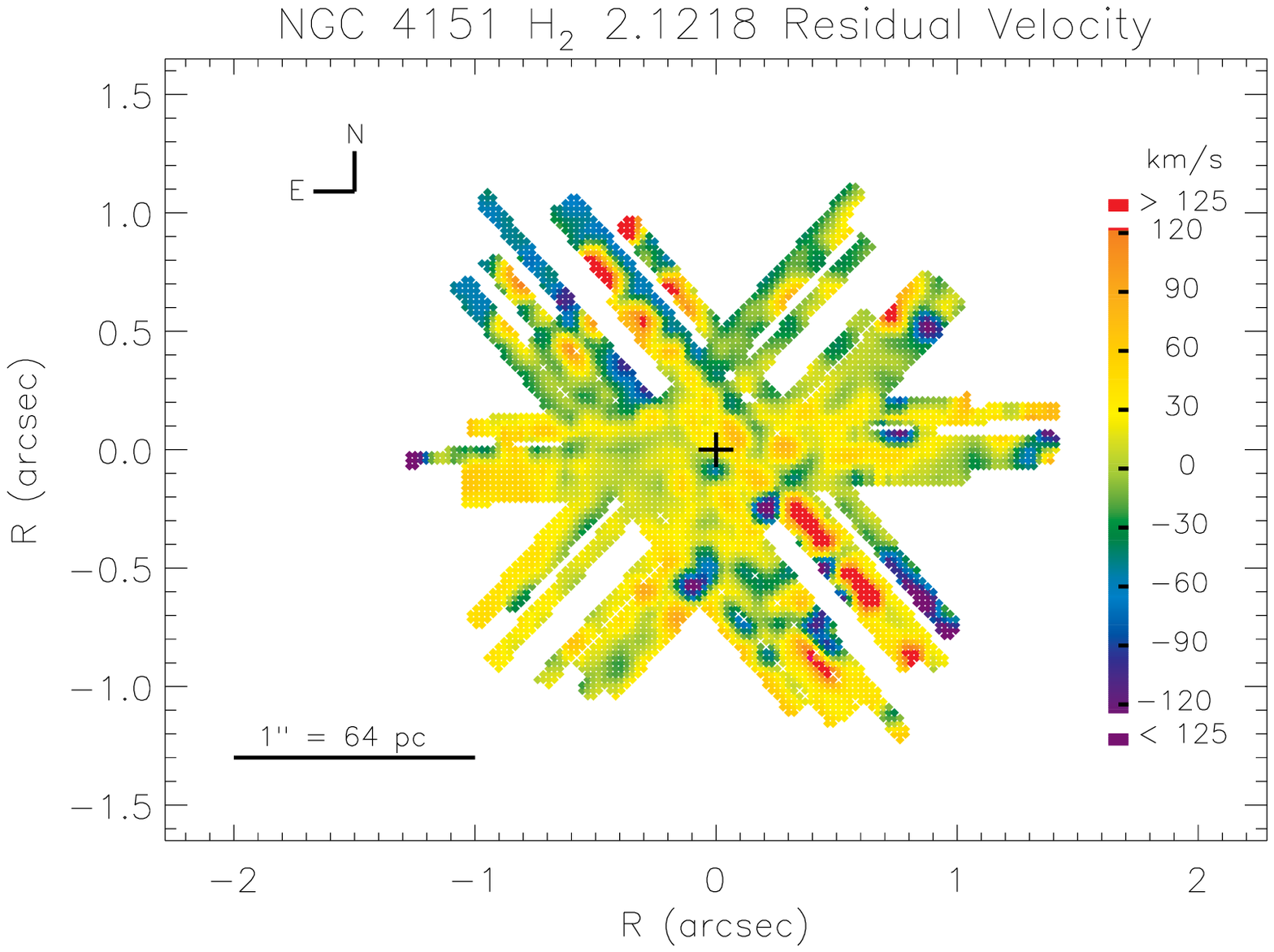}
\caption[NGC 4151: 2-D Map of Model Velocity Field and Residuals]{2-D map of the NGC 4151 \htwo 2.1218 measured velocity field, model velocity field, and the residuals of the model minus the data, as indicated by the labels at the top of each plot.  The model has \mbh= 3.0$\times10^{7}$ \Msun\s with \ml=0.6 \mlu\s and \t=6.5\deg, assuming a fixed \inc=25\deg\s and \rfit=1\as.0 with a S\`{e}rsic \n=3 stellar light distribution. \label{4151_maps}}
\end{figure*}

The relatively low cut off value of PCC$>$0.3 is used for all modeling of NGC 4151 which allows the weaker \htwo emission in the nuclear region to be included in the fit.  For cut off values less than 0.4, beyond which the central 0\as.6 is eliminated from the fit, the best fit model parameters are not dependent on the PCC cut off limit.  A cut off value of 0.3 is conservative in that less data is eliminated, which adds noise to the velocity field resulting in greater error on the model parameter estimates. 

\subsubsection{Summary of \mbh\s Estimate for NGC 4151}
The molecular hydrogen in the nuclear region of NGC 4151 is in systematic rotation, but there is evidence for a second velocity component that is possibly associated with the nuclear radio jet.  The value of \rfit\s is selected to minimize the influence of this second component on the modeling of the rotating gas disk.  The best fit to the \htwo \lam 2.1218 velocity field in NGC 4151 is \mbh= 3.0$^{+0.75}_{-2.2}\times10^{7}$ \Msun\s with \ml=0.6 \mlu, assuming a fixed \inc=25\deg\s and \rfit=1\as.0 with a S\`{e}rsic \n=3 stellar light distribution.  The range of acceptable parameters values are confirmed with the half-sample bootstrap method (see Table \ref{t_results} for a comparison).  If instead a \n=2 light distribution is used, then the error on the best fit \mbh\s increases to 3.0$^{+4.0}_{-2.2}\times10^{7}$ \Msun\s with \ml=0.6 \mlu\s and \inc\s fixed at 25\deg.  2-D maps of the measured and model velocity fields, as well as the residuals, are shown in Fig. \ref{4151_maps}.   

The complicated nuclear kinematics in NGC 4151 has caused previous, lower spatial resolution studies to have difficulty tracing the rotating component of gas down to the spatial scales necessary to estimate the mass of the BH.  \citet{winge99} were, however, able to place an upper limit of \mbh$\leq$5$\times 10^{7}$ \Msun\s based on the kinematics of the highly ionized \oiii\s gas, which is consistent with the estimated mass from the modeling of the molecular hydrogen kinematics.

\subsection{\mbh\s Upper Limits based on Gas Dynamics}
For those galaxies without an organized \htwo or \bg velocity field indicative of rotation, upper limits on the BH mass can be determined.  Assuming the gas is rotating in a disk, an upper limit can be calculated using
\begin{equation}
M_{BH} \, \le \,  \frac{v_{obs}^{2} \, r}{G \, sin(i)}
\end{equation}
\noindent where $v_{obs}$ is the observed velocity, \inc\s is the inclination of the disk (\inc=0 is face-on), {\em r} is the radius at which the velocity is measured, and G is the gravitational constant.  Table \ref{t_uplimits} gives the maximum velocity gradient at a radius of 1\as\s for each of the nine galaxies.  Although the gas dynamical modeling presented ($\S$ 6.1-6.3) places much tighter constraints on \mbh, NGC 3227, NGC 4151, and NGC 7469 are included in Table \ref{t_uplimits} for comparison.  Upper limits are calculated for an assumed disk inclination of 20\deg\s and 45\deg, and the results are listed in Table \ref{t_uplimits}.

\begin{deluxetable*}{lcccccc}
\tabletypesize{\tiny}
\tablecaption{\mbh\s Upper Limits \label{t_uplimits}} 
\tablewidth{0pt}
\tablehead{
\colhead{Galaxy} &
\colhead{$v_{obs}$} &
\colhead{pc/\as} & 
\colhead{\mbh\s (10$^{7}$ \Msun)} &
\colhead{\mbh\s (10$^{7}$ \Msun)} &
\colhead{\mbh\s (10$^{7}$ \Msun)} &
\colhead{\mbh\s (10$^{7}$ \Msun)} \\ 
\colhead{} &
\colhead{(r=1\as)} &
\colhead{} &
\colhead{\inc=20\deg} &
\colhead{\inc=45\deg} &
\colhead{reverb.} &
\colhead{gas dyn.} \\
}
\startdata

NGC 3227	&	70	&	74	&	22	&	10	&	4.2$\pm$2.1	&	2.0$^{+1.0}_{-0.4}$ \\
NGC 3516	&	50	&	172	&	29	&	14	& 	4.3$\pm$1.5	&	\\
NGC 4051	&	30	&	45	&	28	&	14	&	0.2$\pm$0.1	&	\\
NGC 4151	&	60	&	65	&	16	& 	7.7	&	4.6$\pm$0.5	&	3.0$^{+0.75}_{-2.2}$	\\
NGC 4593	&	70	&	175	&	58	&	28	&	1.0$\pm$0.2	&	\\
NGC 5548	&	60	&	335	&	82	&	40	&	6.5$\pm$0.3	&	\\
NGC 6814	&	40	&	102	&	11	&	5.4	&	1.2$\pm$0.5	\\
NGC 7469	&	60	&	318	&	78	&	38	&	1.2$\pm$0.1	&	$\le$5.0	\\
Ark 120	&	75	&	632	&	241	&	117	& 15.0$\pm$1.9	&	\\					
\enddata
\tablecomments{The fourth and fifth columns are upper limits based on equation 5 with an assumed disk inclination of \inc=20\deg\s and \inc=45\deg.  The two rightmost columns are mass estimates based on reverberation mapping (except for NGC 6814; see references in Table \ref{t_sample}) and modeling of the gas dynamics presented in $\S$ 6.}

\end{deluxetable*}

\subsection{Calibration of the Reverberation Mapping Technique}
\begin{figure}[!b] 
\epsscale{1}
\plotone{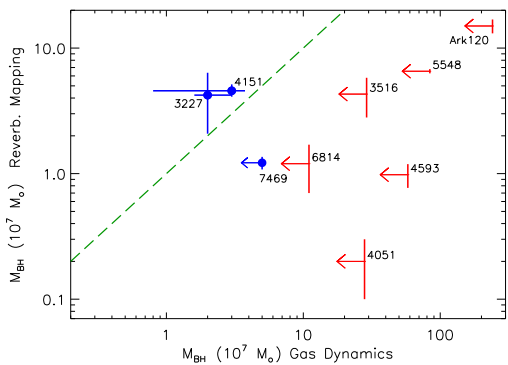}
\caption[Gas Dynamical \mbh\s versus Reverberation Mapping]{ \mbh\s from gas dynamics versus the reverberation mapping estimates.  The filled circles are the three galaxies for which modeling of the \htwo disk was completed and the remaining data points are upper limits based on the gas kinematics.  The line of equality between the direct and indirect \mbh\s estimates is shown by the dashed line.  All galaxies are labeled by their NGC number or full galaxy name. \label{reverb}}
\end{figure}

There is general agreement between the BH masses from reverberation mapping (typically taken from \citealt{peterson04}) and the measurements based on the gas dynamics that have been presented.  For NGC 3227 reverberation mapping gives \mbh=4.2$\pm$2.1 $\times 10^{7}$ \Msun.  This is higher than the estimate given by the gas dynamics, which is \mbh=2.0$^{+1.0}_{-0.4}\times10^{7}$ \Msun, but they agree within the errors of the measurements.  Reverberation mapping gives \mbh=4.6$\pm$0.5$\times10^{7}$ \Msun\s for NGC 4151 (\citealt{bentz07}), which is also consistent with the estimate of \mbh=3.0$^{+0.75}_{-2.2}\times10^{7}$ \Msun\s from the gas kinematics.  Finally, for NGC 7469, reverberation mapping yields a mass of \mbh=1.2$\pm$0.1$\times10^{7}$\Msun, which is consistent with the gas dynamical estimate of \mbh$<5.0\times10^{7}$ \Msun.  A comparison of these mass estimates, along with the \mbh\s upper limits for the rest of the galaxies, is shown in Fig. \ref{reverb}.  The consistency of the mass determinations through the method of reverberation mapping with the masses determined from the gas kinematics demonstrates that the reverberation mapping method is accurate to within a factor of three without indication of any systematic error. 

\subsection{Significance of the Low Inclination Angeles: AGN Unification Theory}

AGN unification theory states that type 1 and type 2 AGNs, such as Seyfert 1 and 2 galaxies, are from a single population and they differ in appearance as a result of the angle at which they are viewed.  In this scenario, Seyfert 1 galaxies are seen with a direct line of sight to the central engine, whereas the AGN is obscured in Seyfert 2 galaxies.  It is speculated that the obscuring material is in a thick disk or torus-like structure such that when the `disk' is viewed face-on (i.e. low inclination angles) the central engine is unobscured and a Seyfert 1 is seen.  Similarly, when the `disk' is viewed at a high inclination angle the line of sight is through the obscuring material and a Seyfert 2 galaxy is detected.

In all three of the Seyfert 1 galaxies for which gas dynamical modeling was completed, the inclination of the nuclear molecular gas disk is found to be \inc\s $\sim$ 20\deg, with a 3\sig limit of \inc\s $\le$ 45\deg\s in all three cases.  If the nuclear disks were oriented at random with respect to the line of sight then there would be only a 0.02 probability of observing three galaxies with a disk inclination of \inc\s $\le$ 25\deg, and a 0.12 probability of observing three galaxies with a disk inclination of \inc\s $\le$ 45\deg.  This supports the theory that Seyfert 1 galaxies preferentially have nuclear gas disks that are oriented nearly face-on with respect to our line of sight.

\section{Conclusions}

The 2-D distribution and kinematics of the molecular, ionized, and highly ionized gas in the nuclear regions of nearby bright Seyfert 1 galaxies have been studied using high spatial resolution, single slit near-infrared spectroscopy from NIRSPEC with AO on the Keck telescope.  Molecular hydrogen is detected at \lam2.1218 in all nine of the Seyfert 1 galaxies in the sample, and in four of the galaxies \htwo is detected in more than one transition.  In six of the galaxies \htwo is spatially resolved on scales of tens of parsecs with a major axis PA consistent with that of larger scale optical emission, and the remaining three galaxies have distributions consistent with that of the compact \kb\s continuum.  In addition, NGC 3227 and NGC 4151 have a complex \htwo emission structure throughout the 2\as\s FOV, with several spatially resolved knots of emission.  Broad \bg emission is also detected in all nine Seyfert 1 galaxies, four of which also have a separable narrow \bg component.  In all four of these galaxies the distribution of the narrow \bg emission matches that of the \kb\s continuum and its peak is coincident with that of the AGN, which is assumed to be located at the peak of the \kb\s continuum.

In five of the nine Seyfert 1 galaxies the \htwo kinematics are consistent with disk rotation.  Three of these galaxies have a velocity gradient of 100 \kms or greater across the central 0\as.5.  The other two galaxies have a similar velocity gradient over 1\as.5.  In the remaining four galaxies, it is possible that insufficient 2-D coverage or a face-on gas disk can explain the lack of detected rotation.  In each of the four galaxies for which a narrow \bg component was measured, the \bg kinematics are similar to that of the rotating molecular hydrogen.  However, in the nuclear region (r $\le$0\as.5) \bg is redshifted in two of the galaxies and blue shifted in one galaxy by more than 75 \kms with respect to \htwons.  Also of note is that in one galaxy, NGC 4151, there is evidence of a second velocity component superimposed on the general rotation, which could be associated with a radial outflow of gas related to the nuclear radio jet.

The 2-D coronal line flux distributions and kinematics are also measured in the sample of Seyfert 1 galaxies.  In all but one case, the [Ca VIII] and [Si VII] coronal emission is found to be consistent with the flux distributions of \bg and the \kb\s continuum.  The only galaxy for which the coronal flux distribution differs from the \kb\s emission is NGC 4151, and in this case the coronal emission does follow the flux distribution of \bgns.  The coronal emission is not measured to be spatially coincident with the \htwo emission in any of the galaxies.  In each galaxy the coronal emission is measured to be red- or blueshifted in the central 0\as.5, and is in agreement with the kinematics of \bgns.  

The velocity dispersions of both the coronal emission and the narrow \bg component are greater than that of \htwo by 1.3$\pm$0.5 and 2.0$\pm$0.7 times, respectively.  In addition, the velocity dispersion of \bg is similar to that reported in the literature for \oiii.  If this dispersion is assumed to be due to rotation, then the \oiii~ and \bg line emitting gas is located closest to the AGN and \htwo furthest, with the coronal gas located at an intermediate radius. 

The 2-D gas kinematics have been fitted with dynamical models that assume a flat thin disk undergoing circular rotation in a gravitational field created by both the stellar population and a central point mass, presumably a massive black hole.  The stellar gravitational field is estimated from archival HST NICMOS near-infrared images.  The stellar light is separated from the Seyfert 1 nucleus by fitting a S\`{e}rsic function plus a point source (represented by a model PSF), and a constant mass-to-light ratio, \ml, is used to convert it to a stellar mass distribution.  The model velocity field for the gas disk is then synthetically observed to fit the model to the observed velocity field.  This takes into account the PSF for each spectroscopic exposure, which is monitored with SCAM.  A method was developed to reliably estimate the PSF from a fit to just the core of the galaxy light profile, which is dominated by the point-like AGN.  The emission line surface brightness distribution, determined from the measured 2-D flux distribution of the line emitting gas, is also taken into account.

Based on the \htwo gas dynamical modeling, \mbh\s has been estimated in three galaxies.  In NGC 3227, the best fit model of the \htwo \lam2.1218 kinematics, assuming \inc=20\deg, gives \mbh=2.0$^{+1.0}_{-0.4}\times10^{7}$ \Msun\s with \ml=0.70$\pm$0.05 \mlu\s and a S\`{e}rsic \n=3 stellar light distribution.  This mass estimate is consistent with the estimate from two additional \htwo transitions at \lam2.4066 and \lam2.4066.  For NGC 4151 the rotating component of the \htwo \lam2.1218 velocity field is best fit by \mbh=3.0$^{+0.75}_{-2.2}\times10^{7}$ \Msun\s with \ml=0.6 \mlu, also with a S\`{e}rsic \n=3 stellar light distribution and assuming \inc=25\deg.  In NGC 7469, the AGN-subtracted light distribution was found to better represent the stellar light distribution, and with this distribution the best fit model of the \htwo \lam2.1218 kinematics gives \mbh$<5.0\times10^{7}$ \Msun\s with \ml=0.9 \mlu\s and assuming \inc=20\deg.  

A very interesting result of the dynamical modeling is that each of the galaxies prefers a near face-on molecular gas disk.  The best fit models have \inc\s $<$ 16\deg, 22\deg, and 25\deg, at the 1\sig level, for NGC 3227, NGC 7469, and NGC 4151, respectively (\inc\s $<$ 22\deg, 40\deg, and 47\deg\s at the 3\sig confidence level).  This is consistent with unification theory, which predicts that Seyfert 1 galaxies should have a near line-of-sight view of the central engine.  Another interesting result is that the stellar surface densities estimated in the inner 2\as\s for all three galaxies are found to be higher than most non-active spiral galaxies.  Finally, a comparison of the direct gas dynamical \mbh\s estimates to estimates based on the technique of reverberation mapping indicates that the latter technique is accurate to a factor of three with no indications of systematic errors.

The high spatial resolutions now available have made it possible to investigate the nuclear regions of active galaxies with a level of accuracy never before achieved.  With a sample of Seyfert 1 galaxies, this research has, for the first time, investigated the distribution and kinematics of the molecular and ionized nuclear gas at spatial scales down to $\sim$10 pc in galaxies with a direct view of the central engine.  Modeling of the gas kinematics in three galaxies has provided a direct measurement of \mbh, which shows that the reverberation mapping technique is capable of accurately determining \mbh.  This method, and others calibrated against it, can now be extended with greater confidence to larger samples of galaxies and to galaxies at greater distances, providing an essential tool for furthering our understanding of the role black holes play in galaxy formation and evolution.

The authors wish to recognize and acknowledge the very significant cultural role and reverence that the summit of Mauna Kea has always had within the indigenous Hawaiian community.  We are most fortunate to have the opportunity to conduct observations from this mountain.  We also thank the referee for comments that helped improve the manuscript.

\appendix
\section{Nuclear and Off-Nuclear Near-Infrared Spectra}
\begin{figure}[!ht]	
\epsscale{0.4}
\plotone{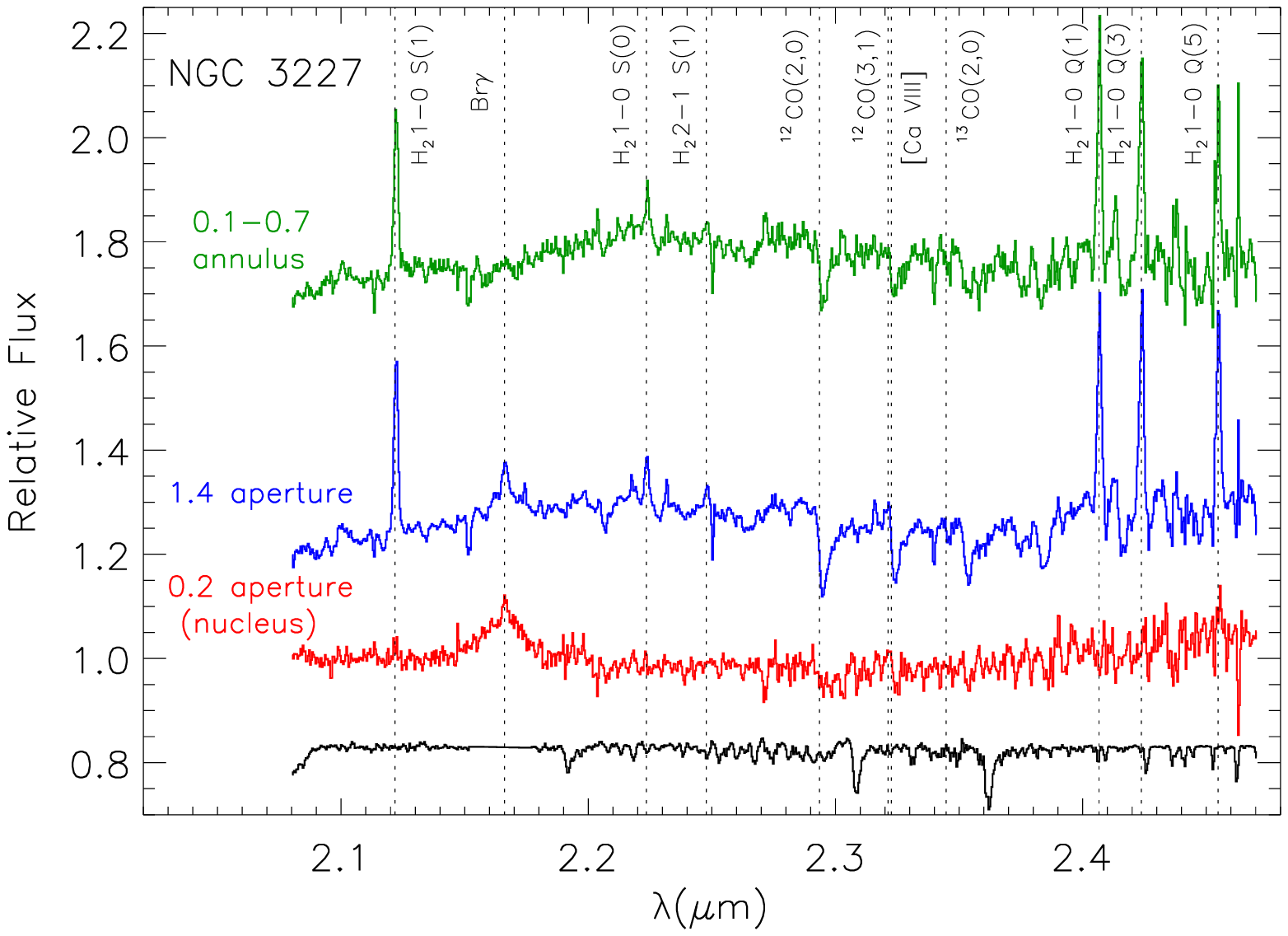}
\plotone{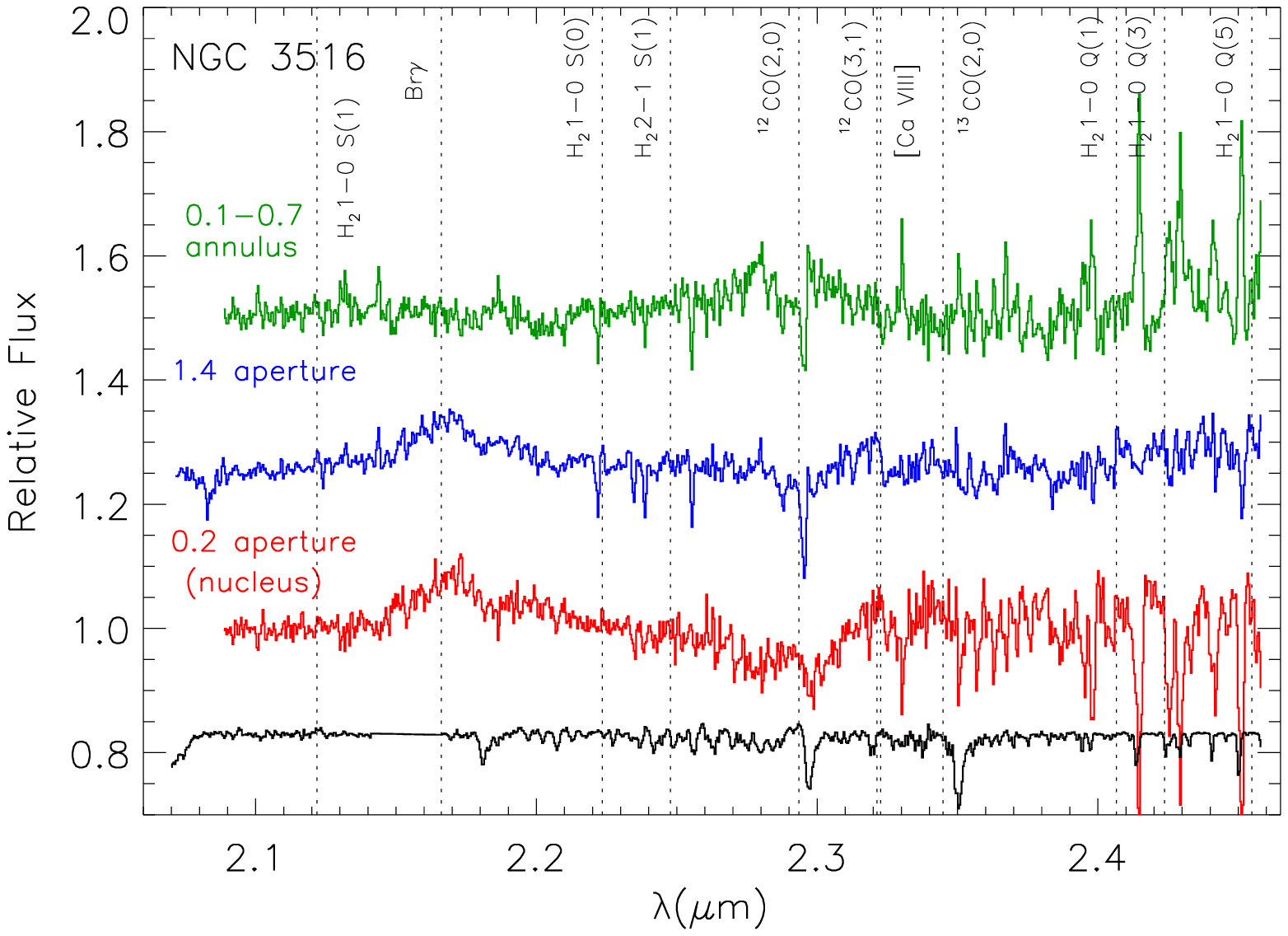}
\plotone{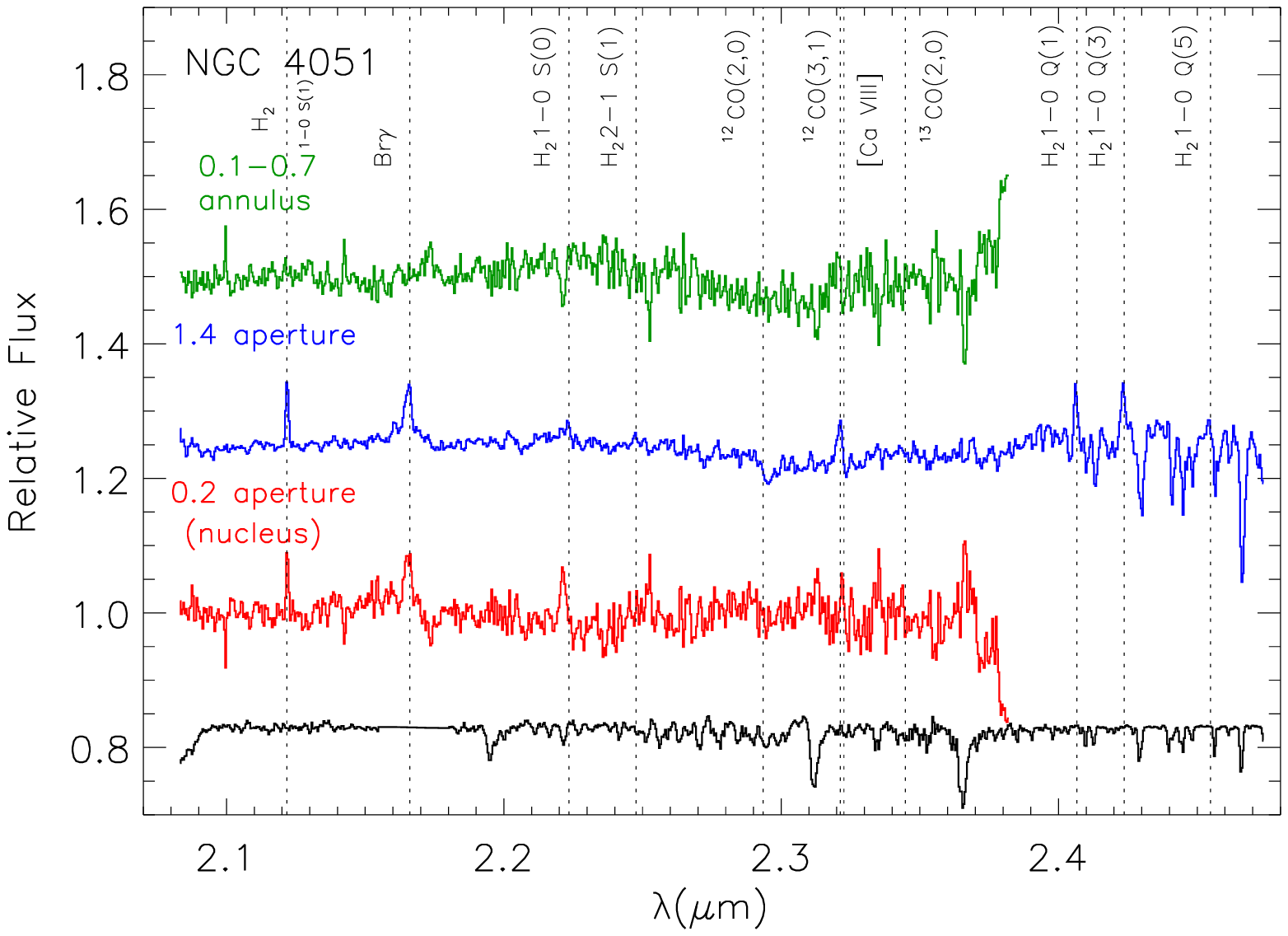}
\plotone{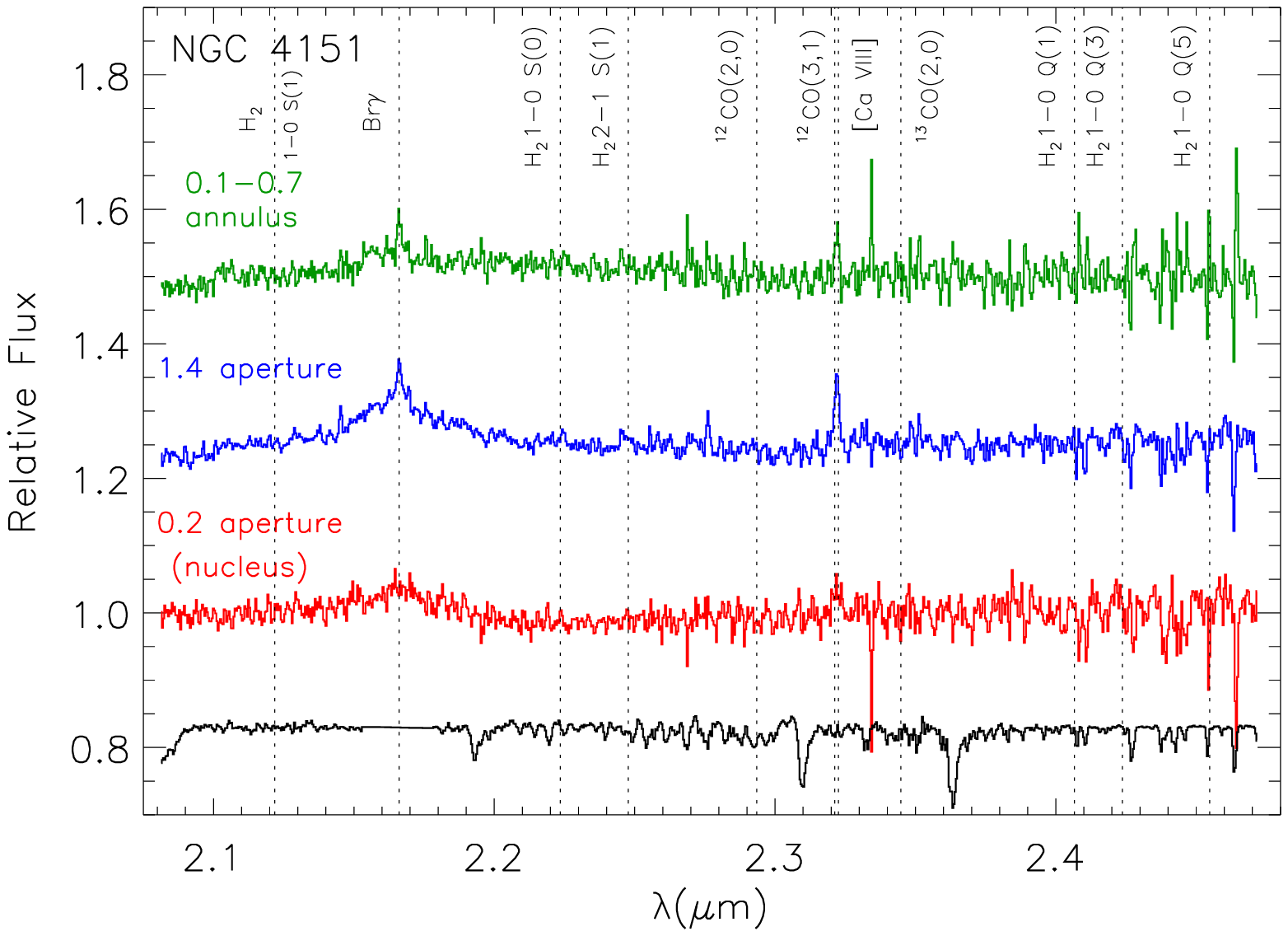}
\plotone{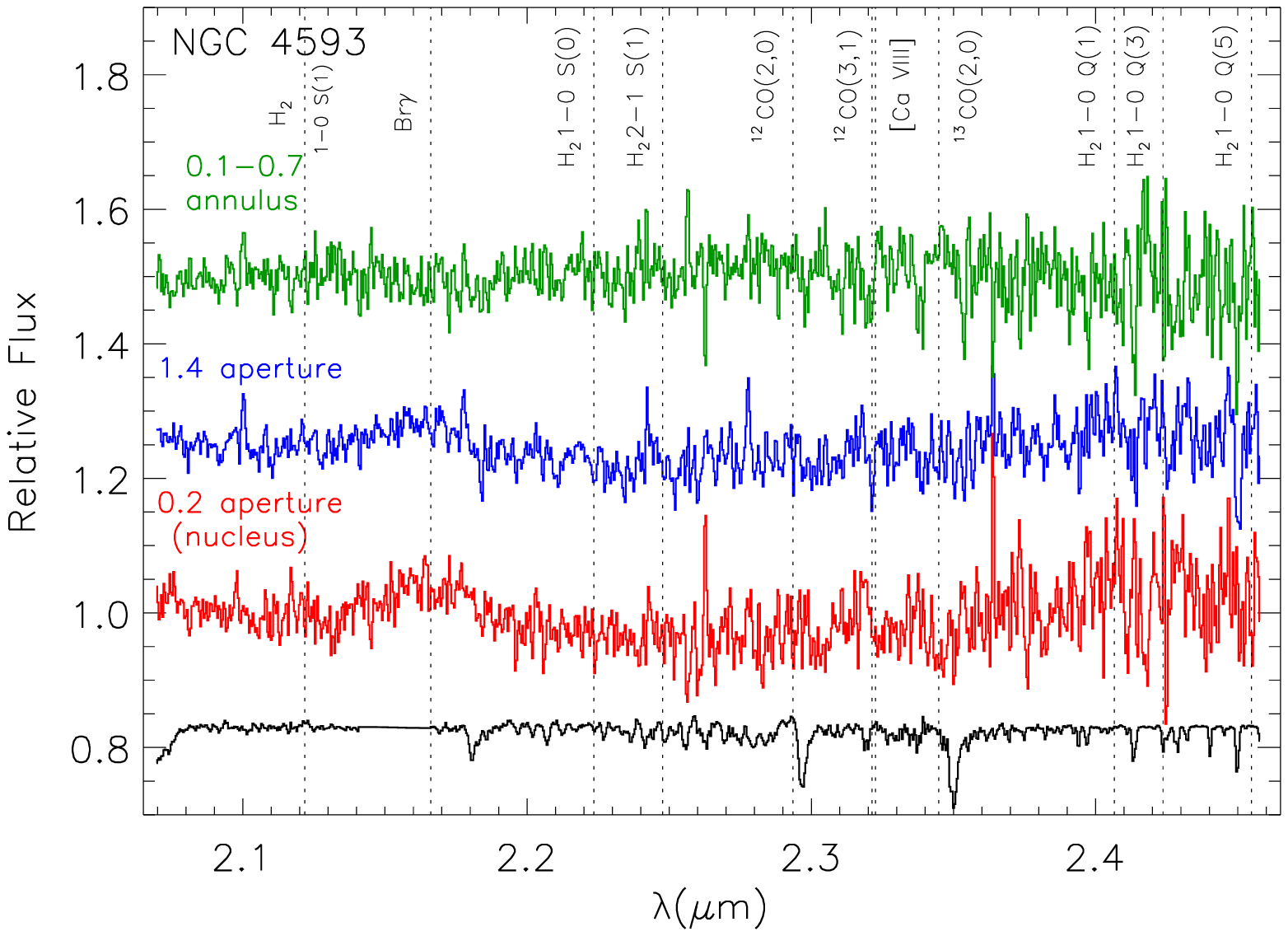}
\plotone{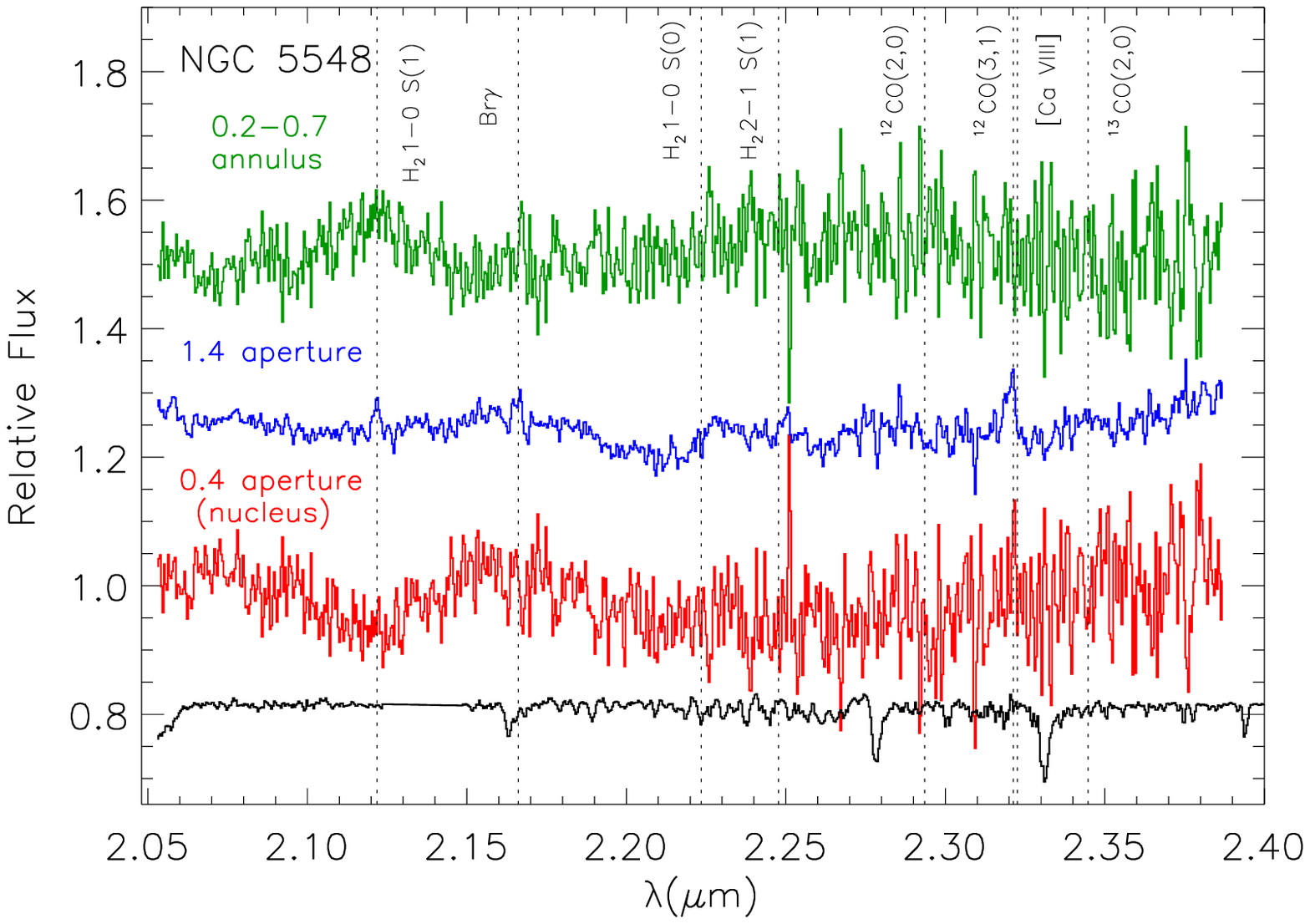}
\plotone{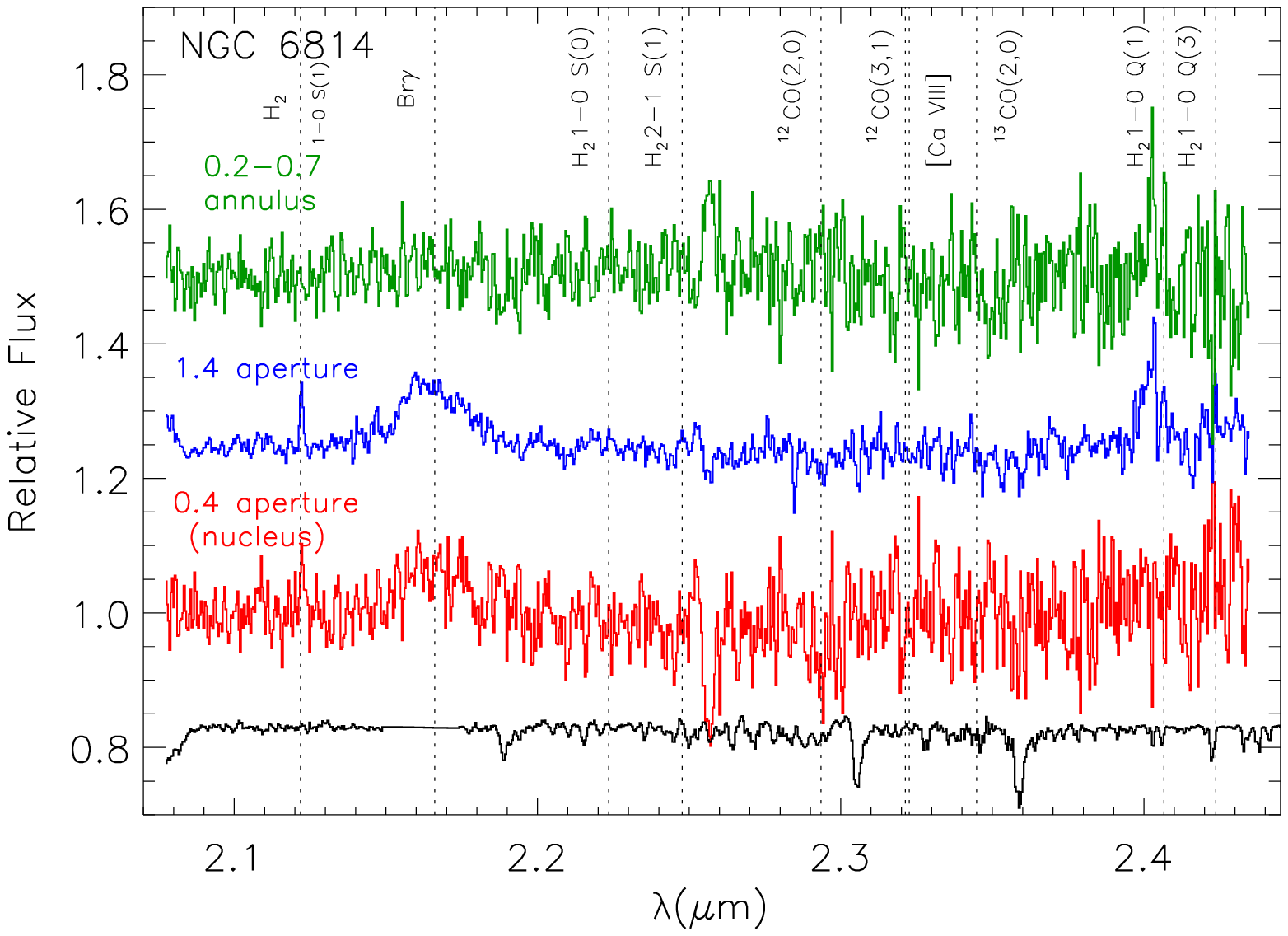}
\plotone{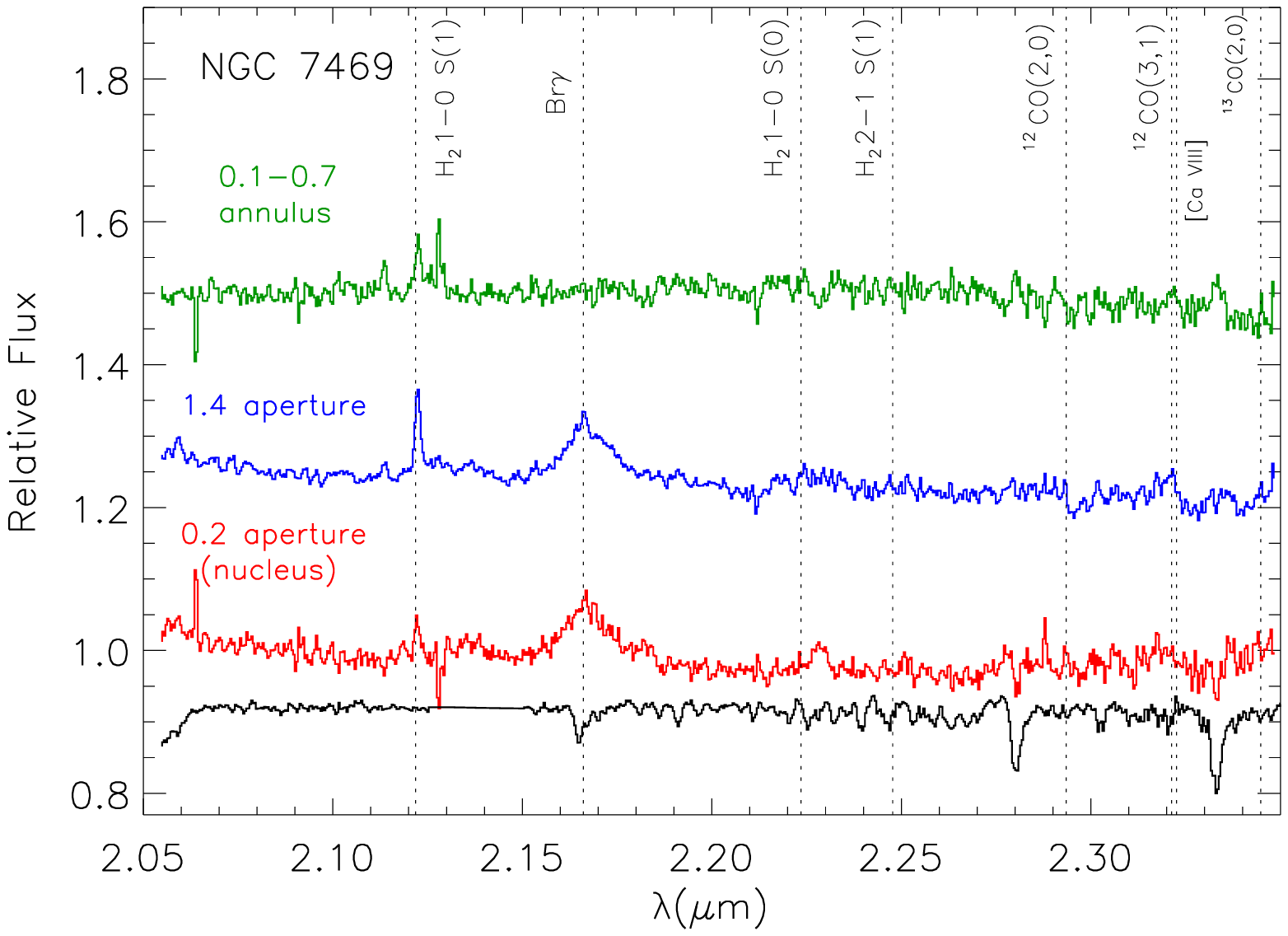}
\plotone{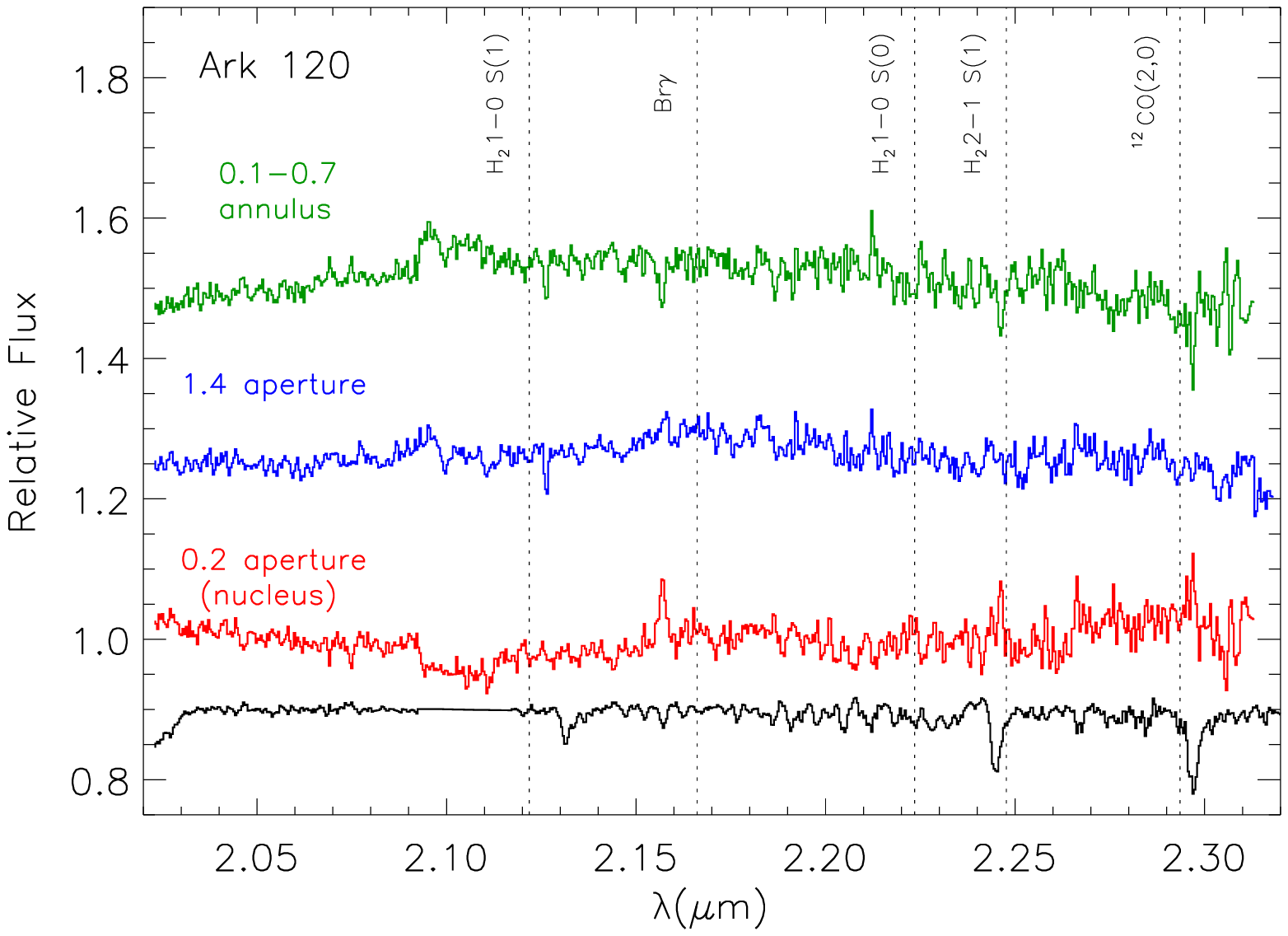}
\caption[Nuclear and Off-Nuclear Spectra]{Nuclear and off-nuclear spectra for each Seyfert 1 galaxy.  Each plot contains the galaxy name in the upper left and the spectra are also labeled on the left.  The nuclear spectra are centered on the AGN and extracted with a 0\as.2 diameter aperture (0\as.4 aperture for NGC 5548 and NGC 6814), and a larger 1\as.4 diameter aperture spectrum was also extracted centered on the AGN.  Also shown is a spectrum from an annulus of radius 0\as.1-0\as.7  (0\as.2-0\as.7 for NGC 5548 and NGC 6814).  For comparison, the bottom spectrum is telluric absorption.  \label{indspec}}
\end{figure}

For comparison with telluric absorption, the spectra extracted for each galaxy are shown in Fig. \ref{indspec}.  As discussed in $\S$ 4.3, the spectra shown are a composite spectrum extracted with a 1\as.4 diameter aperture, a nuclear spectrum from a 0\as.2 diameter aperture (0\as.4 aperture for NGC 5548 and NGC 6814), and the spectrum from an annulus of 0\as.1-0\as.7 (0\as.2-0\as.7 for NGC 5548 and NGC 6814), which is the difference of the other two spectra.  For a comparison of the spectra extracted for the nine galaxies, see Fig. \ref{spectra}.

\normalsize

\section{PSF Prediction Using SCAM Seyfert 1 Images}

To characterize the PSF, SCAM images were acquired of point sources (i.e. stars), with the source placed free of the slit.  A neutral density filter was used to decrease the counts on the wave front sensor to match those measured with the Seyfert 1 galaxies, and the AO system parameters (correction rate, gains, etc.) were also set to be identical.  During each observing run several images were taken resulting in a sample of point source images taken during a range of seeing conditions.  All images are reduced, including flat fielding, as described in $\S$ 3.2.

\begin{figure}[!h] 
\epsscale{0.37}
\plotone{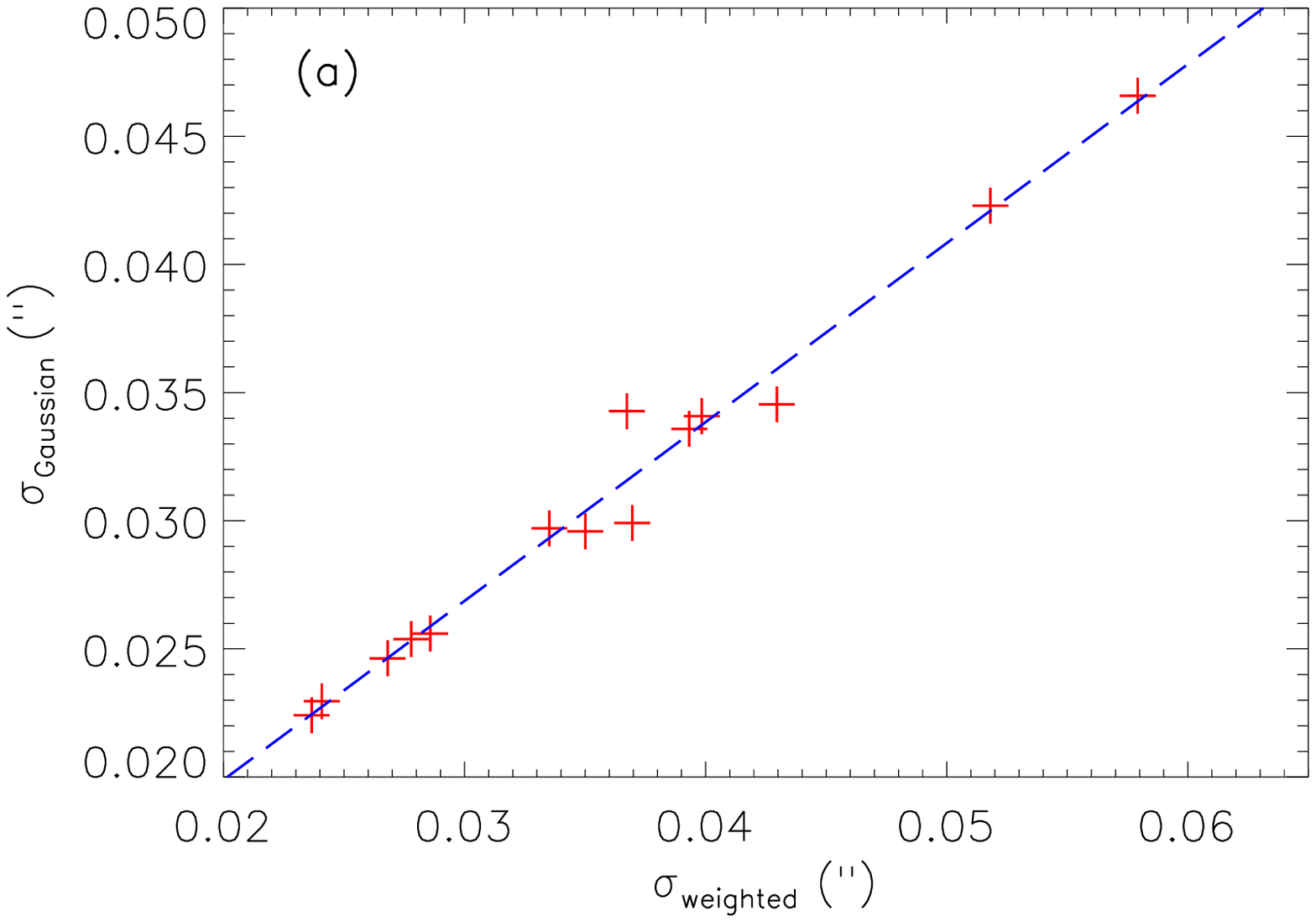}
\plotone{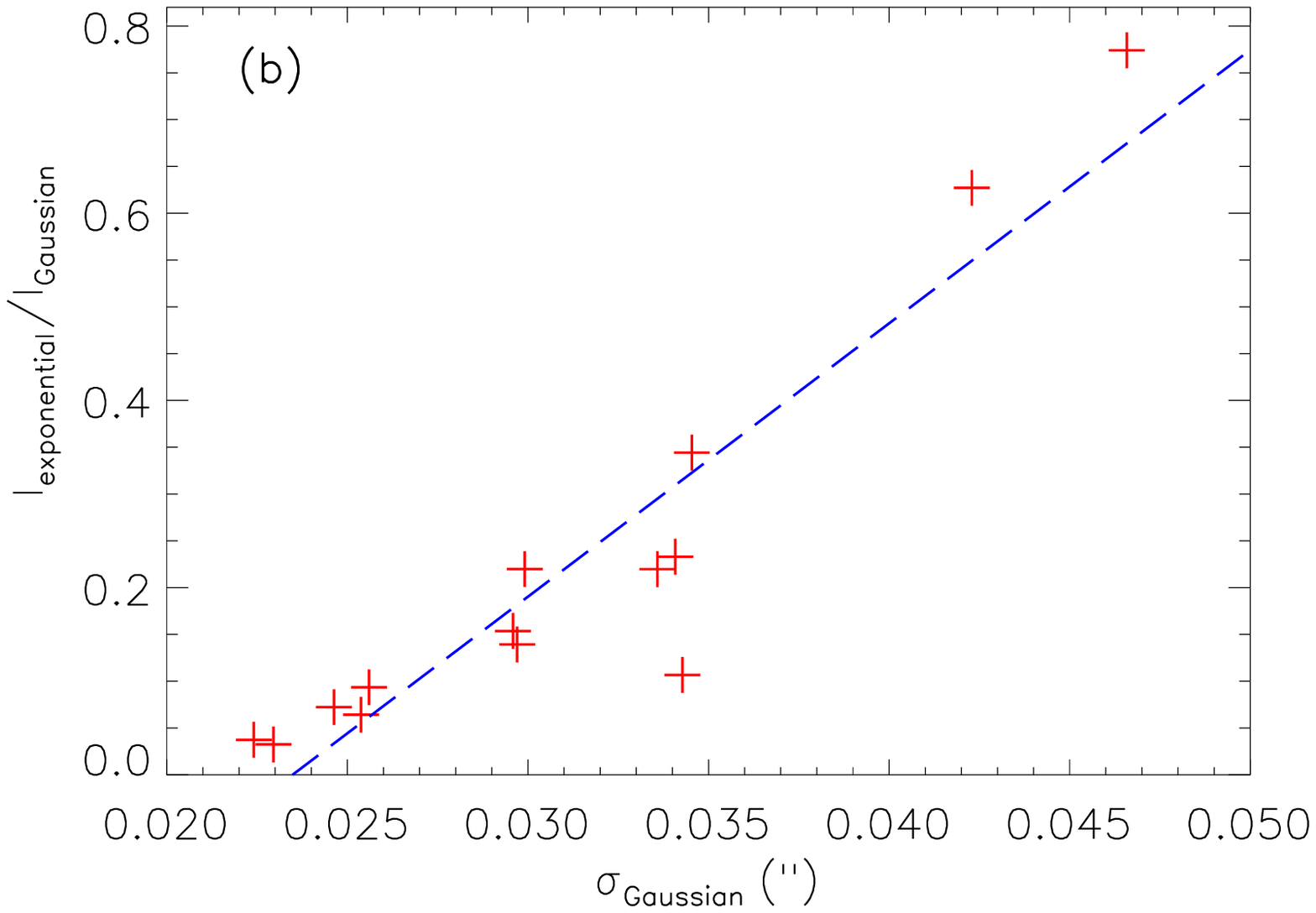}
\plotone{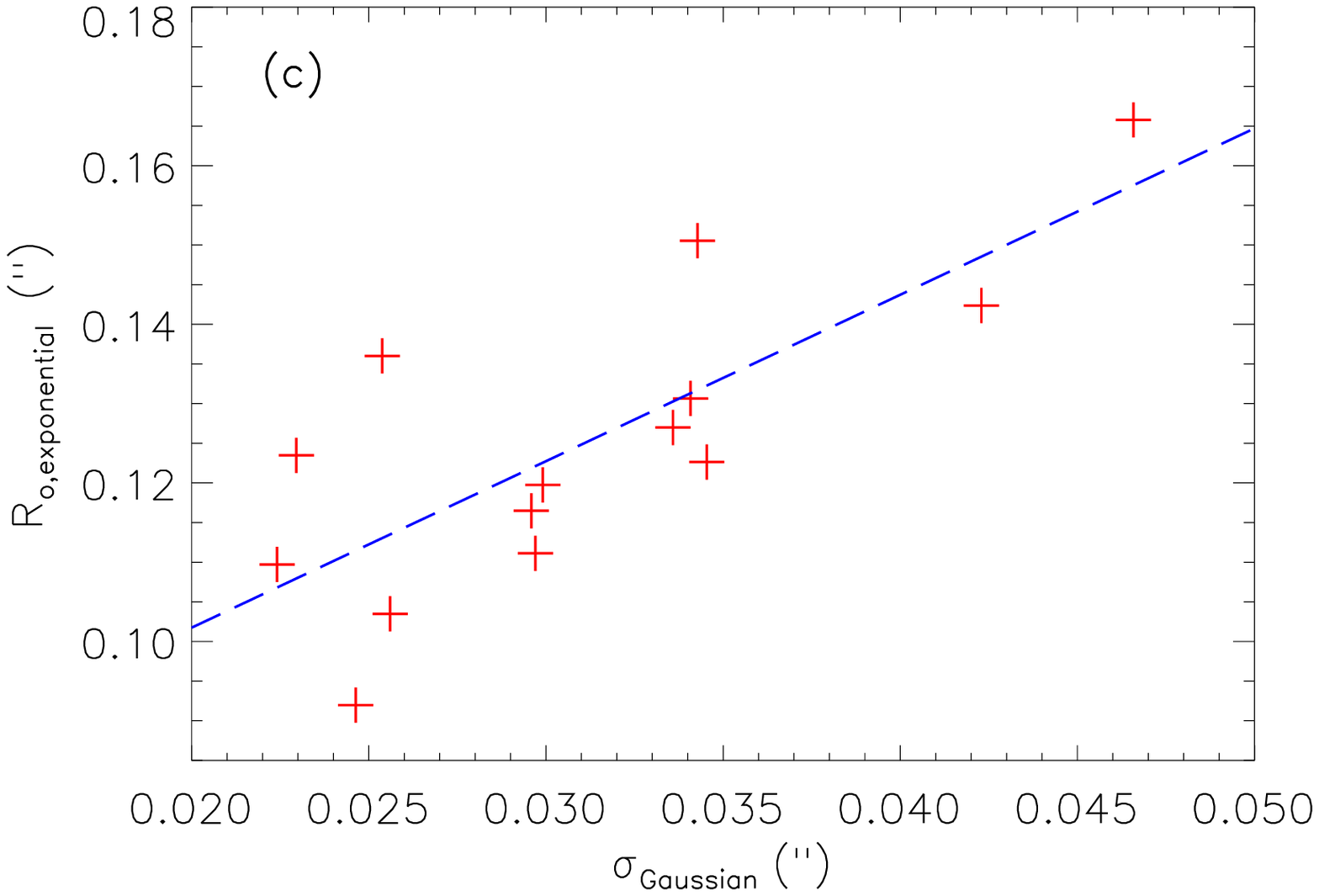}
\caption[PSF Prediction Relationships]{ PSF prediction relationships based on PSF image from known point sources (stars): (a) Gaussian core width from a weighted single Gaussian fit to the central 0\as.08 of the light profile, (b) peak intensity ratio of the exponential wing and Gaussian core from the Gaussian core width, and (c) the exponential width (scale length) from the Gaussian core width.  Measurements from the weighted single Gaussian fit and the best fit two function PSF (Gaussian core plus exponential wings) are shown by the data points (crosses), and the dashed line represents the best straight line fits to the data given by equations B.1-3. \label{psfrel}}
\end{figure}

The PSF, as determined by the stellar images, is well characterized by a circular Gaussian core plus exponential wing (see Fig \ref{spsf}).  Other functions that were attempted, but were found to not fit the PSF images as well, are combinations of Gaussians, Lorentzians, and Moffats.  Prediction of the Gaussian core plus exponential wing PSF function is possible based on the width, or $\sigma$, of a core weighted single Gaussian fit to the central 0\as.08 of the light profile.  Using the PSF images the following relationships (see Fig. \ref{psfrel}), with 1\sig errors, were established between the PSF function and the core weighted fit:
\begin{equation}
\sigma_{Gaussian}(\as) \, = \, (0.698 \pm 0.029) \, \times \, \sigma_{weighted}(\as) \, + 0.006
\end{equation}
\begin{equation}
I_{exponential} \, / \, I_{Gaussian }= \, (29.206 \pm 3.279) \, \times \, \sigma_{Gaussian}(\as) \, - 0.686
\end{equation}
\begin{equation}
R_{o,exponential}(\as) \, = \, (2.099 \pm 0.502) \, \times \, \sigma_{Gaussian}(\as) \, + 0.060
\end{equation}
The correlations between the weighted Gaussian fit and the Gaussian core of the PSF function and between the Gaussian core and the intensity ratio are very good, with reduced chi-squared values of 0.05 and 0.08, respectively.  Intensities ratios below 0.02 were replaced with this value since ratios below this level are unrealistic.  The correlation between the widths of the Gaussian core and the exponential wing is not as strong, however it is not as critical to predict this parameter as accurately because it does not alter the model velocity field significantly. 

To test the impact on the model rotation curves of using a functional form to represent the PSF, and then predicting the parameters of this function, a typical velocity field (smooth S\`{e}rsic n=2 stellar distribution with a peak-to-peak velocity difference of 400 km s$^{-1}$ and a BH of 10$^{7}$ \Msun) was convolved with each PSF.  Convolution with the observe PSF (the stellar image) and the best fit two function model PSF (Gaussian core plus exponential wings) results in an average difference in the velocity field of less than 4 \kms for all PSF images, with the worse case found to be less than 17 \kmsns.  The velocity difference between curves convolved with the actual PSF and the predicted PSF is on average less than 17 \kmsns, and including the 1\sig errors of the prediction relationships, the average difference is still less than 22 \kmsns.  The greatest velocity differences, which were as much as 30 \kms, were found using those stars with the narrowest weighted Gaussian fits, which are those that had the best AO correction, and are not representative of the corrections obtained with the Seyfert 1 nuclei.

The AO correction obtained with the Seyfert 1 nuclei was at times worse then was achieved with the stars, even with the AO parameters manually set to match those used with the galaxies.  As a result, to predict the PSF from the Seyfert 1 nuclei, an extrapolation of the above relationships is required approximately 50\% of the time.  In the most extreme cases the core weighted fit to the center of the galaxy light profile gave a width of \sig$\sim$ 13.4 mas (milliarcseconds = 0\as.001), which an extrapolation of the relationships gives a Gaussian core width of 9.9 mas, an exponential wing scale length of {\em R$_o$} $\sim$ 34.1 mas, and a peak intensity ratio of {\em I$_{Gaussian}$/I$_{exponential}$} $\sim$ 2.2.   

\end{document}